\documentclass[aps,prd]{revtex4-2}
\usepackage{graphicx}
\usepackage{amssymb}
\usepackage{amsmath}
\usepackage{float}
\usepackage{enumerate}
\begin{document}

\title{Gravitational lensing of massive particles in the charged NUT spacetime}

\author{Torben C. Frost}

\affiliation{ZARM, University of Bremen, 28359 Bremen, Germany\\
and Kavli Institute for Astronomy and Astrophysics, Peking University, 100871 Beijing, China\\
e-mail: torben.frost@pku.edu.cn}

\date{December 19th, 2024}

\begin{abstract}
In astronomy, gravitational lensing of light leads to the formation of multiple images, arcs, Einstein rings, and, most important, the shadow of black holes. Analogously in the vicinity of a massive compact object massive particles, following timelike geodesics, are gravitationally lensed. So far gravitational lensing of massive particles was mainly investigated in the weak and strong field limits. In this paper we will, for the first time, investigate exact gravitational lensing of massive particles using the example of the charged Newman-Unti-Tamburino (NUT) metric (and its special cases) which contains three physical parameters, the mass parameter $m$, the electric charge $e$, and the gravitomagnetic charge $n$. We will first discuss and solve the equations of motion for unbound timelike geodesics using elementary and Jacobi's elliptic functions and Legendre's elliptic integrals. Then we will introduce an orthonormal tetrad to relate the $z$ component of the angular momentum and the Carter constant to the energy $E$ of the particles along the timelike geodesics and latitude-longitude coordinates on the celestial sphere of a stationary observer in the domain of outer communication. We will use these relations to derive the angular radius of the particle shadow of the black hole, to formulate an exact lens equation, and to derive the travel time of the particles in terms of the time coordinate and the proper time. Finally, we will discuss the impact of the physical parameters and the energy of the particles on observable lensing features. We will also comment on how we can use these features alone and in a multimessenger context together with the corresponding features for light rays to determine if an astrophysical black hole can be described by the charged NUT metric or one of its special cases.
\end{abstract}

\maketitle
\section{Introduction}
In the recent decades gravitational lensing of light celebrated huge successes. First correctly predicted by Albert Einstein in 1915 in the form of the deflection of light by the Sun, for a thorough version of his original calculation see Ref. \cite{Einstein1916}, and later confirmed by an expedition to Pr\'{i}ncipe lead by Eddington in 1919 \cite{Dyson1920} it served as one of two initial tests of Einstein's theory of general relativity. Nowadays we regularly observe multiple images as well as arcs and Einstein rings. In addition, as a result of recent technological advances in radio astronomy the Event Horizon Telescope Collaboration was able to observe the shadows of the supermassive black holes in the centers of the galaxy M87 \cite{EHTCollaboration2019a} and the Milky Way \cite{EHTCollaboration2022}. Gravitational lensing of massive particles on the other hand has received far less attention so far. This has mainly two reasons. First, for being able to detect particles emitted by distant sources on Earth they have to be stable on long timescales and they should not or only weakly interact with other matter. Second, the weak interaction with other matter has as consequence that they are very difficult to detect. The result is that current particle detectors only have a very low angular resolution of a few square degrees. \\ 
Currently we only know one type of particle which meets both requirements. This type of particle is the neutrino. However, the detection of neutrinos is rather difficult because of two main reasons. The first reason is their weak interaction with other matter and thus their low detection rate in concurrent neutrino detectors. The second reason is much more important. In space, outside the Solar System, currently we can only identify very few strong, individually detectable and characterizable neutrino sources. In addition, the emission events of some of these sources, the so-called burst sources, mainly supernovae, tidal disruption events, and also, predicted although not yet detected, binary neutron star mergers, are rather short-lived  which further limits the probability of detecting lensed neutrino signals. However, within a multimessenger approach the detection of gravitationally lensed massive particles may provide supplementary information to characterize the nature of their source and the lens. In particular in the case that the lens is a black hole it may help to place constraints on the physical parameters characterizing the black hole spacetime and to reduce their uncertainties. Therefore, the main aim of this paper is to investigate gravitational lensing of massive particles by black holes.\\
However, real astrophysical settings are very complicated. Black holes do not exist in isolated environments and, in particular in the vicinity of supermassive black holes, smaller objects, e.g., stars, white dwarfs, neutron stars, or stellar mass black holes, can lead to microlensing. Furthermore, currently we only have an upper bound for the sum of the neutrino masses; see, e.g., Palanque-Delabrouille \emph{et al.} \cite{PalanqueDelabrouille2015}. However, we know that neutrinos travel at velocities close to the speed of light and, using modern neutrino detectors like Super-Kamiokande \cite{Fukuda2003,Ashie2005} and IceCube \cite{Achterberg2006}, we can now detect neutrinos in the TeV and even PeV range \cite{ICCollaboration2013,TheICCollaboration2021}. Therefore, for the remainder of this article we make two simplifying assumptions. First, we assume that we have an isolated, nonrotating black hole; i.e., microlensing does not occur. Second, we assume that the lensed particles can be described by uncharged test particles without spin and with fixed energies $E$ (note that here we \emph{a priori} assume that the particles can have arbitrary energies and we do not limit ourselves to the high-energy range predominantly observed for neutrinos).\\
In this paper we want to approach this problem from a spacetime perspective in general relativity. Therefore, we restrict our discussion to one of the most simple axisymmetric and stationary spacetimes in general relativity, the so-called charged Newman-Unti-Tamburino (NUT) metric. It belongs to the Pleba\'{n}ski-Demia\'{n}ski class of spacetimes of Petrov type D \cite{Plebanski1976} and is an exact solution to Einstein's electrovacuum field equations without cosmological constant. The NUT metric was originally discovered in two steps. In 1951, Taub \cite{Taub1951} derived the nonstationary part, which was originally interpreted as a cosmological solution \cite{Misner1963}. About 12 years later Newman \emph{et al.} \cite{Newman1963} used the Newman-Penrose formalism to derive the stationary part of the spacetime. They also realized that their solution is an extension of Taub's solution. The whole spacetime is commonly referred to as Taub-NUT spacetime; however, because in this paper we focus on investigating gravitational lensing of massive particles in the stationary part we will refer to it as NUT metric throughout the remainder of this paper. The charged NUT metric was first found by Brill \cite{Brill1964} in 1964 (therefore it is sometimes also referred to as Brill solution or Brill spacetime).\\
The charged NUT metric contains three different physical parameters: the mass parameter $m$, the electric charge $e$, and the gravitomagnetic charge $n$. The latter is a gravitational equivalent to a hypothetical magnetic charge in classical electrodynamics. In addition, the charged NUT metric contains a fourth parameter $C$. It was first introduced by Manko and Ruiz \cite{Manko2005} and is related to the existence of conical singularities in the spacetime. Misner \cite{Misner1963} was the first to notice the existence of these singularities in the original NUT metric and concluded that either the metric tensor or the time coordinate $t$ has a singularity on the axis $\vartheta=\pi$. For this historic reason the conical singularities are commonly also referred to as Misner strings. The nature of this singularity was first investigated by Bonnor \cite{Bonnor1969} and later by Sackfield \cite{Sackfield1971}. As a result of his investigations Bonnor drew the conclusion that the singularity can be interpreted as a semi-infinite massless rotating rod which in addition serves as source of angular momentum \cite{Bonnor1969}. While the original form of the NUT metric only contained one conical singularity the parameter $C$ generalizes the original NUT solution and allows one to control the number (one or two) and location ($\vartheta=0$, $\vartheta=\pi$, or on both axes) of the singularities. One important aspect of the existence of the conical singularities is that the charged NUT metric is only asymptotically flat in the sense that for $r\rightarrow\infty$ the Riemann tensor vanishes but it does not become asymptotically Minkowskian \cite{Misner1963}. The charged NUT metric can have two different interpretations. In the standard interpretation it contains up to two horizons and is usually interpreted as a black hole. When the electric charge $e$ exceeds a critical value the spacetime does not contain any horizons and, since the charged NUT metric does not possess a curvature singularity at $r=0$, it can be interpreted as a wormhole; see, e.g., Cl\'{e}ment \emph{et al.} \cite{Clement2016}.\\
The charged NUT metric is usually considered to be a rather exotic solution to Einstein's electrovacuum field equations. This has two main reasons. The first reason is the existence of the Misner strings. The second reason is that close to the Misner strings the spacetime contains closed timelike curves. Closed timelike curves violate causality and thus they are considered to be unphysical. Misner \cite{Misner1963} demonstrated that the conical singularities on the axes can be removed via a periodic coordinate transformation of the type $\tilde{t}=t+2nC\varphi$; however, after the transformation the spacetime contains closed timelike curves everywhere. From the physical aspect this is even less desirable and thus it is a common convention to keep the Misner strings.\\
In fundamental physics the existence of a gravitomagnetic charge is still an open question. Therefore, when we exclude the regions containing closed timelike curves the charged NUT metric may still serve as a good approximation for a spacetime realizing a gravitomagnetic charge. This paper has now two main goals. The first goal is to extend the exact gravitational lensing investigation of Frost \cite{Frost2022} from light rays to massive particles. For this purpose we will first exactly solve the equations of motion for unbound massive test particles using elementary as well as Jacobi's elliptic functions and Legendre's elliptic integrals. In the second part of this paper we will then use the obtained solutions to investigate gravitational lensing of massive particles. The second goal of this paper will then be to discuss how we can observe the electric and gravitomagnetic charges using gravitational lensing of massive particles in three different scenarios (note that commonly one can assume that in astrophysical environments the electric charge of a black hole is negligibly small; however, under the special circumstance that the black hole is embedded in a plasma with strong magnetic fields it can accumulate a significant electric charge \cite{Castellanos2018}). In the first scenario we will only consider particles with a fixed energy. In the second scenario we will consider a spectrum of particles with different energies and in the third scenario we will consider a spectrum of particles in combination with the emission of electromagnetic radiation.\\
For the NUT metric timelike geodesic motion was first discussed by Zimmerman and Shahir \cite{Zimmerman1989}. They first performed a potential analysis and wrote down a time integral for radial timelike geodesics. In addition, Zimmerman and Shahir investigated timelike circular and elliptic bound orbits and showed that in the NUT metric all geodesics, lightlike and timelike, lie on spatial cones. Jefremov and Perlick \cite{Jefremov2016} investigated the positions of circular timelike geodesics with particular focus on the innermost stable and marginally bound circular orbits. The most thorough investigation of lightlike and timelike geodesic motion in the NUT metric was carried out by Kagramanova \emph{et al.} \cite{Kagramanova2010}. In their paper the authors first classified the motion along lightlike and timelike geodesics into five different types of orbits (transit, escape, crossover escape, bound, and crossover bound; for the exact definitions please refer to \cite{Kagramanova2010}). In the second part of their paper Kagramanova \emph{et al.} \cite{Kagramanova2010} used Weierstra\ss' elliptic $\wp$ and Weierstra\ss' $\zeta$ and $\sigma$ functions to solve the equations of motion for lightlike and timelike geodesics. While in the NUT metric (and due to their similar structure also in the charged NUT metric) the types of motion are well known and it is straightforward to calculate the solutions to the equations of motion due to the existence of the Misner strings whether or not the spacetime is geodesically complete is still an open question. While many authors advocate that there is sufficient evidence to assume that the NUT metric is geodesically incomplete \cite{Misner1969,Miller1971,Kagramanova2010}, Cl\'{e}ment \emph{et al.} \cite{Clement2015} investigated this question and concluded that all geodesics can be smoothly continued through the Misner strings. However, recent gravitational lensing results from Frost \cite{Frost2022} show that travel time maps for the NUT metric as well as the charged NUT metric contain discontinuities for lightlike geodesics crossing the Misner strings. Thus for lightlike geodesics crossing the Misner strings the time coordinate is not continuous.\\
Gravitational lensing of light by (charged) NUT black holes has already been investigated by several authors. For a recent summary of the existing literature on gravitational lensing of light we refer the interested reader to Frost \cite{Frost2022}. Here we will only provide a short summary of the main results. As mentioned above in the NUT metric all geodesics lie on spatial cones \cite{Zimmerman1989}. On these cones we can calculate the bending angle of light using the same approach as for spherically symmetric and static spacetimes \cite{Zimmerman1989,LyndenBell1998,NouriZonoz1997,Halla2020,Halla2023,Sharif2016}. Nouri-Zonoz and Lynden-Bell \cite{NouriZonoz1997,LyndenBell1998} showed that in the weak field limit the presence of the gravitomagnetic charge leads to a twist in the lens map. Grenzebach \emph{et al.} \cite{Grenzebach2014,Grenzebach2016} investigated the shadow of Kerr-Newman-NUT--de Sitter black holes and found that the size of the shadow grows with increasing gravitomagnetic charge (an exact analytic formula for the angular radius of the shadow of charged NUT--de Sitter black holes can be found in Frost \cite{Frost2022}). Frost \cite{Frost2022} confirmed the twist for the exact lens map and also found the existence of the aforementioned discontinuities in the travel time maps for lightlike geodesics crossing the Misner strings.\\
In this paper we will see that a lot of the concepts well known from gravitational lensing of light rays can be easily transferred to gravitational lensing of massive particles. Indeed, in spherically symmetric and static spacetimes gravitational lensing of massive particles has already received some interest for quite some time. According to Zakharov \cite{Zakharov2018}, Mielnik and Pleba\'{n}ski \cite{Mielnik1962} were the first to introduce the concept of a particle sphere, the equivalent of the photon sphere for massive particles. Kobialko \emph{et al.} \cite{Kobialko2022} were the first to transfer the concept of a photon surface originally introduced by Claudel \emph{et al.} \cite{Claudel2001} to massive particles. In their work they first defined what they refer to as the \emph{massive particle surface} and then they discussed several examples, among them the Schwarzschild metric, the Reissner-Nordstr\"{o}m metric (as part of the Reissner-Nordstr\"{o}m dyon), and the NUT metric. In the late 1980s Zakharov \cite{Zakharov1988} derived the effective particle capture cross section of a Schwarzschild black hole for particles with arbitrary velocities at spatial infinity. A few years later he revisited his work for the Schwarzschild metric and also extended it to the Reissner-Nordstr\"{o}m metric for particles with $E=1$ \cite{Zakharov1994}. In 2002, Accioly and Ragusa \cite{Accioly2002} calculated the deflection angle of unbound, relativistic, massive particles up to the second post-Newtonian order under the assumption that the gravitational field of the Sun can be described by the Schwarzschild metric. About ten years later Tsupko \cite{Tsupko2014} was the first to investigate gravitational lensing of massive particles in the strong deflection limit in the Schwarzschild metric. He first wrote down the exact deflection angle in terms of Legendre's elliptic integral of the first kind. Then he continued his investigation by deriving four different versions of the strong deflection limit. Two years later Liu \emph{et al.} \cite{Liu2016} derived the bending angle for unbound massive particles in the weak and strong deflection limits for low ($v\ll c$) and ultrarelativistic ($v\approx c$) particle velocities. In the case of ultrarelativistic particles they also calculated the magnification. Crisnejo and Gallo \cite{Crisnejo2018} were the first to apply the Gauss-Bonnet theorem to derive the deflection angle of massive particles in the Schwarzschild metric up to second order using the approach originally developed by Gibbons and Werner \cite{Gibbons2008}. In 2019 Jia and Liu \cite{Jia2019} derived exact and approximative relations for the travel time in the Schwarzschild metric using elementary functions and Legendre's elliptic integrals of the first, second, and third kind. However, their exact result has the disadvantage that it explicitly contains the imaginary unit. In the weak field limit they also wrote down a simple lens equation and derived the time delay between two particle beams with different velocities on the same and opposite sides of the lens. In the same year Pang and Jia \cite{Pang2019} investigated gravitational lensing in the Reissner-Nordstr\"{o}m metric. They first derived the deflection angle in terms of Legendre's elliptic integral of the first kind. In the second step, they derived approximations for the weak and strong field limits and calculated the magnification for particles with velocities close to the speed of light. In the strong deflection limit they also extended their calculations to images of higher orders and estimated the size of the shadow of the supermassive black hole associated with the radio source Sgr $\text{A}^{*}$. One year later He \emph{et al.} \cite{He2020} derived the deflection angle for massive particles in the Schwarzschild--de Sitter metric up to the second post-Minkowskian order.\\
In most of these works the calculation of the deflection angle was limited to an observer and a particle source in the equatorial plane at spatial infinity. Therefore, except for the work of Jia and Liu \cite{Jia2019} in the exact gravitational lensing formalism observers and particle sources at finite distances to the lens have not been considered. In addition, to the best of our knowledge, in the charged NUT metric gravitational lensing of massive particles has not been investigated so far. Therefore, the main goal of this paper will be to set up the mathematical formalism for exact gravitational lensing of massive particles for the charged NUT metrics including the Schwarzschild metric, the Reissner-Nordstr\"{o}m metric, and the NUT metric for observers and particle sources in the domain of outer communication at finite distances to the black hole (note that whenever we use the plural we refer to the metric itself and all its special cases in the following). For this purpose in the first part of this paper we will discuss and analytically solve the equations of motion for the Schwarzschild metric, the Reissner-Nordstr\"{o}m metric, the NUT metric, and the charged NUT metric using elementary and Jacobi's elliptic functions and Legendre's elliptic integrals of the first, second, and third kind following the procedures described in Gralla and Lupsasca \cite{Gralla2020} and Frost \cite{Frost2022}. Note that we can also use Weierstra\ss' elliptic $\wp$ and Weierstra\ss' $\zeta$ and $\sigma$ functions to solve the equations of motion (as described in Kagramanova \emph{et al.} \cite{Kagramanova2010}); however, they have the disadvantage that we have to manually adjust the branches of the natural logarithm in the travel time integrals along the whole geodesic because we do not \emph{a priori} know how many branch cuts occur. Using Jacobi's elliptic functions and Legendre's elliptic integrals on the other hand has the advantage that we only have to consider the turning points and all relations can be written in forms which are explicitly real. In addition, some authors claim that using Jacobi's elliptic functions and Legendre's elliptic integrals has the advantage that their evaluation can be faster than numerical calculations when they are appropriately implemented \cite{Yang2013}. In the second part of the paper we will then investigate gravitational lensing in the charged NUT metrics. For this purpose we will assume that we have a stationary observer and stationary particle sources in the domain of outer communication. At the position of the observer we will then introduce an orthonormal tetrad to define latitude-longitude coordinates on the observer's celestial sphere following the approach of Grenzebach \emph{et al.} \cite{Grenzebach2015}. In the next step we will transfer the approach of Perlick and Tsupko \cite{Perlick2017} for gravitational lensing of light rays in a plasma to unbound timelike geodesics. We relate the constants of motion to the latitude-longitude coordinates on the observer's celestial sphere. In addition, we derive the total energy of a particle as measured by the stationary observer. Then we will derive the angular radius of the shadow for massive particles with constant energy and write down a lens equation. We will discuss the observed lensing features for sources that emit particles with a constant energy, sources that emit a spectrum of particles, and sources that emit a spectrum of particles in combination with electromagnetic radiation. Finally, we will discuss two different travel time measures. The first is the travel time in terms of the time coordinate $t$ and the second is the travel time in terms of the proper time of the particles $\tau$. We will discuss how we can combine travel time differences between "images" of different orders generated by particles with the same and different energies and light rays to determine the physical parameters of the charged NUT metric, namely the electric and gravitomagnetic charges. In addition, we will also discuss the possibility of directly determining the traveled proper time using particle decay or neutrino oscillations.\\
The remainder of this paper is structured as follows. In Sec.~\ref{Sec:CNUT} we will briefly introduce the charged NUT metrics and discuss their physical properties. Then, in Sec.~\ref{Sec:EoM} we will discuss and solve the equations of motion. In Sec.~\ref{Sec:Lensing} we will discuss different lensing features in the charged NUT metrics. For this purpose we will write down a lens equation and discuss three different lensing observables. These are the angular radius of the shadow and the travel times in terms of the time coordinate $t$ and the proper time of the particles $\tau$. We will discuss differences with respect to light rays and how they can be used to identify effects arising from the presence of the electric and gravitomagnetic charges. In addition, we will also comment on how we can combine gravitational lensing of light rays and massive particles in a multimessenger approach. In Sec.~\ref{Sec:Summary} we will summarize our results and conclusions. Throughout the paper we will use geometric units such that $c=G=1$. The metric signature is $\left(-,+,+,+\right)$.\\

\section{The Charged NUT Spacetime} \label{Sec:CNUT}
\begin{figure}[h!]
\centering
\includegraphics[width=0.9\textwidth]{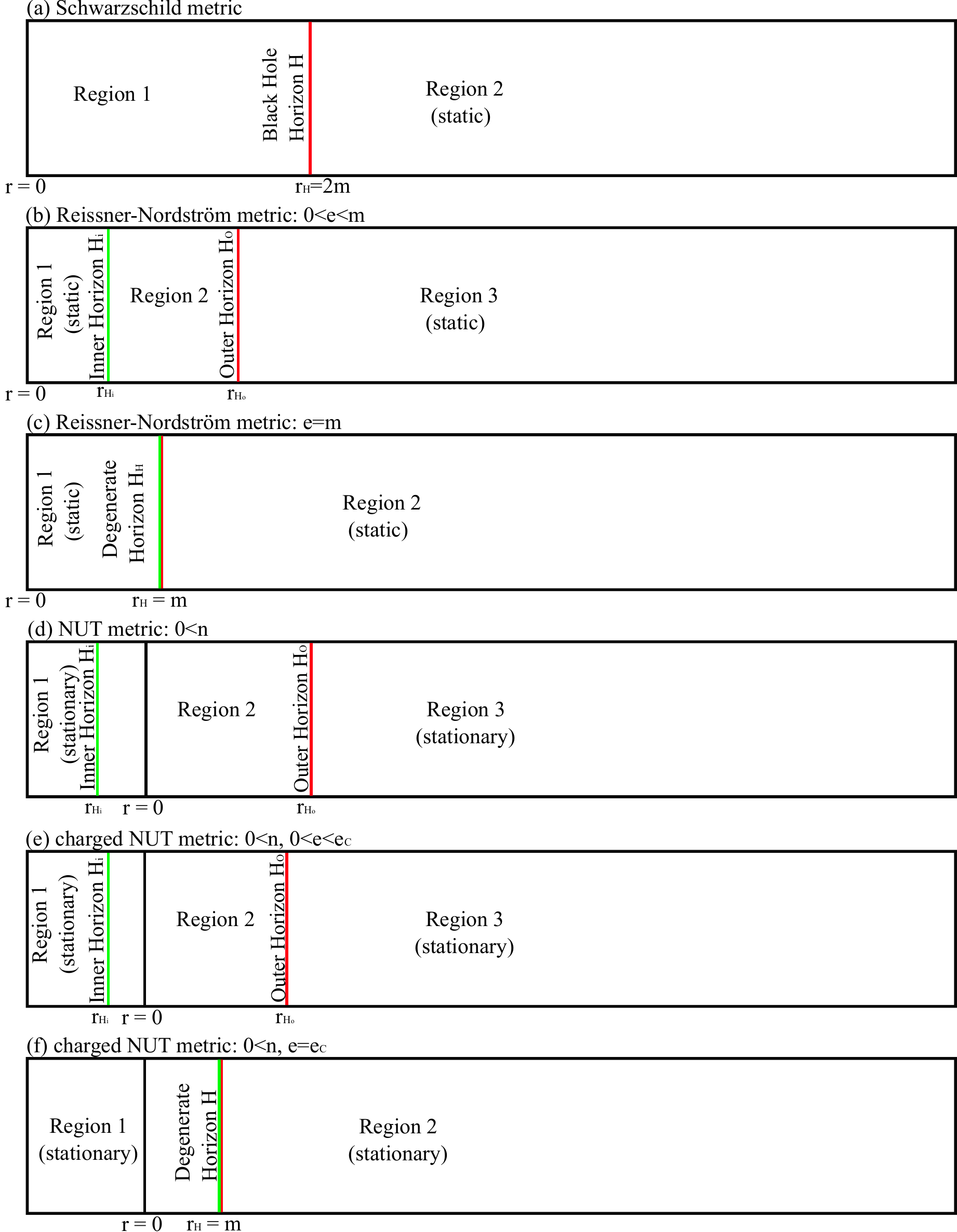}
\caption{Positions of the horizons (coordinate singularities) in (a) the Schwarzschild metric, the Reissner-Nordstr\"{o}m metric with (b) $0<e<m$ and (c) $e=m$, (d) the NUT metric, and the charged NUT metric with (e) $0<e<e_{\text{C}}=\sqrt{m^2+n^2}$ and (f) $e=e_{\text{C}}=\sqrt{m^2+n^2}$. Note that for panels (a)--(c) the spacetimes end at $r=0$ and for panels (d)--(f) large parts of the region $r<0$ are not shown. Also note that the angular coordinates are suppressed and any additional singularities are not shown.}\label{fig:CNUTHor}
\end{figure}

The charged NUT metric is an electrovacuum solution of Einstein's field equation without cosmological constant. It is axisymmetric and stationary and belongs to the Pleba\'{n}ski-Demia\'{n}ski family of spacetimes of Petrov type D \cite{Plebanski1976}. In Boyer-Lindquist-like coordinates the most general form of the line element of the charged NUT metric reads (for a full discussion of the spacetime structure see, e.g., pp.~213--237 in Griffiths and Podolsk\'{y} \cite{Griffiths2009})
\begin{equation}\label{eq:CNUTmetric}
g_{\mu\nu}\text{d}x^{\mu}\text{d}x^{\nu}=-\frac{Q(r)}{\rho(r)}\left(\text{d}t+2n(\cos\vartheta+C)\text{d}\varphi\right)^2+\frac{\rho(r)}{Q(r)}\text{d}r^2+\rho(r)\left(\text{d}\vartheta^2+\sin^2\vartheta\text{d}\varphi^2\right),
\end{equation}
where
\begin{equation}\label{eq:CNUTCoeffQ}
Q(r)=r^2-2mr+e^2-n^2
\end{equation}
and
\begin{equation}
\rho(r)=r^2+n^2.
\end{equation}
The spacetime contains three physical parameters. The first is the mass parameter $m$, the second is the electric charge $e$, and the third is the so-called gravitomagnetic charge $n$, the gravitational equivalent to a hypothetical magnetic charge in classical electrodynamics. In addition, the spacetime also contains a fourth parameter $C$, commonly referred to as Manko-Ruiz parameter \cite{Manko2005}. We will further discuss it below. For $e=n=0$ the metric reduces to the Schwarzschild metric. For $n=0$ it reduces to the Reissner-Nordstr\"{o}m metric and for $e=0$ it reduces to the NUT metric.\\
In theory the four parameters $m$, $e$, $n$, and $C$ can take any arbitrary real value. However, we can limit them using physical reasoning and the symmetries of the spacetime as follows. In nature we only observe positive masses and thus we have $m>0$. In (\ref{eq:CNUTCoeffQ}) and thus the line element (\ref{eq:CNUTmetric}) the electric charge only occurs as square. Thus the sign of the electric charge does not have any effect on the spacetime; only its absolute value has and since in this paper we will only consider uncharged particles we can choose $0\leq e\leq e_{\text{C}}$. Note that here $e_{\text{C}}$ is an upper bound for the electric charge which has to be chosen based on the desired nature of the spacetime. Similarly, we can easily see that when we replace $n\rightarrow -n$ and $C\rightarrow -C$ and perform the coordinate transformation $\vartheta \rightarrow \pi-\vartheta$ the line element (\ref{eq:CNUTmetric}) remains invariant and thus we can choose $0\leq n$ and $C\in \mathbb{R}$.\\
The main structure of the spacetime is strongly influenced by the existence of conical singularities on the axes. While the original Taub-NUT spacetime \cite{Taub1951,Newman1963} only contained one conical singularity at $\vartheta=\pi$ \cite{Misner1963} in its general form given by (\ref{eq:CNUTmetric}) the number of conical singularities and their strength depend on the choice of the Manko-Ruiz parameter. For $C=1$ the spacetime has a conical singularity at $\vartheta=0$. For $C=-1$ the spacetime has a conical singularity at $\vartheta=\pi$. When $\left|C\right|\neq 1$ the spacetime contains conical singularities on both axes; however, only for $C=0$ both singularities have the same strength. Due to the presence of the Misner strings the charged NUT metric is asymptotically flat in the sense that for $r\rightarrow\infty$ the Riemann tensor vanishes but the spacetime does not become asymptotically Minkowskian \cite{Misner1963}. The most common interpretation \cite{Bonnor1969} of the Misner strings is that they represent semi-infinite massless rotating rods which serve as source of angular momentum and give rise to the gravitomagnetic charge $n$. Close to the Misner strings the charged NUT metric possesses regions in which $g_{\varphi\varphi}\leq0$ \cite{Bonnor1969} and thus in these regions closed timelike curves exist. The existence of closed timelike curves is problematic since particles moving on these curves violate causality, which according to our current knowledge is physically not possible. When we look at the line element (\ref{eq:CNUTmetric}) we can see that we can remove the Misner strings by introducing a periodic time coordinate via the periodic transformation $\tilde{t}=t+2nC\varphi$ \cite{Misner1963}. This transformation removes the Misner strings; however, it is only valid locally because we usually assume that the $\varphi$ coordinate is periodic while the time coordinate $t$ is not (as a consequence charged NUT spacetimes with different $C$ are locally isometric \cite{Grenzebach2014}). Furthermore, when we introduce a periodic time coordinate the spacetime contains closed timelike curves everywhere, which is physically even less desirable. Therefore, in this paper we retain the Misner strings.\\
The only question left to answer is the choice of the coordinates $t$, $r$, $\vartheta$, and $\varphi$. As already discussed above in this paper we retain the Misner strings and thus we have $t\in \mathbb{R}$. In addition, we choose the angular coordinates $\vartheta$ and $\varphi$ such that they represent angular coordinates on the two sphere $S^{2}$. Therefore, we have $\vartheta\in\left[0,\pi\right]$ and $\varphi\in[0,2\pi)$. Here, we have to note another interesting aspect of the charged NUT metric. Although the spacetime is only axisymmetric it retains some degree of rotational symmetry. While for the NUT metric spherical symmetry is clearly broken it was shown by Newman \emph{et al.} \cite{Newman1963} and Halla and Perlick \cite{Halla2020} that the spacetime possesses four linearly independent Killing vector fields. Three of these Killing vector fields now generate isometries which are isomorphic to the rotation group $SO(3,\mathbb{R})$. Thus the NUT metric and consequently also the charged NUT metric are symmetric with respect to rotations about any arbitrary radial direction (for more information we would like to refer the interested reader to the original works).\\
Now the only coordinate range left to discuss is that for the $r$ coordinate. For this purpose let us have a look at the roots associated with $Q(r)=0$ and the curvature singularities of the metric tensor. The real roots of $Q(r)=0$ are all coordinate singularities and thus they correspond to horizons. They can be removed using appropriate coordinate transformations. Figure~\ref{fig:CNUTHor} shows the horizon structures of the Schwarzschild metric [Fig. 1(a)], the Reissner-Nordstr\"{o}m metric [Figs. 1(b) and 1(c)], the NUT metric [Fig. 1(d)], and the charged NUT metric [Figs. 1(e) and 1(f)]. We will start our discussion by recapitulating the well known horizon structures of the Schwarzschild and Reissner-Nordstr\"{o}m metrics. Both metrics have a curvature singularity at $r=0$ and thus we have $0<r$. For the Schwarzschild metric this singularity is spacelike while for the Reissner-Nordstr\"{o}m metric it is timelike. In addition, the Schwarzschild metric has a coordinate singularity at $r_{\text{H}}=2m$ which marks the position of a horizon. For $0<r<r_{\text{H}}$ the vector field $\partial_{t}$ is spacelike and the vector field $\partial_{r}$ is timelike and thus this part of the spacetime is nonstatic. For $r_{\text{H}}<r$ the vector field $\partial_{t}$ is timelike and the vector field $\partial_{r}$ is spacelike and thus this part of the spacetime is static. Similarly, for $0<e<m$ the Reissner-Nordstr\"{o}m metric has two coordinate singularities at $r_{\text{H}_{\text{i}}}=m-\sqrt{m^2-e^2}$ and $r_{\text{H}_{\text{o}}}=m+\sqrt{m^2-e^2}$. Here, the first coordinate singularity marks the position of an inner horizon and the second coordinate singularity marks the position of an outer horizon. For $0<r<r_{\text{H}_{\text{i}}}$ and $r_{\text{H}_{\text{o}}}<r$ the spacetime is static while for $r_{\text{H}_{\text{i}}}<r<r_{\text{H}_{\text{o}}}$ the spacetime is nonstatic. For $e=m$ both horizons coincide and we only have one horizon at $r_{\text{H}}=m$ separating two static regions. For $m<e$ the spacetime does not possess coordinate singularities and thus we have a naked singularity (not shown in Fig.~\ref{fig:CNUTHor}). \\
Because we have $\rho(r)>0$ for all $r$ the NUT metric and the charged NUT metric do not possess curvature singularities at $r=0$. Thus for the NUT metric and the charged NUT metric we have $r\in \mathbb{R}$. However, both spacetimes can have up to two coordinate singularities marking the positions of horizons. The horizons of the NUT metric are located at $r_{\text{H}_{\text{i}}}=m-\sqrt{m^2+n^2}$ and $r_{\text{H}_{\text{o}}}=m+\sqrt{m^2+n^2}$. We can easily see that $r_{\text{H}_{\text{i}}}<0<r_{\text{H}_{\text{o}}}$. For $r<r_{\text{H}_{\text{i}}}$ and $r_{\text{H}_{\text{o}}}<r$ the spacetime is stationary while for $r_{\text{H}_{\text{i}}}<r<r_{\text{H}_{\text{o}}}$ the spacetime is nonstationary. For the charged NUT metric the horizons are located at $r_{\text{H}_{\text{i}}}=m-\sqrt{m^2+n^2-e^2}$ and $r_{\text{H}_{\text{o}}}=m+\sqrt{m^2+n^2-e^2}$ and the spacetime structure is similar to that of the NUT metric. For $e=\sqrt{m^2+n^2}$ both horizons coincide and the spacetime has only one horizon at $r_{\text{H}}=m$ separating two stationary regions. For $\sqrt{m^2+n^2}<e$ the spacetime does not possess horizons and we only have a single stationary region (not shown in Fig.~\ref{fig:CNUTHor}). Thus some authors interpret this case as a wormhole; see, e.g., Cl\'{e}ment \emph{et al.} \cite{Clement2016}. Throughout most of the remainder of this paper we want the metrics to represent black hole spacetimes. For the Reissner-Nordstr\"{o}m metric and the charged NUT metric this is only possible when $0<e\leq e_{\text{C}}=m$ and $0<e\leq e_{\text{C}}=\sqrt{m^2+n^2}$, respectively. Furthermore, we will follow the convention to refer to the domain outside the (outer) black hole horizon as domain of outer communication.

\section{Solving the Equations of Motion}\label{Sec:EoM}
We start with writing down the equations of motion for timelike geodesics in the charged NUT metric. For massive particles moving along timelike geodesics we have four constants of motion. These are the Lagrangian $\mathcal{L}=-1/2$, the energy of the particles $E$, the angular momentum about the $z$ axis $L_{z}$, and the Carter constant $K$. Using the constants of motion the equations of motion can be fully separated. They read
\begin{align}
\frac{\text{d}t}{\text{d}\lambda}&=\frac{\rho(r)^2}{Q(r)}E-2n(\cos\vartheta+C)\frac{L_{z}+2n(\cos\vartheta+C)E}{\sin^2\vartheta},\label{eq:EoMt}\\
\left(\frac{\text{d}r}{\text{d}\lambda}\right)^2&=\rho(r)^2E^2-\rho(r)Q(r)-Q(r)K,\label{eq:EoMr}\\
\left(\frac{\text{d}\vartheta}{\text{d}\lambda}\right)^2&=K-\frac{\left(L_{z}+2n(\cos\vartheta+C)E\right)^2}{\sin^2\vartheta},\label{eq:EoMtheta}\\
\frac{\text{d}\varphi}{\text{d}\lambda}&=\frac{L_{z}+2n(\cos\vartheta+C)E}{\sin^2\vartheta},\label{eq:EoMphi}
\end{align}
where the parameter $\lambda$ is the so-called Mino parameter \cite{Mino2003}, sometimes also referred to as Mino time. It is related to the proper time $\tau$ via
\begin{equation}\label{eq:Mino}
\frac{\text{d}\lambda}{\text{d}\tau}=\frac{1}{\rho(r)}.
\end{equation}
In this paper we choose the proper time and the Mino parameter such that they increase for future-directed timelike geodesics and decrease for past-directed timelike geodesics. This requires that the energy parameter $E$ is positive. In this paper we are mainly interested in unbound timelike motion. This requires an even more strict constraint because in the charged NUT metric unbound timelike geodesics only exist for $E>1$. Note that (\ref{eq:EoMt}), (\ref{eq:EoMtheta}), and (\ref{eq:EoMphi}) are structurally the same as for lightlike geodesics; see Eqs.~(5), (7), and (8) in Frost \cite{Frost2022}. In the following we will now proceed to discuss and analytically solve the equations of motion for timelike geodesics using elementary and  Jacobi's elliptic functions as well as Legendre's elliptic integrals of the first, second, and third kind. We will also explicitly derive analytic expressions for the proper time $\tau$. For this purpose we will for now assume that we have arbitrary initial conditions $(x_{i}^{\mu})=(x^{\mu}(\lambda_{i}))=(t_{i},r_{i},\vartheta_{i},\varphi_{i})$ and $\tau(\lambda_{i})=\tau_{i}$. In the next section we will then use the obtained solutions to discuss exact gravitational lensing of massive particles in the Schwarzschild metric, the Reissner-Nordstr\"{o}m metric, the NUT metric, and the charged NUT metric. Therefore, we will limit our discussion to timelike geodesics in the domain of outer communication outside the (outer) black hole horizon. \\
Before we proceed we would like to note that many of the results presented in this section are very similar to the results for lightlike geodesics presented in Frost \cite{Frost2022}. However, we still provide a thorough discussion for two reasons. First, the equations of motion for timelike geodesics, in particular for the $r$ motion and the time coordinate $t$, have some distinct differences which do not occur for lightlike geodesics. Second, for a reader, who is interested in understanding the mathematical technicalities required to solve the equations of motion, it is more convenient to find all results in one paper. In addition, for keeping comparability to our earlier paper we will largely keep the same notation and also heavily lean on the structural outline.

\subsection{The $r$ motion}\label{Sec:EoMSolr}
\subsubsection{Potential and the particle sphere}
We start our discussion with the $r$ motion. Let us first rewrite (\ref{eq:EoMr}) in terms of a potential $V_{E}(r)$. It reads
\begin{equation}\label{eq:EoMrwPot}
-\frac{1}{Q(r)}\left(\frac{\text{d}r}{\text{d}\lambda}\right)^2+V_{E}(r)=K,
\end{equation}
where the potential $V_{E}(r)$ is given by
\begin{equation}\label{eq:EoMrPot}
V_{E}(r)=\frac{\rho(r)(\rho(r)E^2-Q(r))}{Q(r)}.
\end{equation}
Unlike for lightlike geodesics the potential for unbound timelike geodesics depends on the particle energy along the timelike geodesics $E$. We can easily read from (\ref{eq:EoMtheta}) that geodesic motion can only occur for $K\geq0$. Therefore, in the domain of outer communication the left-hand side of (\ref{eq:EoMrwPot}) is required to be positive or zero. From (\ref{eq:EoMrwPot}) we can read that this requirement is only fulfilled when $V_{E}(r)>0$ and thus all unbound timelike geodesics in the domain of outer communication of the charged NUT metrics have to fulfill this criterion. \\
Figure~2 shows plots of the potentials $V_{E}(r)$ for the Schwarzschild metric (top left), the Reissner-Nordstr\"{o}m metric with $e=m$ (top right), the NUT metric with $n=m/2$ (bottom left), and the charged NUT metric with $e=m$ and $n=m/2$ (bottom right) for the three energies $E_{1}=\sqrt{101/100}$ (solid line), $E_{2}=\sqrt{5}/2$ (dotted line), and $E_{3}=\sqrt{2}$ (dashed line). We start our discussion with the potential for $E_{1}=\sqrt{101/100}$ for the Schwarzschild metric. The potential approaches the limit $V_{E}(r)\rightarrow\infty$ for $r\rightarrow r_{\text{H}}$ and $r\rightarrow\infty$. In addition, the potential has a minimum at $r_{\text{pa}_{1}}$. This minimum marks the position of what we will refer to as particle sphere (a more formal definition and discussion of the particle sphere will follow below). When we now increase the energy $E$ the value of the minimum at $r_{\text{pa}}$ increases and the potential's width around the minimum decreases. When we turn on the electric charge $e$ the whole potential structure is shifted to lower radius coordinates $r$ and the minimum values of the potentials at the radius coordinates of the corresponding particle spheres decrease. The contrary happens when we turn on the gravitomagnetic charge $n$. In this case the whole potential structure shifts to larger radius coordinates $r$ while the minimum values of the potentials at the radius coordinates of the corresponding particle spheres increase.\\
\begin{figure}\label{fig:VErpot}
  \begin{tabular}{cc}
    \hspace{-0.5cm}\includegraphics[width=95mm]{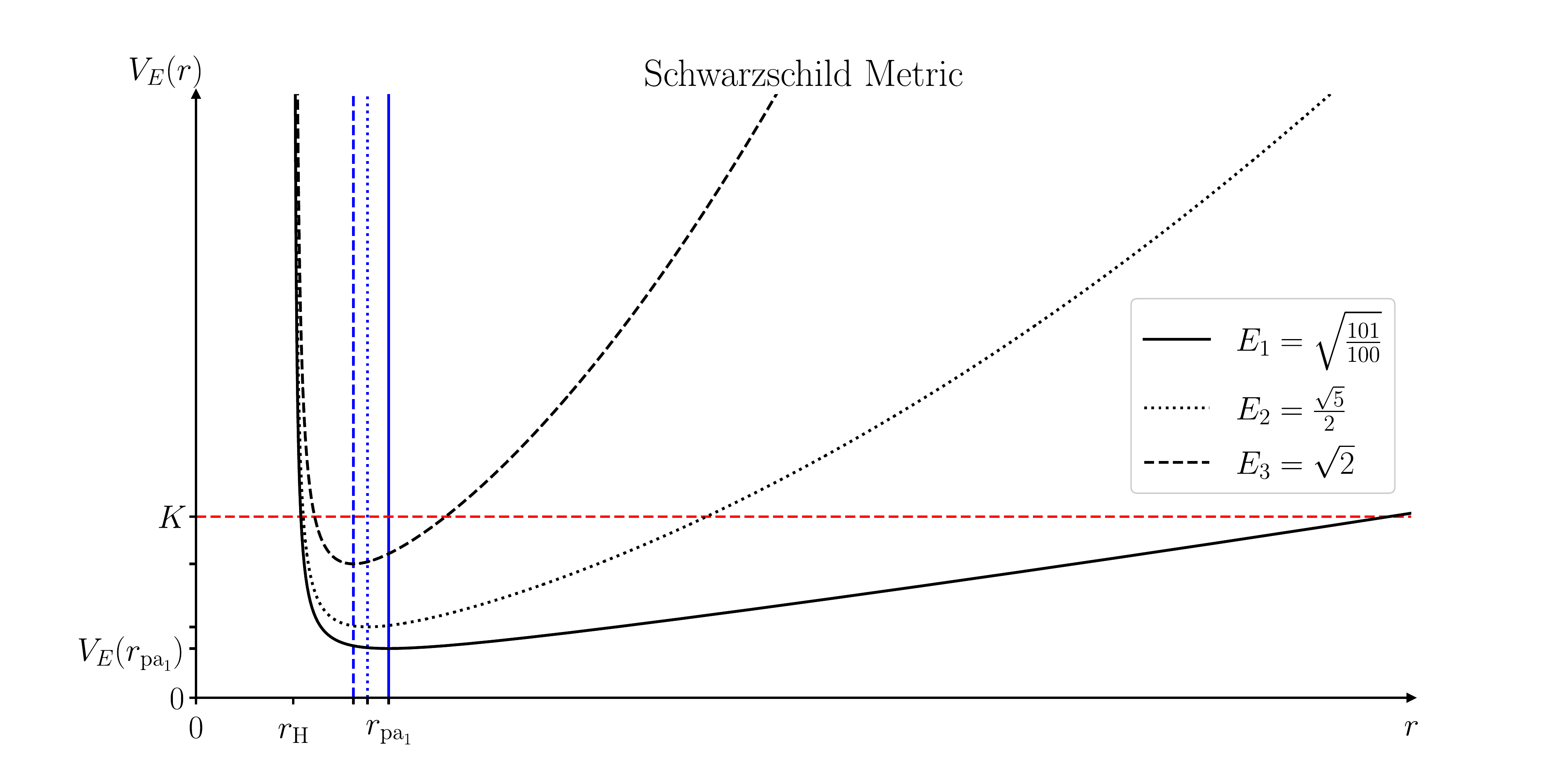} &   \includegraphics[width=95mm]{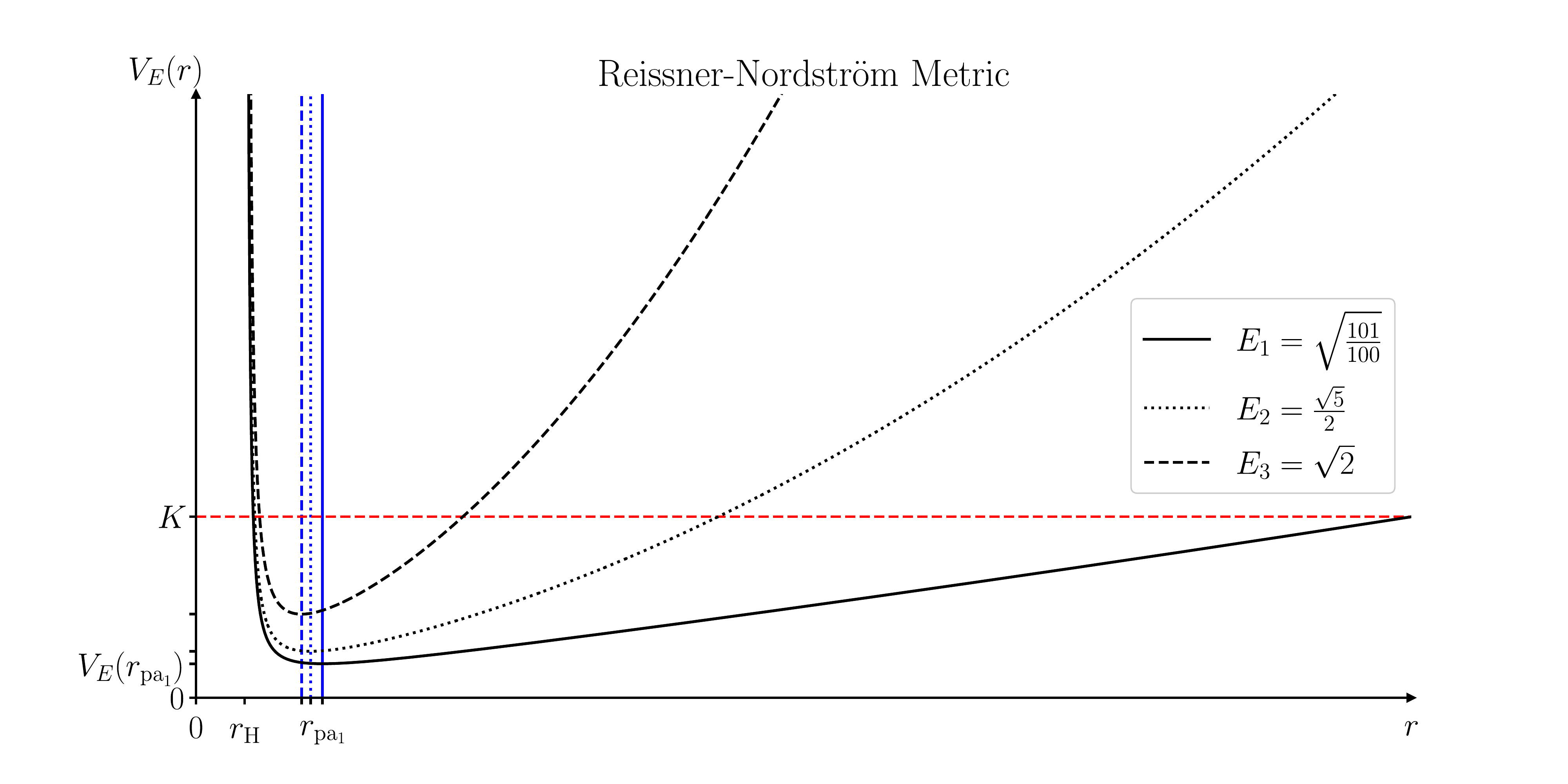} \\
    \hspace{-0.5cm}\includegraphics[width=95mm]{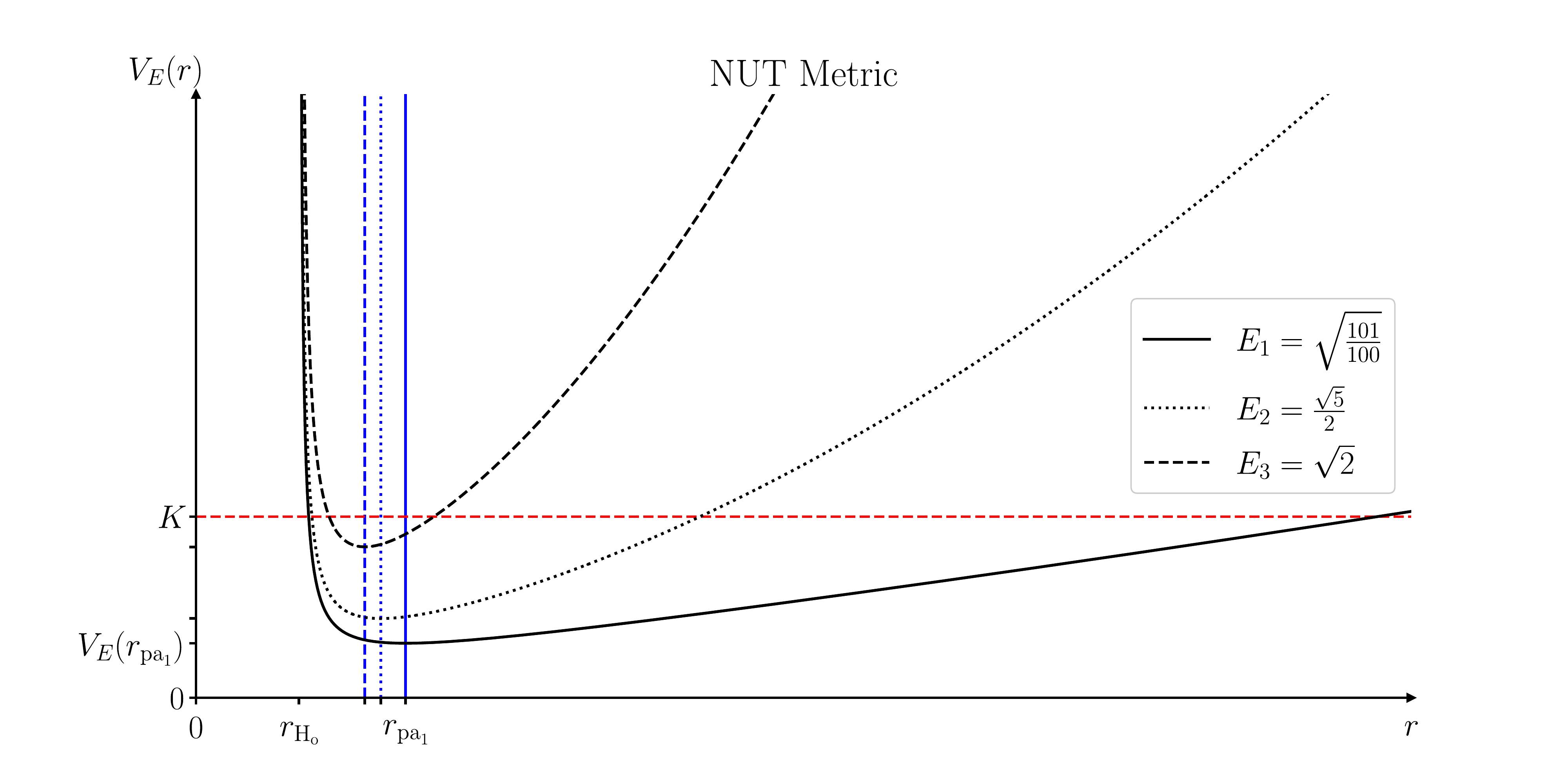} &   \includegraphics[width=95mm]{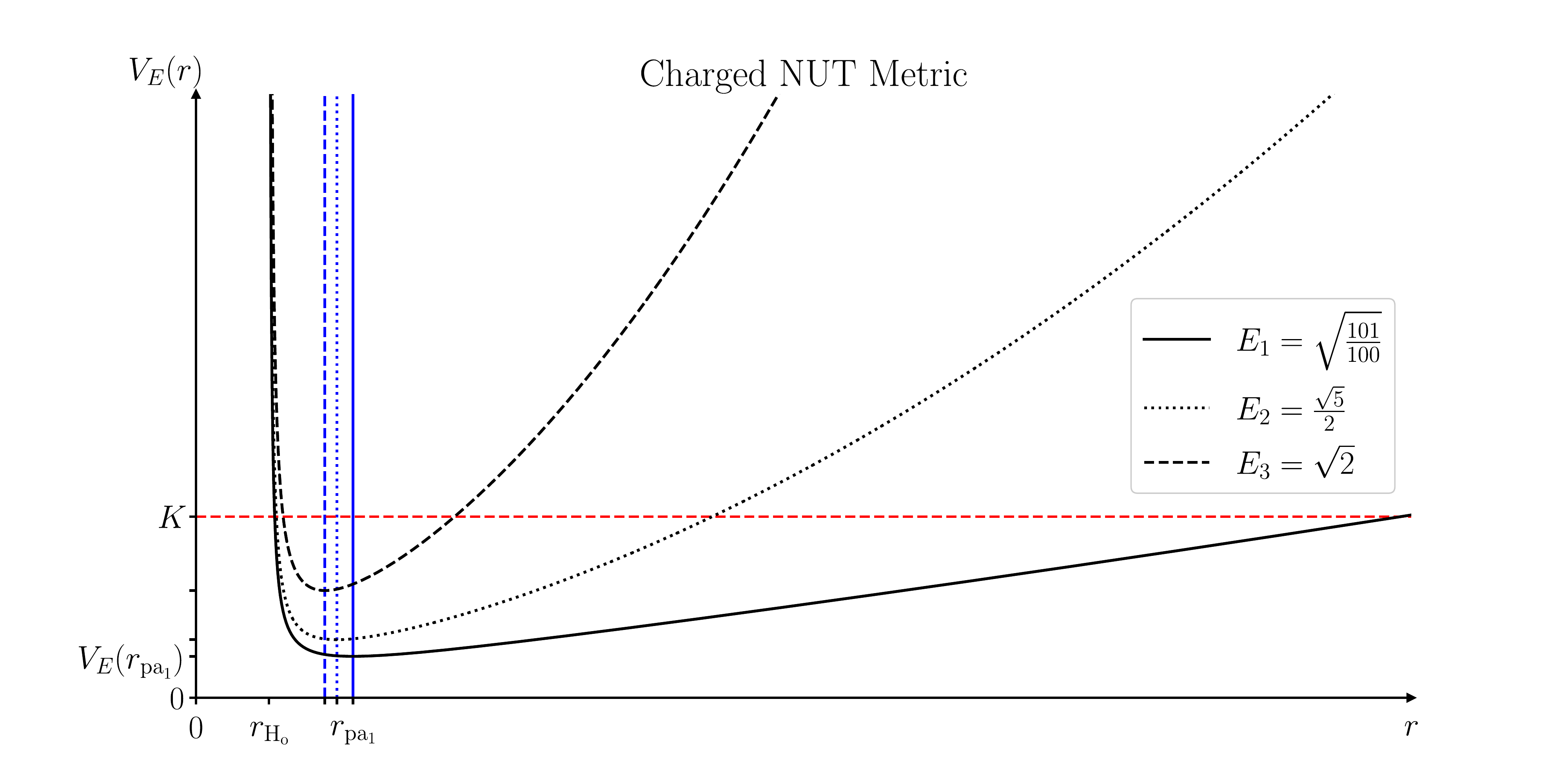} \\
  \end{tabular}
	\caption{Potential $V_{E}(r)$ of the $r$ motion for the Schwarzschild metric (top left), the Reissner-Nordstr\"{o}m metric (top right), the NUT metric (bottom left), and the charged NUT metric (bottom right) for $e=m$ and $n=m/2$ for three different energy values $E_{1}=\sqrt{101/100}$ (solid line), $E_{2}=\sqrt{5}/2$ (dotted line), and $E_{3}=\sqrt{2}$ (dashed line). The vertical lines mark the positions of the particle spheres for the three energy values, respectively. The unlabelled dashes on the horizontal and vertical axes mark the positions of $r_{\text{pa}_{2}}$ and $r_{\text{pa}_{3}}$ as well as $V_{E}(r_{\text{pa}_{2}})$ and $V_{E}(r_{\text{pa}_{3}})$, respectively. The axes have the same scale in all four plots.}
\end{figure}
While we discussed the potential structures for timelike geodesics above we already used the term particle sphere; however, we did not yet explain what it is. We recall that in the charged NUT metrics for lightlike geodesics we always have a photon sphere in the domain of outer communication. For massive particles on timelike geodesics the particle sphere is now the equivalent to the photon sphere for light rays on lightlike geodesics. Like the photon sphere for black hole spacetimes the particle sphere in the domain of outer communication is unstable in the sense that if we slightly perturb a particle moving on a timelike geodesic on the particle sphere in the radial direction the particle either falls into the black hole or escapes to infinity. Note that in this paper only the particle sphere in the domain of outer communication is of particular relevance. However, we will see that for the characterization of the $r$ motion knowledge about the existence of all potentially existing unstable particle spheres is required. Therefore, in the following we will provide a thorough and general discussion of the particle spheres for particles with energies $E>1$. Here, our discussion will mainly focus on the existence and stability of particle spheres in black hole spacetimes; however, for the sake of completeness we will also discuss these points for the horizonless Reissner-Nordstr\"{o}m and charged NUT metrics.\\
We can calculate the radius coordinates of the particle spheres from the conditions $\text{d}r/\text{d}\lambda=\text{d}^2r/\text{d}\lambda^2=0$. Combining both conditions we get as determining equation for the radius coordinates of the particle spheres
\begin{align}\label{eq:rpa}
r^5+\frac{m(1-3(E^2-1))}{E^2-1}r^4+\frac{2((e^2-n^2)(E^2-1)-2m^2)}{E^2-1}r^3+\frac{2m(2(e^2-n^2)-n^2E^2)}{E^2-1}r^2\\
+\frac{2n^2e^2E^2-3n^4E^2-(n^2-e^2)^2}{E^2-1}r+\frac{mn^4E^2}{E^2-1}=0.\nonumber
\end{align}
Equation (\ref{eq:rpa}) can have up to five real solutions. As long as $n\neq0$ (\ref{eq:rpa}) is a polynomial of fifth order and cannot be solved analytically. In the Schwarzschild limit ($e=0$ and $n=0$) Eq.~(\ref{eq:rpa}) effectively reduces to a polynomial of second order and we have two real roots. Only one of these roots is positive and lies outside the black hole horizon. It reads
\begin{equation}\label{eq:rpaS}
r_{\text{pa}_{\text{S}}}(E)=\frac{m(3E^2-4+\sqrt{9E^4-8E^2})}{2(E^2-1)}.
\end{equation}
The radius coordinate of the particle sphere depends on the particle energy $E$ [note that in terms of an (energy-dependent) parameter or the particle velocity at spatial infinity it was already derived in a more complicated way by Zakharov \cite{Zakharov1988} and by Liu \emph{et al.} \cite{Liu2016}, respectively, and that when rewritten in terms of the particle energy the result of Liu \emph{et al.} is equivalent to (\ref{eq:rpaS})]. It decreases with increasing particle energy $E$ and for $E\rightarrow \infty$ we recover the radius coordinate of the photon sphere $r_{\text{ph}_{\text{S}}}=3m$. In the Reissner-Nordstr\"{o}m limit ($n=0$) Eq.~(\ref{eq:rpa}) reduces to a polynomial of fourth order. We can derive the roots using Ferrari's method (note that Pang and Jia \cite{Pang2019} used a slightly more complicated approach to calculate, in terms of the particle velocity at spatial infinity, the radius coordinate of the particle sphere; however, their discussion only focused on the unstable particle sphere in the domain of outer communication and not all potential unstable and stable particle spheres). In the general case we can have up to three real positive roots. Let us label these roots $r_{\text{pa}_{-}}$, $r_{\text{pa}_{+}}$, and $r_{\text{pa}}$ and sort them such that in the case of a black hole spacetime we have $0<r_{\text{pa}_{-}}<r_{\text{H}_{\text{i}}}\leq r_{\text{pa}_{+}}\leq r_{\text{H}_{\text{o}}}<r_{\text{pa}}$. As long as we have a black hole spacetime $r_{\text{pa}_{+}}$ is located between the horizons in the nonstatic region and thus it cannot mark the position of a particle sphere. $r_{\text{pa}_{-}}$ lies in the interior static domain between curvature singularity and inner black hole horizon $r_{\text{H}_{\text{i}}}$. As long as we have a black hole spacetime we have $V_{E}(r_{\text{pa}_{-}})<0$ and thus a negative Carter constant. As a consequence $r_{\text{pa}_{-}}$ does not mark the position of a particle sphere. Thus only $r_{\text{pa}}$ marks the position of a particle sphere and this particle sphere is unstable. When we exceed the critical value $e=m$ both horizons disappear and now we have an unstable particle sphere at $r_{\text{pa}}$ and a stable particle sphere at $r_{\text{pa}_{+}}$. When we keep $E$ constant and further increase the electric charge $e$ $r_{\text{pa}_{+}}$ approaches $r_{\text{pa}}$. When we exceed a second critical value $e_{\text{C}_{2}}$ both particle spheres disappear (for $e=e_{\text{C}_{2}}$ we have $r_{\text{pa}_{+}}=r_{\text{pa}}$). This is analogous to the photon sphere; however, for timelike geodesics the exact value $e_{\text{C}_{2}}$ beyond which the particle spheres disappear does not only depend on the electric charge $e$ but also on the particle energy $E$.\\
In the case of the NUT metric (\ref{eq:rpa}) commonly has three real roots. Similarly, for the charged NUT metric (\ref{eq:rpa}) can have up to five real roots. According to our investigations, when (\ref{eq:rpa}) has five real roots, two of these roots seem to be associated with $V_{E}(r)<0$ and thus a negative Carter constant $K$. As was already mentioned above in this case we cannot have angular motion and thus at these radius coordinates particle spheres cannot exist. Also for the remaining three roots the answer to the question how many of them actually represent real particle spheres depends on the choice for $e$, $n$, and $E$. For analyzing how many particle spheres can exist let us first have a look at (\ref{eq:EoMr}) for geodesics with $K=0$ (we will see below that these are radial timelike geodesics). In this case we can write the right-hand side of (\ref{eq:EoMr}) in terms of $\rho(r)$ and a polynomial of second order. As $\rho(r)=r^2+n^2>0$ the associated roots are complex conjugate and purely imaginary. The roots of the polynomial of second order on the other hand can be real or complex conjugate. For distinguishing both cases we now define a new quantity
\begin{equation}\label{eq:ClassCons}
\Delta=m^2+e^2(E^2-1)-n^2(E^4-1),
\end{equation}
which allows one to classify the nature of the roots and determines the number of particle spheres that can exist. If $\Delta>0$ for the NUT metric and the charged NUT metric the polynomial has two distinct real roots. When we have a black hole spacetime we only have one unstable particle sphere at the radius coordinate $r_{\text{H}_{\text{o}}}<r_{\text{pa}}$. In the case of the charged NUT metric we can have three more cases. For $\sqrt{m^2+n^2}<e<e_{\text{C}_{2}}$ we have a stable particle sphere at $r_{\text{pa}_{+}}$ and an unstable particle sphere at $r_{\text{pa}}>r_{\text{pa}_{+}}$. When we have $e=e_{\text{C}_{2}}$ we have $r_{\text{pa}_{+}}=r_{\text{pa}}$ and when we have $e_{\text{C}_{2}}<e$ the charged NUT metric does not possess particle spheres. When $\Delta=0$ the polynomial has a real double root. When we choose $0\leq e\leq \sqrt{m^2+n^2}$ the NUT metric and the charged NUT metric represent a black hole and we have two unstable particle spheres at the radius coordinates $r_{\text{pa}_{-}}<r_{\text{H}_{\text{i}}}$ and $r_{\text{H}_{\text{o}}}<r_{\text{pa}}$. When $\Delta<0$ the polynomial has a pair of complex conjugate roots. When we choose $0\leq e\leq \sqrt{m^2+n^2}$ for both, the NUT metric and the charged NUT metric, we again have a black hole spacetime. Also in this case we only have two unstable particle spheres at the radius coordinates $r_{\text{pa}_{-}}<r_{\text{H}_{\text{i}}}$ and $r_{\text{H}_{\text{o}}}<r_{\text{pa}}$. Again in the case of the charged NUT metric for both, $\Delta=0$ and $\Delta<0$, we can have three more cases. When we have $\sqrt{m^2+n^2}<e<e_{\text{C}_{2}}$ we have two unstable particle spheres at the radius coordinates $r_{\text{pa}_{-}}<r_{\text{pa}}$ and a stable particle sphere at the radius coordinate $r_{\text{pa}_{+}}$, where we have $r_{\text{pa}_{-}}<r_{\text{pa}_{+}}<r_{\text{pa}}$. For $e=e_{\text{C}_{2}}$ we have $r_{\text{pa}_{+}}=r_{\text{pa}}$. For $e_{\text{C}_{2}}<e$ the particle spheres at $r_{\text{pa}_{+}}$ and $r_{\text{pa}}$ disappear and only the particle sphere at $r_{\text{pa}_{-}}$ remains. Here, the exact value of $e_{\text{C}_{2}}$ depends on $e$, $n$, and $E$. Note that due to the possible number of combinations of $e$, $n$, and $E$ we could not test all potential combinations and it is possible that also other cases may occur. However, because the cases discussed above are similar to the different cases for the photon sphere we are convinced that they represent the majority of all realized cases. In the following we will now restrict our discussion to black hole spacetimes.\\
\begin{figure}\label{fig:rpaE}
  \begin{tabular}{cc}
    \hspace{-0.5cm}\includegraphics[width=95mm]{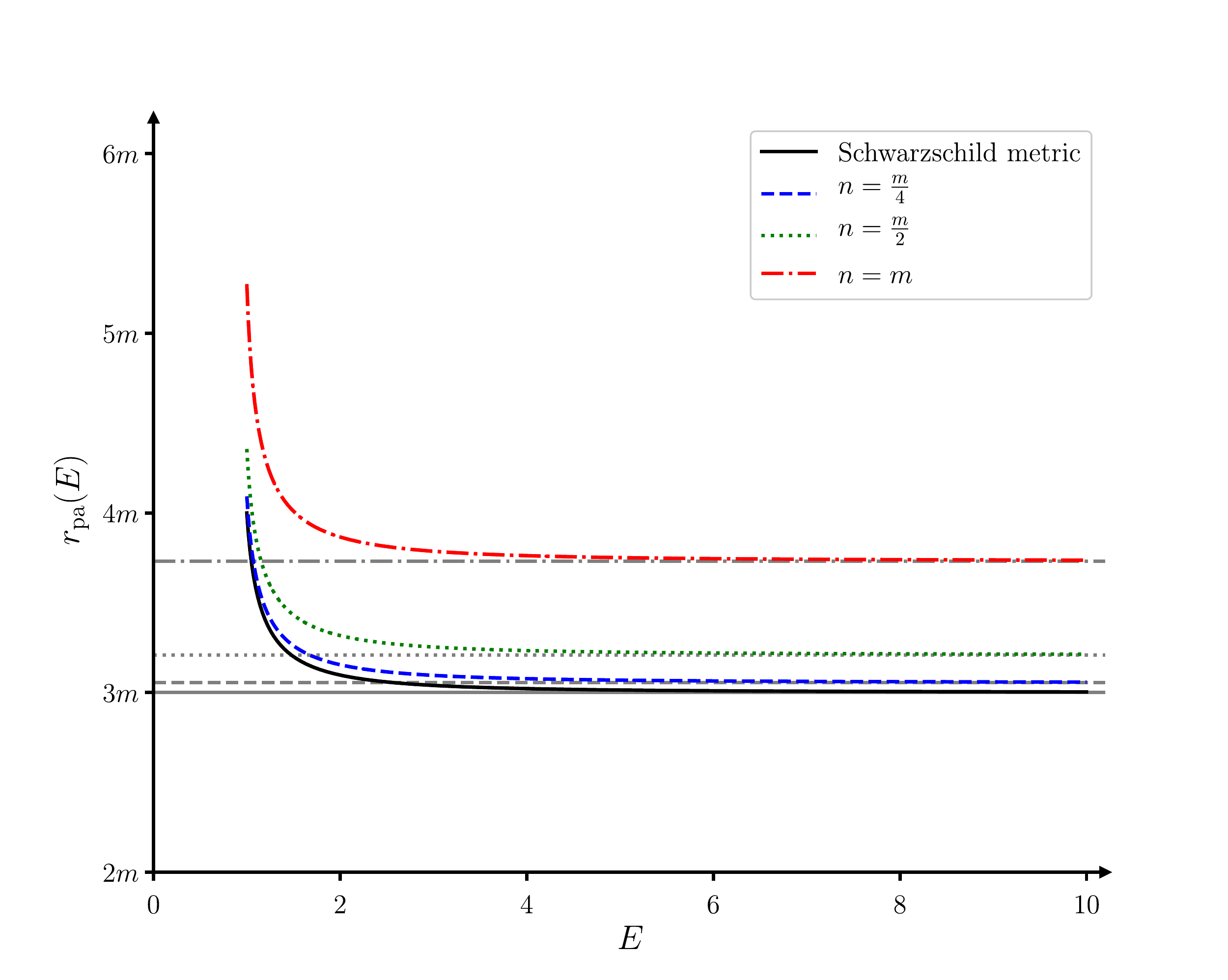} &   \includegraphics[width=95mm]{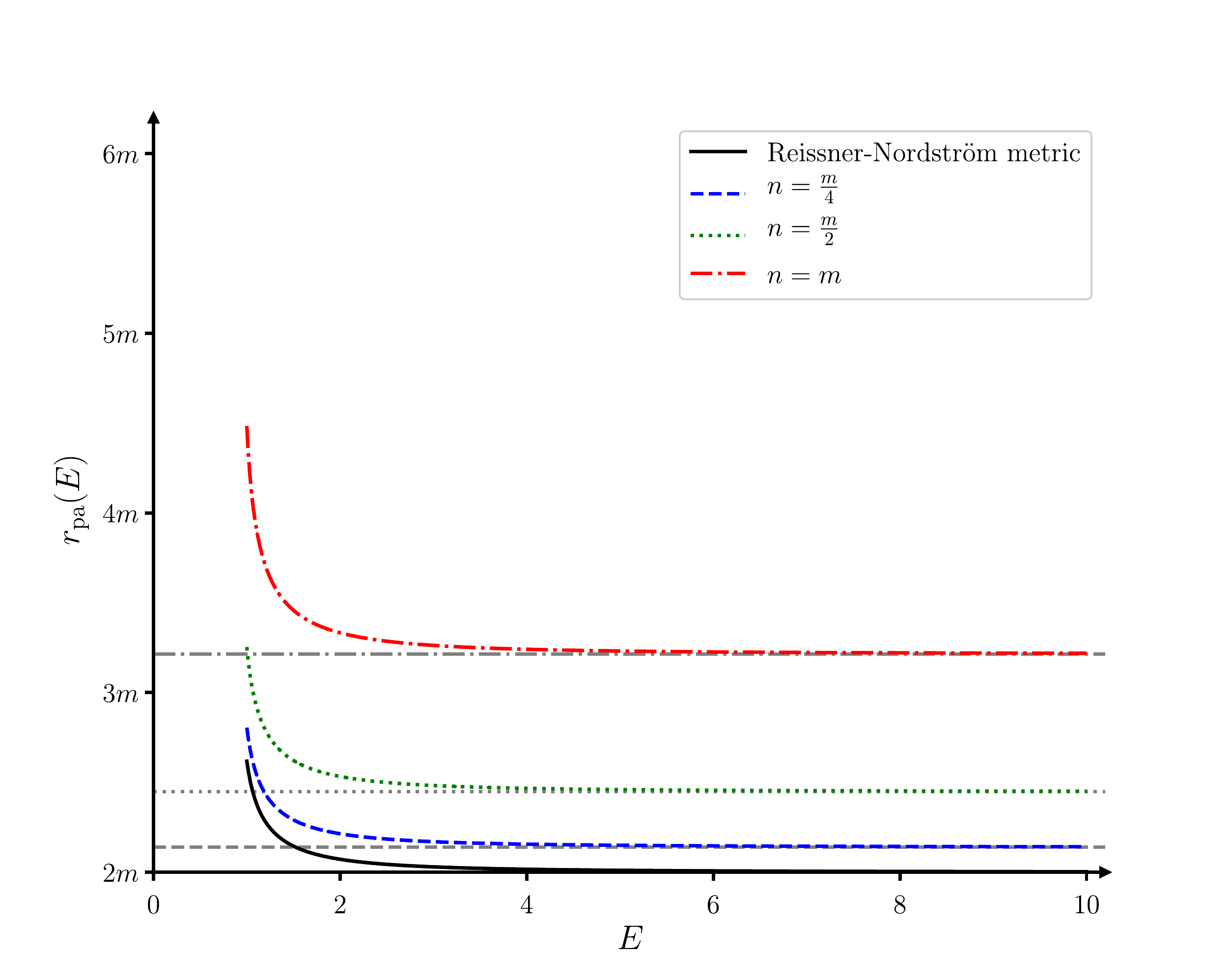}
  \end{tabular}
	\caption{Radius coordinate of the particle sphere at $r_{\text{H}_{\text{o}}}<r_{\text{pa}}$ as function of the energy $E$. Left panel: $r_{\text{pa}}$ as function of $E$ for the Schwarzschild metric (black solid line) and the NUT metric with $n=m/4$ (blue dashed line), $n=m/2$ (green dotted line), and $n=m$ (red dash-dotted line). Right panel: $r_{\text{pa}}$ as function of $E$ for the Reissner-Nordstr\"{o}m metric (black solid line) and the charged NUT metric with $n=m/4$ (blue dashed line), $n=m/2$ (green dotted line), and $n=m$ (red dash-dotted line). In all four cases we have $e=m$. The horizontal gray lines with the same line styles mark the radius coordinates of the corresponding photon spheres.}
\end{figure}
The left panel of Fig.~3 shows the radius coordinate of the particle sphere $r_{\text{pa}}$ as function of $E$ for the Schwarzschild metric (black solid line) and the NUT metric with $n=m/4$ (blue dashed line), $n=m/2$ (green dotted line), and $n=m$ (red dash-dotted line). The horizontal lines with the same line styles mark the radius coordinates of the corresponding photon spheres. We can see that in all four cases for $E\rightarrow \infty$ the radius coordinates of the particle spheres approach the radius coordinates of the corresponding photon spheres. When we increase the gravitomagnetic charge $n$ and keep the energy $E$ constant the radius coordinate of the particle sphere increases. Similarly the right panel of Fig.~3 shows the radius coordinate of the particle sphere $r_{\text{pa}}$ as function of $E$ for the Reissner-Nordstr\"{o}m metric (black solid line) and the charged NUT metric with $n=m/4$ (blue dashed line), $n=m/2$ (green dotted line), and $n=m$ (red dash-dotted line). In all four cases we have $e=m$. It basically shows the same features as the left panel but shifted to lower radius coordinates.\\
\begin{figure}\label{fig:rpan}
  \begin{tabular}{cc}
    \hspace{-0.5cm}\includegraphics[width=95mm]{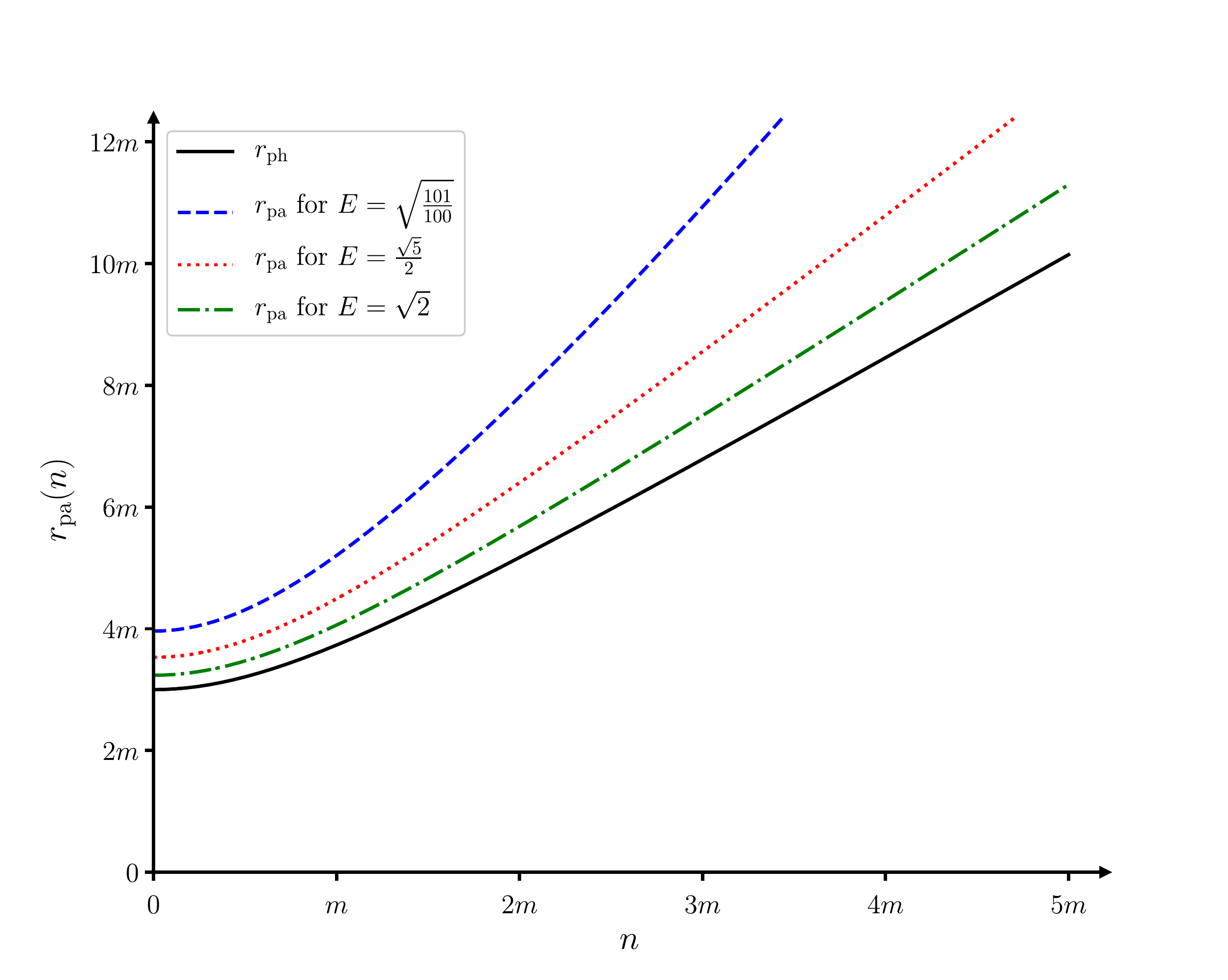} &   \includegraphics[width=95mm]{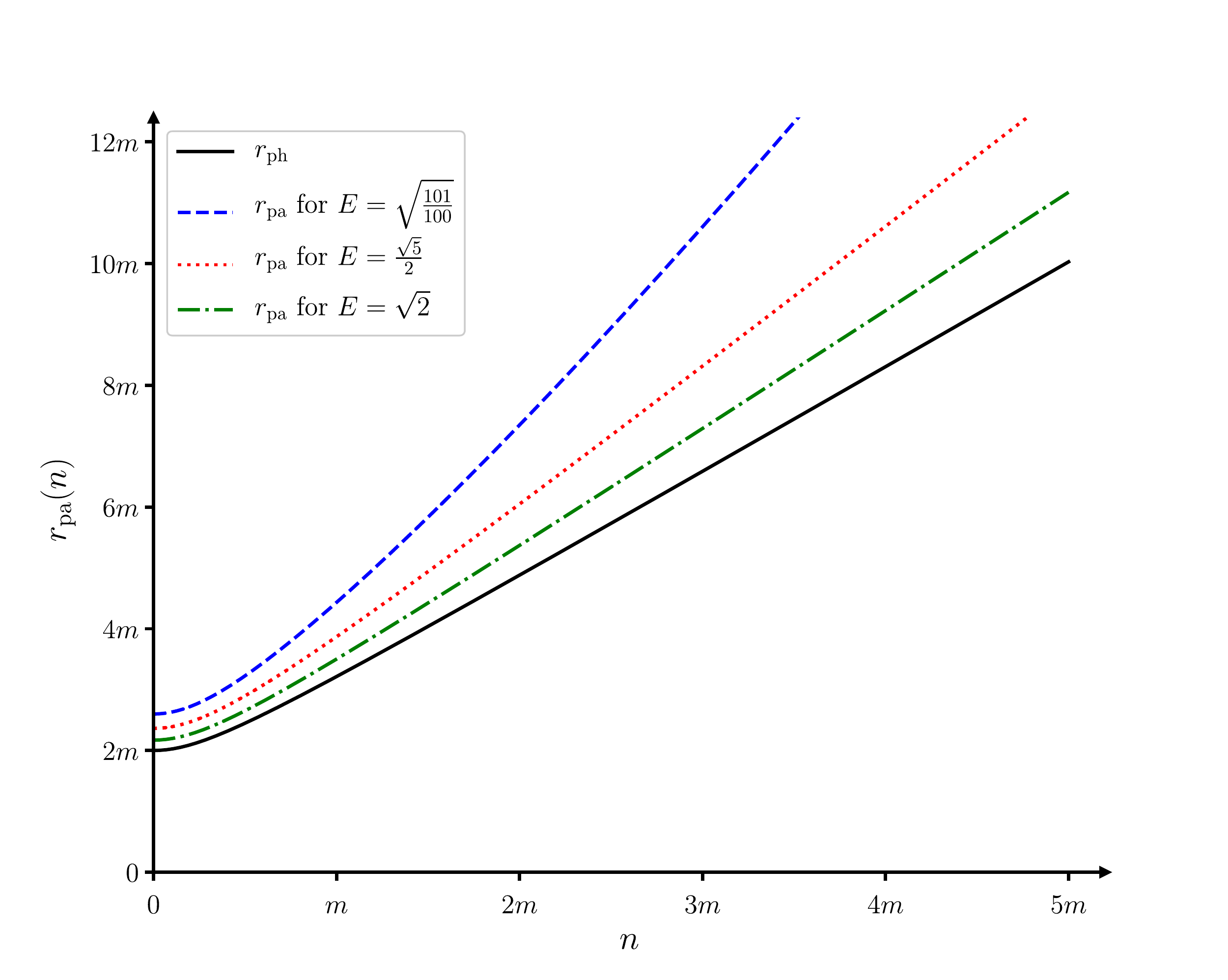}
  \end{tabular}
	\caption{Radius coordinates of the photon sphere $r_{\text{ph}}$ and the particle spheres with $r_{\text{H}_{\text{o}}}<r_{\text{pa}}(E)$ for three different energies $E$ as functions of the gravitomagnetic charge $n$. Left panel: radius coordinate of the photon sphere $r_{\text{ph}}$ (black solid line) and radius coordinates of the particle spheres $r_{\text{pa}}$ for $E=\sqrt{101/100}$ (blue dashed line), $E=\sqrt{5}/2$ (red dotted line), and $E=\sqrt{2}$ (green dash-dotted line) as function of $n$ for the NUT metric. Right panel: radius coordinate of the photon sphere $r_{\text{ph}}$ (black solid line) and radius coordinates of the particle spheres $r_{\text{pa}}$ for $E=\sqrt{101/100}$ (blue dashed line), $E=\sqrt{5}/2$ (red dotted line), and $E=\sqrt{2}$ (green dash-dotted line) as functions of $n$ for the charged NUT metric with $e=m$.}
\end{figure}
The left panel of Fig.~4 shows the radius coordinate of the photon sphere $r_{\text{ph}}$ (black solid line) and the radius coordinates $r_{\text{pa}}$ of the particle spheres for $E=\sqrt{101/100}$ (blue dashed line), $E=\sqrt{5}/2$ (red dotted line), and $E=\sqrt{2}$ (green dash-dotted line) as functions of the gravitomagnetic charge $n$ for the NUT metric. As already indicated in Fig.~3 the radius coordinates of the photon sphere and the particle spheres increase with increasing $n$. In addition the radius coordinates of the particle spheres decrease with increasing $E$. Similarly, the right panel of Fig.~4 shows the radius coordinate of the photon sphere $r_{\text{ph}}$ (black solid line) and the radius coordinates $r_{\text{pa}}$ of the particle spheres for $E=\sqrt{101/100}$ (blue dashed line), $E=\sqrt{5}/2$ (red dotted line), and $E=\sqrt{2}$ (green dash-dotted line) as functions of the gravitomagnetic charge $n$ for the charged NUT metric. We basically see the same features as in the left panel; however, turning on the electric charge shifted them to lower radius coordinates. Note that in the rest of Sec.~\ref{Sec:EoM} we will assume that the reader is aware that $r_{\text{pa}_{-}}$ and $r_{\text{pa}}$ are functions of the energy $E$ and thus we will not explicitly write the energy dependency.\\

\subsubsection{Types of $r$ motion}
We can now use the radius coordinates of the unstable particle spheres $r_{\text{pa}_{-}}$ and $r_{\text{pa}}$, the potential $V_{E}(r)$, and the Carter constant $K$ for classifying seven different types of motion (note that we count radial timelike geodesics in the Schwarzschild metric and the Reissner-Nordstr\"{o}m metric as two different types of motion). These are as follows.
\begin{enumerate}[(i)]
\item \emph{Case 1-S}.--This case covers radial timelike geodesics with $K=0$ in the Schwarzschild metric. In this case the right-hand side of (\ref{eq:EoMr}) has four real roots. Three of these roots are equal and located at $r_{1}=r_{2}=r_{3}=0$. The fourth root is also real and located at $r_{4}=-2m/(E^2-1)$. These geodesics do not have turning points in the domain of outer communication.
\item \emph{Case 1-RN}.--This case covers radial timelike geodesics with $K=0$ in the Reissner-Nordstr\"{o}m metric. In this case the right-hand side of (\ref{eq:EoMr}) has four real roots. The first root is positive and located at $r_{1}=(\sqrt{m^2+e^2(E^2-1)}-m)/(E^2-1)$. Then we have a double root at $r_{2}=r_{3}=0$. The fourth root is negative and located at $r_{4}=-(\sqrt{m^2+e^2(E^2-1)}+m)/(E^2-1)$. Again these geodesics do not have turning points in the domain of outer communication.
\item \emph{Case 1-NUT-a}.--This case covers timelike geodesics with $\Delta<0$ and $0\leq K<V_{E}(r_{\text{pa}_{-}})$ in the NUT metric and the charged NUT metric. In this case the right-hand side of (\ref{eq:EoMr}) has two pairs of complex conjugate roots. We sort and label them such that we have $r_{1}=\bar{r}_{2}=R_{1}+iR_{2}$ and $r_{3}=\bar{r}_{4}=R_{3}+iR_{4}$,
where we choose the real and imaginary parts such that $R_{1}<R_{3}$, and $R_{2}>0$ and $R_{4}>0$, respectively. These geodesics do not have turning points in the domain of outer communication.
\item \emph{Case 1-NUT-b}.--This case covers timelike geodesics with $\Delta\leq0$ and $K=V_{E}(r_{\text{pa}_{-}})$ in the NUT metric and the charged NUT metric. In this case the right-hand side of (\ref{eq:EoMr}) has a real double root and a pair of complex conjugate roots. We label the real roots such that $r_{1}=r_{2}=r_{\text{pa}_{-}}$ and we write the complex conjugate roots in terms of their real and imaginary parts such that $r_{3}=\bar{r}_{4}=R_{3}+iR_{4}$, where we choose $R_{4}>0$. These geodesics do not have turning points in the domain of outer communication.
\item \emph{Case 2}.--This case covers timelike geodesics with three different combinations of $\Delta$ and $K$. For the Schwarzschild metric and the Reissner-Nordstr\"{o}m metric it occurs when we have $0<K<V_{E}(r_{\text{pa}})$ (in this case we always have $0<\Delta$). For the NUT metric and the charged NUT metric it occurs when $0<\Delta$ and $0\leq K<V_{E}(r_{\text{pa}})$ or $\Delta\leq 0$ and $V_{E}(r_{\text{pa}_{-}})<K<V_{E}(r_{\text{pa}})$. In this case the right-hand side of (\ref{eq:EoMr}) has two distinct real roots and a pair of complex conjugate roots. We sort and label the real roots such that $r_{1}>r_{2}$. As above we write the complex conjugate roots in terms of their real and imaginary parts such that $r_{3}=\bar{r}_{4}=R_{3}+iR_{4}$, where we choose $R_{4}>0$. These geodesics do not have turning points in the domain of outer communication.
\item \emph{Case 3}.--This case covers timelike geodesics with $K=V_{E}(r_{\text{pa}})$. In this case the right-hand side of (\ref{eq:EoMr}) has four real roots, two of which are equal. We sort the roots such that $r_{1}=r_{2}=r_{\text{pa}}>r_{3}>r_{4}$. These are timelike geodesics on or asymptotically coming from or going to an unstable particle sphere. They do not have turning points in the domain of outer communication.
\item \emph{Case 4}.--This case covers timelike geodesics with $V_{E}(r_{\text{pa}})<K$. In this case the right-hand side of (\ref{eq:EoMr}) has four distinct real roots. We sort and label them such that $r_{1}>r_{2}>r_{3}>r_{4}$. For $r_{\text{pa}}<r$ these geodesics can have a turning point at $r_{1}=r_{\text{min}}$ (minimum) and for $r_{\text{H}_{\text{o}}}<r<r_{\text{pa}}$ these geodesics can have a turning point at $r_{2}=r_{\text{max}}$ (maximum).
\end{enumerate}
Note that all four metrics have three common types of motion (cases 2--4). For keeping them consistent we classified the remaining four types of motion under case 1 (case 1-S and case 1-RN both cover the same type of motion; however, because their root structure is different we discussed them separately). In the following we will now show how to solve the equation of motion for $r$ using elementary and Jacobi's elliptic functions.

\subsubsection{Solving the equation of motion}
\emph{Case 1-S}.--We start with radial timelike geodesics in the Schwarzschild metric. These geodesics have $K=0$ and thus (\ref{eq:EoMr}) reduces to
\begin{equation}
\left(\frac{\text{d}r}{\text{d}\lambda}\right)^2=(E^2-1)r^4+2mr^3.
\end{equation}
We separate variables and integrate from $r(\lambda_{i})=r_{i}$ to $r(\lambda)=r$. We get
\begin{equation}\label{eq:Case1Sint}
\lambda-\lambda_{i}=i_{r_{i}}\int_{r_{i}}^{r}\frac{\text{d}r'}{r'\sqrt{(E^2-1)r'^2+2mr'}},
\end{equation}
where $i_{r_{i}}=\text{sgn}(\left.\text{d}r/\text{d}\lambda\right|_{r=r_{i}})$. The right-hand side of (\ref{eq:Case1Sint}) has the same structure as $I_{3}$ given by (\ref{eq:I3}) in Appendix~\ref{App:EmInt1}. Now we use (\ref{eq:I3}) to rewrite (\ref{eq:Case1Sint}) in terms of elementary functions and solve for $r$. We get as solution for $r(\lambda)$
\begin{equation}\label{eq:Case1Ssol}
r(\lambda)=\frac{r_{i}}{1-i_{r_{i}}\sqrt{((E^2-1)r_{i}+2m)r_{i}}(\lambda-\lambda_{i})+\frac{mr_{i}}{2}(\lambda-\lambda_{i})^2}.
\end{equation}
\emph{Case 1-RN}.--Now we turn to radial timelike geodesics in the Reissner-Nordstr\"{o}m metric. These geodesics have $K=0$. In this case (\ref{eq:EoMr}) reduces to
\begin{equation}
\left(\frac{\text{d}r}{\text{d}\lambda}\right)^2=(E^2-1)r^4+2mr^3-e^2r^2.
\end{equation}
We again separate variables and integrate from $r(\lambda_{i})=r_{i}$ to $r(\lambda)=r$. This time we get
\begin{equation}\label{eq:Case1RNint}
\lambda-\lambda_{i}=i_{r_{i}}\int_{r_{i}}^{r}\frac{\text{d}r'}{r'\sqrt{(E^2-1)r'^2+2mr'-e^2}}.
\end{equation}
The right-hand side of (\ref{eq:Case1RNint}) has the same structure as $I_{7}$ given by (\ref{eq:I7}) in Appendix~\ref{App:EmInt1}. We use it to rewrite the integral in terms of elementary functions. We solve for $r$ and get as solution for $r(\lambda)$
\begin{equation}\label{eq:Case1RNsol}
r(\lambda)=\frac{e^2}{m-\sqrt{(E^2-1)e^2+m^2}\sin\left(\arcsin\left(\frac{mr_{i}-e^2}{r_{i}\sqrt{(E^2-1)e^2+m^2}}\right)+i_{r_{i}}\left|e\right|\left(\lambda-\lambda_{i}\right)\right)}.
\end{equation}
\emph{Case 1-NUT-a}.--This case covers timelike geodesics with $\Delta<0$ and $0\leq K<V_{E}(r_{\text{pa}_{-}})$. It only occurs for the NUT metric and the charged NUT metric. In this case we first rewrite the right-hand side of (\ref{eq:EoMr}) in terms of the real and imaginary parts of the roots. We integrate from $r(\lambda_{i})=r_{i}$ to $r(\lambda)=r$ and get
\begin{equation}\label{eq:Case1NUTaint}
\lambda-\lambda_{i}=i_{r_{i}}\int_{r_{i}}^{r}\frac{\text{d}r'}{\sqrt{(E^2-1)((R_{1}-r')^2+R_{2}^2)((R_{3}-r')^2+R_{4}^2)}}.
\end{equation}
As second step we use the real and imaginary parts of the complex conjugate roots to define two new constants of motion \cite{Gralla2020,Byrd1954}:
\begin{align}
R=\sqrt{(R_{2}-R_{4})^2+(R_{1}-R_{3})^2},\label{eq:Case1NUTaCoeff1}\\
\bar{R}=\sqrt{(R_{2}+R_{4})^2+(R_{1}-R_{3})^2}.\label{eq:Case1NUTaCoeff2}
\end{align}
Now we use the substitution 
\begin{equation}\label{eq:Case1NUTasub}
r=R_{1}-R_{2}\frac{g_{0}-\tan\chi}{1+g_{0}\tan\chi},
\end{equation}
where $g_{0}$ is given by
\begin{equation}\label{eq:Case1NUTaCoeffg0}
g_{0}=\sqrt{\frac{4R_{2}^2-(R-\bar{R})^2}{(R+\bar{R})^2-4R_{2}^2}},
\end{equation}
to transform the integral (\ref{eq:Case1NUTaint}) into the Legendre form (\ref{eq:LegIntsep}). Note that because of $\tan\chi=\tan(\chi\pm \bar{n}\pi)$, where $\bar{n}$ is a positive integer, (\ref{eq:Case1NUTasub}) is valid for all $\tilde{\chi}=\chi\pm \bar{n}\pi$. Now we follow the steps outlined in Appendix~\ref{App:SolDiff} to derive the solution for $r(\lambda)$. It is given in terms of Jacobi's elliptic $\text{sc}$ function and reads
\begin{equation}\label{eq:Case1NUTasol}
r(\lambda)=R_{1}-R_{2}\frac{g_{0}-\text{sc}\left(i_{r_{i}}\sqrt{E^2-1}\frac{R+\bar{R}}{2}(\lambda-\lambda_{i})+\lambda_{r_{i},k_{1}},k_{1}\right)}{1+g_{0}\text{sc}\left(i_{r_{i}}\sqrt{E^2-1}\frac{R+\bar{R}}{2}(\lambda-\lambda_{i})+\lambda_{r_{i},k_{1}},k_{1}\right)},
\end{equation}
where in our case $\lambda_{r_{i},k_{1}}$, $\chi_{i}$ and the square of the elliptic modulus $k_{1}$ are given, respectively, by
\begin{equation}
\lambda_{r_{i},k_{1}}=F_{L}(\chi_{i},k_{1}),
\end{equation}
\begin{equation}\label{eq:Case1NUTachi}
\chi_{i}=\arctan\left(\frac{r_{i}-R_{1}}{R_{2}}\right)+\arctan\left(g_{0}\right),
\end{equation}
and
\begin{equation}\label{eq:Case1NUTak}
k_{1}=\frac{4R\bar{R}}{(R+\bar{R})^2}.
\end{equation}
\emph{Case 1-NUT-b}.--This case covers timelike geodesics with $\Delta\leq0$ and $K=V_{E}(r_{\text{pa}_{-}})$. They only occur for the NUT metric and the charged NUT metric. We again rewrite (\ref{eq:EoMr}) in terms of the roots and integrate from $r(\lambda_{i})=r_{i}$ to $r(\lambda)=r$. This time we get
\begin{equation}\label{eq:Case1NUTbint}
\lambda-\lambda_{i}=i_{r_{i}}\int_{r_{i}}^{r}\frac{\text{d}r'}{(r'-r_{\text{pa}_{-}})\sqrt{(E^2-1)((R_{3}-r')^2+R_{4}^2)}}.
\end{equation}
We can easily see that the right-hand side has the form of the integral $I_{12}$ given by (\ref{eq:I12}) in Appendix~\ref{App:EmInt2}. Now we set $r_{a}=r_{\text{pa}_{-}}$ in (\ref{eq:I12}), insert, and solve for $r$. We obtain as solution for $r(\lambda)$
\begin{equation}\label{eq:Case1NUTbsol}
r(\lambda)=r_{\text{pa}_{-}}+\frac{(R_{3}-r_{\text{pa}_{-}})^2+R_{4}^2}{R_{3}-r_{\text{pa}_{-}}+R_{4}\text{sinh}\left(a_{-}-i_{r_{i}}\sqrt{(E^2-1)((R_{3}-r_{\text{pa}_{-}})^2+R_{4}^2)}(\lambda-\lambda_{i})\right)},
\end{equation}
where the coefficient $a_{-}$ reads
\begin{equation}
a_{-}=\text{arsinh}\left(\frac{(r_{\text{pa}_{-}}-R_{3})(r_{i}-r_{\text{pa}_{-}})+(R_{3}-r_{\text{pa}_{-}})^2+R_{4}^2}{R_{4}(r_{i}-r_{\text{pa}_{-}})}\right).
\end{equation}
\emph{Case 2}.--This case covers timelike geodesics with $0<K<V_{E}(r_{\text{pa}})$ in the Schwarzschild metric and the Reissner-Nordstr\"{o}m metric and timelike geodesics with $0<\Delta$ and $0\leq K<V_{E}(r_{\text{pa}})$ or $\Delta\leq 0$ and $V_{E}(r_{\text{pa}_{-}})< K<V_{E}(r_{\text{pa}})$ in the NUT metric and the charged NUT metric. We rewrite (\ref{eq:EoMr}) in terms of the roots and integrate from $r(\lambda_{i})=r_{i}$ to $r(\lambda)=r$. We get
\begin{equation}\label{eq:Case2int}
\lambda-\lambda_{i}=i_{r_{i}}\int_{r_{i}}^{r}\frac{\text{d}r'}{\sqrt{(E^2-1)(r'-r_{1})(r'-r_{2})((R_{3}-r')^2+R_{4}^2)}}.
\end{equation}
Now we use the real roots and the real and imaginary parts of the complex conjugate roots to define two new constants of motion. They read \cite{Hancock1917,Gralla2020}
\begin{align}
R=\sqrt{(R_{3}-r_{1})^2+R_{4}^2},\label{eq:Case2Coeff1}\\
\bar{R}=\sqrt{(R_{3}-r_{2})^2+R_{4}^2}.\label{eq:Case2Coeff2}
\end{align}
Now we use the transformation 
\begin{equation}\label{eq:Case2sub}
r=\frac{r_{1}\bar{R}-r_{2}R+(r_{1}\bar{R}+r_{2}R)\cos\chi}{\bar{R}-R+(\bar{R}+R)\cos\chi}
\end{equation}
to transform the integral (\ref{eq:Case2int}) into the Legendre form (\ref{eq:LegIntsep}). Again we follow the steps outlined in Appendix~\ref{App:SolDiff} to derive the solution $r(\lambda)$. This time it is given in terms of Jacobi's elliptic $\text{cn}$ function and reads
\begin{equation}\label{eq:Case2sol}
r(\lambda)=\frac{r_{1}\bar{R}-r_{2}R+(r_{1}\bar{R}+r_{2}R)\text{cn}\left(i_{r_{i}}\sqrt{(E^2-1)R\bar{R}}(\lambda-\lambda_{i})+\lambda_{r_{i},k_{2}},k_{2}\right)}{\bar{R}-R+(\bar{R}+R)\text{cn}\left(i_{r_{i}}\sqrt{(E^2-1)R\bar{R}}(\lambda-\lambda_{i})+\lambda_{r_{i},k_{2}},k_{2}\right)},
\end{equation}
where $\lambda_{r_{i},k_{2}}$, $\chi_{i}$ and the square of the elliptic modulus $k_{2}$ are given, respectively, by
\begin{equation}
\lambda_{r_{i},k_{2}}=F_{L}(\chi_{i},k_{2}),
\end{equation}
\begin{equation}\label{eq:Case2chi}
\chi_{i}=\arccos\left(\frac{(r_{i}-r_{2})R-(r_{i}-r_{1})\bar{R}}{(r_{i}-r_{2})R+(r_{i}-r_{1})\bar{R}}\right),
\end{equation}
\begin{equation}\label{eq:Case2k}
k_{2}=\frac{(R+\bar{R})^2-(r_{1}-r_{2})^2}{4R\bar{R}}.
\end{equation}
\emph{Case 3}.--This case covers timelike geodesics with $K=V_{E}(r_{\text{pa}})$. These are timelike geodesics on or asymptotically coming from or going to an unstable particle sphere. In this case we have a real double root at $r_{1}=r_{2}=r_{\text{pa}}$. For timelike geodesics on an unstable particle sphere we have $r(\lambda_{i})=r_{i}=r_{\text{pa}}$ and thus the solution reads $r(\lambda)=r_{\text{pa}}$. For all other geodesics we now rewrite (\ref{eq:EoMr}) in terms of the roots and integrate from $r(\lambda_{i})=r_{i}$ to $r(\lambda)=r$. We get
\begin{equation}\label{eq:Case3int}
\lambda-\lambda_{i}=i_{r_{i}}\int_{r_{i}}^{r}\frac{\text{d}r'}{\sqrt{(E^2-1)(r'-r_{\text{pa}})^2(r'-r_{3})(r'-r_{4})}}.
\end{equation}
We immediately see that when we pull the term $r'-r_{\text{pa}}$ out of the root this is an elementary integral. Now we substitute
\begin{equation}\label{eq:Case3sub}
r=r_{3}+\frac{3a_{3}}{12y-a_{2}},
\end{equation}
where the two coefficients $a_{2}$ and $a_{3}$ read
\begin{align}
&a_{2}=6(E^2-1)r_{3}^2+6mr_{3}+2n^2E^2-e^2-K,\\
&a_{3}=4(E^2-1)r_{3}^3+6mr_{3}^2+2(2n^2E^2-e^2-K)r_{3}+2m(n^2+K).
\end{align}
Now the integral reads
\begin{equation}\label{eq:Case3yint}
\lambda-\lambda_{i}=-\frac{i_{r_{i}}}{2}\int_{y_{i}}^{y}\frac{\text{d}y'}{\sqrt{(y'-y_{\text{pa}})^2(y'-y_{1})}},
\end{equation}
where $y_{\text{pa}}$ and $y_{1}$ are related to $r_{\text{pa}}$ and $r_{4}$ by (\ref{eq:Case3sub}), respectively. We can easily see that the right-hand side can be rewritten in terms of the integral $I_{14}$ given by (\ref{eq:I14y}) in Appendix~\ref{App:EmInt3}. Now we have to distinguish two different cases. In the first case timelike motion takes place outside the particle sphere and thus we have $r_{\text{pa}}<r$. In this case we use (\ref{eq:I143}) to integrate (\ref{eq:Case3yint}). We solve for $r$ and obtain as solution for $r(\lambda)$
\begin{equation}\label{eq:Case3sol1}
r(\lambda)=r_{3}-\frac{(r_{\text{pa}}-r_{3})(r_{3}-r_{4})}{r_{\text{pa}}-r_{3}-(r_{\text{pa}}-r_{4})\text{tanh}^2\left(\text{artanh}\left(\sqrt{\frac{(r_{i}-r_{4})(r_{\text{pa}}-r_{3})}{(r_{i}-r_{3})(r_{\text{pa}}-r_{4})}}\right)-i_{r_{i}}\sqrt{\frac{a_{3}(r_{\text{pa}}-r_{4})}{4(r_{\text{pa}}-r_{3})(r_{3}-r_{4})}}(\lambda-\lambda_{i})\right)}.
\end{equation}
In the second case we have timelike motion between (outer) black hole horizon and particle sphere $r_{\text{H}_{\text{o}}}<r<r_{\text{pa}}$. Note that in this paper we do not use this solution in our investigation of gravitational lensing of massive particles; however, we still include it for the sake of a complete discussion of the exact solutions of the equations of motion for timelike geodesics. In this case we use (\ref{eq:I142}) to integrate (\ref{eq:Case3yint}). Again we solve for $r$ and get as solution for $r(\lambda)$
\begin{equation}\label{eq:Case3sol2}
r(\lambda)=r_{3}-\frac{(r_{\text{pa}}-r_{3})(r_{3}-r_{4})}{r_{\text{pa}}-r_{3}-(r_{\text{pa}}-r_{4})\text{coth}^2\left(\text{arcoth}\left(\sqrt{\frac{(r_{i}-r_{4})(r_{\text{pa}}-r_{3})}{(r_{i}-r_{3})(r_{\text{pa}}-r_{4})}}\right)+i_{r_{i}}\sqrt{\frac{a_{3}(r_{\text{pa}}-r_{4})}{4(r_{\text{pa}}-r_{3})(r_{3}-r_{4})}}(\lambda-\lambda_{i})\right)}.
\end{equation}

\emph{Case 4}.--This case covers timelike geodesics with $V_{E}(r_{\text{pa}})<K$. The right-hand side of (\ref{eq:EoMr}) has four distinct real roots and we can have turning points in the domain of outer communication. Again we rewrite (\ref{eq:EoMr}) in terms of the roots and integrate from $r(\lambda_{i})=r_{i}$ to $r(\lambda)=r$. We get
\begin{equation}\label{eq:Case4int}
\lambda-\lambda_{i}=i_{r_{i}}\int_{r_{i}}^{r}\frac{\text{d}r'}{\sqrt{(E^2-1)(r'-r_{1})(r'-r_{2})(r'-r_{3})(r'-r_{4})}}.
\end{equation}
Again we have to distinguish two different cases. In the first case we have timelike geodesics with $r_{\text{pa}}<r$. These geodesics can have a turning point at $r_{\text{pa}}<r_{1}=r_{\text{min}}$.
We use the substitution \cite{Hancock1917,Gralla2020}
\begin{equation}\label{eq:Case4sub1}
r=r_{2}+\frac{(r_{1}-r_{2})(r_{2}-r_{4})}{r_{2}-r_{4}-(r_{1}-r_{4})\sin^2\chi}
\end{equation}
to transform the integral (\ref{eq:Case4int}) into the Legendre form (\ref{eq:LegIntsep}). Once more we follow the steps outlined in Appendix~\ref{App:SolDiff} to derive the solution for $r(\lambda)$. This time the solution is given in terms of Jacobi's elliptic $\text{sn}$ function and reads
\begin{equation}\label{eq:Case4sol1}
r(\lambda)=r_{2}+\frac{(r_{1}-r_{2})(r_{2}-r_{4})}{r_{2}-r_{4}-(r_{1}-r_{4})\text{sn}^2\left(\frac{i_{r_{i}}}{2}\sqrt{(E^2-1)(r_{1}-r_{3})(r_{2}-r_{4})}\left(\lambda-\lambda_{i}\right)+\lambda_{r_{i},k_{3}},k_{3}\right)},
\end{equation}
where $\lambda_{r_{i},k_{3}}$, $\chi_{i}$, and the square of the elliptic modulus $k_{3}$ are given, respectively, by
\begin{equation}\label{eq:Case4lambdai}
\lambda_{r_{i},k_{3}}=F_{L}(\chi_{i},k_{3}),
\end{equation}
\begin{equation}\label{eq:Case4chi1}
\chi_{i}=\arcsin\left(\sqrt{\frac{(r_{i}-r_{1})(r_{2}-r_{4})}{(r_{i}-r_{2})(r_{1}-r_{4})}}\right),
\end{equation}
\begin{equation}\label{eq:Case4k}
k_{3}=\frac{(r_{2}-r_{3})(r_{1}-r_{4})}{(r_{1}-r_{3})(r_{2}-r_{4})}.
\end{equation}
In the second case we have timelike geodesics between (outer) black hole horizon and particle sphere $r_{\text{H}_{\text{o}}}<r<r_{\text{pa}}$. As for case 3 these geodesics are not relevant for investigating gravitational lensing and they are only included in the discussion for the sake of completeness. These geodesics can have a turning point at $r_{\text{H}_{\text{o}}}<r_{2}=r_{\text{max}}<r_{\text{pa}}$. In this case we substitute \cite{Hancock1917}
\begin{equation}\label{eq:Case4sub2}
r=r_{1}-\frac{(r_{1}-r_{2})(r_{1}-r_{3})}{r_{1}-r_{3}-(r_{2}-r_{3})\sin^2\chi}
\end{equation}
to transform the integral (\ref{eq:Case4int}) into the Legendre form (\ref{eq:LegIntsep}). Note that on the first view Gralla and Lupsasca \cite{Gralla2020} use a transformation which looks similar to (\ref{eq:Case4sub2}); however, because they sort and label their roots opposite to our convention the transformations are not the same. In addition, it is worth noting that the substitution used by Gralla and Lupsasca has the disadvantage that when we integrate the time coordinate $t$ below Legendre's elliptic integral of the third kind would diverge because we have to integrate over the horizons. The substitution (\ref{eq:Case4sub2}) has the advantage that it completely avoids this problem. \\
One last time we follow the steps outlined in Appendix~\ref{App:SolDiff} to derive the solution for $r(\lambda)$. Again the solution is given in terms of Jacobi's elliptic $\text{sn}$ function; however, this time it reads
\begin{equation}\label{eq:Case4sol2}
r(\lambda)=r_{1}-\frac{(r_{1}-r_{2})(r_{1}-r_{3})}{r_{1}-r_{3}-(r_{2}-r_{3})\text{sn}^2\left(-\frac{i_{r_{i}}}{2}\sqrt{(E^2-1)(r_{1}-r_{3})(r_{2}-r_{4})}\left(\lambda-\lambda_{i}\right)+\lambda_{r_{i},k_{3}},k_{3}\right)},
\end{equation}
where $\lambda_{r_{i},k_{3}}$ and the square of the elliptic modulus $k_{3}$ are given by (\ref{eq:Case4lambdai}) and (\ref{eq:Case4k}), respectively. $\chi_{i}$ on the other hand is related to $r_{i}$ by
\begin{equation}\label{eq:Case4chi2}
\chi_{i}=\arcsin\left(\sqrt{\frac{(r_{2}-r_{i})(r_{1}-r_{3})}{(r_{1}-r_{i})(r_{2}-r_{3})}}\right).
\end{equation}
Note that strictly seen the integral (\ref{eq:Case4int}) is only valid up to the first turning point; however, because of the periodicity of the $\text{sn}$ the solutions (\ref{eq:Case4sol1}) and (\ref{eq:Case4sol2}) are also valid beyond the turning points. As a closing note for this subsection we would like to point out that the cases 1-NUT-a--4 are structurally the same as the cases 2--6 for lightlike geodesics in Frost \cite{Frost2022}.

\subsection{The $\vartheta$ motion}\label{Sec:EoMthetaSol}
Now we turn to the $\vartheta$ motion. From (\ref{eq:EoMtheta}) we can immediately see that for $K=0$ the right-hand side has to vanish. This has as consequence that the right-hand side of (\ref{eq:EoMphi}) has to vanish as well and thus all timelike geodesics with $K=0$ are radial timelike geodesics. As we already noted above the equations of motion for $\vartheta$ and $\varphi$ are the same for lightlike and timelike geodesics. Therefore, the solutions derived in Frost \cite{Frost2022,Frost2022b} for lightlike geodesics can be immediately transferred to timelike geodesics. However, as most readers may not be familiar with them we will briefly summarise the derivations here. \\
We start with transferring the concept of the \emph{individual photon cone}. As for lightlike geodesics timelike geodesics on cones of constant $\vartheta$ have to fulfill the conditions $\text{d}\vartheta/\text{d}\lambda=\text{d}^2\vartheta/\text{d}\lambda^2=0$. When we apply these conditions to (\ref{eq:EoMtheta}) we get from the condition $\text{d}^2\vartheta/\text{d}\lambda^2=0$ for the particle cone
\begin{equation}\label{eq:thetapa}
\vartheta_{\text{pa}}=\arccos\left(-\frac{2nE(2nEC+L_{z})}{K+4n^2E^2}\right).
\end{equation}
Inserting the obtained expression for $\vartheta_{\text{pa}}$ into the condition $\text{d}\vartheta/\text{d}\lambda=0$ we get that for timelike geodesics on the particle cone the constants of motion have to fulfill the relation
\begin{equation}\label{eq:thetapaEoM}
K+4n^2E^2-(2nEC+L_{z})^2=0.
\end{equation}
As for lightlike geodesics these particle cones depend on the constants of motion and thus they are \emph{individual particle cones}. Note that for the Schwarzschild and the Reissner-Nordstr\"{o}m metric ($n=0$) we get $\vartheta_{\text{pa}}=\pi/2$ and $K=L_{z}^2$ which corresponds to timelike geodesics in the equatorial plane. In the following we will now briefly discuss how to solve the equation of motion for $\vartheta$. \\
Although for the Schwarzschild metric and the Reissner-Nordstr\"{o}m metric we can always project the lens equation to the equatorial plane, in Sec.~\ref{Sec:Lensing} we want to compare the lens equations for the NUT metric and the charged NUT metric with the lens equations for the Schwarzschild metric and the Reissner-Nordstr\"{o}m metric. Therefore, in this paper we deviate from this common approach and solve the equations of motion for arbitrary timelike geodesics. In addition, we explicitly distinguish between timelike geodesics in the spherically symmetric and static Schwarzschild and Reissner-Nordstr\"{o}m metrics and the stationary and axisymmetric NUT and charged NUT metrics and discuss them separately.\\
\emph{Case 1: Schwarzschild metric and Reissner-Nordstr\"{o}m metric}.--For the Schwarzschild metric and the Reissner-Nordstr\"{o}m metric (\ref{eq:EoMtheta}) reduces to
\begin{equation}\label{eq:EoMthetaS}
\left(\frac{\text{d}\vartheta}{\text{d}\lambda}\right)^2=K-\frac{L_{z}^2}{\sin^2\vartheta}.
\end{equation}
We can easily read from (\ref{eq:EoMthetaS}) that we have to distinguish three different cases. In the first case we have $K=L_{z}=0$. These are radial timelike geodesics. In the second case we have $\vartheta_{i}=\pi/2$ and $K=L_{z}^2$. These are timelike geodesics in the equatorial plane. In both cases we have $\text{d}\vartheta/\text{d}\lambda=0$ and thus the solution to (\ref{eq:EoMthetaS}) is given by $\vartheta(\lambda)=\vartheta_{i}$. The third case includes all other timelike geodesics. These geodesics oscillate between the two turning points \cite{Frost2022b}
\begin{equation}
\vartheta_{\text{min}}=\arccos\left(\sqrt{1-\frac{L_{z}^2}{K}}\right)~~~\text{and}~~~\vartheta_{\text{max}}=\arccos\left(-\sqrt{1-\frac{L_{z}^2}{K}}\right).
\end{equation}
In this case we first substitute $x=\cos\vartheta$. Then we separate variables and integrate from $x_{i}=\cos\vartheta(\lambda_{i})=\cos\vartheta_{i}$ to $x(\lambda)=\cos\vartheta(\lambda)=\cos\vartheta$. We solve for $\vartheta$ and
get as solution for $\vartheta(\lambda)$ \cite{Frost2022b}
\begin{equation}\label{eq:EoMthetasolS}
\vartheta(\lambda)=\arccos\left(\sqrt{1-\frac{L_{z}^2}{K}}\sin\left(\arcsin\left(\frac{\cos\vartheta_{i}}{\sqrt{1-\frac{L_{z}^2}{K}}}\right)+i_{\vartheta_{i}}\sqrt{K}(\lambda_{i}-\lambda)\right)\right),
\end{equation}
where $i_{\vartheta_{i}}=\text{sgn}(\left.\text{d}\vartheta/\text{d}\lambda\right|_{\vartheta=\vartheta_{i}})$.\\
\emph{Case 2: NUT metric and charged NUT metric}.--For the NUT metric and the charged NUT metric we again have to distinguish three different cases. In the first case we have $K=0$. As discussed above these are radial timelike geodesics. In the second case we have $\vartheta_{i}=\vartheta_{\text{pa}}$ and the constants of motion fulfill (\ref{eq:thetapaEoM}). In both cases we have $\text{d}\vartheta/\text{d}\lambda=0$ and thus the solution to the equation of motion is given by $\vartheta(\lambda)=\vartheta_{i}$. The third case includes all other timelike geodesics. These geodesics oscillate between the turning points \cite{Frost2022,Frost2022b}
\begin{align}
x_{\text{min}}=\cos\vartheta_{\text{min}}=&\frac{\sqrt{K(K+4n^2E^2-(2nEC+L_{z})^2)}-2nE(2nEC+L_{z})}{K+4n^2E^2},\\
x_{\text{max}}=\cos\vartheta_{\text{max}}=&-\frac{\sqrt{K(K+4n^2E^2-(2nEC+L_{z})^2)}+2nE(2nEC+L_{z})}{K+4n^2E^2}.
\end{align}
Again we first substitute $x=\cos\vartheta$. In the next step we separate variables and integrate from $x_{i}=\cos\vartheta(\lambda_{i})=\cos\vartheta_{i}$ to $x(\lambda)=\cos\vartheta(\lambda)=\cos\vartheta$ using the elementary integral
\begin{equation}\label{eq:EMInttheta1}
I_{\vartheta_{1}}=\int\frac{\text{d}x}{\sqrt{b_{1}x^2+b_{2}x+b_{3}}}=-\frac{1}{\sqrt{-b_{1}}}\arcsin\left(\frac{2b_{1}x+b_{2}}{\sqrt{b_{2}^2-4b_{1}b_{3}}}\right),
\end{equation}
where $b_{1}=-(K+4n^2E^2)$, $b_{2}=-4nE(2nEC+L_{z})$, and $b_{3}=K-(2nEC+L_{z})^2$. We solve for $\vartheta$ and obtain as solution for $\vartheta(\lambda)$ \cite{Frost2022,Frost2022b}
\begin{align}\label{eq:EoMthetaNUT}
\vartheta(\lambda)=&\arccos\left(\frac{\sqrt{K(K+4n^2E^2-(2nEC+L_{z})^2)}}{K+4n^2E^2}\sin\Biggl(\arcsin\left(\frac{(K+4n^2E^2)\cos\vartheta_{i}+2nE(2nEC+L_{z})}{\sqrt{K(K+4n^2E^2-(2nEC+L_{z})^2)}}\right)\right.\\
&\left.+i_{\vartheta_{i}}\sqrt{K+4n^2E^2}(\lambda_{i}-\lambda)\Biggr)-\frac{2nE(2nEC+L_{z})}{K+4n^2E^2}\right).\nonumber
\end{align}
Note that also here strictly seen the original integrals we solved to derive (\ref{eq:EoMthetasolS}) and (\ref{eq:EoMthetaNUT}) are only valid up to the first turning point; however, due to the periodicity of the sine the solutions are also valid beyond the first turning point. 

\subsection{The $\varphi$ motion}\label{Sec:EoMphiSol}
As for the $\vartheta$ motion the equations of motion for lightlike and timelike geodesics have the same structure. As a consequence the solutions to the equations of motion for timelike geodesics have the same structure as the solutions for lightlike geodesics derived in Frost \cite{Frost2022,Frost2022b}. Again we will only briefly outline how to derive them.\\
\emph{Case 1: Schwarzschild metric and Reissner-Nordstr\"{o}m metric}.--We start with the Schwarzschild metric and the Reissner-Nordstr\"{o}m metric. In this case we have $n=0$ and the equation of motion for $\varphi$ (\ref{eq:EoMphi}) reduces to
\begin{equation}\label{eq:EoMphiS}
\frac{\text{d}\varphi}{\text{d}\lambda}=\frac{L_{z}}{\sin^2\vartheta}.
\end{equation}
Now we have to distinguish four different cases. In the first case we have $K=L_{z}=0$. These geodesics are radial timelike geodesics and thus we have $\text{d}\varphi/\text{d}\lambda=0$. Therefore, the solution to the equation of motion is given by $\varphi(\lambda)=\varphi_{i}$. In the second case we have $K\neq 0$ and $L_{z}=0$. These are timelike geodesics crossing the axes. Since the Schwarzschild metric and the Reissner-Nordstr\"{o}m metric are both spherically symmetric and static these geodesics form grand circles; however, because we have $\text{d}\varphi/\text{d}\lambda=0$ and the $\varphi$ coordinate has a discontinuity when the geodesics cross the axes (\ref{eq:EoMphiS}) cannot be analytically solved. However, we can define a solution such that $\varphi(\lambda)=\varphi_{i}+n(\lambda)\pi$, where $n(\lambda)$ is the number of axis crossings. The third case covers timelike geodesics with $\vartheta_{i}=\vartheta_{\text{pa}}=\pi/2$ and $K=L_{z}^2$. These are timelike geodesics in the equatorial plane. Since $\vartheta$ is constant the solution is easy to derive and reads
\begin{equation}
\varphi(\lambda)=\varphi_{i}+L_{z}(\lambda-\lambda_{i}).
\end{equation}
The last case includes all remaining geodesics with arbitrary constants of motion $L_{z}$ and $K$. In this case we first replace $\sin^2\vartheta=1-\cos^2\vartheta$ and integrate from $\varphi(\lambda_{i})=\varphi_{i}$ to $\varphi(\lambda)=\varphi$. We insert (\ref{eq:EoMthetasolS}) and evaluate the integral. The obtained solution reads \cite{Frost2022b}
\begin{equation}\label{eq:EoMphisolS}
\varphi(\lambda)=\varphi_{i}+i_{\vartheta_{i}}\left(\arctan\left(\frac{L_{z}\cos\vartheta_{i}}{\sqrt{K-L_{z}^2-K\cos^2\vartheta_{i}}}\right)-\arctan\left(\frac{L_{z}}{\sqrt{K}}\tan\left(a_{\varphi}+i_{\vartheta_{i}}\sqrt{K}\left(\lambda_{i}-\lambda\right)\right)\right)\right),
\end{equation}
where the coefficient $a_{\varphi}$ is given by
\begin{equation}
a_{\varphi}=\arcsin\left(\frac{\cos\vartheta_{i}}{\sqrt{1-\frac{L_{z}^2}{K}}}\right).
\end{equation}
Note that for the explicit evaluation of (\ref{eq:EoMphisolS}) the multivaluedness of the $\arctan$ has to be taken into account.\\
\emph{Case 2: NUT metric and charged NUT metric}.--For the NUT metric and the charged NUT metric we only have to distinguish three different cases. In the first case we have $K=0$. These are radial timelike geodesics. In this case we have $\text{d}\varphi/\text{d}\lambda=0$. Thus the solution reads $\varphi(\lambda)=\varphi_{i}$. In the second case we have $\vartheta_{i}=\vartheta_{\text{pa}}$ and $K+4n^2E^2-(2nEC+L_{z})^2=0$. These are timelike geodesics on \emph{individual particle cones}. The right-hand side of (\ref{eq:EoMphi}) is constant. A simple integration gives for the solution $\varphi(\lambda)$ 
\begin{equation}
\varphi(\lambda)=\varphi_{i}+\frac{L_{z}+2n(\cos\vartheta_{\text{pa}}+C)E}{\sin^2\vartheta_{\text{pa}}}(\lambda-\lambda_{i}).
\end{equation}
The third case includes all other timelike geodesics with arbitrary constants of motion. In this case we first rewrite the right-hand side of (\ref{eq:EoMphi}) in terms of $\cos\vartheta$. Then we perform a partial fraction decomposition and rewrite the result such that only terms with $\cos\vartheta$ in the denominator remain. We insert (\ref{eq:EoMthetaNUT}) and integrate from $\varphi(\lambda_{i})=\varphi_{i}$ to $\varphi(\lambda)=\varphi$. Here, we rewrite the integrals over $\lambda$ such that they have the form
\begin{equation}\label{eq:EMIntPhi}
\int\frac{\text{d}\tilde{\lambda}}{1+a\sin\tilde{\lambda}}=\frac{2}{\sqrt{1-a^2}}\arctan\left(\frac{\tan\left(\frac{\tilde{\lambda}}{2}\right)+a}{\sqrt{1-a^2}}\right).
\end{equation}
We integrate and rearrange all terms. We then obtain as solution for $\varphi(\lambda)$ \cite{Frost2022,Frost2022b}
\begin{align}\label{eq:EoMphisolNUT}
\varphi(\lambda)=&\varphi_{i}+i_{\vartheta_{i}}\left(\arctan\left(\frac{c_{\vartheta_{1}}}{c_{\vartheta_{2}}c_{\vartheta_{5}}}\left(\tan\left(\frac{\tilde{\lambda}(\lambda_{i})}{2}\right)-\frac{c_{\vartheta_{6}}}{c_{\vartheta_{1}}}\right)\right)-\arctan\left(\frac{c_{\vartheta_{1}}}{c_{\vartheta_{2}}c_{\vartheta_{5}}}\left(\tan\left(\frac{\tilde{\lambda}(\lambda)}{2}\right)-\frac{c_{\vartheta_{6}}}{c_{\vartheta_{1}}}\right)\right)\right.\\
&\left.+\arctan\left(\frac{c_{\vartheta_{3}}}{c_{\vartheta_{4}}c_{\vartheta_{5}}}\left(\tan\left(\frac{\tilde{\lambda}(\lambda)}{2}\right)+\frac{c_{\vartheta_{6}}}{c_{\vartheta_{3}}}\right)\right)-\arctan\left(\frac{c_{\vartheta_{3}}}{c_{\vartheta_{4}}c_{\vartheta_{5}}}\left(\tan\left(\frac{\tilde{\lambda}(\lambda_{i})}{2}\right)+\frac{c_{\vartheta_{6}}}{c_{\vartheta_{3}}}\right)\right)\right),\nonumber
\end{align}
where the six coefficients $c_{\vartheta_{1}}$, $c_{\vartheta_{2}}$, $c_{\vartheta_{3}}$, $c_{\vartheta_{4}}$, $c_{\vartheta_{5}}$, and $c_{\vartheta_{6}}$ read
\begin{align}
c_{\vartheta_{1}}=K+4n^2E^2+2nE(2nEC+L_{z}),~~~c_{\vartheta_{2}}=2nE(1+C)+L_{z},\label{eq:Coeffphi}\\
c_{\vartheta_{3}}=K+4n^2E^2-2nE(2nEC+L_{z}),~~~c_{\vartheta_{4}}=2nE(1-C)-L_{z},\label{eq:Coeffphi2}\\
c_{\vartheta_{5}}=\sqrt{K+4n^2E^2},~~~c_{\vartheta_{6}}=\sqrt{K(K+4n^2E^2-(2nEC+L_{z})^2)}.\label{eq:Coeffphi3}
\end{align}
The quantity $\tilde{\lambda}(\lambda')$ is given by
\begin{equation}\label{eq:Coeffphilambdatilde}
\tilde{\lambda}(\lambda')=\arcsin\left(\frac{(K+4n^2E^2)\cos\vartheta_{i}+2nE(2nEC+L_{z})}{\sqrt{K(K+4n^2E^2-(2nEC+L_{z})^2)}}\right)+i_{\vartheta_{i}}\sqrt{K+4n^2E^2}(\lambda_{i}-\lambda').
\end{equation}
Note that also in this case for the explicit evaluation of (\ref{eq:EoMphisolNUT}) the multivaluedness of the $\arctan$ has to be taken into account.

\subsection{The time coordinate $t$}\label{Sec:EoMSolt}
Now we turn to the time coordinate $t$. The right-hand side of (\ref{eq:EoMt}) has an $r$-dependent and a $\vartheta$-dependent part. In the first step we integrate from $t(\lambda_{i})=t_{i}$ to $t(\lambda)=t$. Now the integral reads
\begin{equation}\label{eq:timecoordinate}
t(\lambda)=t_{i}+t_{r}(\lambda)+t_{\vartheta}(\lambda),
\end{equation}
where $t_{r}(\lambda)$ and $t_{\vartheta}(\lambda)$ are given, respectively, by
\begin{align}
t_{r}(\lambda)=\int_{\lambda_{i}}^{\lambda}\frac{\rho(r(\lambda'))^2E\text{d}\lambda'}{Q(r(\lambda'))},\label{eq:tr1}\\
t_{\vartheta}(\lambda)=-2n\int_{\lambda_{i}}^{\lambda}(\cos\vartheta(\lambda')+C)\frac{L_{z}+2n(\cos\vartheta(\lambda')+C)E}{\sin^2\vartheta(\lambda')}\text{d}\lambda'.\label{eq:ttheta1}
\end{align}
Note that these equations are structurally identical to those for lightlike geodesics. We will now briefly outline how to evaluate them.

\subsubsection{Evaluating $t_{\vartheta}(\lambda)$}
We start with evaluating $t_{\vartheta}(\lambda)$. For the Schwarzschild metric and the Reissner-Nordstr\"{o}m metric we have $n=0$ and the right-hand side vanishes. Thus in this case we always have $t_{\vartheta}(\lambda)=0$. For the NUT metric and the charged NUT metric we have to distinguish the same three cases as for the $\varphi$ motion. In the first case we have $K=0$ (radial timelike geodesics) and the right-hand side of (\ref{eq:ttheta1}) vanishes. Thus in this case we have $t_{\vartheta}(\lambda)=0$. In the second case we have $\vartheta_{i}=\vartheta_{\text{pa}}$ and $K+4n^2E^2-(2nEC+L_{z})^2=0$. These are timelike geodesics on \emph{individual particle cones}. In this case the right-hand side of (\ref{eq:ttheta1}) is constant. A simple integration gives for $t_{\vartheta}(\lambda)$
\begin{equation}
t_{\vartheta}(\lambda)=-2n(\cos\vartheta_{\text{pa}}+C)\frac{L_{z}+2n(\cos\vartheta_{\text{pa}}+C)E}{\sin^2\vartheta_{\text{pa}}}(\lambda-\lambda_{i}).
\end{equation}
The third case covers all remaining timelike geodesics with arbitrary constants of motion. In this case we first rewrite the integrand in terms of $\cos\vartheta$ and perform a partial fraction decomposition such that only a constant term and two terms with $\cos\vartheta$ in the denominator remain. We insert the solution for $\vartheta(\lambda)$ given by (\ref{eq:EoMthetaNUT}) and evaluate the obtained integrals using (\ref{eq:EMIntPhi}). The final result reads \cite{Frost2022,Frost2022b}
\begin{align}\label{eq:EoMtthetasol}
t_{\vartheta}(\lambda)=&4n^2E(\lambda-\lambda_{i})\\
&+i_{\vartheta_{i}}2n\left((1+C)\left(\arctan\left(\frac{c_{\vartheta_{1}}}{c_{\vartheta_{2}}c_{\vartheta_{5}}}\left(\tan\left(\frac{\tilde{\lambda}(\lambda)}{2}\right)-\frac{c_{\vartheta_{6}}}{c_{\vartheta_{1}}}\right)\right)-\arctan\left(\frac{c_{\vartheta_{1}}}{c_{\vartheta_{2}}c_{\vartheta_{5}}}\left(\tan\left(\frac{\tilde{\lambda}(\lambda_{i})}{2}\right)-\frac{c_{\vartheta_{6}}}{c_{\vartheta_{1}}}\right)\right)\right)\right.\nonumber\\
&\left.+(1-C)\left(\arctan\left(\frac{c_{\vartheta_{3}}}{c_{\vartheta_{4}}c_{\vartheta_{5}}}\left(\tan\left(\frac{\tilde{\lambda}(\lambda)}{2}\right)+\frac{c_{\vartheta_{6}}}{c_{\vartheta_{3}}}\right)\right)-\arctan\left(\frac{c_{\vartheta_{3}}}{c_{\vartheta_{4}}c_{\vartheta_{5}}}\left(\tan\left(\frac{\tilde{\lambda}(\lambda_{i})}{2}\right)+\frac{c_{\vartheta_{6}}}{c_{\vartheta_{3}}}\right)\right)\right)\right),\nonumber
\end{align}
where the six coefficients $c_{\vartheta_{1}}$--$c_{\vartheta_{6}}$ and $\tilde{\lambda}(\lambda')$ are given by (\ref{eq:Coeffphi})--(\ref{eq:Coeffphi3}) and (\ref{eq:Coeffphilambdatilde}), respectively. Note that again for the explicit evaluation of (\ref{eq:EoMtthetasol}) the multivaluedness of the $\arctan$ has to be taken into account.

\subsubsection{Evaluating $t_{r}(\lambda)$}
Now we turn to $t_{r}(\lambda)$. Here, we first rewrite (\ref{eq:tr1}) using the root of (\ref{eq:EoMr}) and obtain
\begin{equation}\label{eq:tr2}
t_{r}(\lambda)=\int_{r_{i}...}^{...r(\lambda)}\frac{\rho(r')^2E\text{d}r'}{Q(r')\sqrt{\rho(r')^2E^2-\rho(r')Q(r')-Q(r')K}},
\end{equation}
where the sign of the root in the denominator has to be chosen in agreement with the direction of the $r$ motion and the dots in the limits shall indicate that we have to split the integral at potential turning points. For the evaluation of this integral we have to consider the same seven different types of motion as for $r$. Note that due to the large number of cases and the length of the resulting terms in the following we will only explicitly present a few selected solutions, namely those for radial timelike geodesics in the Schwarzschild and Reissner-Nordstr\"{o}m spacetimes and those for timelike geodesics on a particle sphere. In all other cases we will only briefly outline the steps necessary for the derivation.\\
\emph{Case 1-S}.--In this case we have radial timelike geodesics in the Schwarzschild metric. These geodesics have $K=0$ and thus (\ref{eq:tr2}) reduces to
\begin{equation}\label{eq:tr2S}
t_{r}(\lambda)=i_{r_{i}}\int_{r_{i}}^{r(\lambda)}\frac{r'^3E\text{d}r'}{Q(r')\sqrt{r'^2E^2-Q(r')}}.
\end{equation}
Now we rewrite the factor $r'^3/Q(r')$ in (\ref{eq:tr2S}) such that only terms with $r'$ in the nominator or the denominator remain. The result contains three elementary integrals of the form of the integrals $I_{1}$, $I_{2}$, and $I_{4}$ given by (\ref{eq:I1}), (\ref{eq:I2}), and (\ref{eq:I4}) in Appendix~\ref{App:EmInt1}. We integrate and obtain as result for $t_{r}(\lambda)$
\begin{align}
t_{r}(\lambda)=&i_{r_{i}}E\left(\frac{\sqrt{r(\lambda)((E^2-1)r(\lambda)+2m)}-\sqrt{r_{i}((E^2-1)r_{i}+2m)}}{E^2-1}\right.\\
&\left.+2m\frac{3-2E^2}{(E^2-1)^{\frac{3}{2}}}\ln\left(\frac{\sqrt{(E^2-1)r_{i}}+\sqrt{(E^2-1)r_{i}+2m}}{\sqrt{(E^2-1)r(\lambda)}+\sqrt{(E^2-1)r(\lambda)+2m}}\right)\right.\nonumber\\
&\left.+\frac{4m}{E}\left(\text{arcoth}\left(E\sqrt{\frac{r_{i}}{(E^2-1)r_{i}+2m}}\right)-\text{arcoth}\left(E\sqrt{\frac{r(\lambda)}{(E^2-1)r(\lambda)+2m}}\right)\right)\right).\nonumber
\end{align}
\emph{Case 1-RN}.--In this case we have radial timelike geodesics in the Reissner-Nordstr\"{o}m metric. Again these geodesics have $K=0$ and (\ref{eq:tr2}) reduces to (\ref{eq:tr2S}). Now, using a partial fraction decomposition if necessary, we rewrite the factor $r'^3/Q(r')$ in (\ref{eq:tr2S}) such that only terms with $r'$ in the nominator or the denominator remain. For $0<e<m$ the resulting expression contains three elementary integrals of the form of $I_{5}$, $I_{6}$, and $I_{8}$ given by (\ref{eq:I5}), (\ref{eq:I6}), and (\ref{eq:I8}) in Appendix~\ref{App:EmInt1}. We integrate and obtain for $t_{r}(\lambda)$
\begin{align}
&t_{r}(\lambda)=i_{r_{i}}E\left(\frac{\sqrt{P(r(\lambda))}-\sqrt{P(r_{i})}}{E^2-1}+\frac{(E^2-1)(r_{\text{H}_{\text{o}}}+r_{\text{H}_{\text{i}}})-m}{(E^2-1)^{\frac{3}{2}}}\ln\left(\frac{\sqrt{(E^2-1)P(r(\lambda))}+(E^2-1)r(\lambda)+m}{\sqrt{(E^2-1)P(r_{i})}+(E^2-1)r_{i}+m}\right)\right.\\
&\left.+\frac{r_{\text{H}_{\text{o}}}^3}{(r_{\text{H}_{\text{o}}}-r_{\text{H}_{\text{i}}})\sqrt{P(r_{\text{H}_{\text{o}}})}}\left(\ln\left(\frac{r(\lambda)-r_{\text{H}_{\text{o}}}}{r_{i}-r_{\text{H}_{\text{o}}}}\right)+\ln\left(\frac{P(r_{\text{H}_{\text{o}}})+((E^2-1)r_{\text{H}_{\text{o}}}+m)(r_{i}-r_{\text{H}_{\text{o}}})+\sqrt{P(r_{\text{H}_{\text{o}}})P(r_{i})}}{P(r_{\text{H}_{\text{o}}})+((E^2-1)r_{\text{H}_{\text{o}}}+m)(r(\lambda)-r_{\text{H}_{\text{o}}})+\sqrt{P(r_{\text{H}_{\text{o}}})P(r(\lambda))}}\right)\right)\right.\nonumber\\
&\left.+\frac{r_{\text{H}_{\text{i}}}^3}{(r_{\text{H}_{\text{o}}}-r_{\text{H}_{\text{i}}})\sqrt{P(r_{\text{H}_{\text{i}}})}}\left(\ln\left(\frac{r_{i}-r_{\text{H}_{\text{i}}}}{r(\lambda)-r_{\text{H}_{\text{i}}}}\right)+\ln\left(\frac{P(r_{\text{H}_{\text{i}}})+((E^2-1)r_{\text{H}_{\text{i}}}+m)(r(\lambda)-r_{\text{H}_{\text{i}}})+\sqrt{P(r_{\text{H}_{\text{i}}})P(r(\lambda))}}{P(r_{\text{H}_{\text{i}}})+((E^2-1)r_{\text{H}_{\text{i}}}+m)(r_{i}-r_{\text{H}_{\text{i}}})+\sqrt{P(r_{\text{H}_{\text{i}}})P(r_{i})}}\right)\right)\right),\nonumber
\end{align}
where we defined
\begin{equation}\label{eq:CoeffPr}
P(r')=(E^2-1)r'^2+2mr'-e^2.
\end{equation}
For $e=m$ we proceed analogously. In this case the right-hand side of (\ref{eq:tr2S}) contains four elementary integrals of the form of $I_{5}$, $I_{6}$, $I_{8}$, and $I_{9}$ given by (\ref{eq:I5}), (\ref{eq:I6}), (\ref{eq:I8}), and (\ref{eq:I9}) in Appendix~\ref{App:EmInt1}. Again we integrate and obtain as result for $t_{r}(\lambda)$
\begin{align}
t_{r}(\lambda)=&i_{r_{i}}E\left(\frac{\sqrt{P(r(\lambda))}-\sqrt{P(r_{i})}}{E^2-1}+\frac{r_{\text{H}}^3}{P(r_{\text{H}})}\left(\frac{\sqrt{P(r_{i})}}{r_{i}-r_{\text{H}}}-\frac{\sqrt{P(r(\lambda))}}{r(\lambda)-r_{\text{H}}}\right)\right.\\
&\left.+\frac{2r_{\text{H}}(E^2-1)-m}{(E^2-1)^{\frac{3}{2}}}\ln\left(\frac{\sqrt{(E^2-1)P(r(\lambda))}+(E^2-1)r(\lambda)+m}{\sqrt{(E^2-1)P(r_{i})}+(E^2-1)r_{i}+m}\right)\right.\nonumber\\
&\left.+\frac{3r_{\text{H}}^2P(r_{\text{H}})-r_{\text{H}}^3((E^2-1)r_{\text{H}}+m)}{P(r_{\text{H}})^{\frac{3}{2}}}\Biggl(\ln\left(\frac{r(\lambda)-r_{\text{H}}}{r_{i}-r_{\text{H}}}\right)\right.\nonumber\\
&\left.+\ln\left(\frac{P(r_{\text{H}})+((E^2-1)r_{\text{H}}+m)(r_{i}-r_{\text{H}})+\sqrt{P(r_{\text{H}})P(r_{i})}}{P(r_{\text{H}})+((E^2-1)r_{\text{H}}+m)(r(\lambda)-r_{\text{H}})+\sqrt{P(r_{\text{H}})P(r(\lambda))}}\right)\Biggr)\right).\nonumber
\end{align}
\emph{Case 1-NUT-a}.--In this case we have timelike geodesics with $\Delta<0$ and $0\leq K< V_{E}(r_{\text{pa}_{-}})$ in the NUT metric and the charged NUT metric. We recall that in this case the right-hand side of (\ref{eq:EoMr}) has two pairs of distinct complex conjugate roots. For $0\leq e<\sqrt{m^2+n^2}$ we first use a partial fraction decomposition and rewrite $\rho(r')^2/Q(r')$ such that only terms with $r'$ in the nominator or the denominator remain. Then we substitute using (\ref{eq:Case1NUTasub}) and rewrite the obtained expression in terms of Legendre's elliptic integral of the first kind and the two nonstandard elliptic integrals given by $J_{L_{1}}(\chi_{i},\chi(\lambda),k_{1},\tilde{n})$ and $J_{L_{2}}(\chi_{i},\chi(\lambda),k_{1},\tilde{n})$ given by (\ref{eq:J1}) and (\ref{eq:J2}) in Appendix~\ref{App:EllInt}. Now we use (\ref{eq:J1R})--(\ref{eq:JL}) to rewrite (\ref{eq:tr2}) in terms of elementary functions and Legendre's elliptic integrals of the first, second, and third kind. For $e=\sqrt{m^2+n^2}$ we proceed analogously; however, we have to explicitly distinguish the case $K=0$ from all other cases. In this case we have $r_{\text{H}}=m=R_{2}/g_{0}+R_{1}$ and thus we rewrite (\ref{eq:tr2}) in terms of Legendre's elliptic integral of the first kind and the four nonstandard (pseudo)elliptic integrals $J_{L_{1}}(\chi_{i},\chi(\lambda),k_{1},\tilde{n})$, $J_{L_{2}}(\chi_{i},\chi(\lambda),k_{1},\tilde{n})$, $J_{L_{3}}(\chi_{i},\chi(\lambda),k_{1})$, and $J_{L_{4}}(\chi_{i},\chi(\lambda),k_{1})$ given by (\ref{eq:J1})--(\ref{eq:J4}) in Appendix~\ref{App:EllInt}. In this case we use Eqs.~(\ref{eq:J1R})--(\ref{eq:J4R}) to rewrite (\ref{eq:tr2}) in terms of elementary functions and Legendre's elliptic integrals of the first, second, and third kind. For $0<K<V_{E}(r_{\text{pa}_{-}})$, with the exception of the partial fraction decomposition, the integration procedure is the same as before.\\
\emph{Case 1-NUT-b}.--In this case we have timelike geodesics with $\Delta\leq0$ and $K=V_{E}(r_{\text{pa}_{-}})$ in the NUT metric and the charged NUT metric. We recall that in this case the right-hand side of (\ref{eq:EoMr}) has a real double root at $r_{1}=r_{2}=r_{\text{pa}_{-}}$ and a pair of complex conjugate roots. We again use a partial fraction decomposition and rewrite $\rho(r')^2/Q(r')$ such that only terms with $r'$ in the nominator or the denominator remain. For $0\leq e<\sqrt{m^2+n^2}$ we rewrite the resulting integrals in terms of the elementary integrals $I_{10}$, $I_{11}$, and $I_{12}$ given by (\ref{eq:I10})--(\ref{eq:I12}) in Appendix~\ref{App:EmInt2}. We integrate and insert the results to rewrite $t_{r}(\lambda)$ in terms of elementary functions. For $e=\sqrt{m^2+n^2}$ we follow the same procedure. The only difference is that this time we use $I_{10}$, $I_{11}$, $I_{12}$, and $I_{13}$ given by (\ref{eq:I10})--(\ref{eq:I13}) in Appendix~\ref{App:EmInt2} to rewrite $t_{r}(\lambda)$ in terms of elementary functions.\\
\emph{Case 2}.--In this case we either have timelike geodesics with $0<K<V_{E}(r_{\text{pa}})$ in the Schwarzschild metric or the Reissner-Nordstr\"{o}m metric or timelike geodesics with $0<\Delta$ and $0\leq K<V_{E}(r_{\text{pa}})$ or $\Delta\leq 0$ and $V_{E}(r_{\text{pa}_{-}})<K<V_{E}(r_{\text{pa}})$ in the NUT metric or the charged NUT metric. We recall that in this case the right-hand side of (\ref{eq:EoMr}) has two distinct real roots and a pair of complex conjugate roots. In this case the $r$ motion does not have a turning point in the domain of outer communication. We again rewrite $\rho(r')^2/Q(r')$ such that only terms with $r'$ in the nominator or the denominator remain. Now we substitute using (\ref{eq:Case2sub}). For $0\leq e<\sqrt{m^2+n^2}$ we rewrite the obtained expressions in terms of Legendre's elliptic integral of the first kind and the two nonstandard elliptic integrals $J_{L_{5}}(\chi_{i},\chi(\lambda),k_{2},\tilde{n})$ and $J_{L_{6}}(\chi_{i},\chi(\lambda),k_{2},\tilde{n})$ given by (\ref{eq:J5}) and (\ref{eq:J6}) in Appendix~\ref{App:EllInt}. We use (\ref{eq:J5R}), (\ref{eq:J6R}), and (\ref{eq:JTL}) to rewrite $t_{r}(\lambda)$ in terms of elementary functions and Legendre's elliptic integrals of the first, second, and third kind. Note that again we have to consider a special case. When $e=\sqrt{m^2+n^2}$ and $K=0$ we have $r_{\text{H}}=m=(r_{1}\bar{R}-r_{2}R)/(\bar{R}-R)$. In this case we rewrite the right-hand side of (\ref{eq:tr2}) in terms of Legendre's elliptic integral of the first kind and the four nonstandard (pseudo)elliptic integrals $J_{L_{5}}(\chi_{i},\chi(\lambda),k_{2},\tilde{n})$, $J_{L_{6}}(\chi_{i},\chi(\lambda),k_{2},\tilde{n})$, $J_{L_{7}}(\chi_{i},\chi(\lambda),k_{2})$, and $J_{L_{8}}(\chi_{i},\chi(\lambda),k_{2})$ given by (\ref{eq:J5})--(\ref{eq:J8}) in Appendix~\ref{App:EllInt}. We now use (\ref{eq:J5R}) and (\ref{eq:J6R})--(\ref{eq:J8R}) to rewrite $t_{r}(\lambda)$ in terms of elementary functions and Legendre's elliptic integrals of the first, second, and third kind. In all other cases with $e=\sqrt{m^2+n^2}$ the integration procedure is the same as before.\\
\emph{Case 3}.--In this case we have timelike geodesics with $K=V_{E}(r_{\text{pa}})$. We recall that in this case the right-hand side of (\ref{eq:EoMr}) has four real roots. Two of these roots form a double root at $r_{1}=r_{2}=r_{\text{pa}}$. These are timelike geodesics on or asymptotically coming from or going to an unstable particle sphere. They do not have turning points in the domain of outer communication. In the case that we have a timelike geodesic on an unstable particle sphere we have $r(\lambda)=r_{\text{pa}}$ and thus the integrand on the right-hand side of (\ref{eq:tr1}) is constant. We integrate and get
\begin{equation}
t_{r}(\lambda)=\frac{\rho(r_{\text{pa}})^2E}{Q(r_{\text{pa}})}(\lambda-\lambda_{i}).
\end{equation}
In all other cases we again rewrite $\rho(r')^2/Q(r')$ in (\ref{eq:tr2}) such that only terms with $r'$ in the nominator or the denominator remain. In the next step we substitute using (\ref{eq:Case3sub}). We rewrite the right-hand side of (\ref{eq:tr2}) in terms of the two elementary integrals given by $I_{14}$ (\ref{eq:I14y}) and $I_{15}$ (\ref{eq:I15y}) in Appendix~\ref{App:EmInt3}. Now we rewrite them in terms of elementary functions using $I_{14_{1}}$, $I_{14_{2}}$ (for the term with $r_{a}=r_{\text{pa}}$ and only for $r_{\text{H}_{\text{o}}}<r<r_{\text{pa}}$), $I_{14_{3}}$ (for the term with $r_{a}=r_{\text{pa}}$ only for $r_{\text{pa}}<r$), $I_{15_{1}}$, and $I_{15_{2}}$ given by (\ref{eq:I141})--(\ref{eq:I152}) in Appendix~\ref{App:EmInt3} and obtain $t_{r}(\lambda)$.\\
\emph{Case 4}.--In this case we have timelike geodesics with $V_{E}(r_{\text{pa}})<K$. We recall that in this case the right-hand side of (\ref{eq:EoMr}) has four distinct real roots. For $r_{\text{H}_{\text{o}}}<r<r_{\text{pa}}$ these timelike geodesics can have a turning point at $r_{2}=r_{\text{max}}$ (maximum). For $r_{\text{pa}}<r$ these timelike geodesics can have a turning point at $r_{1}=r_{\text{min}}$ (minimum). One last time we rewrite $\rho(r')^2/Q(r')$ such that only terms with $r'$ in the nominator or the denominator remain. For timelike geodesics with $r_{\text{H}_{\text{o}}}<r<r_{\text{pa}}$ we now substitute using (\ref{eq:Case4sub2}). For timelike geodesics with $r_{\text{pa}}<r$ we substitute using (\ref{eq:Case4sub1}). We rewrite the right-hand side of (\ref{eq:tr2}) in terms of Legendre's elliptic integrals of the first and third kind and the nonstandard elliptic integral $J_{L_{9}}(\chi_{i},\chi(\lambda),k_{3},\tilde{n})$ given by (\ref{eq:J9}) in Appendix~\ref{App:EllInt}. We now use (\ref{eq:J9R}) from Appendix~\ref{App:EllInt} to rewrite $t_{r}(\lambda)$ in terms of elementary functions and Legendre's elliptic integrals of the first, second, and third kind. Note that for timelike geodesics passing through a turning point we have to split the integral at the turning point into two terms and for the second term we have to change $i_{r_{i}}$ to $-i_{r_{i}}$.

\subsection{The proper time $\tau$}\label{Sec:EoMtau}
The last quantity related to the equations of motion we have to discuss is the proper time. While for lightlike geodesics the affine parameter $s$ is not a physically measurable quantity, this is different for the proper time $\tau$. It measures the time a clock moving with a particle along a timelike geodesic would show and thus is another time measure which in general differs from the time coordinate $t$. We can easily derive it from (\ref{eq:Mino}). We write $r$ as function of the Mino parameter $\lambda$ and separate variables. We integrate from $\tau(\lambda_{i})=\tau_{i}$ to $\tau(\lambda)=\tau$. Then we use the root of (\ref{eq:EoMr}) to rewrite the integral over $\lambda$ as integral over $r$ (note that we can also integrate using two fixed limits for $r$; however, in the most general case, which we consider here, the upper limit still depends on $\lambda$ and thus the result for $\tau$ is also a function of $\lambda$). The result reads
\begin{equation}\label{eq:PTint}
\tau(\lambda)=\tau_{i}+\int_{r_{i}...}^{...r(\lambda)}\frac{\rho(r')\text{d}r'}{\sqrt{\rho(r')^2E^2-\rho(r')Q(r')-Q(r')K}}.
\end{equation}
Again we have to choose the sign of the root in the denominator such that it agrees with the direction of the $r$ motion and the dots in the limits shall indicate that we have to split the integral at potential turning points. Again we have to distinguish the same seven different types of motion as for $r$. Again we will only explicitly present the solutions for radial timelike geodesics in the Schwarzschild metric and the Reissner-Nordstr\"{o}m metric and for timelike geodesics on a particle sphere while in all other cases we will only outline the steps of their derivation.\\
\emph{Case 1-S}.--Again we start with radial timelike geodesics in the Schwarzschild metric. These geodesics have $K=0$ and thus (\ref{eq:PTint}) reduces to
\begin{equation}\label{eq:PTCase1Sint}
\tau(\lambda)=\tau_{i}+i_{r_{i}}\int_{r_{i}}^{r(\lambda)}\frac{r'\text{d}r'}{\sqrt{(E^2-1)r'^2+2mr'}}.
\end{equation}
We notice that the integral on the right-hand side has the form of $I_{1}$ given by (\ref{eq:I1}) in Appendix~\ref{App:EmInt1}. We integrate and obtain for $\tau(\lambda)$
\begin{align}\label{eq:PTCase1Ssol}
\tau(\lambda)=\tau_{i}+i_{r_{i}}\left(\frac{\sqrt{r(\lambda)((E^2-1)r(\lambda)+2m)}-\sqrt{r_{i}((E^2-1)r_{i}+2m)}}{E^2-1}\right.\\
\left.+\frac{2m}{(E^2-1)^{\frac{3}{2}}}\ln\left(\frac{\sqrt{(E^2-1)r_{i}}+\sqrt{(E^2-1)r_{i}+2m}}{\sqrt{(E^2-1)r(\lambda)}+\sqrt{(E^2-1)r(\lambda)+2m}}\right)\right).\nonumber
\end{align}
\emph{Case 1-RN}.--We have radial timelike geodesics in the Reissner-Nordstr\"{o}m metric. Again we have $K=0$. This time (\ref{eq:PTint}) reduces to
\begin{equation}\label{eq:PTCase1RNint}
\tau(\lambda)=\tau_{i}+i_{r_{i}}\int_{r_{i}}^{r(\lambda)}\frac{r'\text{d}r'}{\sqrt{(E^2-1)r'^2+2mr'-e^2}}.
\end{equation}
The integral over $r'$ has the same form as $I_{5}$ given by (\ref{eq:I5}) in Appendix~\ref{App:EmInt1}. We use $I_{5}$ to integrate the right-hand side of (\ref{eq:PTCase1RNint}) and get
for $\tau(\lambda)$
\begin{align}\label{eq:PTCase1RNsol}
\tau(\lambda)=\tau_{i}+i_{r_{i}}\left(\frac{\sqrt{P(r(\lambda))}-\sqrt{P(r_{i})}}{E^2-1}+\frac{m}{(E^2-1)^{\frac{3}{2}}}\ln\left(\frac{\sqrt{(E^2-1)P(r_{i})}+(E^2-1)r_{i}+m}{\sqrt{(E^2-1)P(r(\lambda))}+(E^2-1)r(\lambda)+m}\right)\right),
\end{align}
where $P(r(\lambda))$ and $P(r_{i})$ are given by (\ref{eq:CoeffPr}).\\
\emph{Case 1-NUT-a}.--In this case we have timelike geodesics with $\Delta<0$ and $0\leq K<V_{E}(r_{\text{pa}_{-}})$ in the NUT metric and the charged NUT metric. We recall that in this case the right-hand side of (\ref{eq:EoMr}) has two distinct pairs of complex conjugate roots and no turning points in the domain of outer communication. We substitute using (\ref{eq:Case1NUTasub}) and rewrite the obtained expression in terms of Legendre's elliptic integral of the first kind and the two nonstandard elliptic integrals $J_{L_{1}}(\chi_{i},\chi(\lambda),k_{1},\tilde{n})$ and $J_{L_{2}}(\chi_{i},\chi(\lambda),k_{1},\tilde{n})$ given by (\ref{eq:J1}) and (\ref{eq:J2}) in Appendix~\ref{App:EllInt}. Then we use (\ref{eq:J1R})--(\ref{eq:JL}) to rewrite (\ref{eq:PTint}) in terms of elementary functions and Legendre's elliptic integrals of the first, second, and third kind.\\
\emph{Case 1-NUT-b}.--In this case we have timelike geodesics with $\Delta\leq0$ and $K=V_{E}(r_{\text{pa}_{-}})$ in the NUT metric and the charged NUT metric. We recall that in this case the right-hand side of (\ref{eq:EoMr}) has a real double root at $r_{1}=r_{2}=r_{\text{pa}_{-}}$ and two complex conjugate roots. We rewrite the right-hand side of (\ref{eq:PTint}) in terms of the integrals $I_{10}$, $I_{11}$, and $I_{12}$ given by (\ref{eq:I10})--(\ref{eq:I12}) in Appendix~\ref{App:EmInt2} and integrate. We use the obtained expressions to write down $\tau(\lambda)$ in terms of elementary functions.\\
\emph{Case 2}.--In this case we either have timelike geodesics with $0<K<V_{E}(r_{\text{pa}})$ in the Schwarzschild metric or the Reissner-Nordstr\"{o}m metric or timelike geodesics with $0<\Delta$ and $0\leq K<V_{E}(r_{\text{pa}})$ or $\Delta\leq 0$ and $V_{E}(r_{\text{pa}_{-}})<K<V_{E}(r_{\text{pa}})$ in the NUT metric or the charged NUT metric. We recall that in this case the right-hand side of (\ref{eq:EoMr}) has two distinct real roots and a pair of complex conjugate roots but no turning points in the domain of outer communication. We substitute using (\ref{eq:Case2sub}) and rewrite the obtained expression in terms of Legendre's elliptic integral of the first kind and the two nonstandard elliptic integrals $J_{L_{5}}(\chi_{i},\chi(\lambda),k_{2},\tilde{n})$ and $J_{L_{6}}(\chi_{i},\chi(\lambda),k_{2},\tilde{n})$ given by (\ref{eq:J5}) and (\ref{eq:J6}) in Appendix~\ref{App:EllInt}. Now we use (\ref{eq:J5R}), (\ref{eq:J6R}), and (\ref{eq:JTL}) to rewrite (\ref{eq:PTint}) in terms of elementary functions and Legendre's elliptic integrals of the first, second, and third kind.\\
\emph{Case 3}.--In this case we have timelike geodesics with $K=V_{E}(r_{\text{pa}})$. We recall that in this case the right-hand side of (\ref{eq:EoMr}) has four real roots, two of which form a double root at $r_{1}=r_{2}=r_{\text{pa}}$. These are timelike geodesics on or asymptotically coming from or going to an unstable particle sphere. When we have a timelike geodesic on an unstable particle sphere we have $r(\lambda)=r_{\text{pa}}$ and thus the right-hand side of (\ref{eq:Mino}) is constant. In this case we deviate from the standard procedure and directly integrate over $\lambda$ and get
\begin{equation}
\tau(\lambda)=\tau_{i}+\rho(r_{\text{pa}})(\lambda-\lambda_{i}).
\end{equation}
In all other cases we substitute using (\ref{eq:Case3sub}) and rewrite the obtained expression in terms of the elementary integrals $I_{14}$ and $I_{15}$ given by (\ref{eq:I14y}) and (\ref{eq:I15y}) in Appendix~\ref{App:EmInt3}. We rewrite the right-hand side of (\ref{eq:PTint}) in terms of the expressions $I_{14_{1}}$, $I_{14_{2}}$ (for $r_{\text{H}_{\text{o}}}<r<r_{\text{pa}}$), $I_{14_{3}}$ (for $r_{\text{pa}}<r$), and $I_{15_{1}}$ given by (\ref{eq:I141})--(\ref{eq:I151}) in Appendix~\ref{App:EmInt3} and obtain $\tau(\lambda)$ in terms of elementary functions.\\
\emph{Case 4}.--In this case we always have timelike geodesics with $V_{E}(r_{\text{pa}})<K$. We recall that in this case the right-hand side of (\ref{eq:EoMr}) has four distinct real roots. Timelike geodesics with $r_{\text{H}_{\text{o}}}<r<r_{\text{pa}}$ can have a turning point at $r_{2}=r_{\text{max}}$ (maximum) while timelike geodesics with $r_{\text{pa}}<r$ can have a turning point at $r_{1}=r_{\text{min}}$ (minimum). For the former we substitute using (\ref{eq:Case4sub2}), while for the latter we substitute using (\ref{eq:Case4sub1}). We rewrite the resulting expressions in terms of Legendre's elliptic integrals of the first and third kind and the nonstandard elliptic integral $J_{L_{9}}(\chi_{i},\chi(\lambda),k_{3},\tilde{n})$ given by (\ref{eq:J9}) in Appendix~\ref{App:EllInt}. In the next step we use (\ref{eq:J9R}) from Appendix~\ref{App:EllInt} to rewrite $\tau(\lambda)$ in terms of elementary functions and Legendre's elliptic integrals of the first, second, and third kind. Note that analogously to the $r$ component of the time coordinate $t_{r}(\lambda)$ for timelike geodesics passing through a turning point we have to split the integral at the turning point into two terms and for the second term we have to change $i_{r_{i}}$ to $-i_{r_{i}}$.

\section{Gravitational Lensing}\label{Sec:Lensing}
\subsection{The observer's celestial sphere and the particle's energy}
\subsubsection{The observer's celestial sphere: The orthonormal tetrad and the celestial coordinates}
\begin{figure}[h]
\centering
\includegraphics[width=0.6\textwidth]{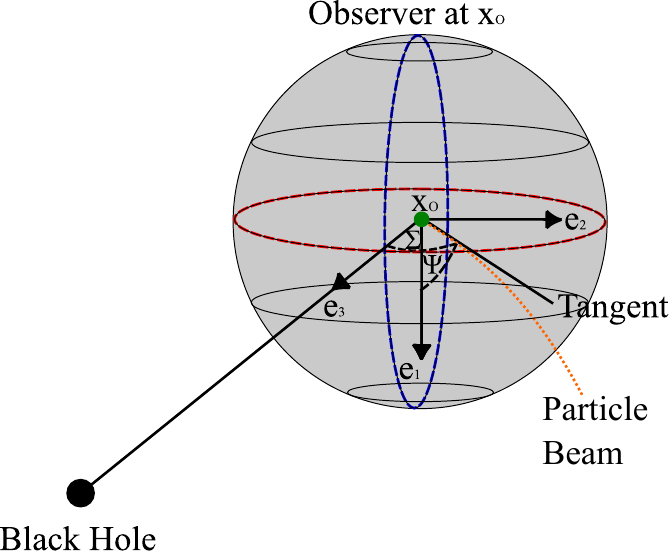}
\caption{Illustration of the celestial sphere of a stationary observer located in the domain of outer communication (green dot) and the three tetrad vectors $e_{1}$, $e_{2}$, and $e_{3}$. The observer detects a particle beam (orange dotted line) at the event $x_{O}$. The celestial latitude $\Sigma$ is measured from $e_{3}$ and the celestial longitude is measured from $e_{1}$ in the direction of $e_{2}$.}\label{fig:ObsGeom}
\end{figure}
When we observe a source on the sky we usually mark its position using latitude-longitude coordinates on our celestial sphere (or what is visible thereof). However, \emph{a priori} the coordinates on the celestial sphere are not fixed. In astronomy the most common convention is that we take a known reference source on the sky and then measure the positions of all other sources in relation to this reference. We now transfer this approach to gravitational lensing of massive particles by a black hole. We place a stationary
observer in the domain of outer communication at the coordinates $(x_{O}^{\mu})=(t_{O},r_{O},\vartheta_{O},\varphi_{O})$ and take the black hole as reference direction. As in Frost \cite{Frost2022} for lightlike geodesics, we now follow the approach of Grenzebach \emph{et al.} \cite{Grenzebach2015} and introduce an orthonormal tetrad. The four tetrad vectors $e_{0}$, $e_{1}$, $e_{2}$, and $e_{3}$ read
\begin{eqnarray}\label{eq:tetrad0}
e_{0}=\left.\sqrt{\frac{\rho(r)}{Q(r)}}\partial_{t}\right|_{(x_{O}^{\mu})},
\end{eqnarray}
\begin{eqnarray}\label{eq:tetrad1}
e_{1}=\left.\frac{1}{\sqrt{\rho(r)}}\partial_{\vartheta}\right|_{(x_{O}^{\mu})},
\end{eqnarray}
\begin{eqnarray}\label{eq:tetrad2}
e_{2}=-\left.\frac{\partial_{\varphi}-2n(\cos\vartheta+C)\partial_{t}}{\sqrt{\rho(r)}\sin\vartheta}\right|_{(x_{O}^{\mu})},
\end{eqnarray}
\begin{eqnarray}\label{eq:tetrad3}
e_{3}=-\left.\sqrt{\frac{Q(r)}{\rho(r)}}\partial_{r}\right|_{(x_{O}^{\mu})}.
\end{eqnarray}
Here, the tetrad vector $e_{0}$ is also the four-velocity of the observer. Having introduced the orthonormal tetrad we now define the celestial latitude $\Sigma$ and the celestial longitude $\Psi$ on the observer's celestial sphere as depicted in Fig.~\ref{fig:ObsGeom}. The celestial latitude $\Sigma$ is measured from the direction of the tetrad vector $e_{3}$. The celestial longitude $\Psi$ on the other hand is measured from the tetrad vector $e_{1}$ in the direction of the tetrad vector $e_{2}$. Now let us consider the tangent vector to a timelike geodesic of a particle detected by the observer:
\begin{eqnarray}\label{eqn:tanvec}
  \frac{\text{d}\eta}{\text{d}\lambda}=\frac{\text{d}t}{\text{d}\lambda}\partial_{t}+\frac{\text{d}r}{\text{d}\lambda}\partial_{r}+\frac{\text{d}\vartheta}{\text{d}\lambda}\partial_{\vartheta}+\frac{\text{d}\varphi}{\text{d}\lambda}\partial_{\varphi}.
\end{eqnarray}
We now use the approach of Perlick and Tsupko \cite{Perlick2017} for lightlike geodesics in a plasma and transfer it to our test particles. Using their notation and transferring it to Mino parametrization the tangent vector to a timelike geodesic ending at the observer reads in terms of the tetrad vectors, the celestial angles $\Sigma$ and $\Psi$, and two normalization constants $\alpha$ and $\beta$ 
\begin{eqnarray}\label{eqn:tanvecloc}
\left.\frac{\text{d}\eta}{\text{d}\lambda}\right|_{(x_{O}^{\mu})}=-\alpha e_{0}+\beta\left(\sin\Sigma\cos\Psi e_{1}+\sin\Sigma\sin\Psi e_{2}+\cos\Sigma e_{3}\right).
\end{eqnarray}
We now use (\ref{eqn:tanvecloc}) to rewrite $\beta$ in terms of $\alpha$ via
\begin{eqnarray}
g\left(\left.\frac{\text{d}\eta}{\text{d}\lambda}\right|_{(x_{O}^{\mu})},\left.\frac{\text{d}\eta}{\text{d}\lambda}\right|_{(x_{O}^{\mu})}\right)=-\alpha^2+\beta^2=-\rho(r_{O})^2.
\end{eqnarray}
Solving for $\beta$ we get
\begin{eqnarray}
\beta=-\sqrt{\alpha^2-\rho(r_{O})^2}.
\end{eqnarray}
Note that here we originally had a sign ambiguity for the root; however, we already fixed it such that for $n=0$ the sign of $L_{z}$ is consistent with the sign for lightlike geodesics in Frost \cite{Frost2022}. The remaining normalization constant $\alpha$ can now be calculated via
\begin{equation}
\alpha=g\left(\left.\frac{\text{d}\eta}{\text{d}\lambda}\right|_{(x_{O}^{\mu})},e_{0}\right).
\end{equation}
Inserting (\ref{eq:tetrad0}) and (\ref{eqn:tanvec}) we get for $\alpha$ and $\beta$
\begin{align}
\alpha&=-\sqrt{\frac{\rho(r_{O})^3}{Q(r_{O})}}E,\label{eq:NCA}\\
\beta&=-\rho(r_{O})\sqrt{\frac{\rho(r_{O})E^2-Q(r_{O})}{Q(r_{O})}}.\label{eq:NCB}
\end{align}
Now we insert the tetrad vectors as well as the normalization constants $\alpha$ and $\beta$ in (\ref{eqn:tanvecloc}) and sort all terms. A comparison of coefficients with (\ref{eqn:tanvec}) now allows us to rewrite $L_{z}$ and $K$ in terms of the energy $E$ and the celestial latitude $\Sigma$ and longitude $\Psi$. We get
\begin{align}
L_{z}&=\sqrt{\frac{\rho(r_{O})(\rho(r_{O})E^2-Q(r_{O}))}{Q(r_{O})}}\sin\vartheta_{O}\sin\Sigma\sin\Psi-2n\left(\cos\vartheta_{O}+C\right)E,\label{eq:EoMCelesLz}\\
K&=\frac{\rho(r_{O})(\rho(r_{O})E^2-Q(r_{O}))}{Q(r_{O})}\sin^2\Sigma.\label{eq:EoMCelesK}
\end{align}

\subsubsection{Particle energy measured by the observer}
We solved the equations of motion and rewrote the $z$ component of the angular momentum as well as the Carter constant in terms of the particle energy $E$ along the geodesic. However, the energy $E$ is not the energy a stationary observer (or better a detector) would measure when the particle is detected. Let us for now assume a particle with four-momentum $(p_{\text{part}\mu})$ and a stationary observer with four-velocity $(u_{\text{O}}^{\mu})$ at the coordinates $(x_{O}^{\mu})=(t_{O},r_{O},\vartheta_{O},\varphi_{O})$. In terms of $(p_{\text{part}\mu})$ and $(u_{\text{O}}^{\mu})$ the total energy $E_{\text{tot}}$ of the particle measured by the observer can now be calculated via (see, e.g., \cite{Wald1984} p.~69)
\begin{equation}
E_{\text{tot}}=-p_{\text{part}\mu}u_{O}^{\mu}.
\end{equation}
In our case the four-velocity $u_{O}$ is given by $e_{0}$ and thus we have $u_{O}^{r}=u_{O}^{\vartheta}=u_{O}^{\varphi}=0$. Consequentially, the only nonvanishing component is
\begin{equation}
u_{O}^{t}=\sqrt{\frac{\rho(r_{O})}{Q(r_{O})}}.
\end{equation}
We now rewrite the four-momentum in terms of the metric coefficients and the components of the tangent vector (\ref{eqn:tanvec}) using the relation
\begin{equation}
p_{\mu}=g_{\mu\nu}\dot{x}^{\nu}=g_{\mu\nu}\frac{\text{d}\lambda}{\text{d}\tau}\frac{\text{d}x^{\nu}}{\text{d}\lambda}.
\end{equation}
Using this relation and (\ref{eq:Mino}) we now get for $E_{\text{tot}}$
\begin{equation}
E_{\text{tot}}=-\frac{g_{\mu\nu}}{\rho(r_{O})}\left.\frac{\text{d}x_{\text{part}}^{\mu}}{\text{d}\lambda}\right|_{(x_{O}^{\rho})}u_{O}^{\nu}.
\end{equation}
In the next step we insert the metric coefficients, the four-velocity of the particle, and the four-velocity of the observer and obtain as relation between the total energy $E_{\text{tot}}$ of the particle measured by the observer at $(x_{O}^{\mu})$ and the energy $E$ along the geodesic
\begin{equation}\label{eq:EtotE}
E_{\text{tot}}=\sqrt{\frac{\rho(r_{O})}{Q(r_{O})}}E.
\end{equation}
Analogously we can also calculate the total energy of the particle at the position of a stationary particle source. Note that for simplicity we continue to use the energy $E$ along the timelike geodesic to parametrize the timelike geodesics in the following. The corresponding energy $E_{\text{tot}}$ in the frame of the observer can simply be calculated using Eq.~(\ref{eq:EtotE}).

\subsection{Angular radius of the particle shadow}
\begin{figure}[h]
\centering
\includegraphics[width=0.8\textwidth]{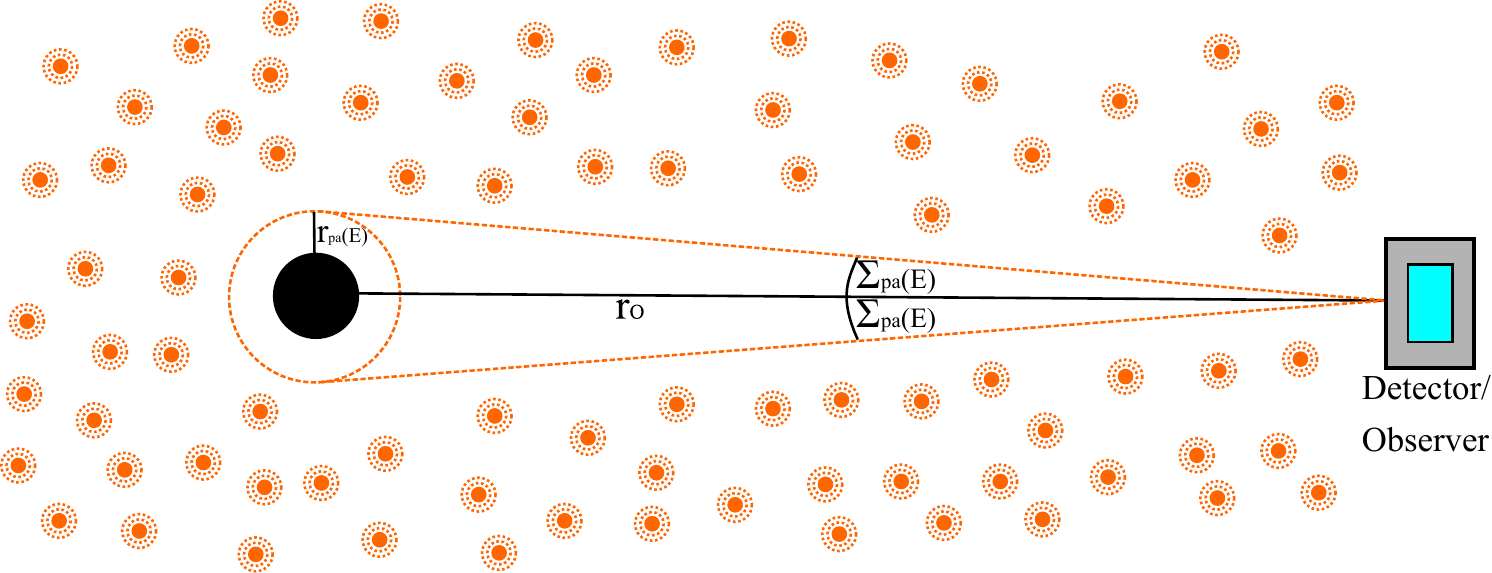}
\caption{Illustration of the construction of the shadow of a black hole for massive particles with a fixed energy $E$. The black circle marks the region hidden behind the (outer) horizon of the black hole. The orange dotted circle around the black hole marks the position of the particle sphere with radius coordinate $r_{\text{pa}}(E)$ at the energy $E$. The orange dots with the dashed circles mark particle sources. The orange dotted lines ending at the detector mark timelike geodesics asymptotically coming from the particle sphere.}\label{fig:IllShad}
\end{figure}
In analogy to the shadow of a black hole for light rays we can also define a shadow for massive particles. To distinguish between both in the following we will call the former \emph{photon shadow} while we call the latter \emph{particle shadow} (ignoring to some degree the particle-wave duality for photons). For massive particles we construct the shadow as depicted in Fig.~6. Let us assume that we have a black hole and that we place a stationary observer (in our case with a particle detector) in the domain of outer communication outside the particle sphere. Then we distribute particle sources everywhere except between black hole and observer. Now, for timelike geodesics with constant energy $E$, the directions on the observer's sky with particle sources are associated with signals in the particle detector. On the other hand the directions on the sky without particle sources are associated with silence in the particle detector. The boundary between both areas marks timelike geodesics asymptotically coming from a particle sphere. Note that since we have a particle sphere for each value of the energy $E$ the definition of the particle shadow is only valid for particles with constant energy $E$. However, we can define a \emph{collective particle shadow}. It is simply the shadow which remains when we superpose the shadows for all particle energies and its boundary is infinitesimally close to the photon shadow.\\
For each value of the energy $E$ we can now derive the angular radius of the particle shadow analogously to the angular radius of the photon shadow. For this purpose we use the fact that timelike geodesics asymptotically coming from a particle sphere have the same constants of motion as timelike geodesics on this particle sphere. We recall that these geodesics have a double root at $r_{1}=r_{2}=r_{\text{pa}}(E)$. We evaluate (\ref{eq:EoMr}) at $r=r_{\text{pa}}(E)$ and insert $K$ given by (\ref{eq:EoMCelesK}). We solve for $\Sigma$ and get as result for the angular radius of the particle shadow (note that $r_{\text{pa}}$ is still energy dependent and we only omit the energy $E$ as argument for brevity)
\begin{equation}\label{eq:PartSDWCN}
\Sigma_{\text{pa}}(E)=\arcsin\left(\sqrt{\frac{\rho(r_{\text{pa}})(\rho(r_{\text{pa}})E^2-Q(r_{\text{pa}}))Q(r_{O})}{\rho(r_{O})(\rho(r_{O})E^2-Q(r_{O}))Q(r_{\text{pa}})}}\right).
\end{equation}
For $n=0$ the obtained expression reduces to the angular radius of the particle shadow for the Reissner-Nordstr\"{o}m metrics. It reads
\begin{equation}\label{eq:PartSDWRN}
\Sigma_{\text{pa}}(E)=\arcsin\left(\frac{r_{\text{pa}}}{r_{O}}\sqrt{\frac{(E^2-\tilde{Q}(r_{\text{pa}}))\tilde{Q}(r_{O})}{(E^2-\tilde{Q}(r_{O}))\tilde{Q}(r_{\text{pa}})}}\right),
\end{equation}
where we defined $\tilde{Q}(r)=Q(r)/r^2$. For $E\rightarrow\infty$ (\ref{eq:PartSDWCN}) and (\ref{eq:PartSDWRN}) reduce to the angular radii of the photon shadows for the charged NUT metrics \cite{Frost2022,Frost2022b}
\begin{equation}\label{eq:PhotSDWCN}
\Sigma_{\text{ph}_{\text{CN}}}=\arcsin\left(\frac{\rho(r_{\text{ph}})}{\rho(r_{O})}\sqrt{\frac{Q(r_{O})}{Q(r_{\text{ph}})}}\right)
\end{equation}
and the Reissner-Nordstr\"{o}m metrics
\begin{equation}\label{eq:PhotSDWRN}
\Sigma_{\text{ph}_{\text{RN}}}=\arcsin\left(\frac{r_{\text{ph}}}{r_{O}}\sqrt{\frac{\tilde{Q}(r_{O})}{\tilde{Q}(r_{\text{ph}})}}\right),
\end{equation}
respectively.\\
\begin{figure}\label{fig:SigmapaE}
  \begin{tabular}{cc}
    \hspace{-0.5cm}\includegraphics[width=85mm]{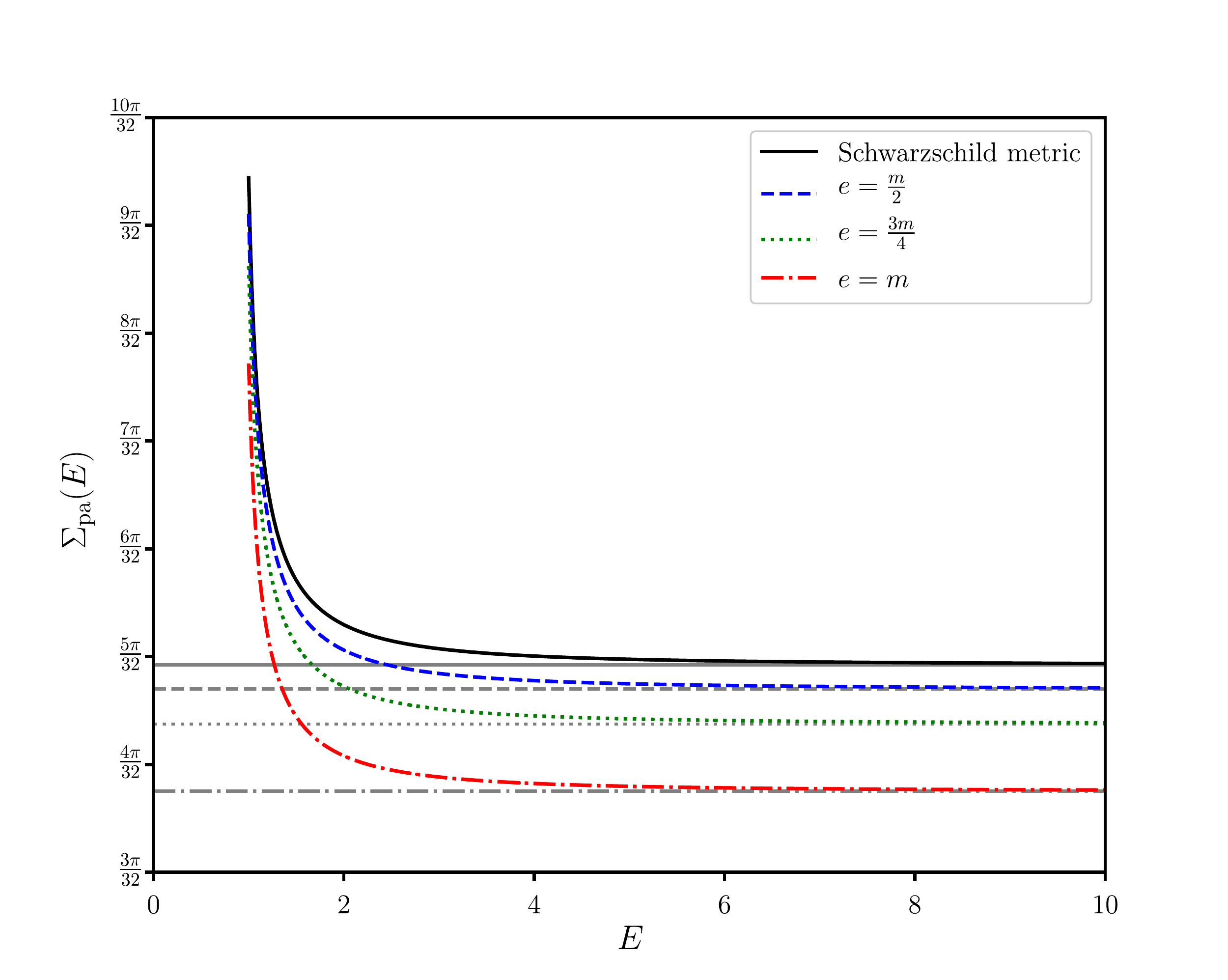} &   \includegraphics[width=85mm]{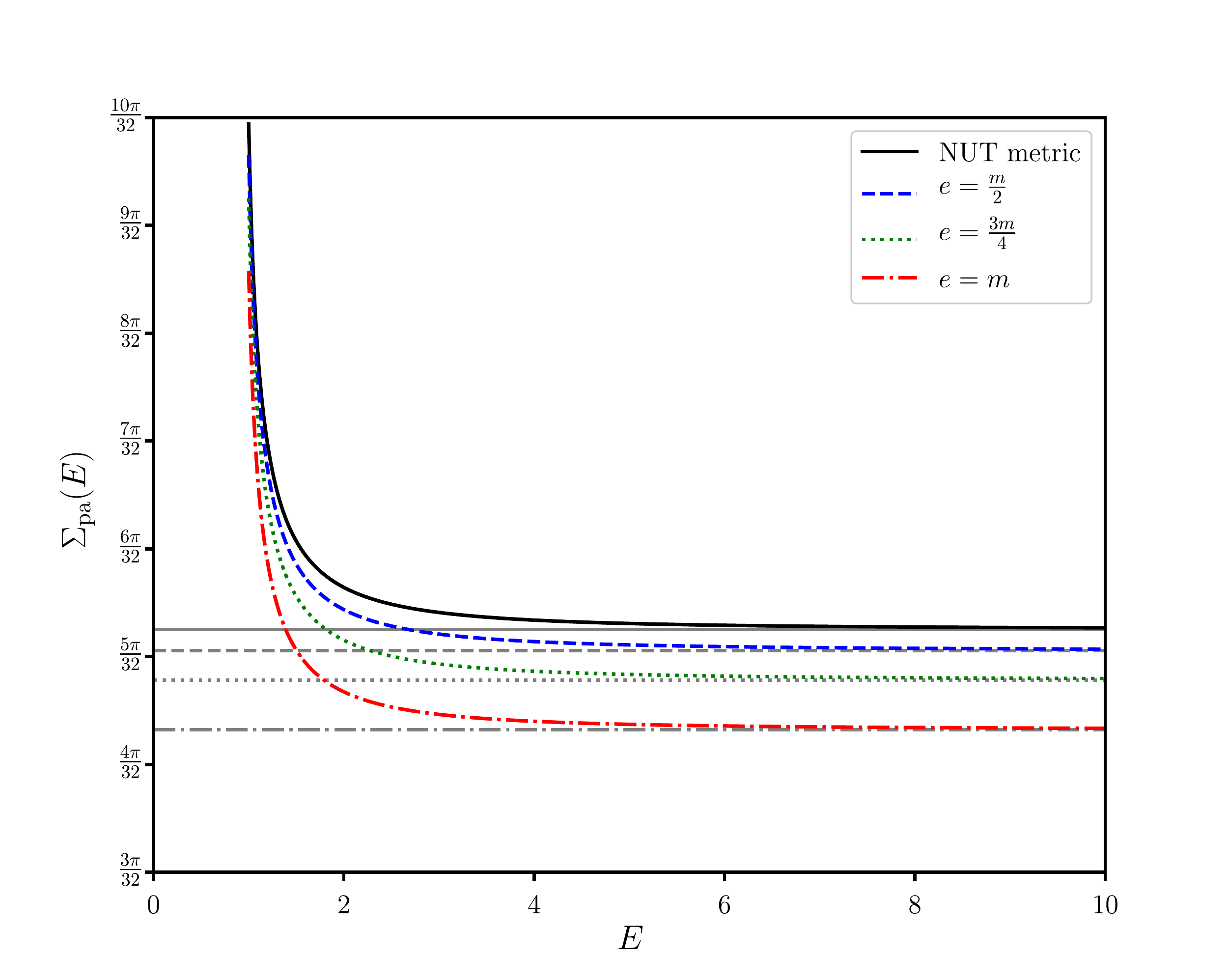}
  \end{tabular}
	\caption{Angular radius of the particle shadow $\Sigma_{\text{pa}}$ as function of the energy $E$. Left panel: Schwarzschild metric (black solid line) and Reissner-Nordstr\"{o}m metric with $e=m/2$ (blue dashed line), $e=3m/4$ (green dotted line), and $e=m$ (red dash-dotted line). Right panel: NUT metric (black solid line) and charged NUT metric with $e=m/2$ (blue dashed line), $e=3m/4$ (green dotted line), and $e=m$ (red dash-dotted line) for $n=m/2$. For both panels the observers are located at $r_{O}=10m$.  The gray horizontal lines with the same line styles mark the corresponding angular radii of the photon shadows $\Sigma_{\text{ph}}$.}
\end{figure}
Figure~7 shows the angular radius of the particle shadow as function of the energy $E$. In the left panel we have the results for the Schwarzschild metric (black solid line) and the Reissner-Nordstr\"{o}m metric with $e=m/2$ (blue dashed line), $e=3m/4$ (green dotted line), and $e=m$ (red dash-dotted line). In the right panel we have the results for the NUT metric (black solid line) and the charged NUT metric with $e=m/2$ (blue dashed line), $e=3m/4$ (green dotted line), and $e=m$ (red dash-dotted line) for $n=m/2$. For both panels the observers are located at $r_{O}=10m$ and the gray horizontal lines mark the angular radii of the corresponding photon shadows. As already mentioned above in both panels we see that for $E\rightarrow \infty$ the angular radii of the particle shadows asymptotically approach the angular radii of the photon shadows.\\
\begin{figure}\label{fig:Sigmapan}
  \begin{tabular}{cc}
    \hspace{-0.5cm}\includegraphics[width=85mm]{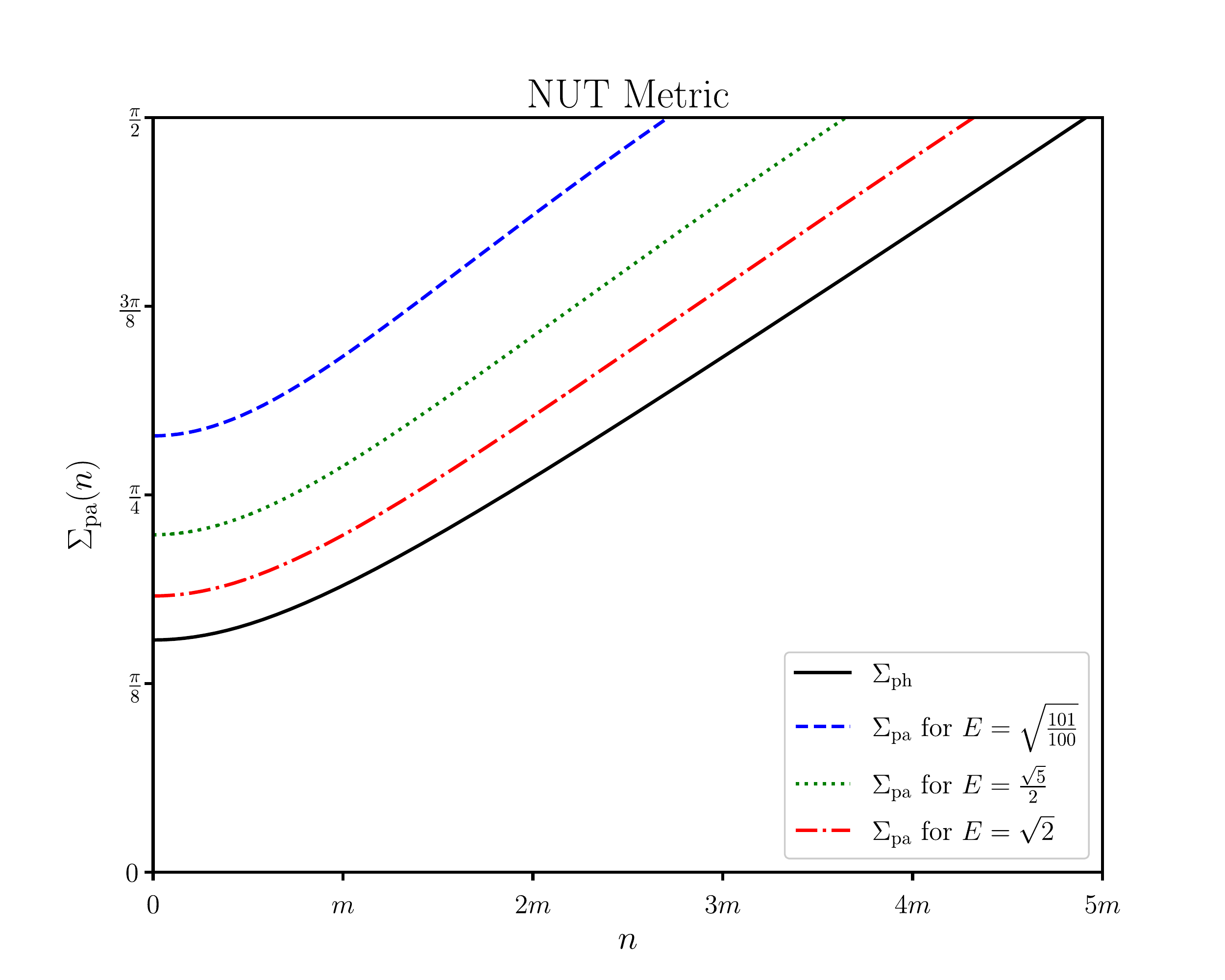} &   \includegraphics[width=85mm]{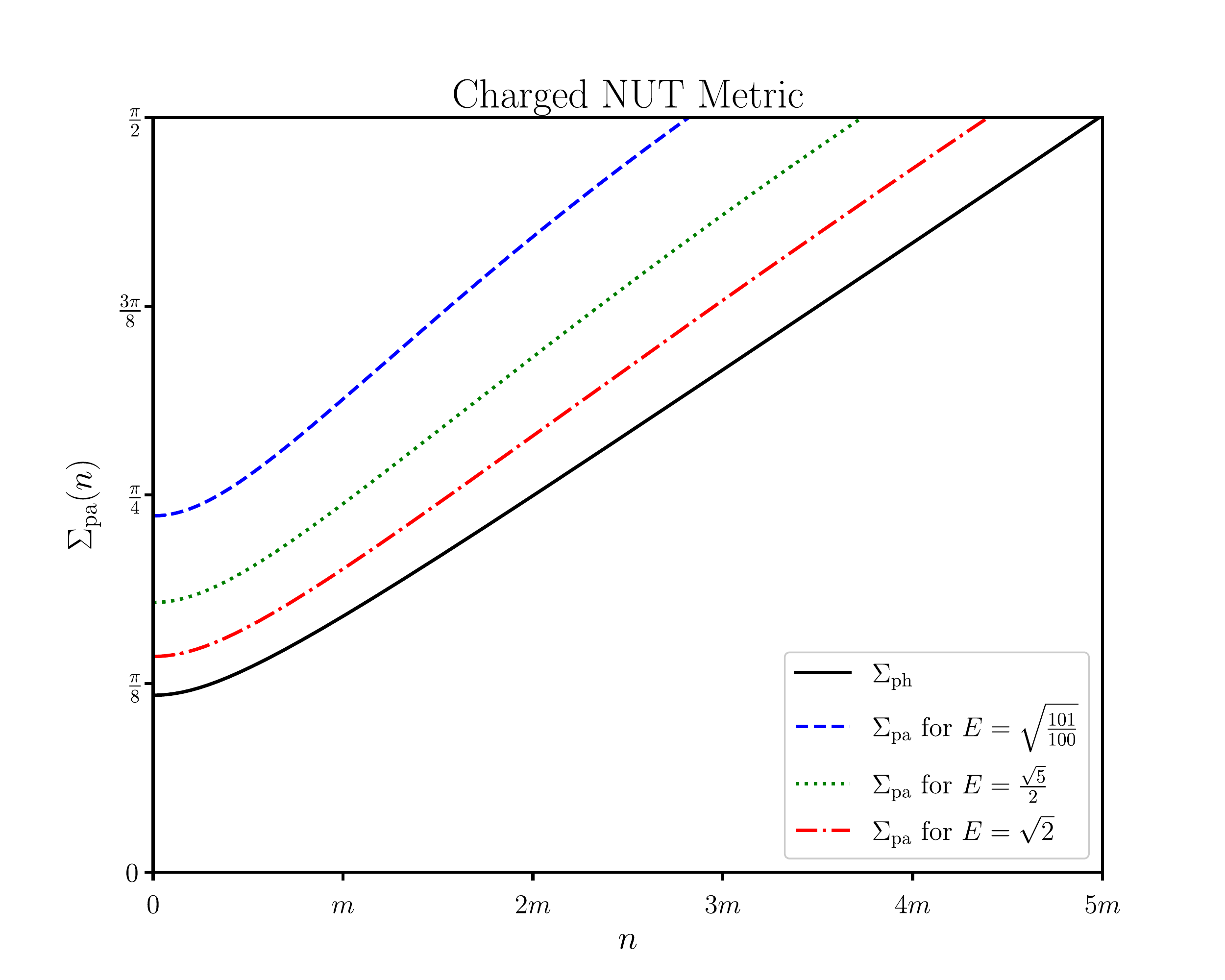} \\
  \end{tabular}
  \caption{Angular radius of the photon shadow $\Sigma_{\text{ph}}$ and the particle shadow $\Sigma_{\text{pa}}$ as function of the gravitomagnetic charge $n$ for the NUT metric (left panel) and the charged NUT metric with $e=m$ (right panel). For both panels the observers are located at $r_{O}=10m$. The black solid line represents the angular radius of the photon shadow. The blue dashed, green dotted, and red dash-dotted lines represent the angular radii of the particle shadows for particles with the energies $E=\sqrt{101/100}$, $E=\sqrt{5}/2$, and $E=\sqrt{2}$, respectively.}
\end{figure}
Figure~8 shows the angular radii of the photon shadows (black solid lines) and the particle shadows for $E=\sqrt{101/100}$ (blue dotted lines), $E=\sqrt{5}/2$ (green dotted lines), and $E=\sqrt{2}$ (red dash-dotted lines) for the NUT metric (left panel) and the charged NUT metric with $e=m$ (right panel) as function of the gravitomagnetic charge $n$. Again for both panels the observers are located at $r_{O}=10m$. For $n=0$ we have the Schwarzschild (left panel) and Reissner-Nordstr\"{o}m (right panel) limits. When we turn on and slowly increase the gravitomagnetic charge $n$ the photon shadows and the particle shadows expand. The reason behind this is simply that for increasing $n$ the photon spheres and the particle spheres expand and approach the position of the observer. When we have $r_{\text{ph}}=r_{O}$ or $r_{\text{pa}}(E)=r_{O}$ the corresponding shadows cover half of the observer's sky and thus we have $\Sigma_{\text{ph}}=\pi/2$ or $\Sigma_{\text{pa}}(E)=\pi/2$, respectively. When we turn on the electric charge and, in the case of the particle spheres, keep the energy $E$ constant the angular radii of the photon shadow and the particle shadows decrease. Therefore, when we increase the gravitomagnetic charge $n$ the photon shadow and the particle shadows approach $\Sigma_{\text{ph}}=\pi/2$ or $\Sigma_{\text{pa}}(E)=\pi/2$ for slightly larger $n$.\\
When we compare the photon shadow with the particle shadow for fixed spacetime parameters and particle energies we see that both are circular and that the particle shadow is always larger than the photon shadow. However, when we have real astrophysical settings we usually do not only have particles at one fixed energy but rather a large number of different particles each of which can have a different energy distribution. While the energy distributions vary for each type of particle we can tendentially say that particles with particularly high energies are less likely to occur than particles with lower energies (note though that some particles like neutrinos always move at velocities close to the speed of light). As a consequence we will detect less particles with particularly high energies and thus toward the photon shadow we will observe a fading out effect of particle detections; i.e., toward the photon shadow the particle shadow becomes darker. Thus from the theoretical perspective the size of the observable particle shadow is determined by the particles with the highest energy.\\
Unfortunately, in astrophysics observing the particle shadow has two caveats. On one hand currently the only particles which could be effectively used to observe the particle shadow are neutrinos. However, the observation of neutrino emission events with an identified and well-characterized source is quite rare. In particular currently we do not have the technical abilities to identify and continuously observe individual sources with a steady emission of neutrinos and characteristic neutrino energy spectra outside the Solar System. In addition, due to their weak interaction neutrinos are difficult to detect and neutrino detectors only have an angular resolution of a few square degrees. Thus even with the recent advances in technology it is questionable if we will be able to resolve the particle shadow in the near future. In addition, when we want to use the particle shadow to distinguish between the Schwarzschild metric, the Reissner-Nordstr\"{o}m metric, the NUT metric, and the charged NUT metric we encounter the same difficulties as for the photon shadow. In all four cases the particle shadow is circular. Its size does not only depend on the spacetime parameters and the energy of the particles but also on the spacetime coordinate of the observer $r_{O}$. As discussed in Frost \cite{Frost2022,Frost2022b} the photon shadow shrinks with increasing distance between black hole and observer, and the same holds for the particle shadow created by particles with a fixed energy $E$ (not shown). Therefore, as long as we do not know the distance between observer and black hole we have a degeneracy between the particle shadows for different $r_{O}$ and different electric and gravitomagnetic charges. The only effect which to some degree may be able to help us to distinguish between different black hole spacetimes is the darkening effect. When we have a closer look at Fig.~7 we can see small variations in how fast the angular radius of the particle shadow approaches the angular radius of the photon shadow with increasing energy $E$ for different combinations of the electric charge $e$ and the gravitomagnetic charge $n$. However, while these variations may allow one to distinguish different black hole spacetimes, for the actual detection of these effects we need particle detectors with very high angular resolution and particle types which can have a broad energy spectrum.

\subsection{The lens equation}\label{Sec:LensEq}
For lightlike geodesics a general exact lens map or also lens equation was first brought forward by Frittelli and Newman \cite{Frittelli1999}. It was later adapted to spherically symmetric and static spacetimes by Perlick \cite{Perlick2004} and to the charged NUT--de Sitter metrics by Frost \cite{Frost2022,Frost2022b}. While in its original form it was only applied to lightlike geodesics it can be easily transferred to timelike geodesics. \\
In this paper we set up the lens map as illustrated in Fig.~9. We place a stationary observer with a particle detector in the domain of outer communication at the coordinates $(x^{\mu}(\lambda_{O}))=(x_{O}^{\mu})=(t_{O},r_{O},\vartheta_{O},\varphi_{O})$ outside the particle sphere. Then we distribute stationary particle sources on a two-sphere $S_{P}^{2}$ at a radius coordinate $r_{O}<r_{P}$. We follow particles moving on timelike geodesics ending at the observer backward in time. Here, we have to distinguish two different types of past-directed timelike geodesics. The first type of past-directed timelike geodesics will intersect with the particle sphere and thus these geodesics will end in the black hole. The second type of past-directed timelike geodesics will intersect with the two-sphere of particle sources $S_{P}^{2}$. The latter now constitute a map from the angular coordinates on the celestial sphere of the observer to the spacetime coordinates on the two-sphere of particle sources $S_{P}^{2}$:
\begin{equation}\label{eq:LensMap}
(\Sigma,\Psi)\rightarrow(\vartheta_{P}(\Sigma,\Psi),\varphi_{P}(\Sigma,\Psi)).
\end{equation}
Because of the symmetries of the charged NUT spacetimes and the fact that the Mino parameter is defined up to an affine transformation we will choose $\lambda_{O}=0$ and $\varphi_{O}=0$ in the following. We now want to use the exact solutions to the equations of motion for $\vartheta$ and $\varphi$ derived in Secs.~\ref{Sec:EoMthetaSol} and \ref{Sec:EoMphiSol} to derive the lens maps. For this purpose we still have to determine $\lambda_{P}<\lambda_{O}=0$. In our setting we know the radius coordinates of the observer and the two-sphere of particle sources $r_{O}$ and $r_{P}$ and thus we can determine $\lambda_{P}$ by separating variables in (\ref{eq:EoMr}), inserting (\ref{eq:EoMCelesK}), and integrating. The result reads
\begin{equation}\label{eq:Minor}
\lambda_{P}=\int_{r_{O}...}^{...r_{P}}\frac{\text{d}r'}{\sqrt{\rho(r')^2E^2-\rho(r')Q(r')-Q(r')\frac{\rho(r_{O})(\rho(r_{O})E^2-Q(r_{O}))}{Q(r_{O})}\sin^2\Sigma}},
\end{equation}
where again the dots in the limits of the integral indicate that we have to split the integral at potential turning points into two terms. The sign of the root in the denominator has to be chosen according to the direction of the $r$ motion. We now rewrite the integral in terms of elementary functions or Legendre's elliptic integral of the first kind. The computational evaluation of the lens equations for massive particles (as well as the travel time measures, which will be discussed in Sec.~\ref{Sec:TravTime}) was carried out in the programming language Julia \cite{Bezanson2017}. The computational evaluation of the lens equations (and the travel times) for lightlike geodesics was carried out following the procedure described in Frost \cite{Frost2022,Frost2022b}.\\
\begin{figure}[h]\label{fig:CelSphere}
\centering
\includegraphics[width=0.8\textwidth]{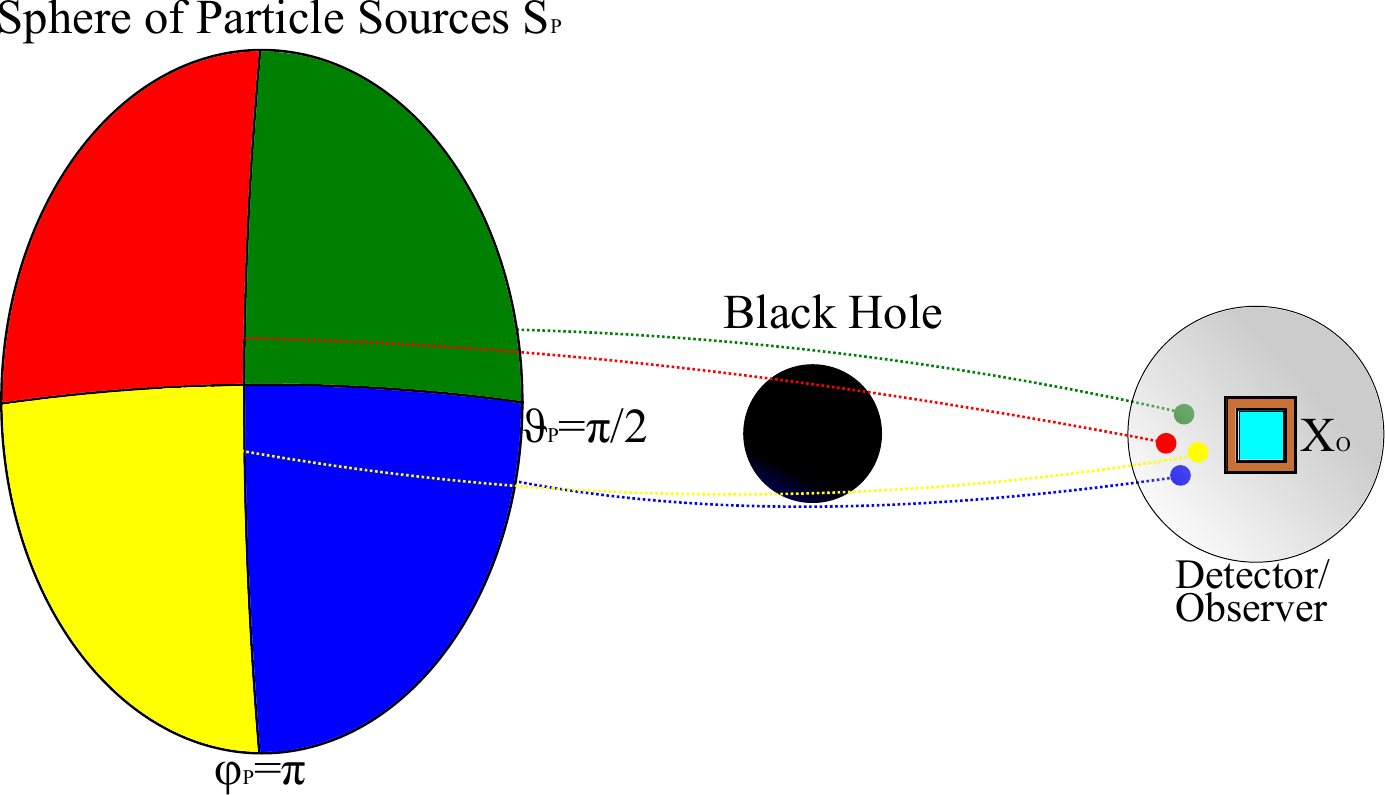}
\caption{Illustration of the lens map. The black hole, in the figure depicted by a black dot, acts as gravitational lens. We place a stationary observer with a detector at the coordinates $(x_{O}^{\mu})=(t_{O},r_{O},\vartheta_{O},\varphi_{O})$ in the domain of outer communication between particle sphere and spatial infinity. The celestial sphere is marked by the gray-shaded sphere around it. We then distribute stationary particle sources on a two-sphere $S_{P}^2$ with $r_{O}<r_{P}$. We divide the two-sphere in four different quadrants
and assign to each quadrant a color following the color convention of Bohn \emph{et al.} \cite{Bohn2015}: $0\leq\vartheta_{P}\leq\pi/2$ and $0\leq\varphi_{P}<\pi$, green; $\pi/2<\vartheta_{P}\leq\pi$ and $0\leq\varphi_{P}<\pi$, blue; $0\leq\vartheta_{P}\leq\pi/2$ and $\pi\leq\varphi_{P}<2\pi$, red; $\pi/2<\vartheta_{P}\leq\pi$ and $\pi\leq\varphi_{P}<2\pi$, yellow. The colored dotted lines represent particles on timelike geodesics emitted by particle sources on the corresponding quadrant of the two-sphere.}
\end{figure}
For the visualization of the lens maps we divide the two-sphere of particle sources $S_{P}^2$ into four quadrants as shown in Fig.~9. To each quadrant we assign a color following a slightly adapted version of the color convention of Bohn \emph{et al.} \cite{Bohn2015}. Here, we assign the colors as follows. The first quadrant with $0\leq\vartheta_{P}\leq\pi/2$ and $0\leq\varphi_{P}<\pi$ is colored green. The second quadrant with $\pi/2<\vartheta_{P}\leq\pi$ and $0\leq\varphi_{P}<\pi$ is colored blue. The third quadrant with $0\leq\vartheta_{P}\leq\pi/2$ and $\pi\leq\varphi_{P}<2\pi$ is colored red. The fourth quadrant with $\pi/2<\vartheta_{P}\leq\pi$ and $\pi\leq\varphi_{P}<2\pi$ is colored yellow. Here, we adapt the colour convention of Bohn \emph{et al.} \cite{Bohn2015} as follows. When we plot the lens maps we have images of different orders. Here, we define images of first order as timelike geodesics for which the covered angle $\Delta \varphi_{P}$ is $0\leq\left|\Delta\varphi_{P}\right|<\pi$. Analogously, we define images of second order as timelike geodesics with $\pi\leq\left|\Delta\varphi_{P}\right|<2\pi$ and so on. In the visual representation of the lens maps we now plot images of even order in slightly fainter colors than images of odd order. In addition, for the discussion of the lens maps and the travel time maps in Sec.~\ref{Sec:TravTime} we agree on the following conventions. We divide the visible part of the observer's celestial sphere into two halves separated by the longitudinal lines $\Psi=\pi/2$ and $\Psi=3\pi/2$. We refer to this line as \emph{celestial equator}. In addition, we agree to refer to all longitudes $\pi/2<\Psi<3\pi/2$ as \emph{northern hemisphere} and to all longitudes $0\leq\Psi<\pi/2$ and $3\pi/2<\Psi$ as \emph{southern hemisphere}. Note that because we divide along longitudes this terminology is actually not correct; however, it will strongly simplify the discussion of the lens and travel time maps. In addition, we divide the observer's celestial sphere along the lines marked by the longitudes $\Psi=0$ and $\Psi=\pi$. For simplicity we will call the former \emph{meridian} and the latter \emph{antimerdian} in the following. In addition, we will refer to the longitudes $0<\Psi<\pi$ as \emph{western hemisphere} and to the longitudes $\pi<\Psi<2\pi$ as \emph{eastern hemisphere}. As last convention we agree on referring to particle detections as images on the celestial sphere although technically this is not correct since we cannot really see these images.\\
\begin{figure}\label{fig:LESchwarzschild}
  \begin{tabular}{cc}
    $E=\sqrt{101/100}$&$E=\sqrt{5}/2$\\
    \hspace{-0.5cm}\includegraphics[width=85mm]{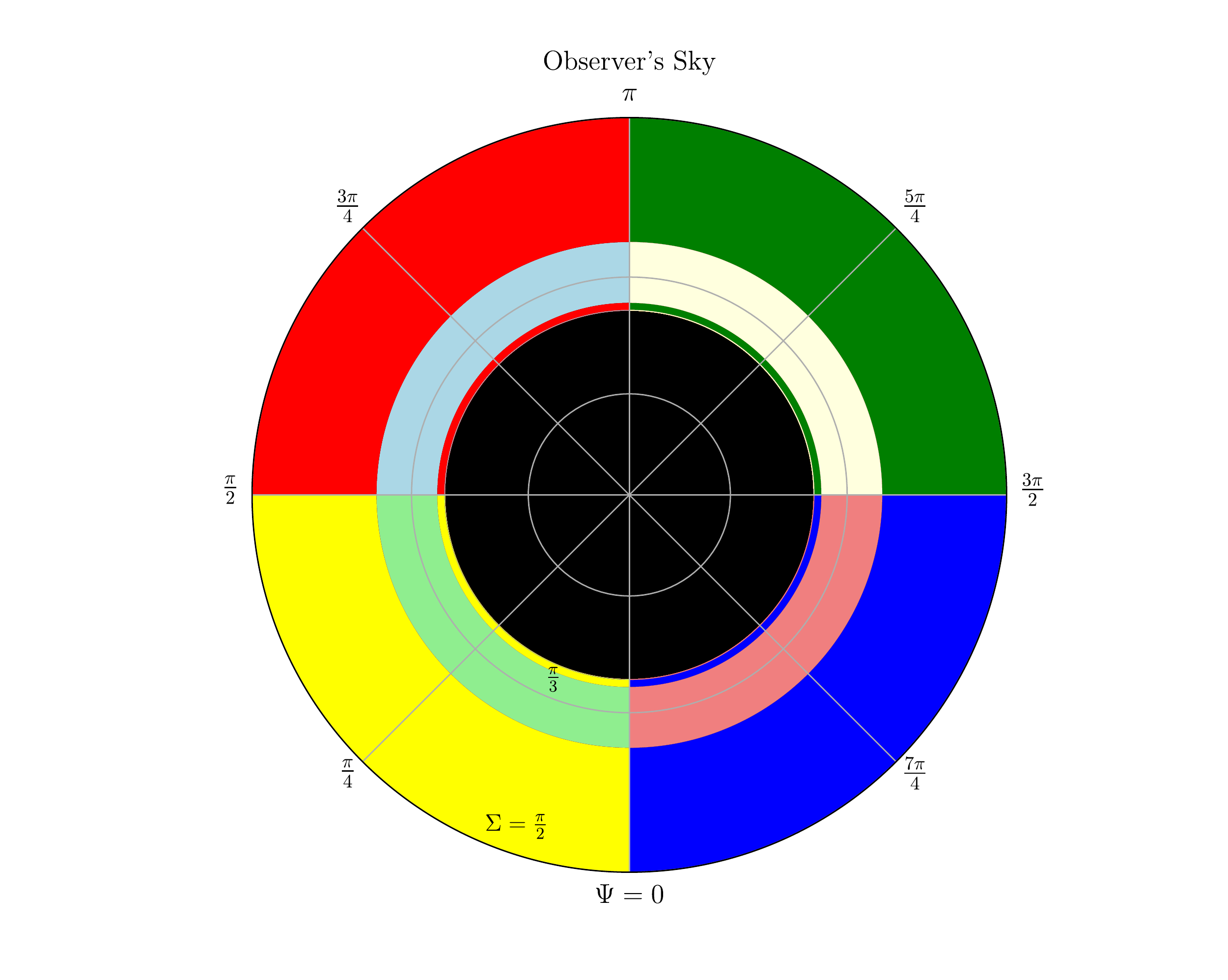} &   \includegraphics[width=85mm]{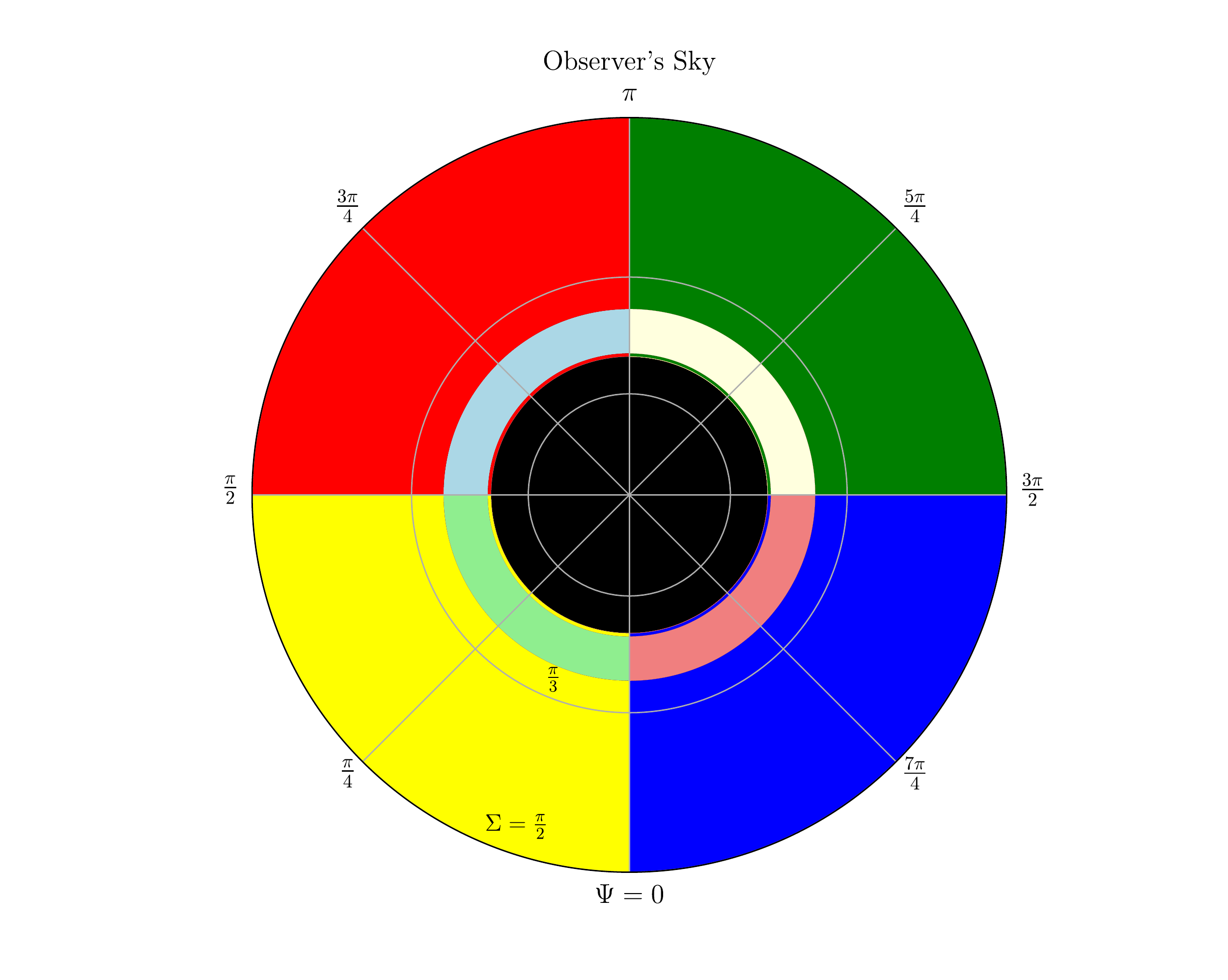} \\
    $E=\sqrt{2}$ & Light Rays\\
    \hspace{-0.5cm}\includegraphics[width=85mm]{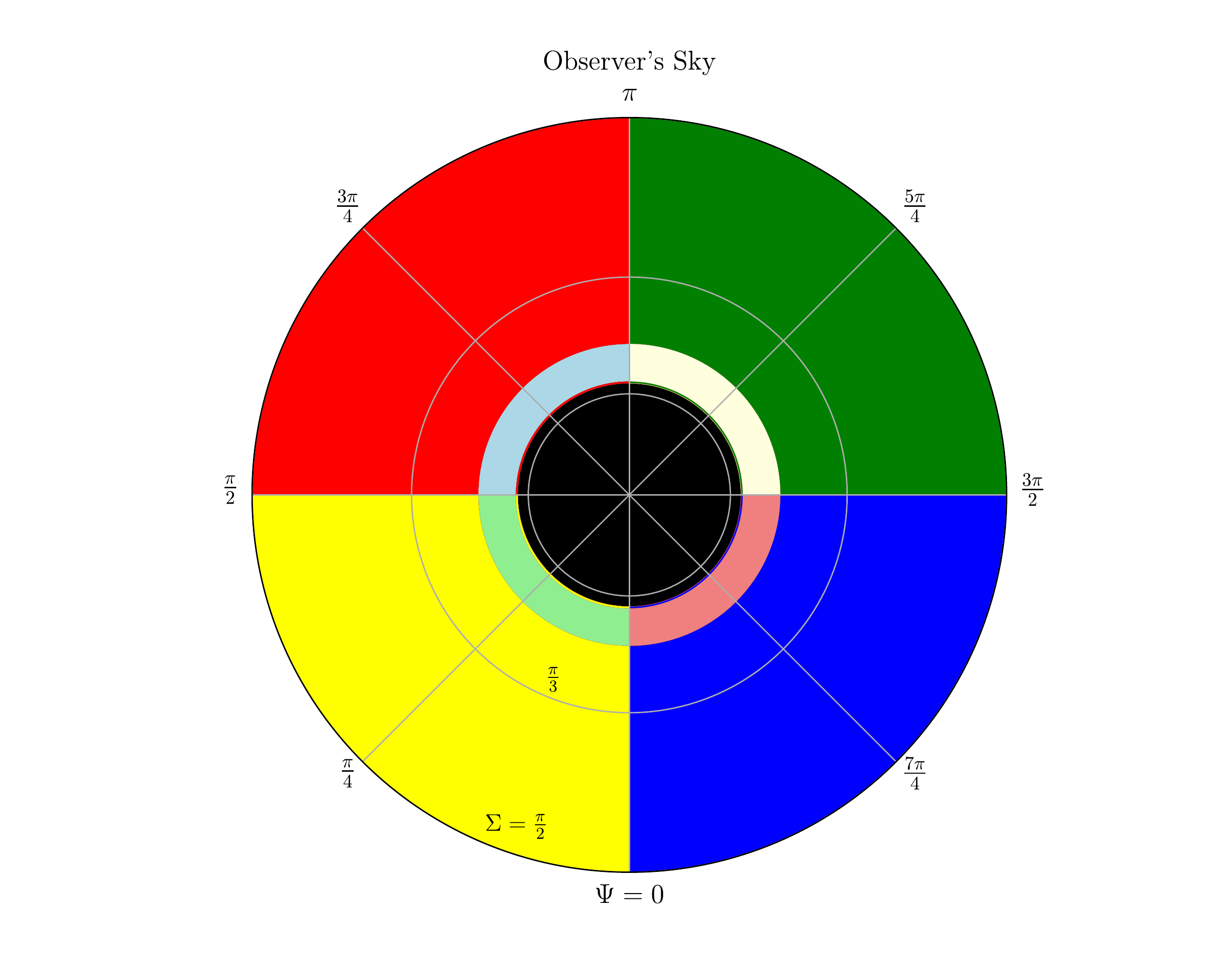} &   \includegraphics[width=85mm]{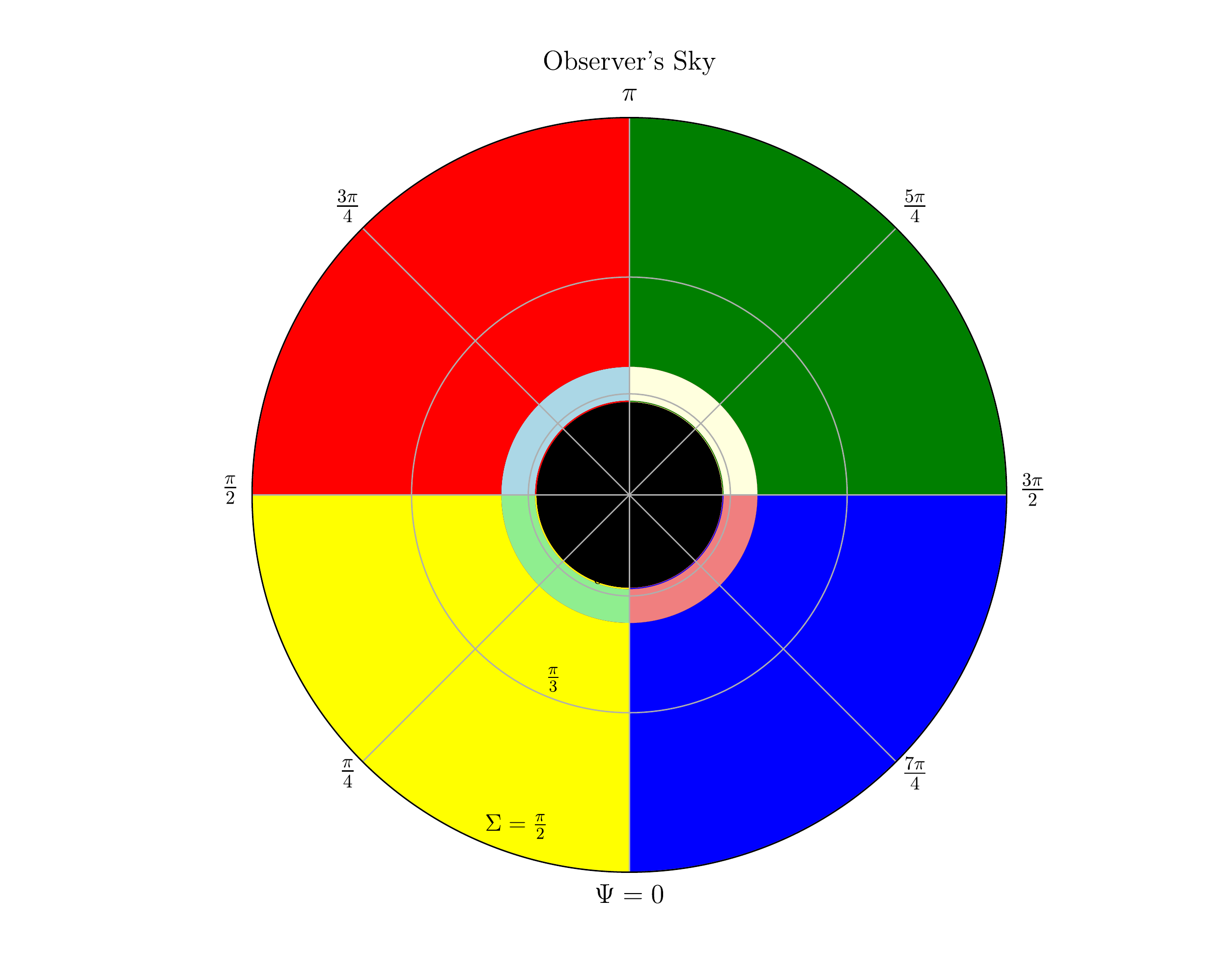} \\
  \end{tabular}
	\caption{Lens maps for massive particles and light rays in the Schwarzschild spacetime. For all four maps the observers are located at the coordinates $r_{O}=10m$ and $\vartheta_{O}=\pi/2$. The first three panels show the lens maps for massive particles (timelike geodesics) with $E=\sqrt{101/100}$ (top left), $E=\sqrt{5}/2$ (top right), and $E=\sqrt{2}$ (bottom left). The bottom right panel shows a reference lens map for light rays (lightlike geodesics). For all four plots the corresponding two-spheres of particle sources $S_{P}^2$ and the two-sphere of light sources $S_{L}^2$ are located at $r_{P}=r_{L}=15m$.}
\end{figure}
We start with discussing the lens map for the Schwarzschild metric. Figure~10 shows lens maps for the Schwarzschild metric for observers at $r_{O}=10m$ and $\vartheta_{O}=\pi/2$. The first three maps are for massive particles with $E=\sqrt{101/100}$ (top left), $E=\sqrt{5}/2$ (top right), and $E=\sqrt{2}$ (bottom left). The two-spheres of particle sources $S_{P}^2$ are located at $r_{P}=15m$. The bottom right panel shows a reference lens map for light rays emitted by light sources distributed on a two-sphere of light sources $S_{L}^2$ at $r_{L}=15m$. The observers look in the direction of the black hole. At the center of each panel we see a black area which is the shadow of the black hole (in the case of massive particles for the respective energy $E$). We start with the image in the top left panel with $E=\sqrt{101/100}$. Images of particle sources from the different quadrants on the two-sphere of particle sources are clearly separated and form rings around the shadow. At the outer boundary of the image we have images of first order, on the eastern hemisphere in green and blue and on the western hemisphere in red and yellow. Then further in we have images of second order, on the eastern hemisphere in faint yellow and faint red, and on the western hemisphere in faint blue and faint green. Further in we also have, well visible, images of third and fourth order. In addition, when we zoom in we can also see images of fifth and, in very close proximity to the shadow, very faint and hard to recognize, images of sixth order. The boundaries between the images of different orders mark the positions of the critical curves. Because of the ring structure of the images of different orders these are circular. When we increase the energy to $E=\sqrt{5}/2$ the area covered by the shadow shrinks. The overall structure remains the same but the rings with images of second, third, and higher order shift to lower latitudes. In addition, the apparent latitudinal width of the rings of images of second, third, and higher order decreases. The images of fifth order close to the shadow are now only barely visible and the images of sixth order seem to not be visible anymore (it is not clear if, close to the shadow, what might be an image of sixth order is real or a plotting effect). When we increase the energy to $E=\sqrt{2}$ the shadow shrinks further. Again the overall structure remains the same. The rings of images of second, third, and higher order shift to lower latitudes and the apparent latitudinal width of the rings also appears to shrink; however, the differences are not as pronounced as before. In addition, images of fifth order close to the shadow are now barely visible or, at least on three of the four quadrants, not visible at all (note that the latter is very likely a plotting effect since the images were created with the same latitude range for all four quadrants). When we now compare the lens maps for massive particles with the lens map for light rays, we immediately see that the overall structure is the same. The only difference is that for lightlike geodesics the shadow is smaller and the images of second, third, and higher order can be found at lower latitudes. In addition, images of fifth order or higher are much more difficult to recognize than for massive particles.\\
\begin{figure}\label{fig:LENUTMetric}
  \begin{tabular}{cc}
    $E=\sqrt{101/100}$&$E=\sqrt{5}/2$\\
    \hspace{-0.5cm}\includegraphics[width=85mm]{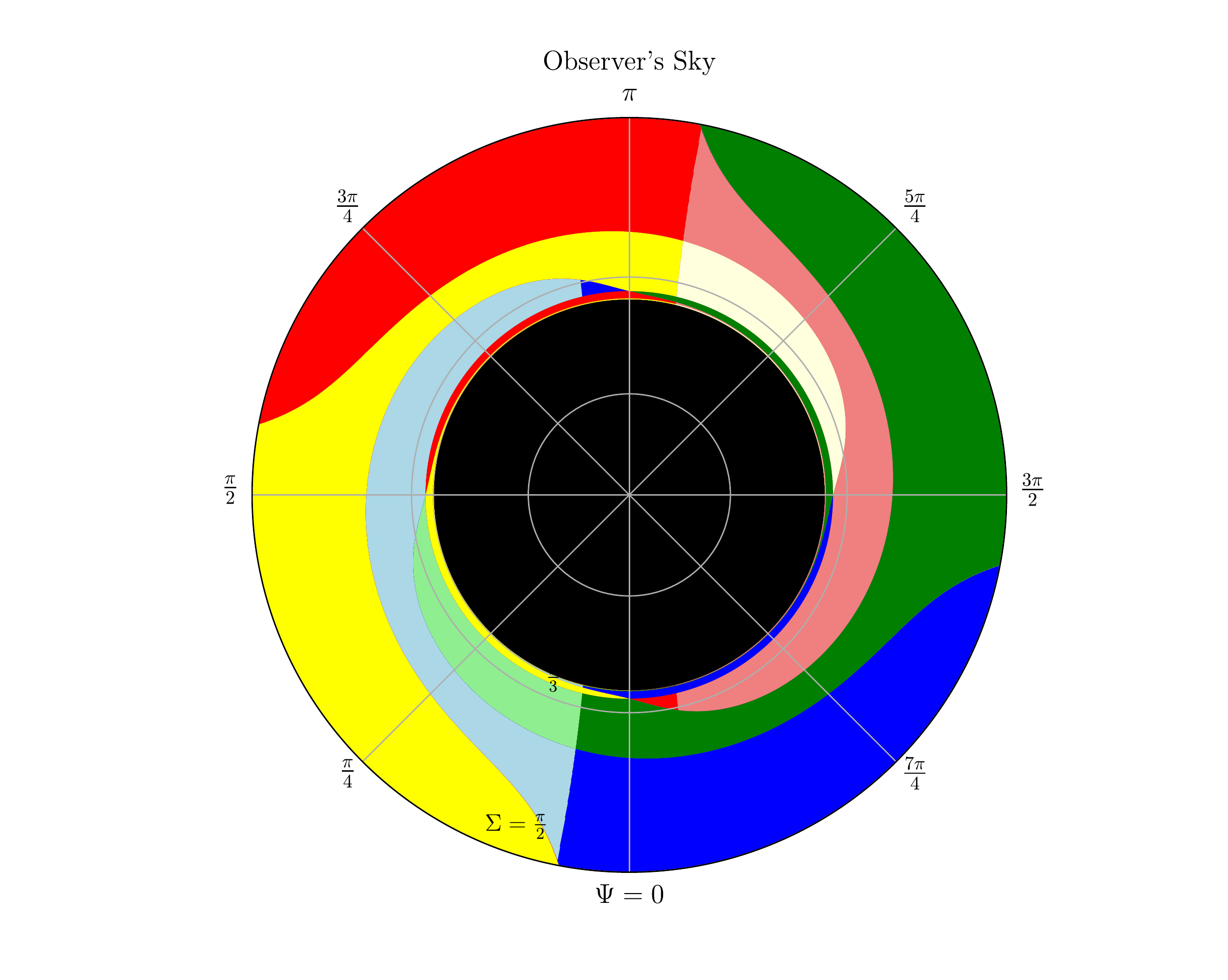} &   \includegraphics[width=85mm]{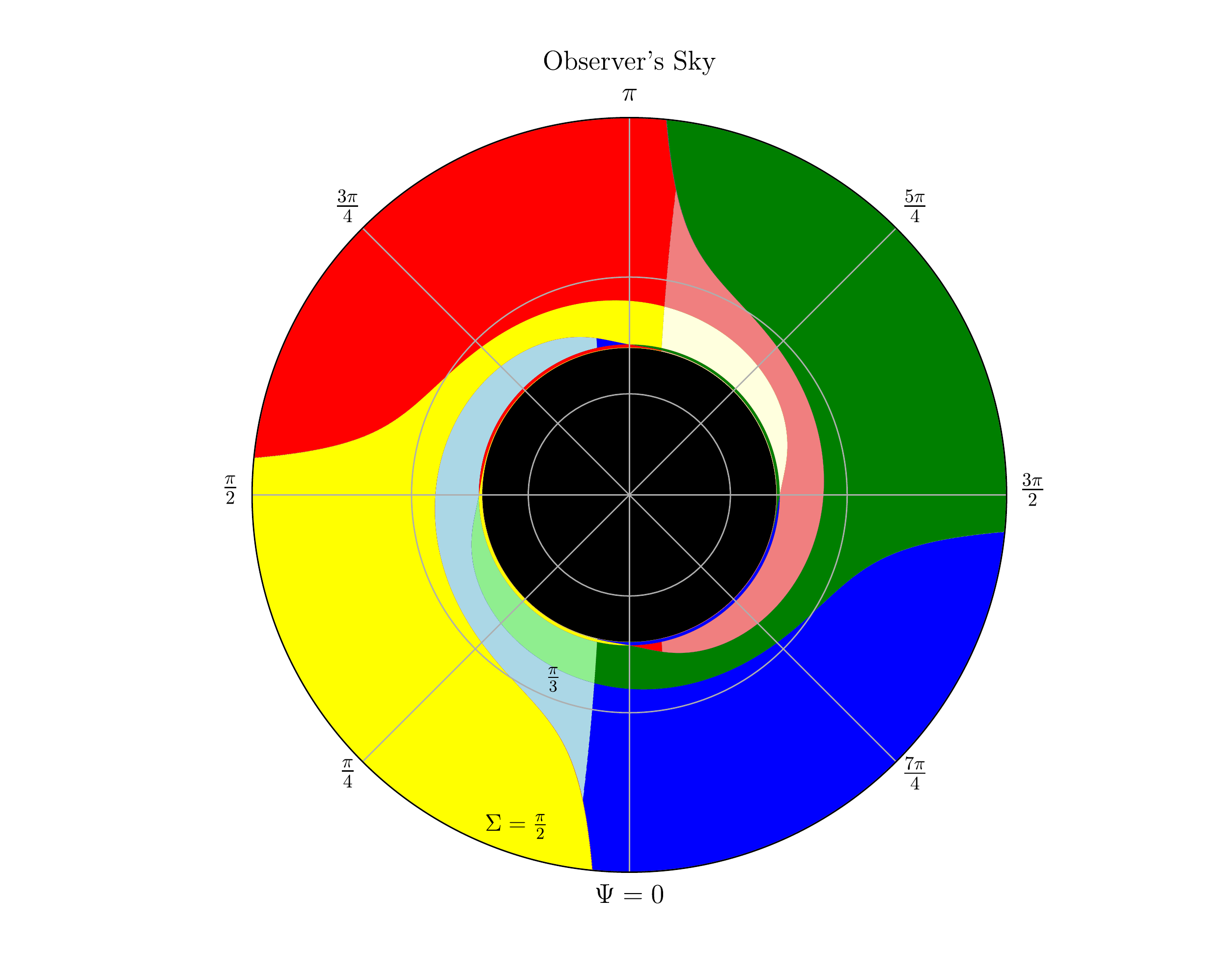} \\
    $E=\sqrt{2}$ & Light Rays\\
    \hspace{-0.5cm}\includegraphics[width=85mm]{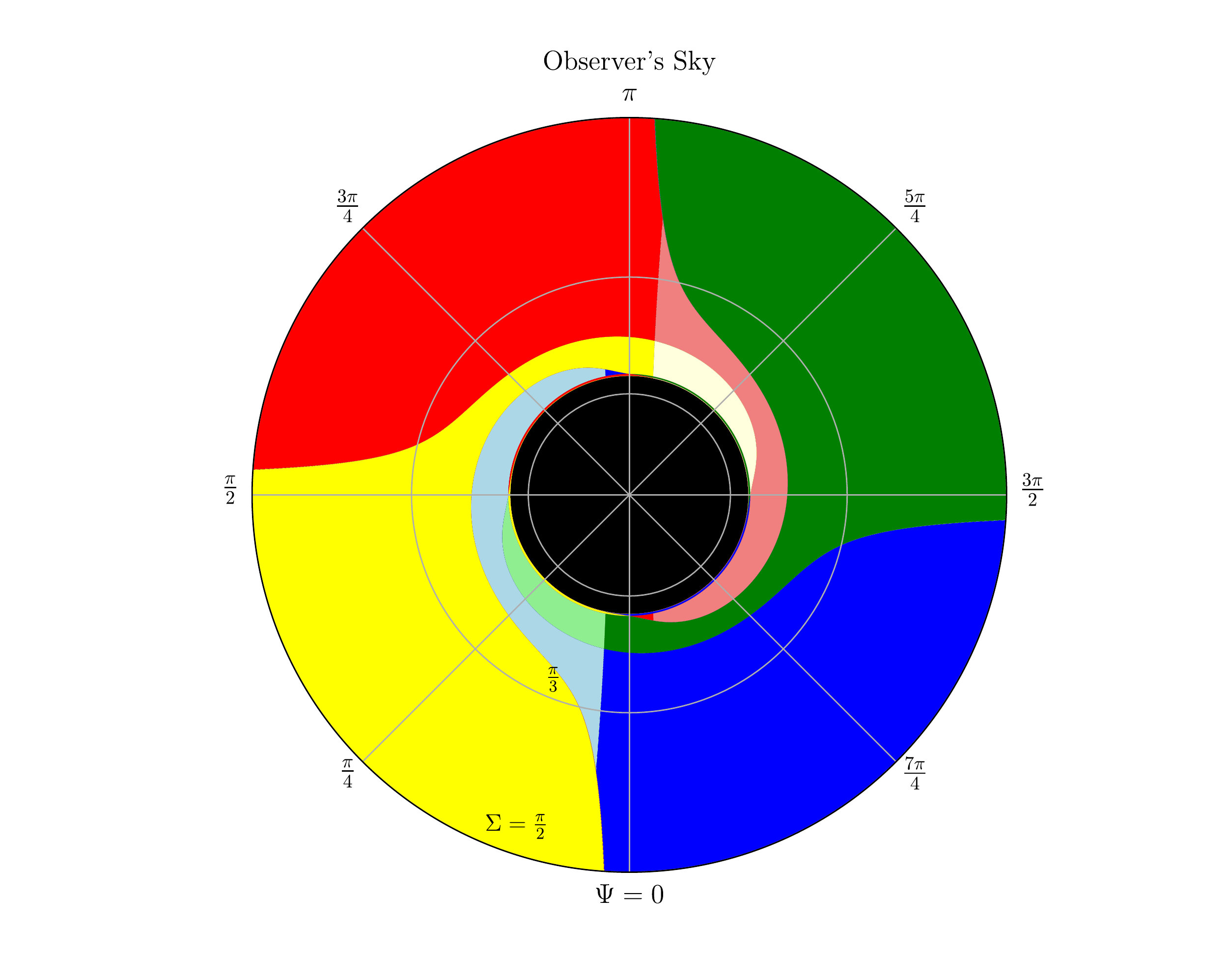} &   \includegraphics[width=85mm]{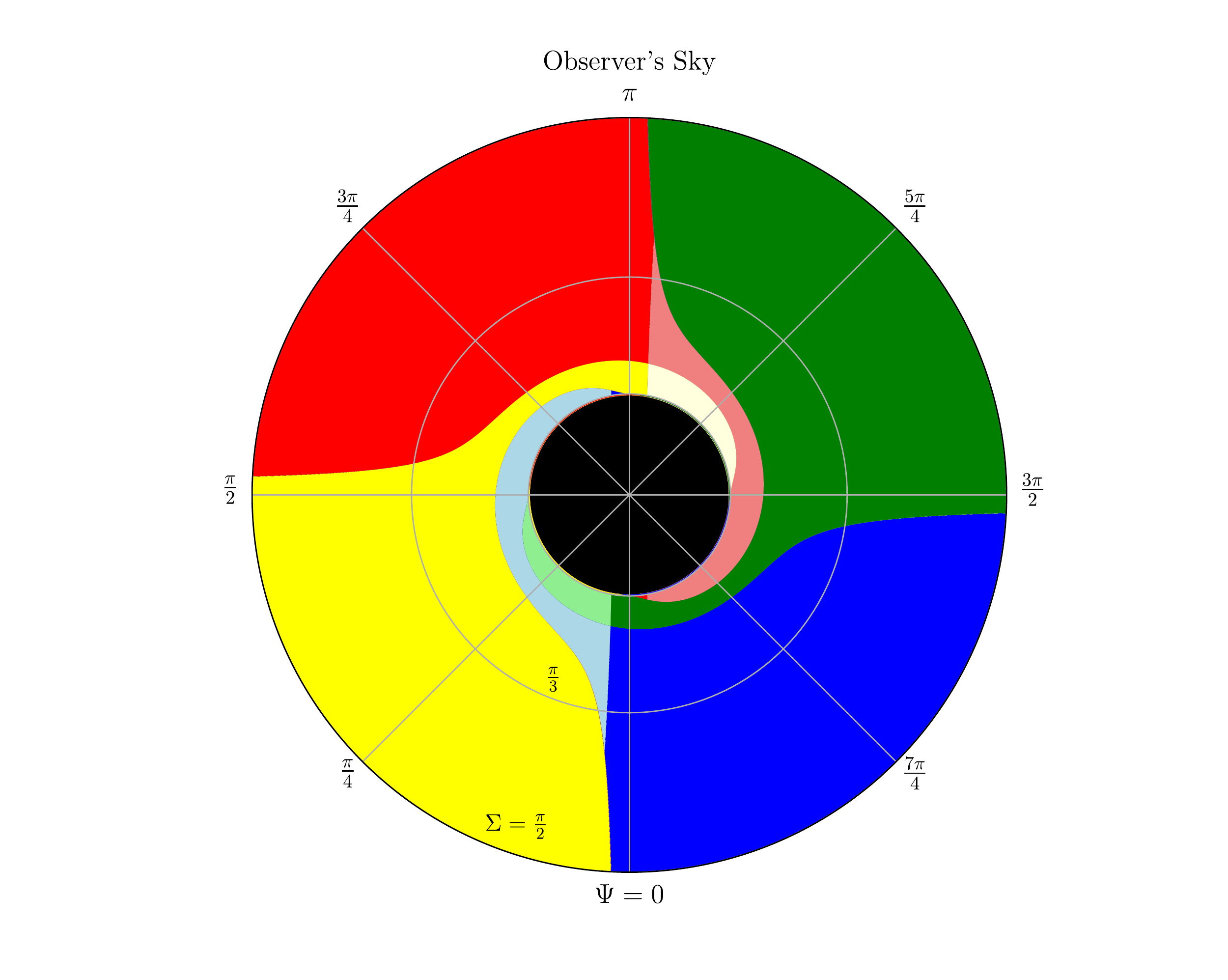} \\
  \end{tabular}
	\caption{Lens maps for massive particles and light rays for the NUT spacetime with $n=m/2$. For all four maps the observers are located at the coordinates $r_{O}=10m$ and $\vartheta_{O}=\pi/2$. The first three panels show the lens maps for massive particles (timelike geodesics) with $E=\sqrt{101/100}$ (top left), $E=\sqrt{5}/2$ (top right), and $E=\sqrt{2}$ (bottom left). The bottom right panel shows a reference lens map for light rays (lightlike geodesics). For all four plots the corresponding two-spheres of particle sources $S_{P}^2$ and the two-sphere of light sources $S_{L}^2$ are located at $r_{P}=r_{L}=15m$. The Manko-Ruiz parameter is $C=1$ and thus we have a Misner string at $\vartheta=0$.}
\end{figure}
Now let us turn to the NUT metric. Figure~11 shows lens maps for the NUT metric with $n=m/2$ for observers located at $r_{O}=10m$ and $\vartheta_{O}=\pi/2$. The Manko-Ruiz parameter is $C=1$ and thus the Misner string is located at $\vartheta=0$. Again the first three maps are for massive particles with $E=\sqrt{101/100}$ (top left), $E=\sqrt{5}/2$ (top right), and $E=\sqrt{2}$ (bottom left). For these maps the corresponding two-spheres of particle sources $S_{P}^2$ are located at $r_{P}=15m$. The bottom right panel shows a reference lens map for light rays emitted by light sources distributed on a two-sphere $S_{L}^2$ at $r_{L}=15m$.\\
Again the black circles at the centers of the maps are the shadows of the black holes. As we can easily see the lens maps for massive particles with fixed energy $E$ and for light rays show the same features. When we turn on the gravitomagnetic charge $n$ and compare the top left panel of Fig.~11 with the top left panel of Fig.~10 ($E=\sqrt{101/100}$) we can easily see that the formerly disconnected patches with images of first and second order from the same quadrants on the two-sphere of particle sources connect resulting in a twist in the lens map. In addition, we see that close to the meridian and the antimeridian images of first and second order from the same quadrant on the two-sphere of particle sources are separated by a sharp line. The longer lines east of the antimeridian and west of the meridian mark timelike geodesics, which cross the axes at $\vartheta=0$ (this is the position of the Misner string as we have $C=1$) and $\vartheta=\pi$, respectively. As already discussed in Frost \cite{Frost2022}, although we see a sharp transition between images of first and second order, across the transition the $\varphi$ coordinate is indeed continuous (in another context this was also discussed in Cl\'{e}ment \emph{et al.} \cite{Clement2015}).\\
West of the antimeridian and east of the meridian we can find regions with images of odd order. As already discussed in Frost \cite{Frost2022} for lightlike geodesics these are formally images of first order; however, they are rather special images. The particles associated with these timelike geodesics move on cones not enclosing one of the axes. Thus for these particles the direction of the $\varphi$ motion reverses. Closer to the shadow we can also see images of third and fourth as well as, when we zoom in, barely visible images of fifth and sixth order (in Fig.~11 these are rather difficult to distinguish). Images of first and second order are separated from the images of third and fourth order by a circular boundary (the same seems to hold for the boundary between images of third and fourth order and images of fifth and sixth order). As discussed for lightlike geodesics in Frost \cite{Frost2022} the circular boundaries between images of first and second and images of third and fourth order as well as images of third and fourth and images of fifth and sixth order are very likely indicating the positions of critical curves. This is also supported by the fact that the spacetime retains an $SO(3,\mathbb{R})$ symmetry and therefore it is very likely that also for timelike geodesics the critical curves are still circular. However, as for light rays from the lens map alone it is rather difficult to determine the critical curves between images of first and second order, between images of third and fourth order, and between images of fifth and sixth order. When we increase the energy to $E=\sqrt{5}/2$ (top right panel) the shadow shrinks and all lens map features shift to lower latitudes. As already observed for the Schwarzschild metric the apparent latitudinal width of the single features decreases. In addition, the lines marking the axes crossings shift to longitudes closer to the meridian and the antimeridian. When we increase the energy to $E=\sqrt{2}$ the effects become even more pronounced. As already noted in Frost \cite{Frost2022} the strength of the twist serves as a good indicator for the magnitude of the gravitomagnetic charge; however, because it is very difficult to construct a full lens map for massive particles it will be very difficult to observe.\\
Figure~12 shows lens maps for the Reissner-Nordstr\"{o}m metric (left column) and the charged NUT metric with $n=m/2$ (right column) for $e=m/2$ (top row), $e=3m/4$ (middle row), and $e=m$ (bottom row) for massive particles with $E=\sqrt{5}/2$ for the same particle source-observer geometry as Figs.~10 and 11. For the charged NUT metric the Manko-Ruiz parameter is $C=1$ and thus the Misner string is located at $\vartheta=0$. When we compare the top right panel of Fig.~10 with the top left panel of Fig.~12 we immediately see that turning on the electric charge has the effect that the particle shadow shrinks and the rings with images of second, third, and higher order shift to lower latitudes. When we now increase the electric charge first to $e=3m/4$ and then to $e=m$ the particle shadow continues to shrink and the rings with images of second, third, and higher order shift to lower latitudes. However, we also observe that with increasing electric charge the rings of images of second, third, and higher order appear to broaden in latitudinal direction. This effect was already observed for light rays in Frost \cite{Frost2022b} and, although this is not surprising, our results show that it also occurs for massive particles. When we now turn on the gravitomagnetic charge the lens maps show the same basic features as for the NUT metric. In addition, when we increase the electric charge from $e=m/2$ to $e=m$ we observe the same latitudinal broadening effect for the features around the shadow which we already observed for the Reissner-Nordstr\"{o}m metric.\\
So far we only discussed the lens maps for massive particles for distinct energies $E$ separately. Now the interesting question is what happens when we consider a spectrum of particles following an energy distribution. Let us first discuss this for the spherically symmetric and static Schwarzschild and Reissner-Nordstr\"{o}m spacetimes. Let us assume for now that we only have a single particle source (point source) on the sphere of particle sources. In this case for a distinct energy $E$ we have a very well-defined image on the observer's sky. The image is simply a dot. When we increase the energy $E$ this dot shifts toward lower latitudes and thus when we combine the images for different energies $E$ the dots form a line along a constant longitude and the innermost end of this line is marked by an image generated by light rays (or when we have a source that also emits gravitational waves and a supermassive black hole as lens we will detect a gravitational wave signal from this direction). Along this line the maximal particle flux can be found for the image which is associated with the particle energy at which the particle density distribution has its maximum. The same effect also occurs for images of higher order just that the overall particle flux at the position of these images is lower. When we turn on the gravitomagnetic charge and consider the (charged) NUT metric the overall pattern for images of point sources is the same as for spherically symmetric and static spacetimes. The images of particles with low energy can be found at high latitudes while images of particles with high energy can be found at lower latitudes. The main difference will come from the twist. Since the gravitomagnetic charge leads to a twist in the lens maps we can assume that when we superpose the images for particles with different energies $E$ emitted by the same point source we will observe a similar effect on the combined image. The images will stretch along different longitudes and most likely have an arcletlike shape. Note that a line stretching along different latitudes and longitudes is also possible but with view on the overall changes due to the gravitomagnetic charge highly unlikely. Besides the general lensing pattern for particles with constant energies the exact shape of this arclet may also be a good indicator for the strength of the gravitomagnetic charge. \\
In addition, when we superpose the lens maps for an energy spectrum of particles we observe another effect. For illustrating this effect let us take the example of three particle energies $E_{1}=\sqrt{101/100}<E_{2}=\sqrt{5}/2<E_{3}=\sqrt{2}$ as shown in Figs.~10 and 11 for the Schwarzschild metric and the NUT metric. When we compare the three lens maps for massive particles for the Schwarzschild metric we immediately see that in general the boundaries between images of successive orders (e.g., order one and order two) of particle sources on different quadrants on the two-sphere of particle sources $S_{P}^2$ overlap. In addition, when the maximal particle energy is high enough images of third order generated by particles with low energies can overlap with images of first order generated by particles with higher energies, both emitted by particle sources on the same quadrant on the two-sphere of particle sources $S_{P}^2$. Under very special circumstances these can even be images of the same particle source. For the NUT metric we observe a similar pattern. However, because of the twist it is rather unlikely that images of third and first order generated by low- and high-energy particles emitted by the same particle source, respectively, overlap.

\begin{figure}\label{fig:LERNCNUTMetric}
  \begin{tabular}{cc}
    Reissner-Nordstr\"{o}m Metric $e=m/2$& Charged NUT Metric $e=m/2$\\
    \hspace{-0.5cm}\includegraphics[width=85mm]{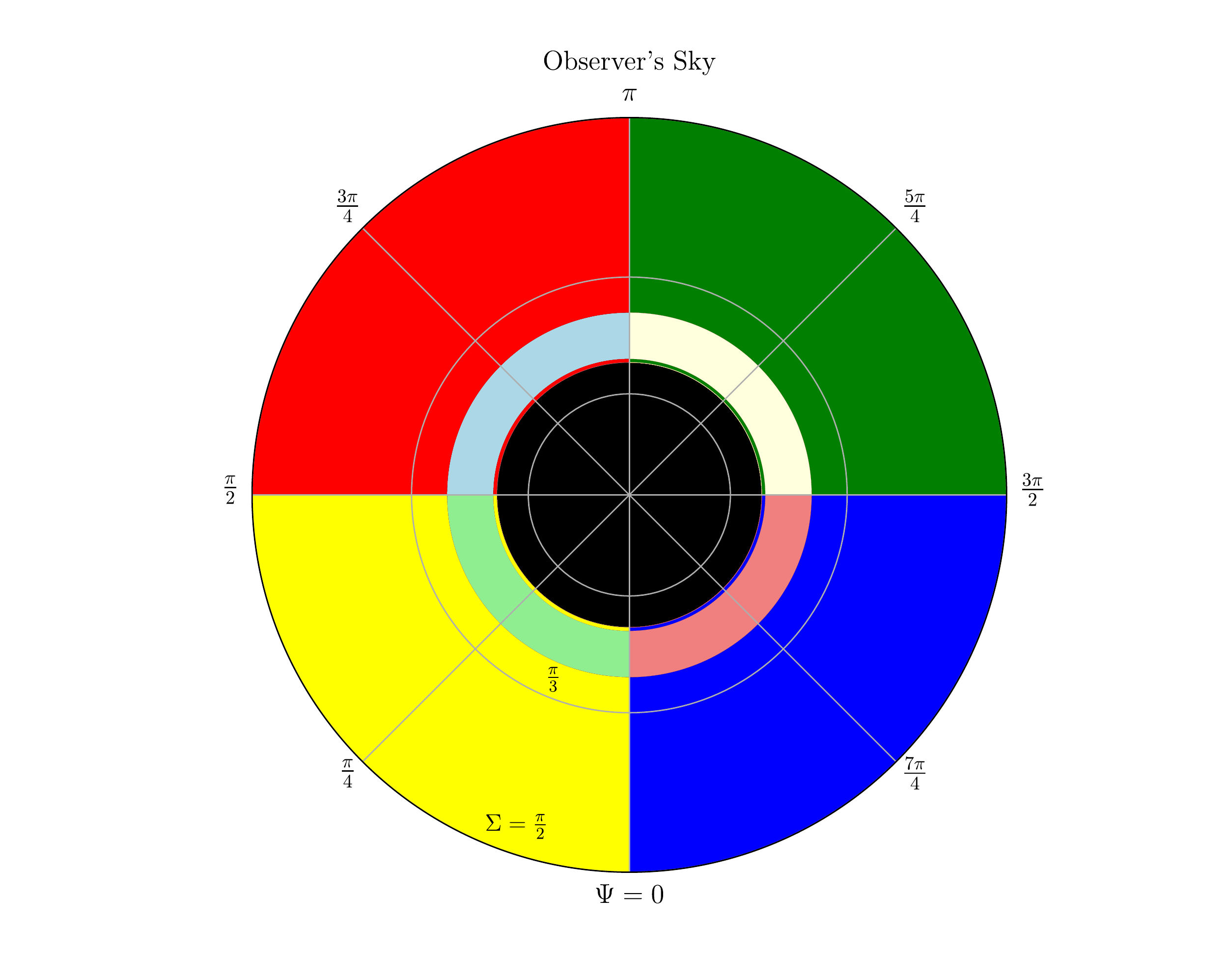} &   \includegraphics[width=85mm]{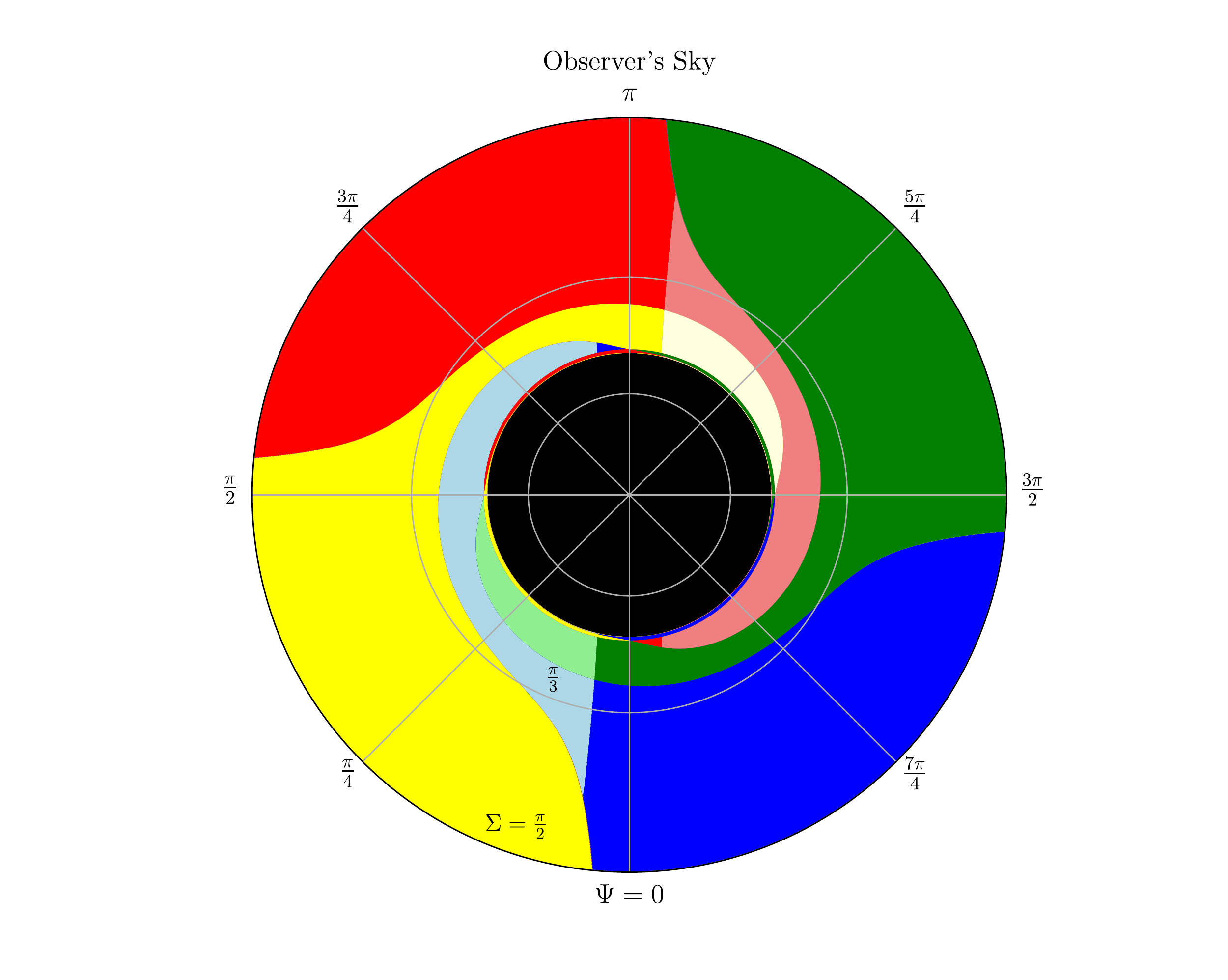} \\
    Reissner-Nordstr\"{o}m Metric $e=3m/4$& Charged NUT Metric $e=3m/4$\\
    \hspace{-0.5cm}\includegraphics[width=85mm]{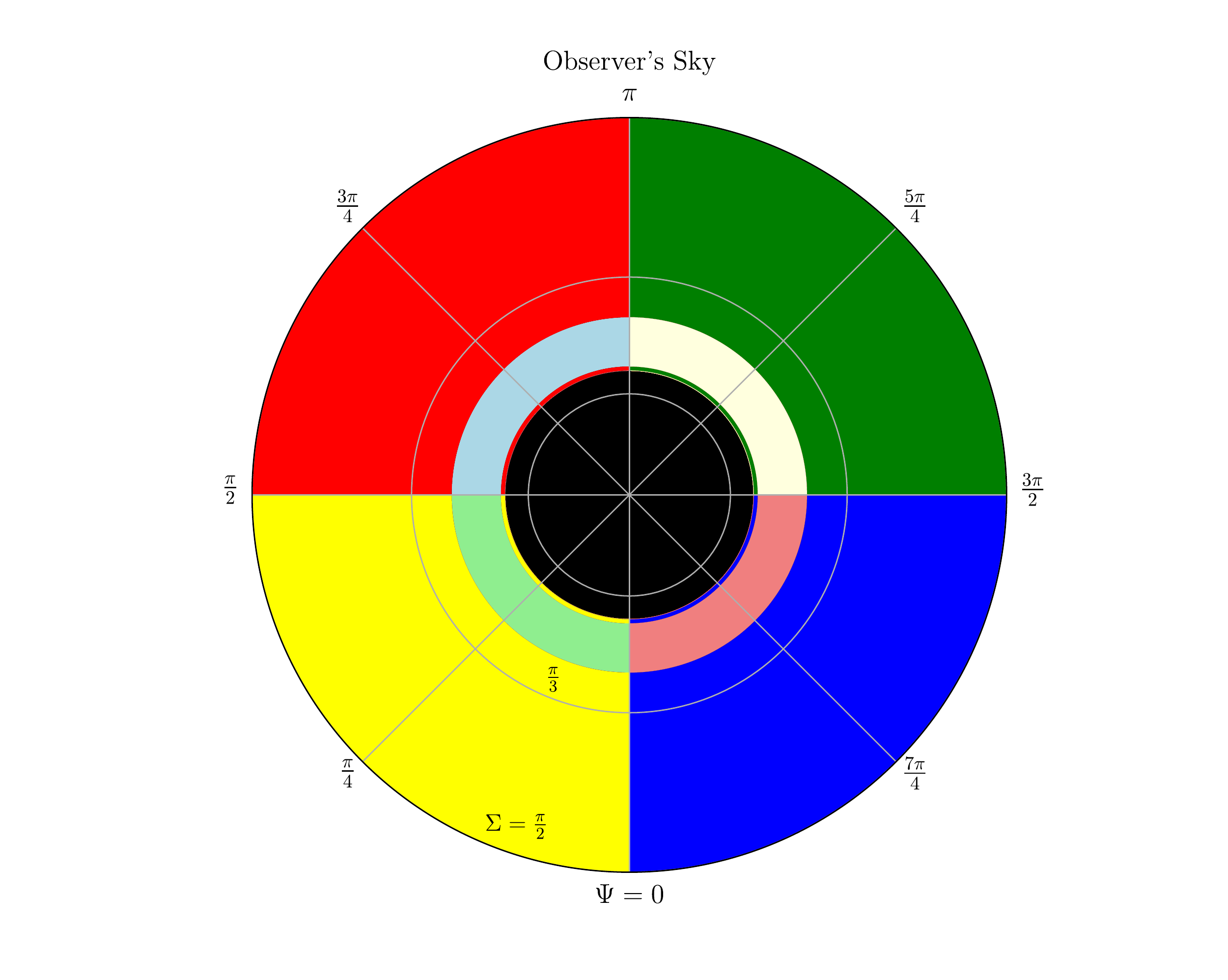} &   \includegraphics[width=85mm]{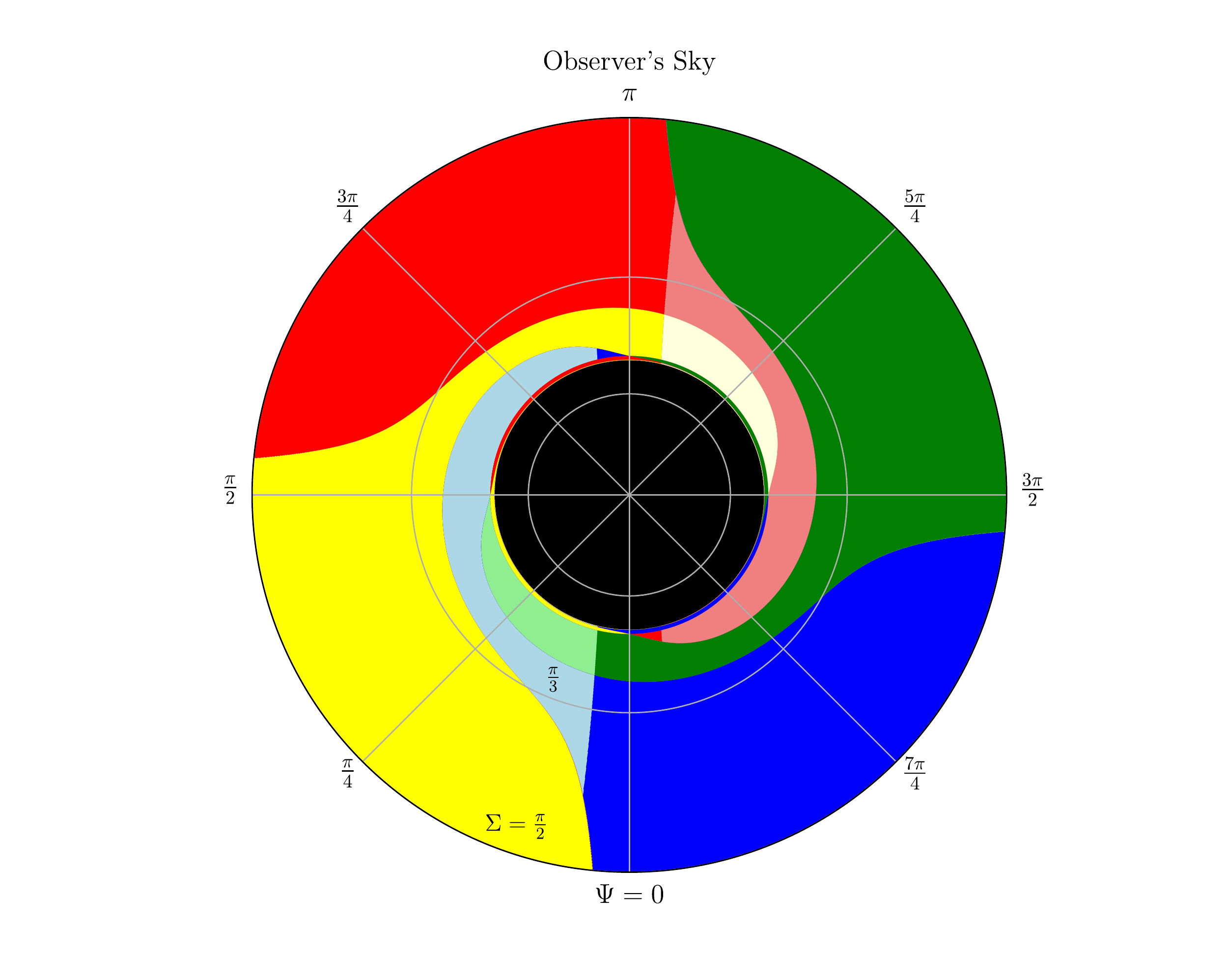} \\
    Reissner-Nordstr\"{o}m Metric $e=m$& Charged NUT Metric $e=m$\\
    \hspace{-0.5cm}\includegraphics[width=85mm]{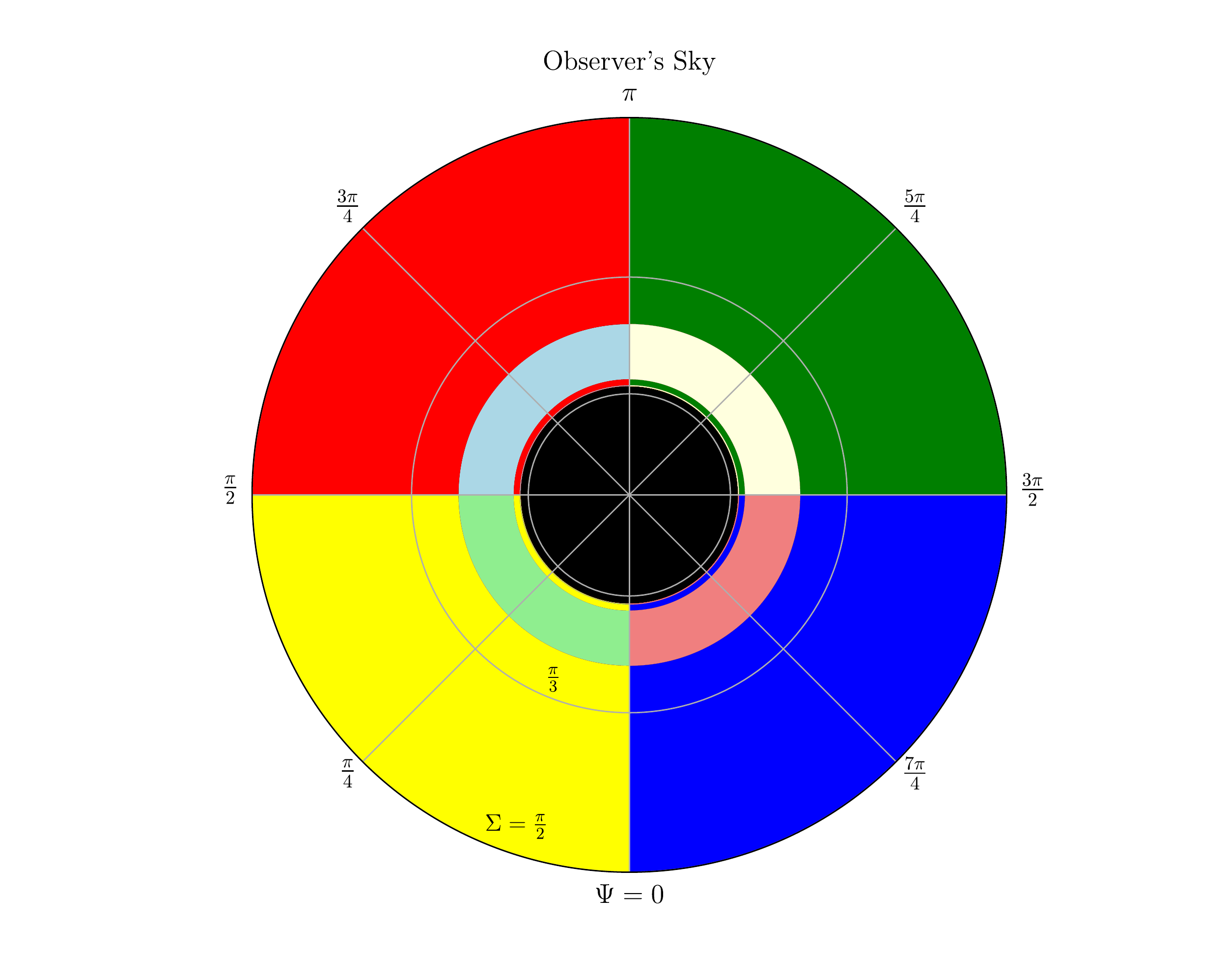} &   \includegraphics[width=85mm]{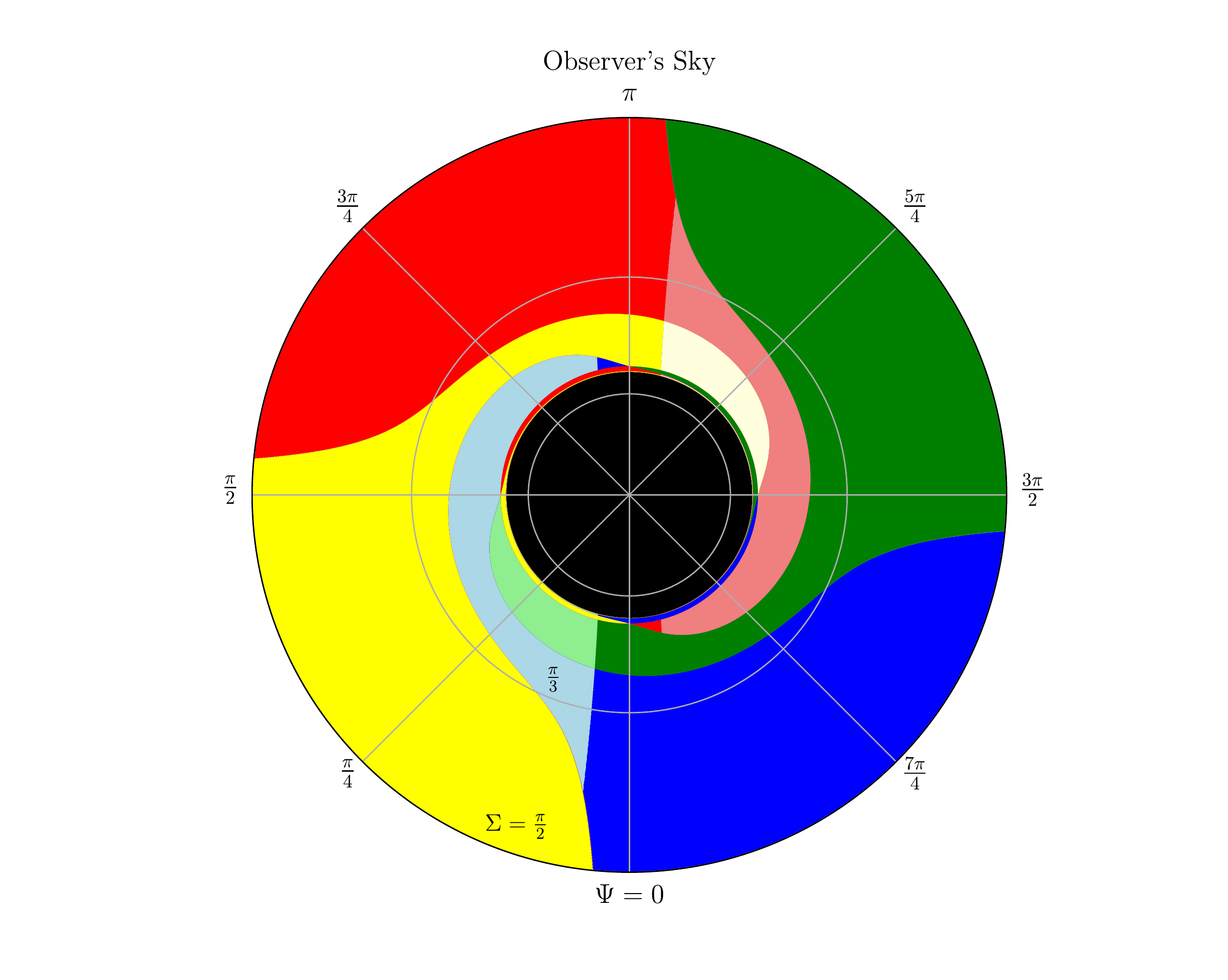} \\
  \end{tabular}
	\caption{Lens maps for massive particles with $E=\sqrt{5}/2$ in the Reissner-Nordstr\"{o}m spacetime (left column) and the charged NUT spacetime with $n=m/2$ (right column) for $e=m/2$ (top row), $e=3m/4$ (middle row), and $e=m$ (bottom row). For the charged NUT spacetime the Manko-Ruiz parameter is $C=1$ and thus the Misner string is located at $\vartheta=0$. For all six maps the observers are located at the coordinates $r_{O}=10m$ and $\vartheta_{O}=\pi/2$ and the two-spheres of particle sources $S_{P}^2$ are located at $r_{P}=15m$.}
\end{figure}

\subsection{Travel time measures}\label{Sec:TravTime}
In this subsection we will discuss two different travel time measures for gravitationally lensed massive particles. The first is derived from the time coordinate $t$ and the second is derived from the proper time $\tau$. Both measure the time a particle needs to travel from the particle source to an observer with a detector. For distinguishing between them we will refer to the former as \emph{travel time} using the standard terminology from gravitational lensing of light rays. The second travel time measure is derived from the proper time and thus we will refer to it as \emph{traveled proper time} in the following.

\subsubsection{Travel time}
The travel time measures in terms of the time coordinate $t$ the time a particle needs to travel from the particle source to an observer with a detector. In terms of the time coordinate $t_{P}$ at which the particle is emitted by the particle source and the time coordinate $t_{O}$ at which it is detected by the observer with the detector it reads
\begin{equation}
T(\Sigma,\Psi)=t_{O}-t_{P}(\Sigma,\Psi).
\end{equation}
For the explicit calculation of the travel time we now assume that $t(\lambda_{O})=t_{O}=0$. We insert the relations between the constants of motion $L_{z}$ and $K$ and the coordinates on the celestial sphere of the observer (\ref{eq:EoMCelesLz}) and (\ref{eq:EoMCelesK}) into (\ref{eq:ttheta1}) and (\ref{eq:tr2}). We evaluate both integrals and insert them into (\ref{eq:timecoordinate}) and get
\begin{align}
&T(\Sigma,\Psi)=\int_{r_{O}...}^{...r_{P}}\frac{\rho(r')^2E\text{d}r'}{Q(r')\sqrt{\rho(r')^2E^2-\rho(r')Q(r')-Q(r')\frac{\rho(r_{O})(\rho(r_{O})E^2-Q(r_{O}))}{Q(r_{O})}\sin^2\Sigma}}\\
&-2n\int_{0}^{\lambda_{P}}\frac{(\cos\vartheta(\lambda')+C)\left(\sqrt{\frac{\rho(r_{O})(\rho(r_{O})E^2-Q(r_{O}))}{Q(r_{O})}}\sin\vartheta_{O}\sin\Sigma\sin\Psi+2n(\cos\vartheta(\lambda')-\cos\vartheta_{O})E\right)\text{d}\lambda'}{1-\cos^2\vartheta(\lambda')},\nonumber
\end{align}
where the dots in the limits of the first term indicate that we have to split the integral into two terms when we have a turning point for $r$ and that we have to evaluate each term separately. The sign of the root in the denominator has to be chosen according to the direction of the $r$ motion. Note that $\lambda_{P}$ also depends on the celestial latitude. We only dropped it for brevity. In the case of the Schwarzschild spacetime and the Reissner-Nordstr\"{o}m spacetime the second term vanishes and we only have to evaluate the first term. We now evaluate the travel time integrals as described in Sec.~\ref{Sec:EoMSolt} using (\ref{eq:Minor}) for calculating the Mino parameter $\lambda_{P}$.\\
\begin{figure}\label{fig:TTSchwarzschild}
  \begin{tabular}{cc}
    $E=\sqrt{101/100}$&$E=\sqrt{5}/2$\\
    \hspace{-0.5cm}\includegraphics[width=95mm]{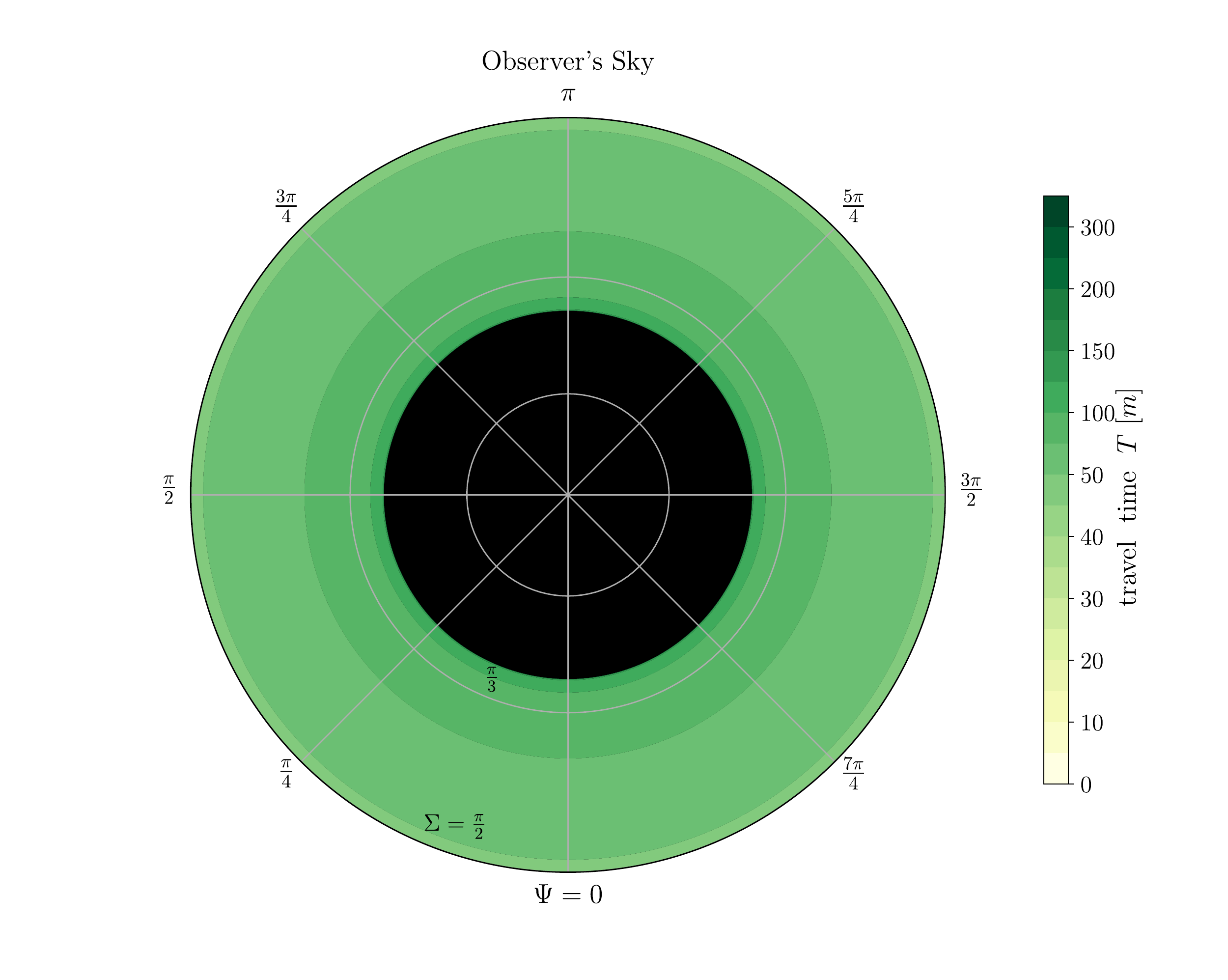} &   \includegraphics[width=95mm]{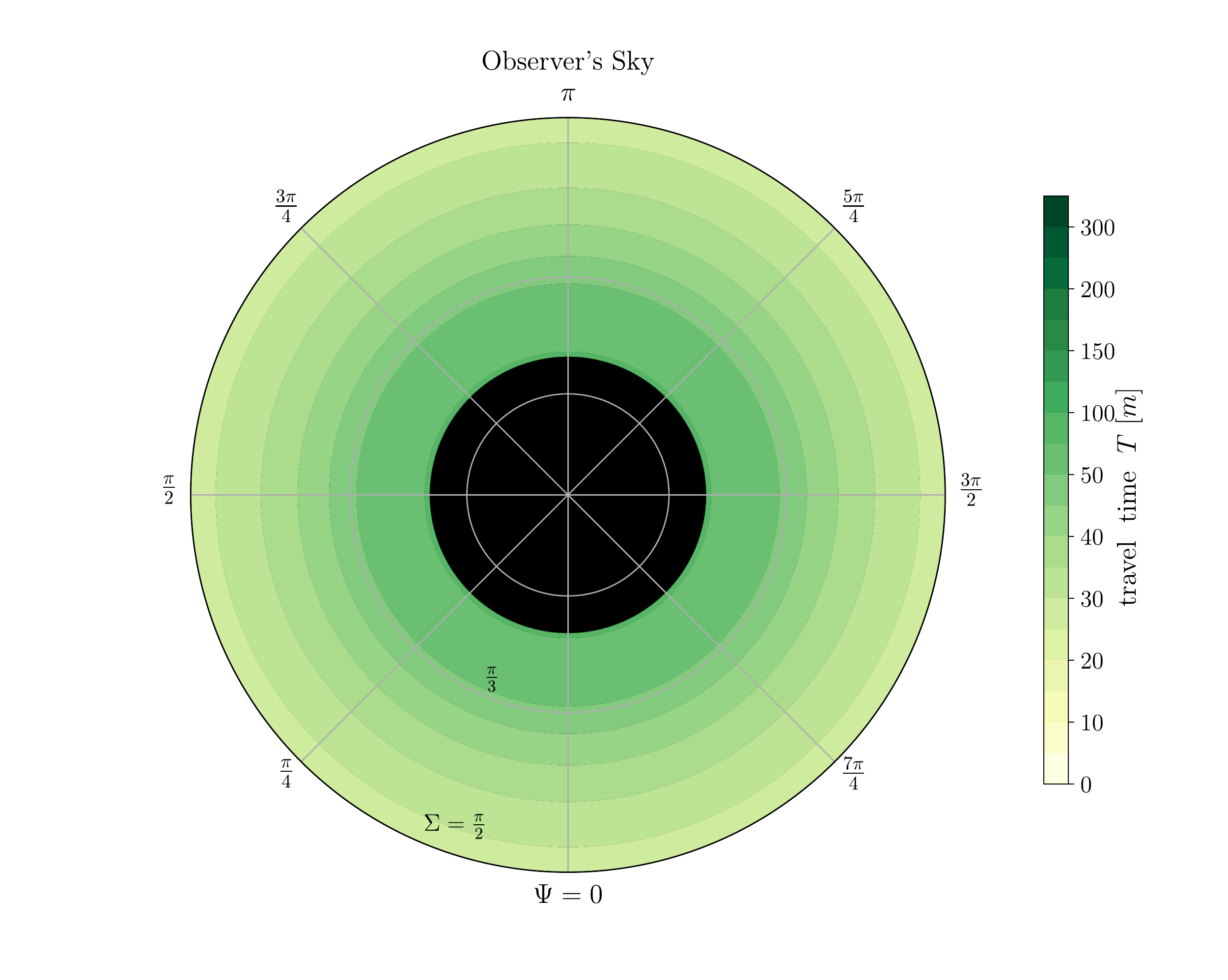} \\
    $E=\sqrt{2}$ & Light Rays\\
    \hspace{-0.5cm}\includegraphics[width=95mm]{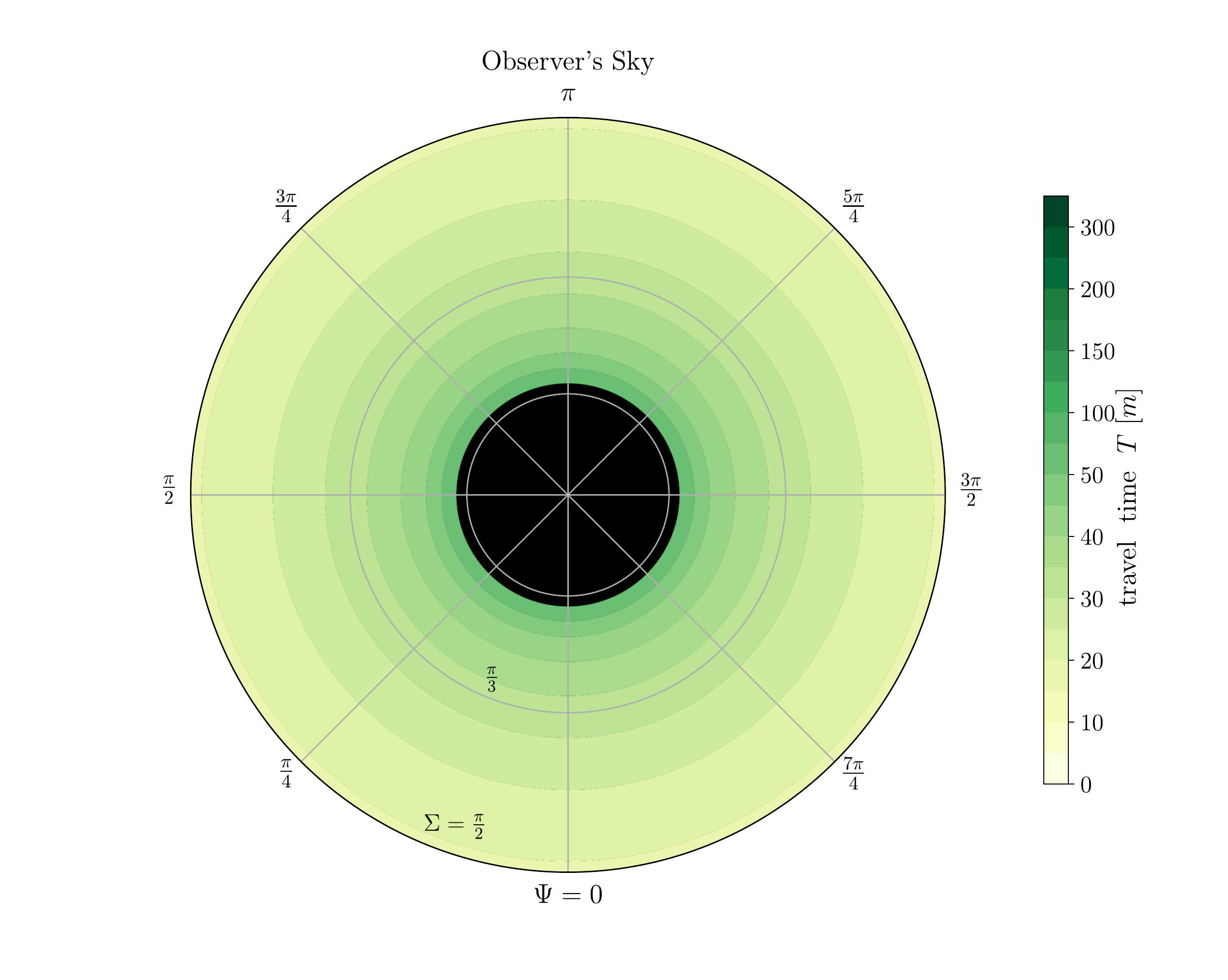} &   \includegraphics[width=95mm]{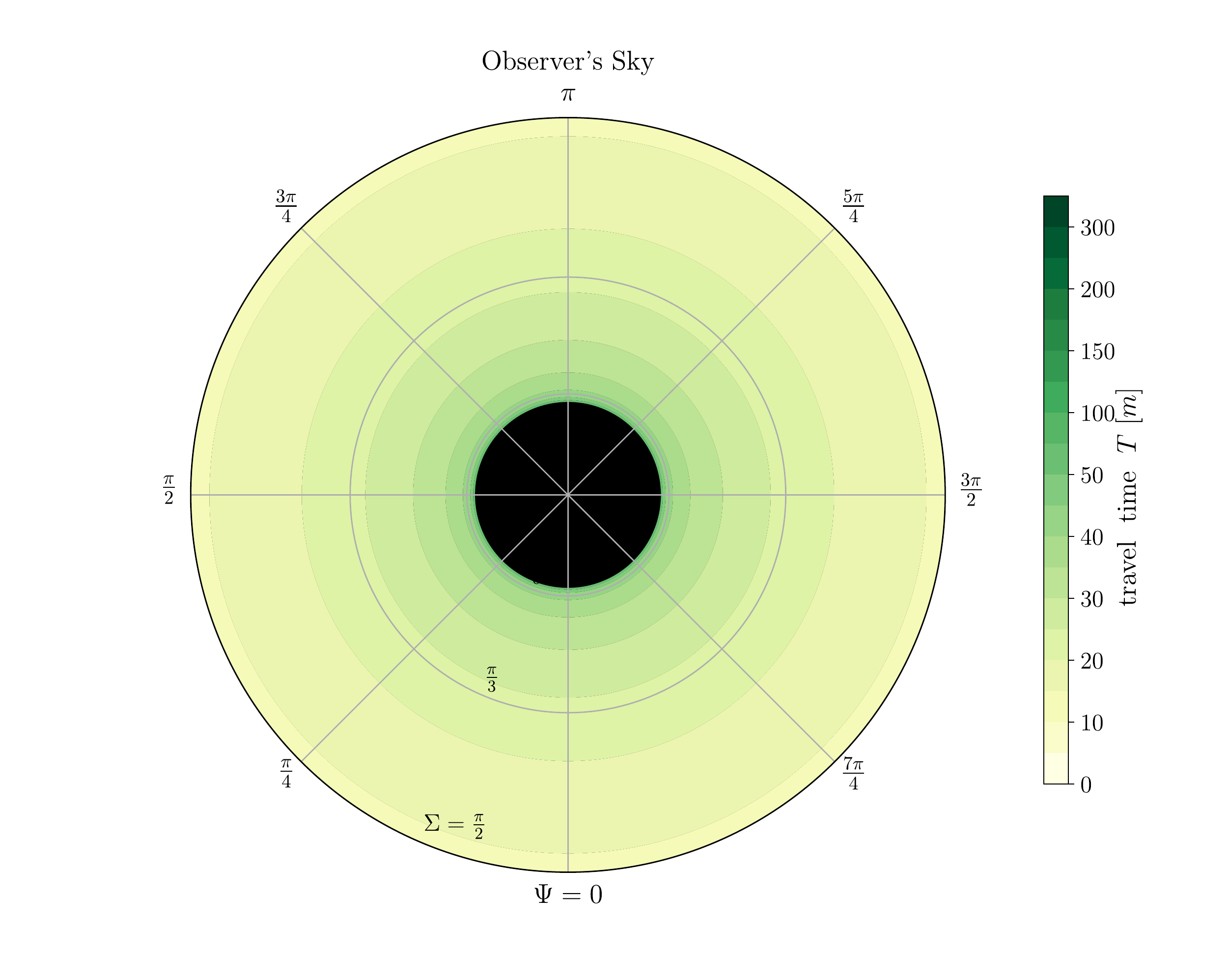} \\
  \end{tabular}
	\caption{Travel time maps for massive particles and light rays in the Schwarzschild spacetime. For all four maps the observers are located at the coordinates $r_{O}=10m$ and $\vartheta_{O}=\pi/2$. The first three panels show travel time maps for massive particles (timelike geodesics) with $E=\sqrt{101/100}$ (top left), $E=\sqrt{5}/2$ (top right), and $E=\sqrt{2}$ (bottom left). The bottom right panel shows a reference travel time map for light rays (lightlike geodesics). For all four plots the corresponding two-spheres of particle sources $S_{P}^2$ and the two-sphere of light sources $S_{L}^2$ are located at $r_{P}=r_{L}=15m$.}
\end{figure}
Figure~13 shows travel time maps for massive particles with $E=\sqrt{101/100}$ (top left panel), $E=\sqrt{5}/2$ (top right panel), and $E=\sqrt{2}$ (bottom left panel) as well as a reference travel time map for light rays (bottom right panel) on the observer's celestial sphere for the Schwarzschild metric. For all four maps the observers are located at the coordinates $r_{O}=10m$ and $\vartheta_{O}=\pi/2$. The two-spheres of particle sources $S_{P}^2$ and the two-sphere of light sources $S_{L}^2$ are located at $r_{P}=r_{L}=15m$. Again the observers look into the direction of the black hole and the black circles at the centers of the maps are the shadows of the black holes. Let us start our discussion with the travel time map for massive particles with $E=\sqrt{101/100}$. Levels of constant travel time form circles around the shadow. The travel time increases with decreasing latitude and diverges for timelike geodesics asymptotically coming from the particle sphere. When we compare the travel time of massive particles with the travel time for light rays (bottom right panel) we see that it is significantly longer. When we increase the particle energy to $E=\sqrt{5}/2$ (top right panel) the shadow shrinks and the travel time decreases because the particles have a higher velocity. Overall levels of constant travel time shift to lower latitudes. This effect is even more pronounced when we increase the energy of the particles to $E=\sqrt{2}$. In addition, with increasing particle energy the travel time approaches the travel time of light rays.\\
\begin{figure}\label{fig:TTNUTMetric}
  \begin{tabular}{cc}
    $\vartheta_{O}=\pi/2$ and $E=\sqrt{101/100}$ & $\vartheta_{O}=\pi/2$ and $E=\sqrt{5}/2$\\
    \hspace{-0.5cm}\includegraphics[width=85mm]{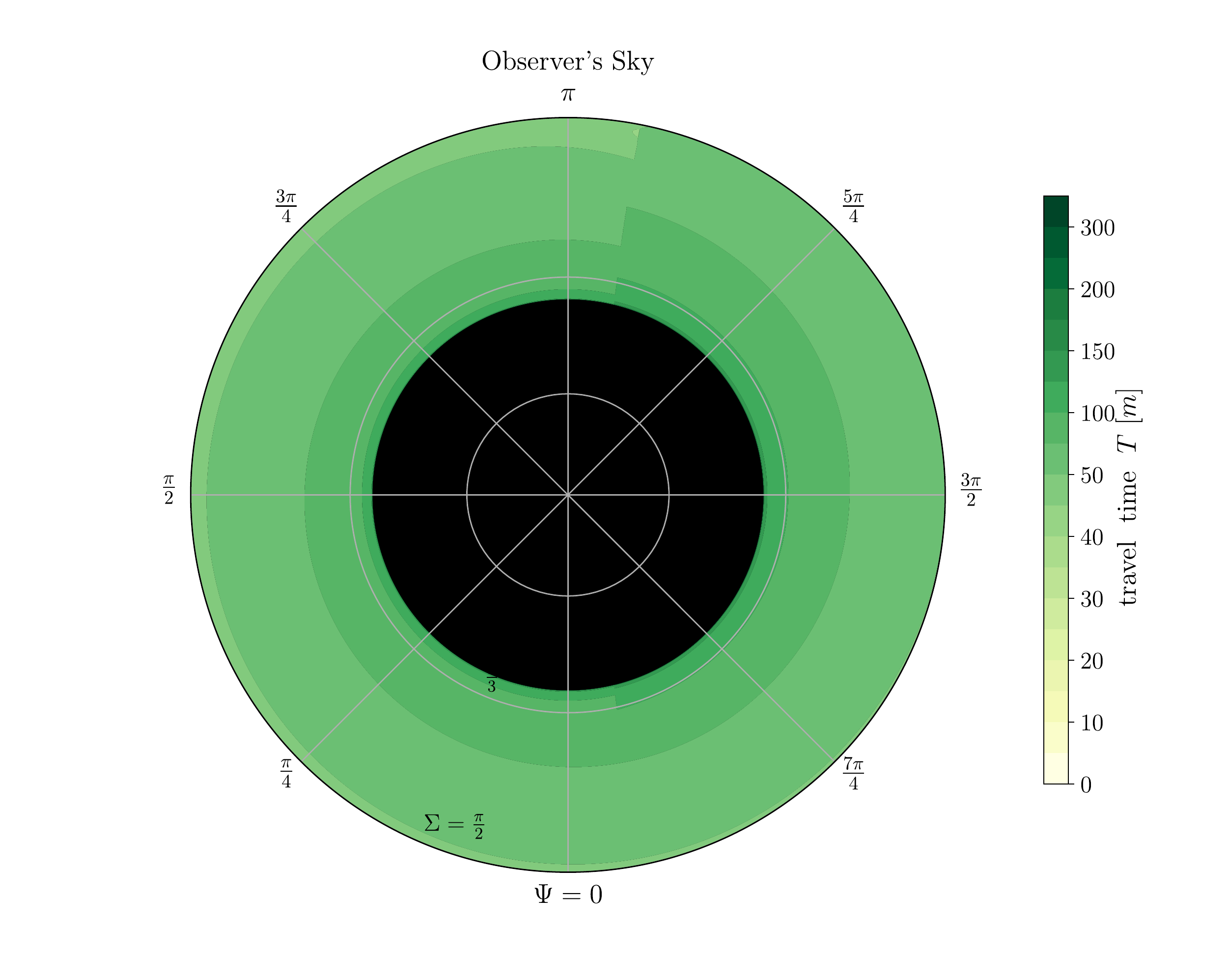} &   \includegraphics[width=85mm]{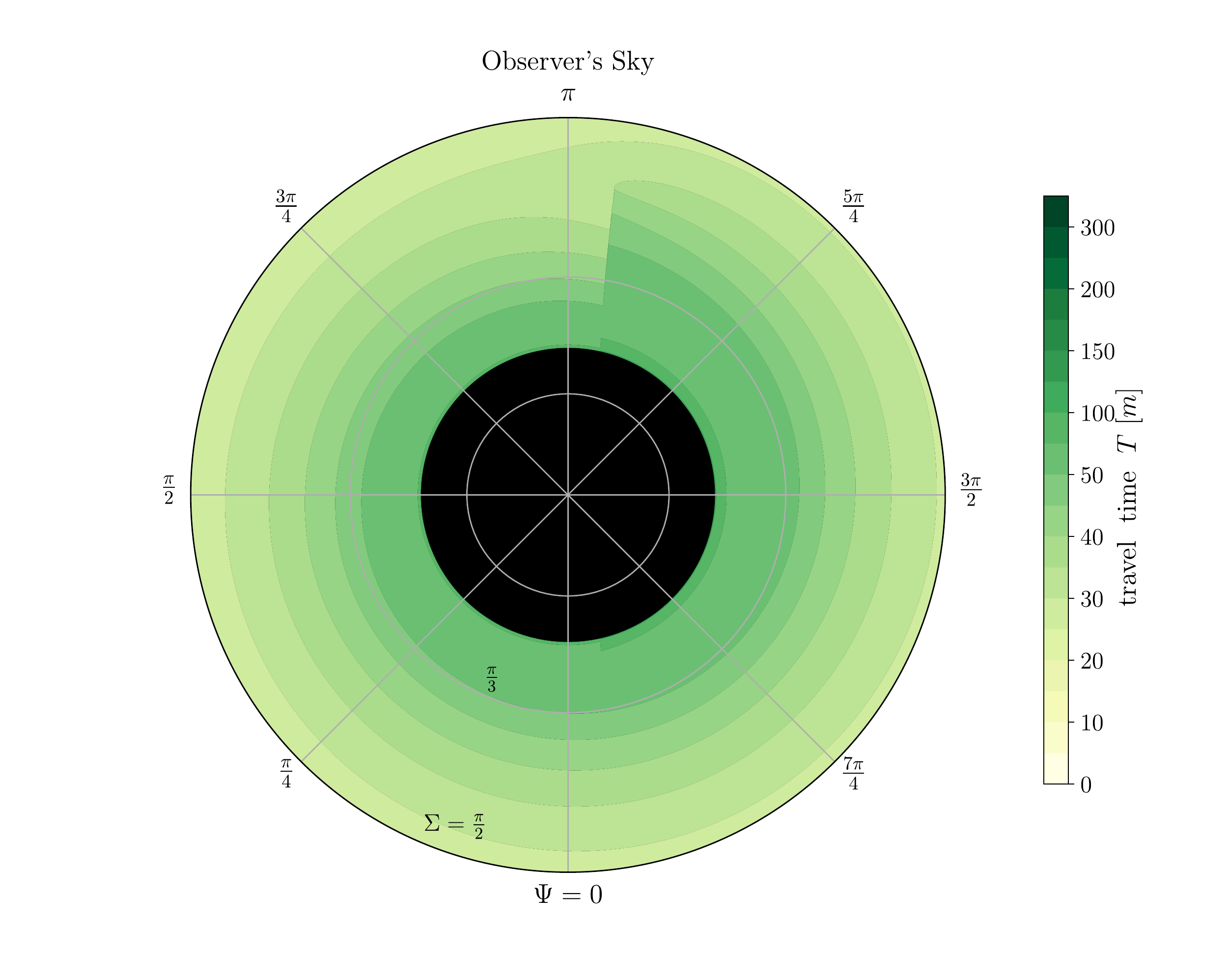} \\
    $\vartheta_{O}=\pi/2$ and $E=\sqrt{2}$ & Light Rays for $\vartheta_{O}=\pi/2$\\
    \hspace{-0.5cm}\includegraphics[width=85mm]{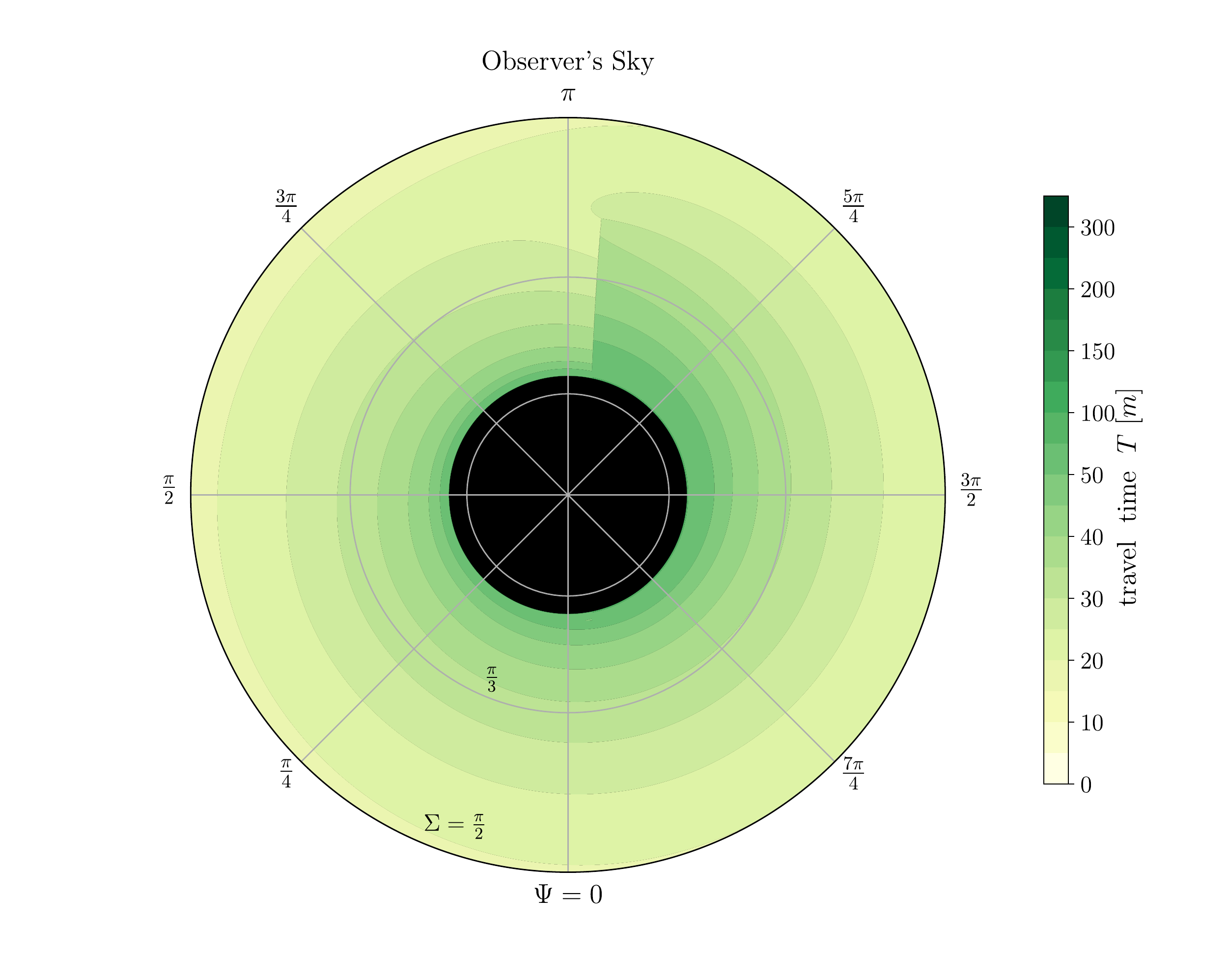} &   \includegraphics[width=85mm]{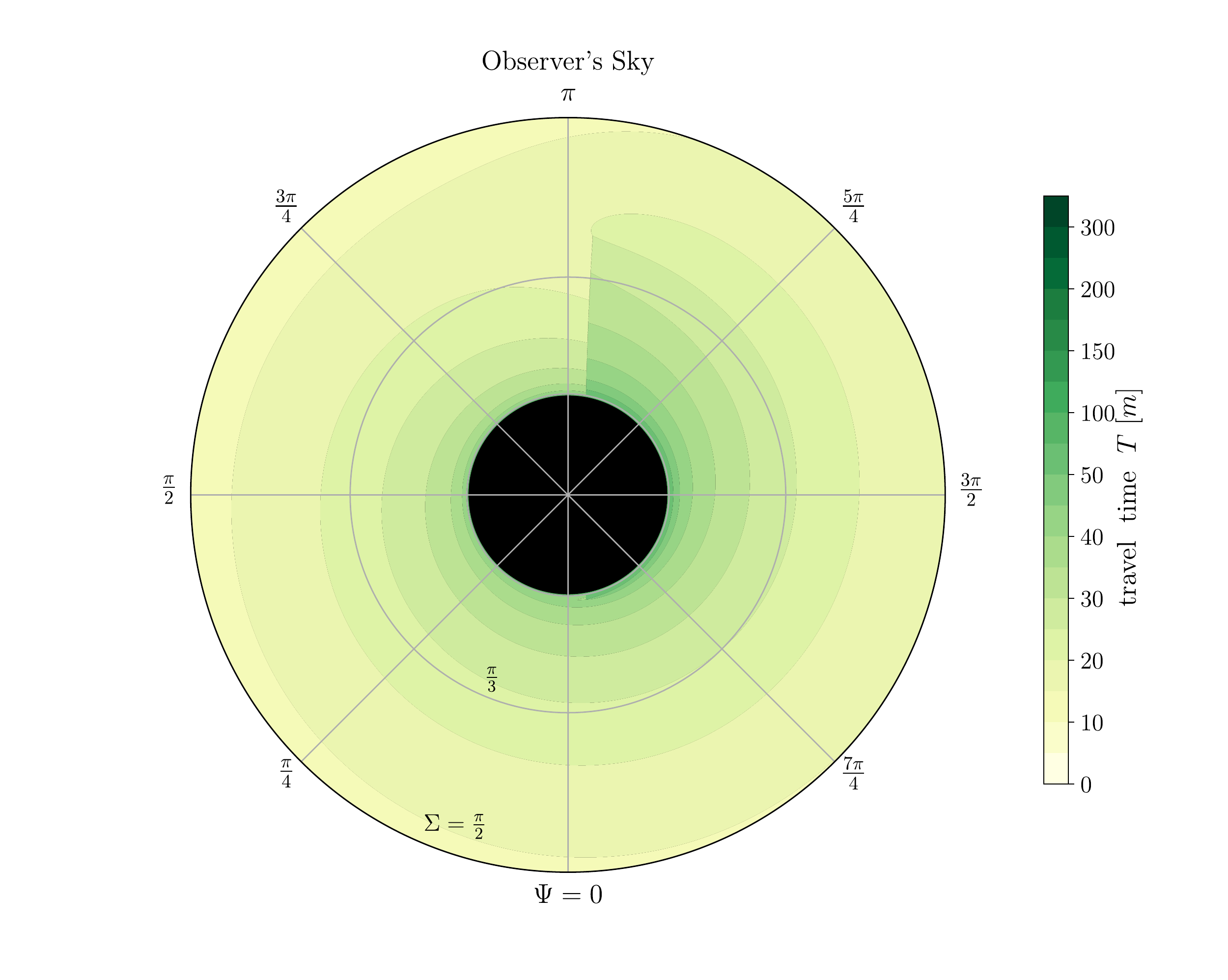} \\
    $\vartheta_{O}=\pi/4$ and $E=\sqrt{5}/2$ & $\vartheta_{O}=3\pi/4$ and $E=\sqrt{5}/2$\\
    \hspace{-0.5cm}\includegraphics[width=85mm]{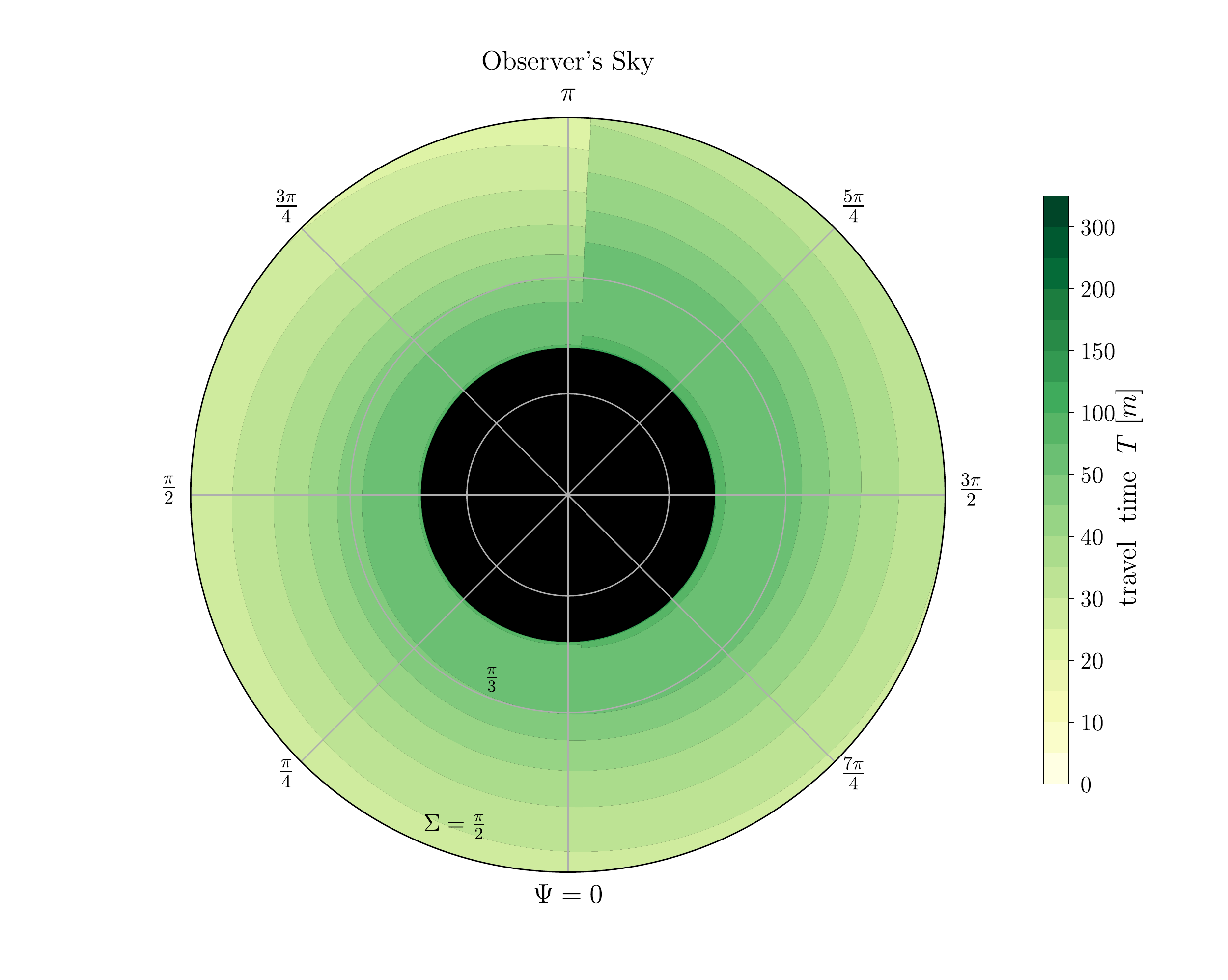} &   \includegraphics[width=85mm]{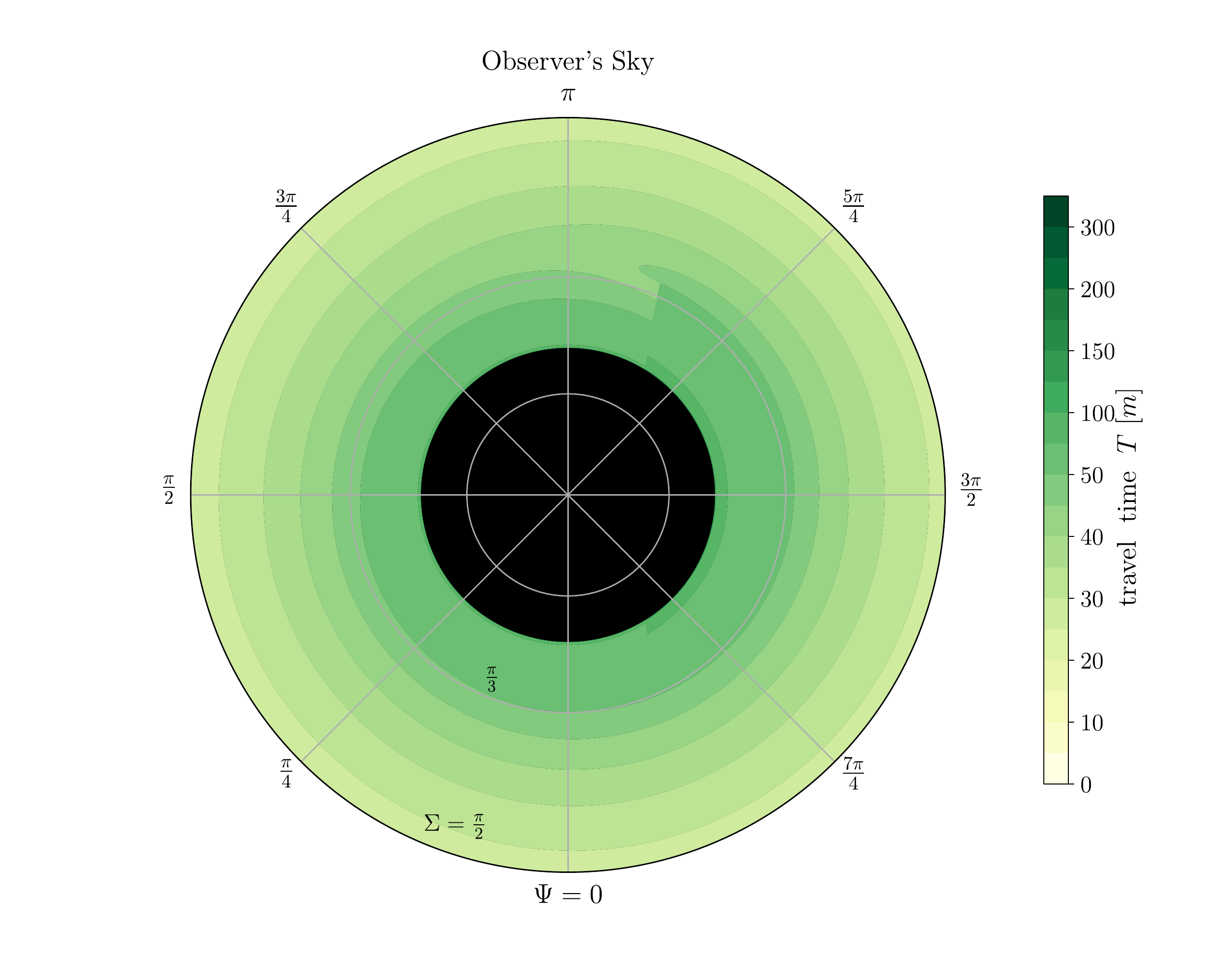} \\
  \end{tabular}
	\caption{Travel time maps for massive particles and light rays in the NUT spacetime with $n=m/2$. In the first four panels the observers are located at the coordinates $r_{O}=10m$ and $\vartheta_{O}=\pi/2$. The first three panels show travel time maps for massive particles (timelike geodesics) with $E=\sqrt{101/100}$ (top left), $E=\sqrt{5}/2$ (top right), and $E=\sqrt{2}$ (middle left). The middle right panel shows a reference travel time map for light rays (lightlike geodesics). The panels in the bottom row show travel time maps for massive particles with $E=\sqrt{5}/2$ for observers at $r_{O}=10m$ and $\vartheta_{O}=\pi/4$ (bottom left) and $\vartheta_{O}=3\pi/4$ (bottom right). For all six plots the corresponding two-spheres of particle sources $S_{P}^2$ and the two-sphere of light sources $S_{L}^2$ are located at $r_{P}=r_{L}=15m$. The Manko-Ruiz parameter is $C=1$ and thus we have a Misner string at $\vartheta=0$.}
\end{figure}
Now we turn to the NUT metric. Figure~14 shows travel time maps for massive particles with $E=\sqrt{101/100}$ (top left panel), $E=\sqrt{5}/2$ (top right panel), $E=\sqrt{2}$ (middle left panel), and light rays (middle right panel) for the NUT metric for the same observer-source geometry as in Fig.~13. The gravitomagnetic charge is $n=m/2$ and the Manko-Ruiz parameter is $C=1$ (Misner string at $\vartheta=0$). In addition, the bottom panels show travel time maps for massive particles with $E=\sqrt{5}/2$ for two observers located at $r_{O}=10m$ and $\vartheta_{O}=\pi/4$ (left panel) and $\vartheta_{O}=3\pi/4$ (right panel) and two-spheres of particle sources $S_{P}^2$ at $r_{P}=15m$.\\
Again the observers look into the direction of the black hole and the black circle in the center of each map is the particle shadow (or photon shadow for light rays) of the corresponding black hole. We again start with the travel time map for $E=\sqrt{101/100}$. Overall the travel time of the particles has the same magnitude as for the Schwarzschild metric. However, there are also a few differences. While for the Schwarzschild metric the travel time was symmetric with respect to rotations about the direction $\Sigma=0$ for the NUT metric this is not the case anymore. The steps in the travel time east to the meridian and the antimeridian indicate that at these positions the travel time has two discontinuities. These mark the positions of timelike geodesics crossing the Misner string (for lightlike geodesics this was already noted in Frost \cite{Frost2022,Frost2022b} and thus it is not surprising that it also occurs for timelike geodesics). When we start at the first discontinuity close to the antimeridian and move along a constant latitude in clockwise direction the travel time decreases. We observe a similar feature when we start at the discontinuity close to the meridian and move along a constant latitude in counterclockwise direction. However, the effect is much weaker and ends after a short angular distance. In addition the discontinuity close to the antimeridian extends nearly to the real celestial equator at $\Sigma=\pi/2$. When we increase the energy to $E=\sqrt{5}/2$ the shadow shrinks and overall the travel time decreases. The discontinuities are now located closer to the meridian and the antimeridian and are more pronounced and thus better visible. However, they are now confined to lower latitudes. Similarly, the spiral-shaped pattern shifts to lower latitudes and is more pronounced. When we increase the particle energy to $E=\sqrt{2}$ we basically observe the same effects. The discontinuities shift to longitudes closer to the meridian and the antimeridian and are confined to lower latitudes. In addition, the spiral pattern becomes even more pronounced. When we further increase the particle energy the pattern in the travel time map slowly approaches the pattern in the travel time map for light rays (middle right panel in Fig.~14). \\
When we shift the observer to $\vartheta_{O}=\pi/4$ (for $E=\sqrt{5}/2$ the travel time map is shown in the bottom left panel of Fig.~14) the discontinuity close to the antimeridian extends well beyond the real celestial equator at $\Sigma=\pi/2$ up to about $\Sigma=7\pi/12$ (not shown) and the spiral pattern becomes much more pronounced. The discontinuity at the meridian on the other hand is confined to latitudes much closer to the particle shadow. In addition, both discontinuities are now observed at longitudes closer to the meridian and the antimeridian. When we move the observer to $\vartheta_{O}=3\pi/4$ the discontinuity close to the antimeridian is confined to latitudes much closer to the particle shadow. The discontinuity close to the meridian appears to extend to roughly the same latitude as for $\vartheta_{O}=\pi/2$. However, it was already shown by Frost \cite{Frost2022,Frost2022b} that for lightlike geodesics the discontinuity extends to higher latitudes and thus this is most likely only an effect due to the choice of the color scale and the discontinuity does indeed extend to higher latitudes. In addition, now both discontinuities are located at longitudes further away from the meridian and the antimeridian.\\
The only question we did not address so far is how the electric charge affects the travel time. Figure~15 shows travel time maps for the Reissner-Nordstr\"{o}m metric (left column) and the charged NUT metric (right column) for three different electric charges $e=m/2$ (top row), $e=3m/4$ (middle row), and $e=m$ (bottom row). The energy of the particles is $E=\sqrt{5}/2$. For the charged NUT metric the gravitomagnetic charge is $n=m/2$ and the Manko-Ruiz parameter is $C=1$ (Misner string at $\vartheta=0$). The observers are located at the coordinates $r_{O}=10m$ and $\vartheta_{O}=\pi/2$ and the two-spheres of particle sources $S_{P}^{2}$ are located at $r_{P}=15m$. When we turn on the electric charge and slowly increase it the shadow shrinks and all features shift to lower latitudes. In addition, as for the lens maps we observe an apparent latitudinal broadening when we approach the shadow.\\
\begin{figure}\label{fig:TTRNCNUTMetric}
  \begin{tabular}{cc}
    Reissner-Nordstr\"{o}m Metric $e=m/2$& Charged NUT Metric $e=m/2$\\
    \hspace{-0.5cm}\includegraphics[width=85mm]{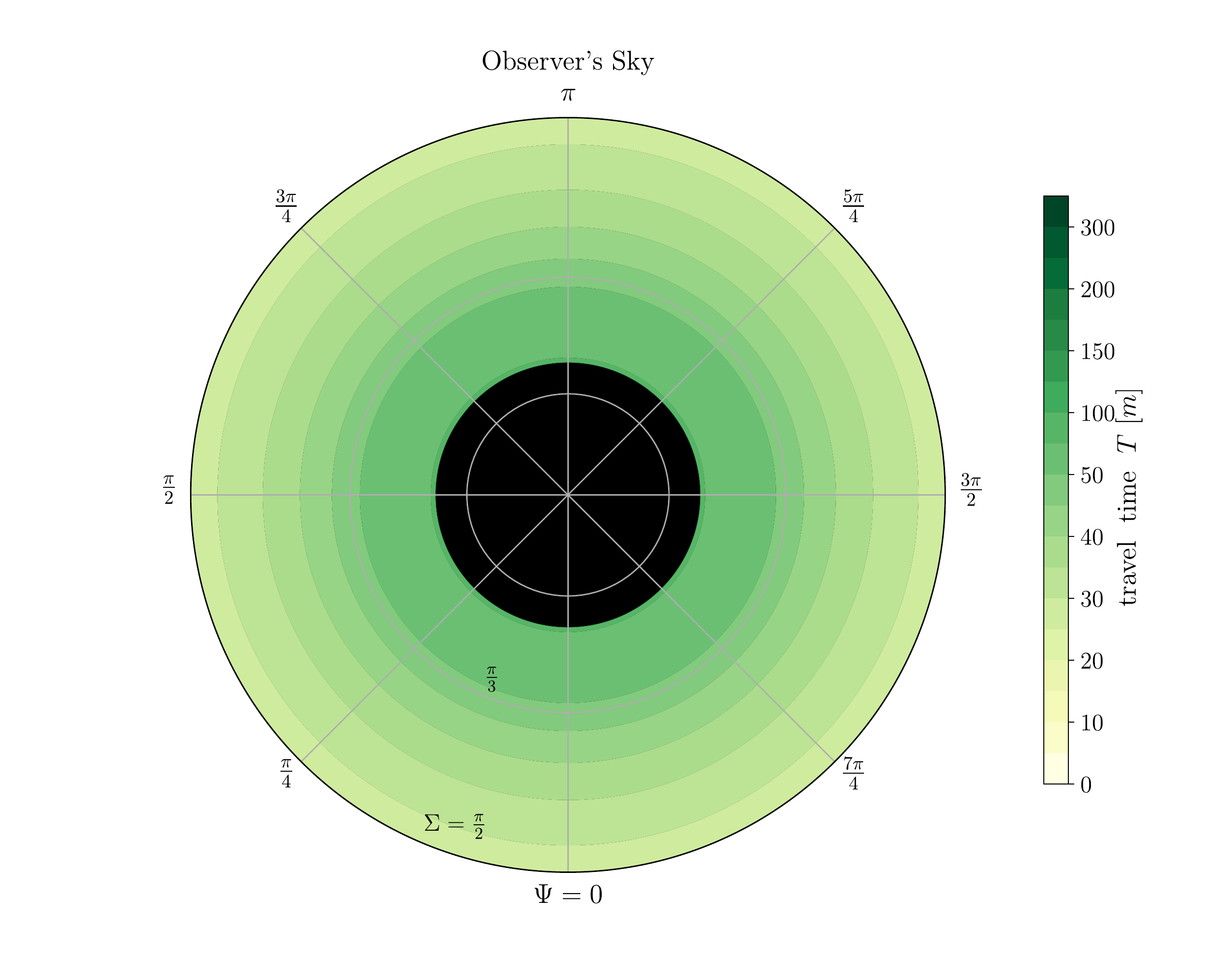} &   \includegraphics[width=85mm]{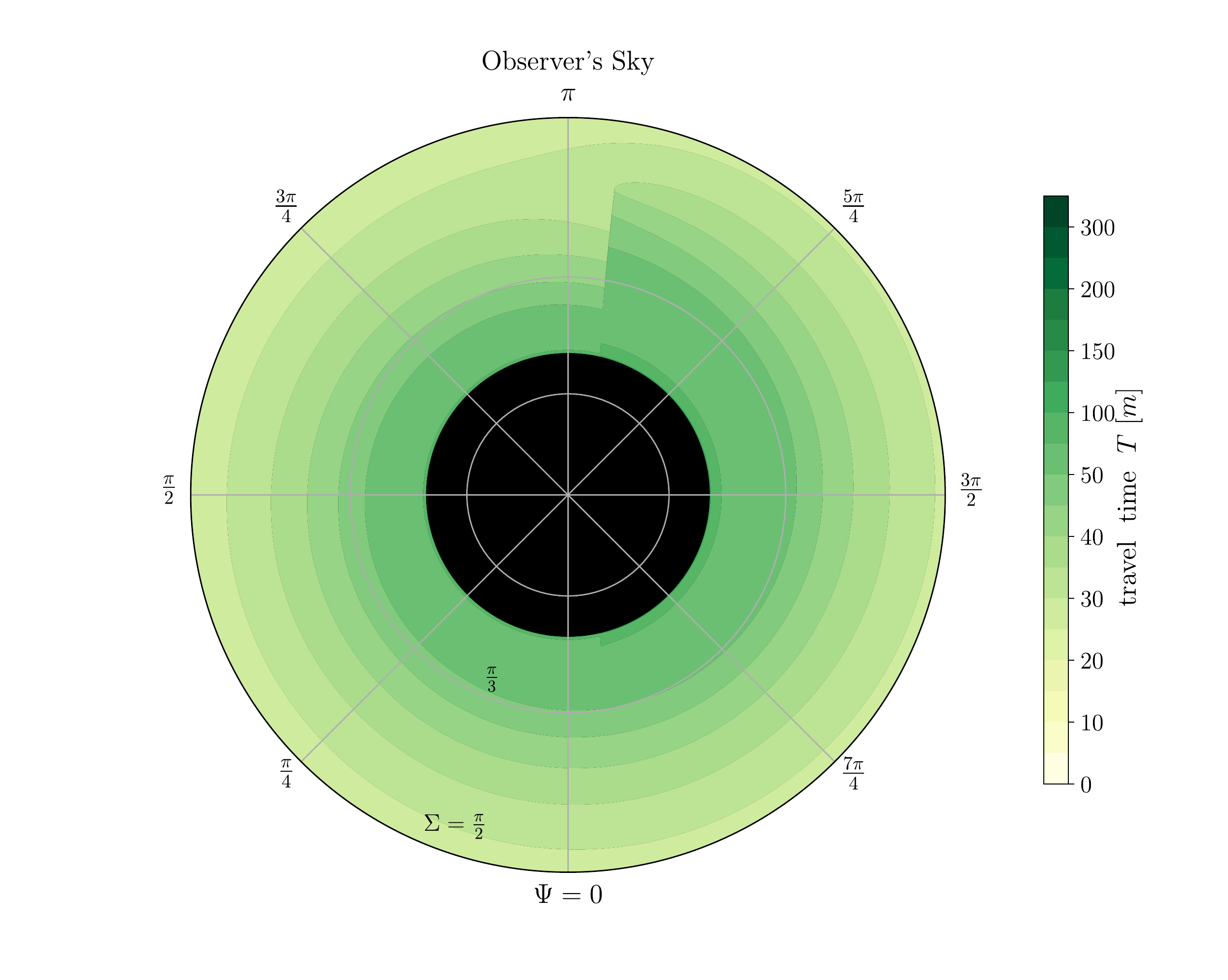} \\
    Reissner-Nordstr\"{o}m Metric $e=3m/4$& Charged NUT Metric $e=3m/4$\\
    \hspace{-0.5cm}\includegraphics[width=85mm]{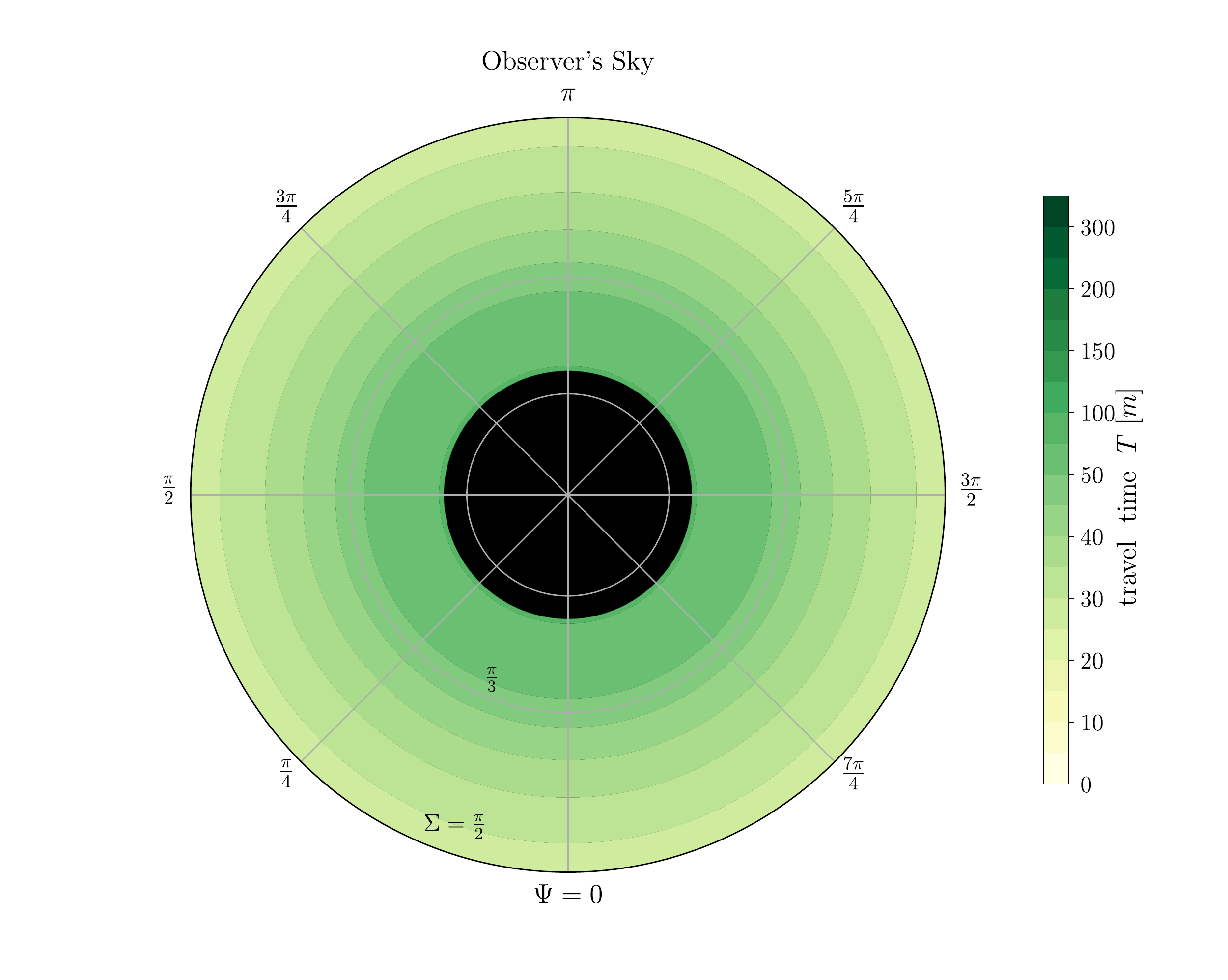} &   \includegraphics[width=85mm]{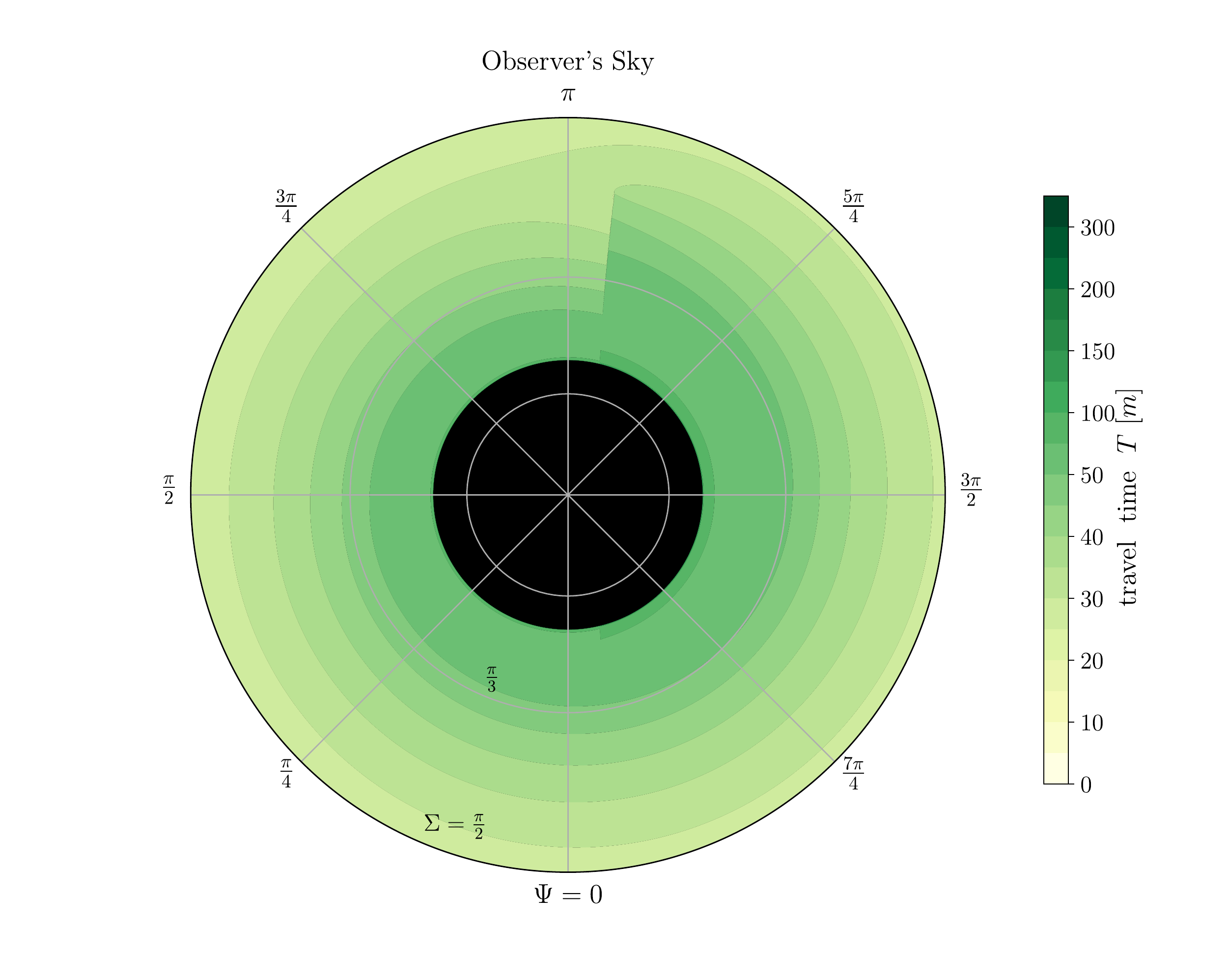} \\
    Reissner-Nordstr\"{o}m Metric $e=m$& Charged NUT Metric $e=m$\\
    \hspace{-0.5cm}\includegraphics[width=85mm]{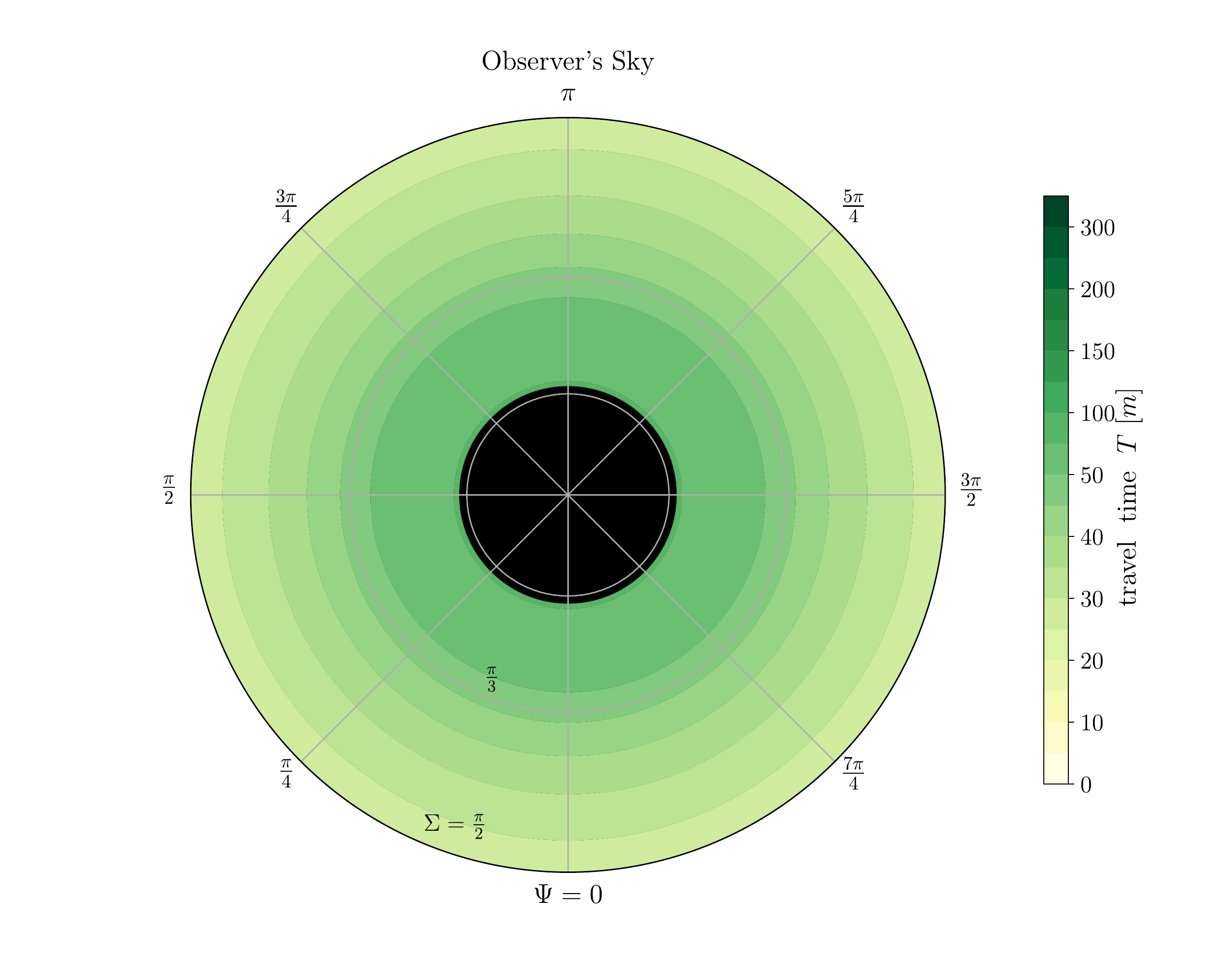} &   \includegraphics[width=85mm]{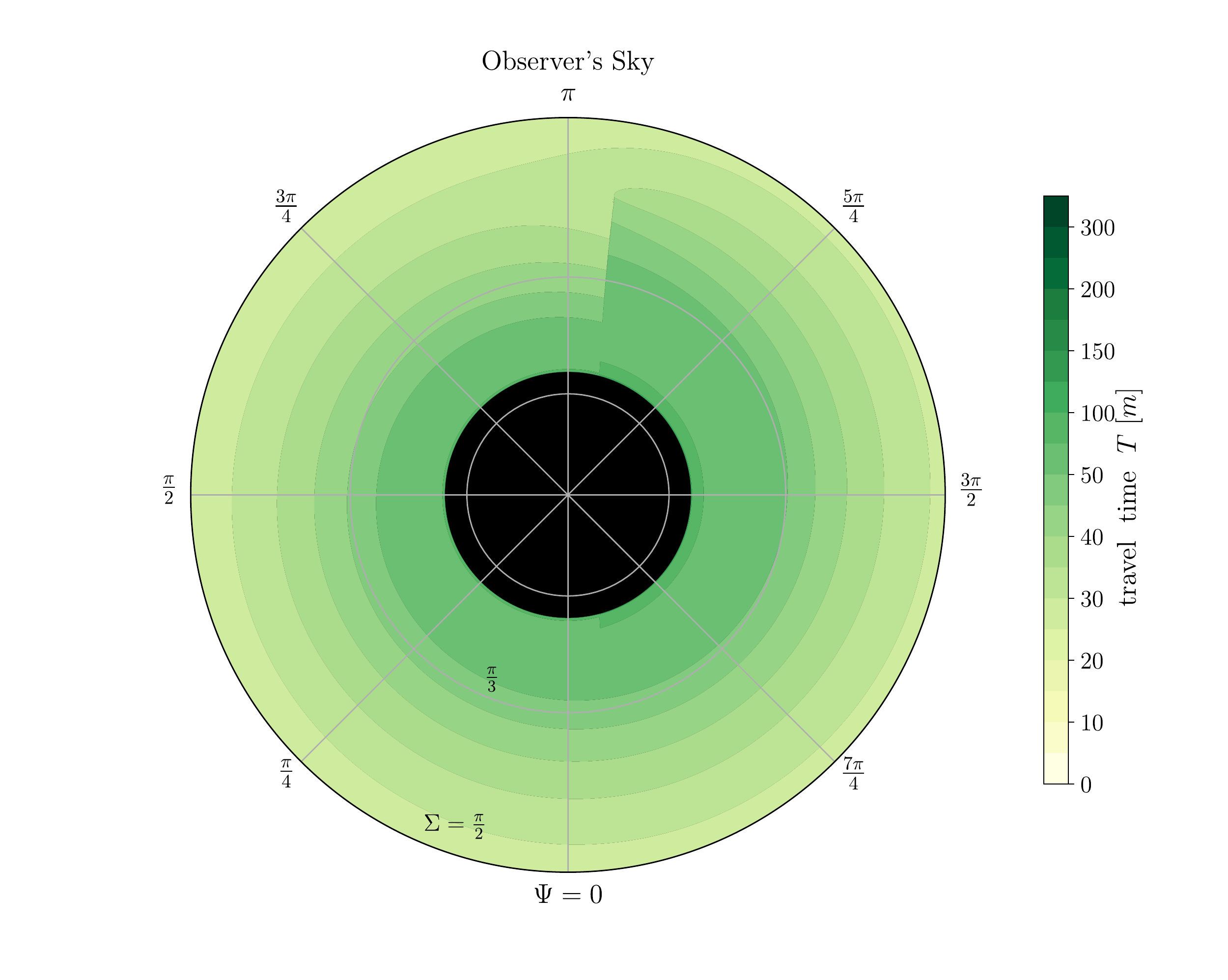} \\
  \end{tabular}
	\caption{Travel time maps for massive particles with $E=\sqrt{5}/2$ in the Reissner-Nordstr\"{o}m spacetime (left column) and the charged NUT spacetime with $n=m/2$ (right column) for $e=m/2$ (top row), $e=3m/4$ (middle row), and $e=m$ (bottom row). For the charged NUT spacetime the Manko-Ruiz parameter is $C=1$ and thus the Misner string is located at $\vartheta=0$. For all six maps the observers are located at the coordinates $r_{O}=10m$ and $\vartheta_{O}=\pi/2$ and the two-spheres of particle sources $S_{P}^2$ are located at $r_{P}=15m$.}
\end{figure}

\subsubsection{Traveled proper time}
In addition to the travel time $T$ for massive particles we can also define another travel time measure using the proper time along a timelike geodesic. It is the \emph{traveled proper time}. Similar to the travel time the traveled proper time measures in terms of the proper time $\tau$ the time a particle needs to travel from the particle source by which it was emitted to the observer. It can be interpreted as the time we could measure with a clock if it moved together with the particle along a timelike geodesic. Initially the traveled proper time does not offer any advantage over the travel time as it is much more difficult to measure. However, when we have particle decay or neutrino oscillations and know the number of particles emitted by the source (in the case of neutrinos with a specific flavor) it can be directly inferred by using an appropriate model for the particle decay or the neutrino oscillations. We can then compare the detected number of particles with the number of particles which should be detected according to the model estimates. We can then vary the proper time until both agree (note that in the most general case we have to also consider the spacetime in which the particles travel). In terms of the proper time $\tau_{P}$ at which the particle is emitted by the particle source and the proper time $\tau_{O}$  at which it is detected by the observer with the detector it reads
\begin{equation}
\tau(\Sigma)=\tau_{O}-\tau_{P}(\Sigma).
\end{equation}
Because the proper time is defined up to an affine parametrization in our case we
choose $\tau(\lambda_{O})=\tau_{O}=0$. We insert (\ref{eq:EoMCelesK}) in (\ref{eq:PTint})
and obtain
\begin{equation}\label{eq:TPT}
\tau(\Sigma)=\int_{r_{O}...}^{...r_{P}}\frac{\rho(r')\text{d}r'}{\sqrt{\rho(r')^2E^2-\rho(r')Q(r')-Q(r')\frac{\rho(r_{O})(\rho(r_{O})E^2-Q(r_{O}))}{Q(r_{O})}\sin^2\Sigma}},
\end{equation}
where again the dots in the limits of the integral shall indicate that we have to split the integral at potential turning points into two terms and the sign of the root in the denominator has to be chosen according to the direction of the $r$ motion. When we use the celestial coordinates to parametrize the timelike geodesics we can immediately read from (\ref{eq:TPT}) that unlike the travel time the traveled proper time only depends on the celestial latitude $\Sigma$ and not on the celestial longitude $\Psi$. We now evaluate the traveled proper time as described in Sec.~\ref{Sec:EoMtau}.\\
\begin{figure}\label{fig:TPTSchwarzschild}
  \begin{tabular}{cc}
    Schwarzschild Metric & Reissner-Nordstr\"{o}m Metric\\
    \hspace{-0.5cm}\includegraphics[width=95mm]{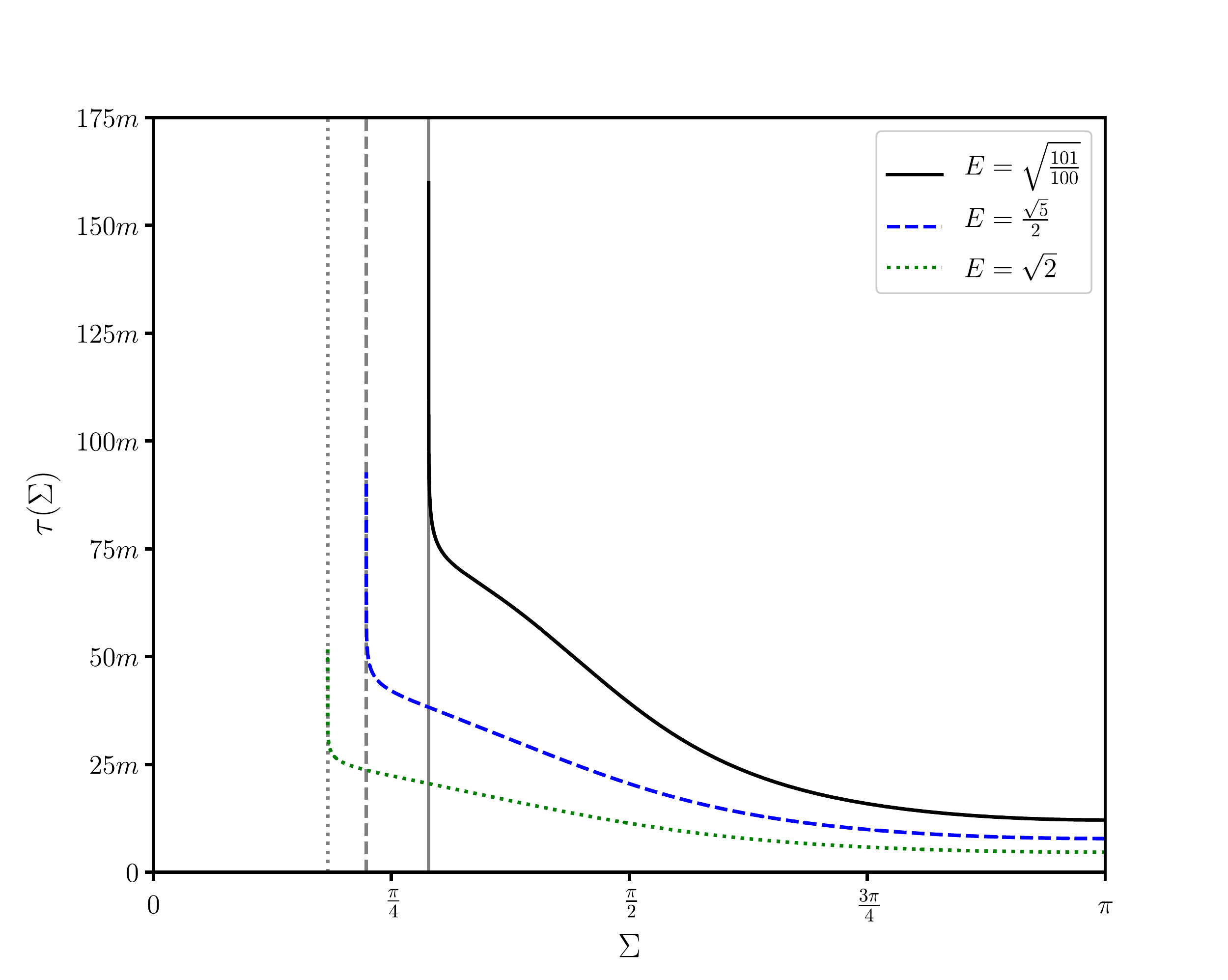} &   \includegraphics[width=95mm]{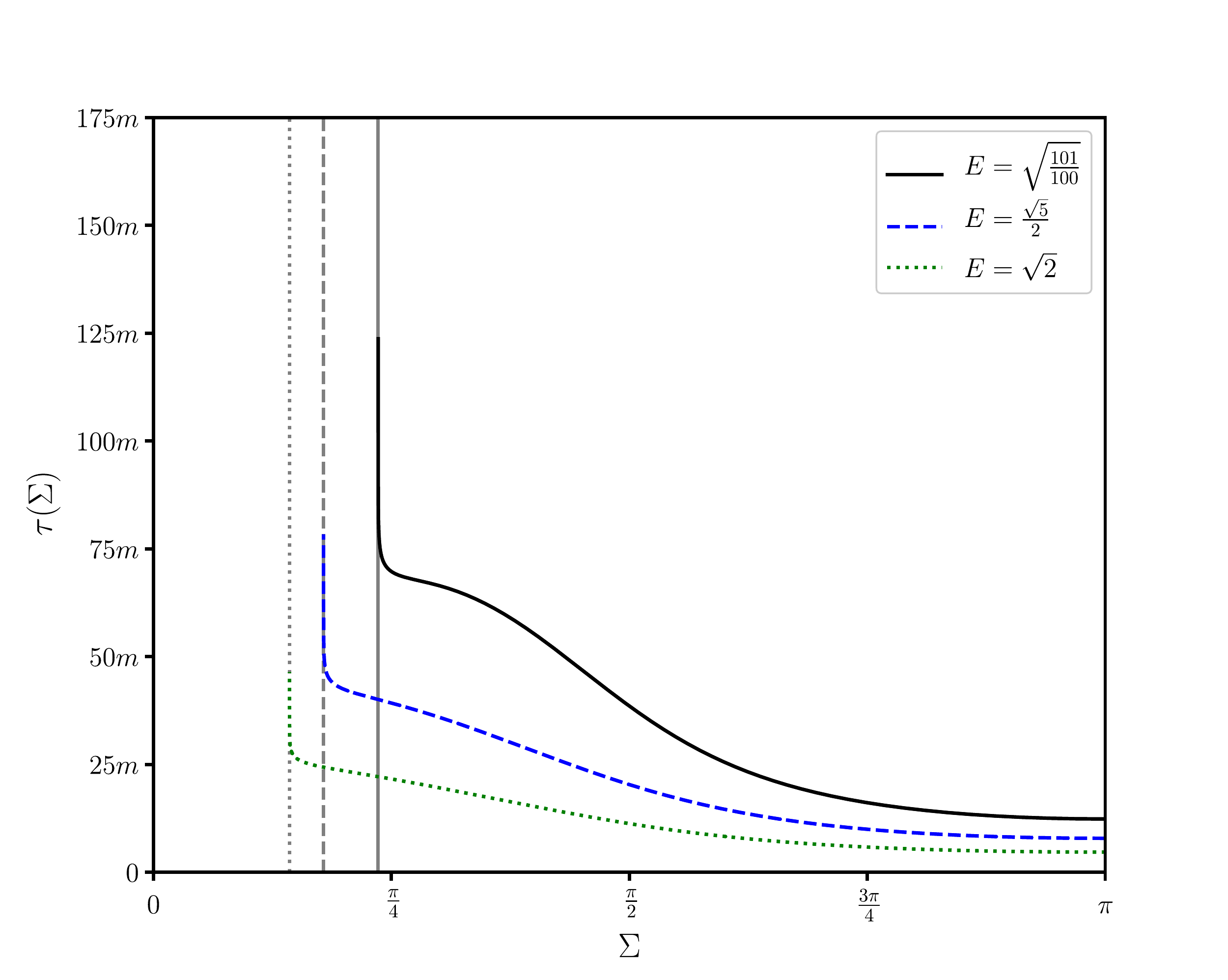} \\
    NUT Metric & Charged NUT Metric\\
    \hspace{-0.5cm}\includegraphics[width=95mm]{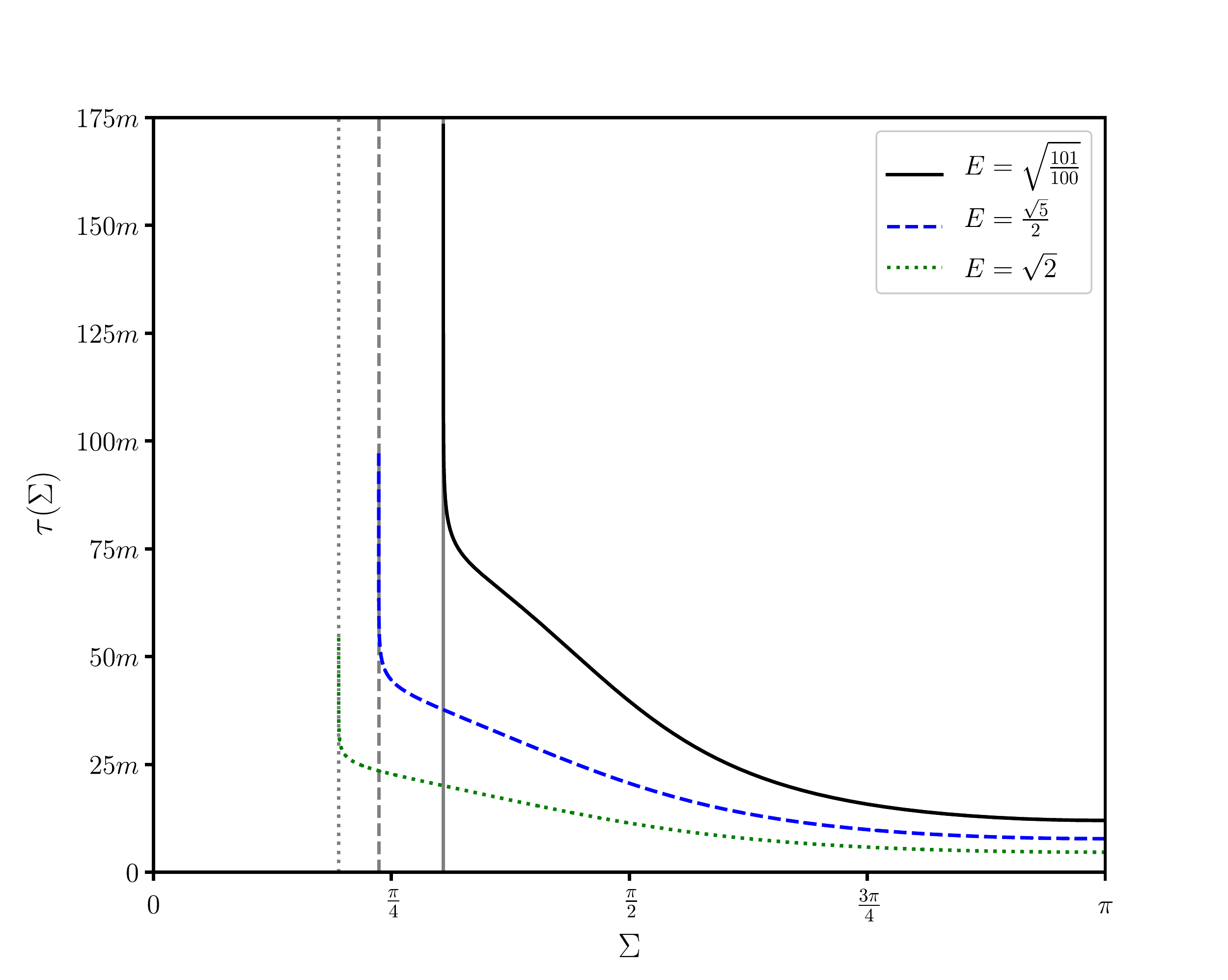} &   \includegraphics[width=95mm]{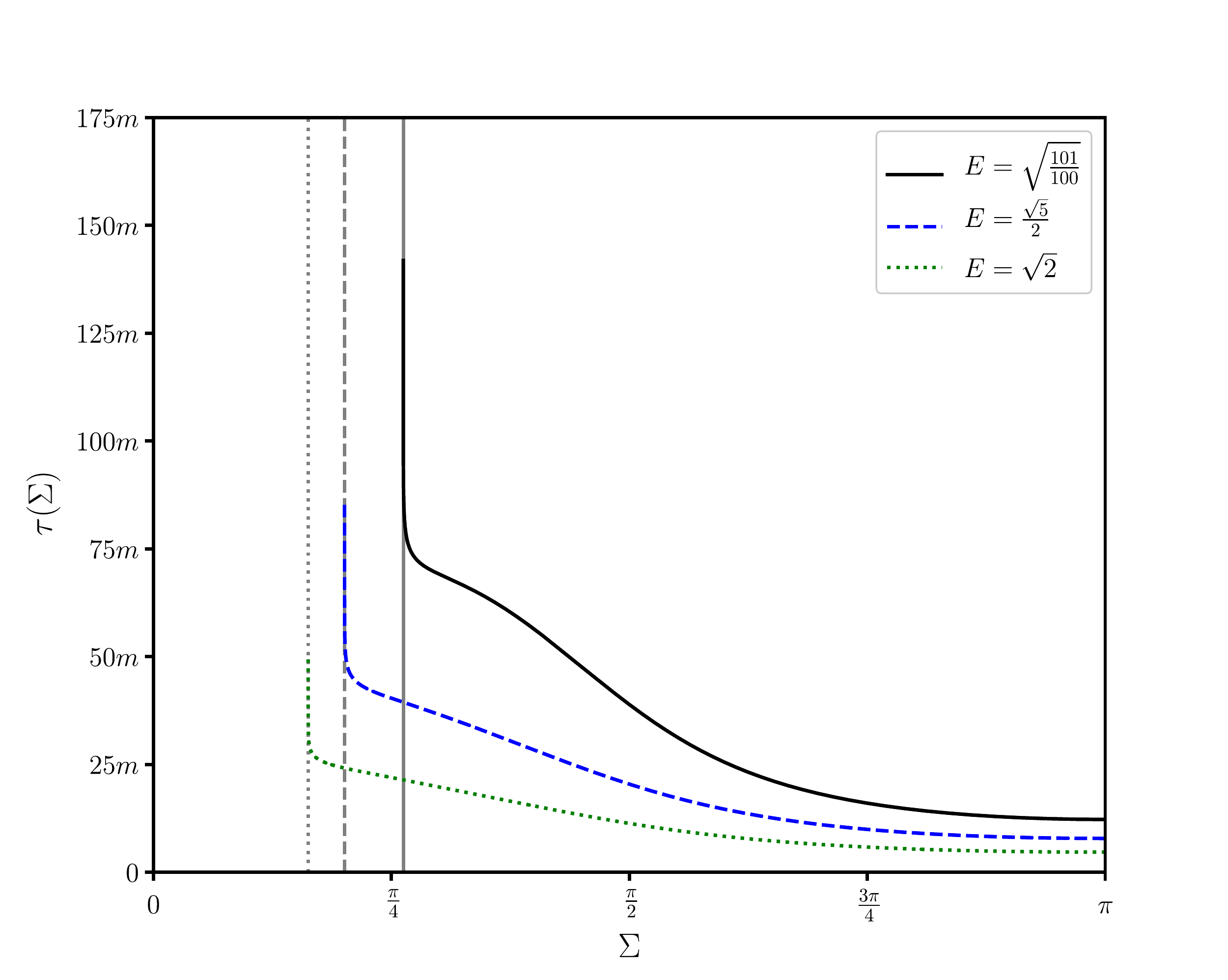} \\
  \end{tabular}
	\caption{Traveled proper time as function of the celestial latitude $\Sigma$ for massive particles in the Schwarzschild spacetime (top left), the Reissner-Nordstr\"{o}m spacetime (top right), the NUT spacetime (bottom left), and the charged NUT spacetime (bottom right) for $E=\sqrt{101/100}$ (black solid line), $E=\sqrt{5}/2$ (blue dashed line), and $E=\sqrt{2}$ (green dotted line). The electric and gravitomagnetic charges are $e=m$ and $n=m/2$, respectively. The observers are located at $r_{O}=10m$ and the particle sources are located at $r_{P}=15m$. The gray vertical lines mark the positions of the corresponding particle shadows.}
\end{figure}
Figure~16 shows the traveled proper time for an observer at the radius coordinate $r_{O}=10m$ and a particle source at the radius coordinate $r_{P}=15m$ for the Schwarzschild metric (top left panel), the Reissner-Nordstr\"{o}m metric with $e=m$ (top right panel), the NUT metric with $n=m/2$ (bottom left panel), and the charged NUT metric with $e=m$ and $n=m/2$ (bottom right panel) for $E=\sqrt{101/100}$ (black solid line), $E=\sqrt{5}/2$ (blue dashed line), and $E=\sqrt{2}$ (green dotted line).\\
Let us again start the discussion with the Schwarzschild metric. We start our discussion with the traveled proper time for $E=\sqrt{101/100}$. We see that the traveled proper time increases from $\Sigma=\pi$ to $\Sigma=\Sigma_{\text{pa}}(E)$. When $\Sigma$ approaches $\Sigma_{\text{pa}}(E)$ the particles make more and more turns around the particle sphere and thus the traveled proper time diverges. When we increase the energy to $E=\sqrt{5}/2$ the angular radius of the particle shadow $\Sigma_{\text{pa}}(E)$ shrinks and the traveled proper time decreases. Further increasing the energy to $E=\sqrt{2}$ has the same effect. When we turn on the electric charge and consider the Reissner-Nordstr\"{o}m metric (top right panel) we do not see any major changes for $\pi/2<\Sigma$ and the traveled proper time is roughly the same. For $\Sigma \rightarrow \Sigma_{\text{pa}}(E)$ the divergences shift to lower latitudes and the traveled proper time generally decreases. However, the larger $E$ the smaller the effect. In addition, for particles with $E=\sqrt{101/100}$ for the Reissner-Nordstr\"{o}m metric close to the divergence the increase in traveled proper time is lower. Now we turn on the gravitomagnetic charge (bottom panels). Again for $\pi/2<\Sigma$ we do not observe any major changes and the traveled proper time remains roughly constant. For $\Sigma<\pi/2$ the divergences shift to higher latitudes and overall the traveled proper time increases for the NUT metric and the charged NUT metric in comparison to the Schwarzschild metric and the Reissner-Nordstr\"{o}m metric, respectively.

\subsubsection{Implications for astrophysical observations}
As discussed above for massive particles we can calculate two different travel time measures:
on one hand the travel time in terms of the time coordinate $t$ and on the other hand the traveled proper time $\tau(\Sigma)$. However, the travel time is not directly accessible to observations. We can only observe travel time differences. In our case at hand we assumed that we have a stationary observer that detects a signal at the event $(x_{O}^{\mu})$ and that the corresponding proper time along the timelike geodesic is $\tau_{O}=0$. The observed travel time differences are all recorded in terms of the proper time in the observer's frame. Therefore, we now have to relate the proper time in the observer's frame to the time coordinate $t$. For this purpose let us first denote the coordinates along the worldline of the observer as $(\bar{x}^{\mu})$ and the corresponding proper time as $\bar{\tau}$. Because we only consider a stationary observer we have $\text{d}\bar{r}/\text{d}\bar{\tau}=\text{d}\bar{\vartheta}/\text{d}\bar{\tau}=\text{d}\bar{\varphi}/\text{d}\bar{\tau}=0$ and thus also $\bar{r}(\lambda)=r_{O}$, $\bar{\vartheta}(\lambda)=\vartheta_{O}$ and $\bar{\varphi}(\lambda)=\varphi_{O}$. Now we only have to calculate the relations between the proper time in the observer's reference frame and the time coordinate $\bar{t}$. The calculation is straightforward. Let us first denote the constants of motion along the worldline of the observer as $E_{O}$, $L_{zO}$, and $K_{O}$. Now we first evaluate (\ref{eq:EoMphi}) and see that 
\begin{equation}\label{eq:CoMStatObsLzE}
L_{zO}+2n(\cos\vartheta_{O}+C)E_{O}=0.
\end{equation}
In the next step we insert this relation in (\ref{eq:EoMtheta}) and get $K_{O}=0$. Then we insert the latter in (\ref{eq:EoMr}) and solve for $E_{O}$. We obtain
\begin{equation}\label{eq:CoMStatObsE}
E_{O}=\sqrt{\frac{Q(r_{O})}{\rho(r_{O})}}.
\end{equation} 
Now we use (\ref{eq:CoMStatObsLzE}) and (\ref{eq:CoMStatObsE}) in (\ref{eq:EoMt}) after reparametrizing it in terms of the proper time $\bar{\tau}$ and get 
\begin{equation}
\frac{\text{d}\bar{t}}{\text{d}\bar{\tau }}=\sqrt{\frac{\rho(r_{O})}{Q(r_{O})}}.
\end{equation}
We integrate from $\bar{t}(\bar{\tau}_{i})=\bar{t}_{i}$ to $\bar{t}(\bar{\tau})=\bar{t}$ and obtain as relation between the time coordinate $\bar{t}$ and the proper time $\bar{\tau}$ along the timelike curve of the observer 
\begin{equation}\label{eq:TtauObs}
\bar{t}(\bar{\tau})=\bar{t}_{i}+\sqrt{\frac{\rho(r_{O})}{Q(r_{O})}}(\bar{\tau}-\bar{\tau}_{i}),
\end{equation}
where in our case the proper time $\bar{\tau}_{i}$ and the corresponding time coordinate $\bar{t}_{i}$ mark the time at which the first particle signal is detected. When we assume that the observer detects the first particle signal at $\bar{\tau}_{1}=\tau_{O}$ and the second signal at $\bar{\tau}_{2}>\bar{\tau}_{1}$ we obtain for the relation between the travel time difference in the observer's frame and the travel time difference in terms of the time coordinate $\bar{t}$
\begin{equation}
\Delta \bar{T}=\bar{t}_{2}-\bar{t}_{1}=\sqrt{\frac{\rho(r_{O})}{Q(r_{O})}}(\bar{\tau}_{2}-\tau_{O}).
\end{equation}
Note that this relation is valid for travel time differences between arbitrary combinations of particle and light signals.\\
When we observe two different signals gravitationally lensed by a black hole and suspect that they were emitted by the same source, we now have to use \emph{a priori} estimates to iteratively determine the travel time along each geodesic (note that here we iterate with respect to the celestial coordinates). When both geodesics end at the same spatial position $(x_{P}^{i})$ and we get $\Delta T=T(\Sigma_{2},\Psi_{2})-T(\Sigma_{1},\Psi_{1})=\Delta \bar{T}$ as travel time difference we know they really come from the same source. However, we have to be cautious to draw conclusions on the underlying spacetime describing the black hole because we still may have a degeneracy with respect to the spacetime's physical parameters as well as the radius coordinates of the observer and the particle or light source. \\
Transforming the proper time along the worldline of the observer to the proper time along the worldline of a particle is much more difficult. We cannot derive a direct relation between the proper time which elapsed for the observer and the traveled proper time. However, when we know the nature of the source and the exact particle spectrum it emits, we can directly determine the traveled proper time for two different astrophysical scenarios. In the first scenario the particle source emits particles which decay along the timelike geodesic. The half-life of the particles is then measured in terms of the proper time $\tau$. In the case that the interaction of the particles with their environment can be neglected, we know the underlying decay processes, and are able to write down the corresponding decay laws, we can determine the traveled proper time $\tau(\Sigma)$ from the detected number of particles and the estimated number of emitted particles determined using numerical particle emission models for the observed source. In the second scenario we have neutrino emission by the source. In the case that we know the source and the flavor composition of the neutrinos it emits, again from, e.g., an appropriate numerical model, and we can observe the flavor composition of the detected neutrinos, we can use an appropriate model for the propagation of the neutrinos along the geodesic under consideration of their flavor oscillations to determine the proper time from the difference between the detected neutrino flavor composition and the estimated neutrino flavor composition of the emission of the source which we should detect according to the source and propagation models. However, in this case we have to make an initial guess for the spacetime describing the black hole lens. \\
Now the most important question is how we can use these two travel time measures for astronomical observations. As long as we only consider one particle source we can have two different scenarios. In the first scenario we only consider particles with the same energy $E$. In this case we have the same situation as for light rays discussed in Frost \cite{Frost2022,Frost2022b}. The travel time increases for images of the same particle source with increasing order of the image. Turning on the electric charge reduces this increase (however, at latitudes close to the shadow for which the geodesics, along which the particles travel, closely approach the particle sphere the travel time of the particles still increases rapidly). When we turn on the gravitomagnetic charge we observe a discontinuity in the travel time for timelike geodesics crossing the Misner string. As already discussed in Frost \cite{Frost2022,Frost2022b} for light rays, this is a real discontinuity and together with the spiral-shaped pattern it is the most distinct indicator for the presence of a gravitomagnetic charge. When we have a multiple-imaging situation with four images we can expect that two of these images are located relatively close to the discontinuity. For these images we then observe a larger travel time difference than for all other image combinations. The magnitude of the travel time differences, each calculated for two particle signals with the same energy $E$, for massive particles with different energies may now allow us to place a tighter constraint on the magnitude of the gravitomagnetic charge than we could obtain with electromagnetic radiation alone. Moreover, when the particle source also emits electromagnetic radiation we can combine the travel time differences for massive particles with the travel time differences for light rays. In this case the obtained estimate for the magnitude of the gravitomagnetic charge will be even more precise. However, we have to note here that even for light rays multiple images created by gravitational lensing by black holes have not been observed so far and thus it is rather unlikely that we will observe multiple images for massive particles in the near future. \\
Let us now assume that the particle source emits a particle spectrum with a particle flux density with a peak at an unspecified energy $E_{\text{max}}$. We already saw that the travel time decreases with increasing particle energy $E$. However, only taking the travel time maps and the travel time differences into account limits the information we can infer. Luckily, when we include the lens maps we get a much clearer picture. We recall that for particles with high energies images of first order are located at lower latitudes than for particles with lower energies. In the same context we also saw that for the same particle source images of first order generated by particles with high energy and images of third order generated by particles with low energy can have a very low angular separation or even overlap (but also remember that the latter is only the case for the spherically symmetric and static Schwarzschild and Reissner-Nordstr\"{o}m spacetimes and not for the rotationally symmetric NUT and charged NUT spacetimes). Let us assume that we have a particle source with a known particle emission spectrum and that this source also emits electromagnetic radiation. Now we can determine the travel time differences between images of first and/or third order at different particle energies and also between the images generated by particle signals and the images created by light rays. In general these will vary depending on the nature of the black hole spacetime and may allow one to distinguish different types of black holes in the framework of general relativity. In the case of the charged NUT metric comparing the calculated travel time differences with measured travel time differences between light and particle signals lensed by an astrophysical black hole will allow us to place constraints on the electric and gravitomagnetic charges of the black hole. However, \emph{a priori} degeneracies with respect to the radius coordinates of the observer and the particle source $r_{O}$ and $r_{P}$ as well as $e$ and $n$ cannot be excluded. Investigating this is beyond the scope of this paper and may be part of future work.\\

\section{Summary}\label{Sec:Summary}
In this paper we performed a detailed and thorough investigation of gravitational lensing of massive particles in the charged NUT metrics. For this purpose we first discussed and solved the equations of motion with focus on unbound timelike geodesics. In analogy to the photon sphere for light rays we rewrote the equation of motion for $r$ in terms of an energy-dependent potential and derived the radius coordinates of the unstable particle spheres in the Schwarzschild metric, the Reissner-Nordstr\"{o}m metric, the NUT metric, and the charged NUT metric. Contrary to the radius coordinates of the unstable photon spheres they depend on the energy $E$ of the particles. That spherically symmetric and static spacetimes possess a particle sphere is nothing new. The concept of a particle sphere was first introduced by Mielnik and Pleba\'{n}ski \cite{Mielnik1962} in 1962. For the Schwarzschild metric and the Reissner-Nordstr\"{o}m metric in terms of an (energy-dependent) parameter or the particle velocity at infinity the radius coordinates of the particle spheres have already been derived by Zakharov \cite{Zakharov1988} and Liu \emph{et al.} \cite{Liu2016}, and Pang and Jia \cite{Pang2019}, respectively. However, to our knowledge this paper is the first to derive the radius coordinates of the particle spheres using the potential without introducing a parameter or the particle velocity at infinity and an impact parameter and to directly present a general form in terms of the particle energy along the geodesic (although in the works mentioned above we can easily restore the energy dependency). For the Schwarzschild metric and the Reissner-Nordstr\"{o}m metric we derived these solutions analytically while for the NUT metric and the charged NUT metric the solutions could only be determined numerically (note that although Kobialko \emph{et al.} \cite{Kobialko2022} applied the concept of a particle surface to the NUT metric they did not explicitly derive the radius coordinate of the unstable particle sphere in the domain of outer communication).\\
Then we solved the equations of motion using elementary and Jacobi's elliptic functions as well as Legendre's elliptic integrals of the first, second, and third kind. Structurally they are very similar or even identical to the solutions for lightlike geodesics derived in Frost \cite{Frost2022,Frost2022b}. In the second part of this paper we then used these solutions to investigate gravitational lensing of massive particles in the Schwarzschild metric, the Reissner-Nordstr\"{o}m metric, the NUT metric, and the charged NUT metric for stationary observers and stationary particle sources distributed on two-spheres $S_{P}^2$  in the domain of outer communication outside the particle sphere. For this purpose we first used the tetrad approach of Grenzebach \emph{et al.} \cite{Grenzebach2015} and adapted it to timelike geodesics by modifying the plasma lensing approach of Perlick and Tsupko \cite{Perlick2017}. We related the constants of motion $L_{z}$ and $K$ to the particle energy $E$ and to latitude-longitude coordinates on the observer's celestial sphere. This effectively allowed us to parametrize the timelike geodesics along which the particles travel in terms of the particle energy $E$ and the celestial latitude and longitude on the observer's celestial sphere. We then derived the relation between the particle energy along a timelike geodesic and the particle energy as detected in the reference frame of a stationary observer with a particle detector.\\
As first part of our lensing analysis we derived the angular radius of the \emph{particle shadow} of a black hole. For all four metrics it depends on the energy $E$ and the particle shadow is circular. When we increase the energy of the particles the angular radius of the particle shadow shrinks and for $E\rightarrow \infty$ it approaches the angular radius of the \emph{photon shadow}. With respect to the electric and gravitomagnetic charges the particle shadow shows the same behavior as the photon shadow. Its size shrinks with increasing electric charge and it grows when we increase the gravitomagnetic charge. When we have only particles with the same energy or a particle distribution with a very narrow peak it is unfortunate that the particle shadow is circular because we have a degeneracy with respect to up to four different parameters, the particle energy $E$, the radius coordinate of the observer $r_{O}$, the electric charge $e$, and the gravitomagnetic charge $n$. We also addressed the question how we can combine the photon shadow and the particle shadow in a multimessenger approach to determine the underlying spacetime. We came to the conclusion that when we have accurate observations of the photon shadow we also know the boundary of the \emph{collective particle shadow}. We concluded that the darkening (or fading-out) effect of the particle shadow, in particular the details how the particle shadow darkens for particles with higher energies may allow us to draw conclusions about the nature of the spacetime and its physical parameters. Thus, theoretically this effect can be used to characterize the spacetime of an astrophysical black hole we observe. However, the angular resolution of current particle detectors like Super-Kamiokande \cite{Fukuda2003,Ashie2005} and IceCube \cite{Achterberg2006} is far too low to observe this effect, in particular since their main targets are neutrinos, which move at velocities close to the speed of light and interact only very weakly with other matter.\\
We also wrote down, to our knowledge for the first time, an exact lens equation for massive particles for a stationary and axisymmetric spacetime, namely the charged NUT metric. For this purpose we transferred the approach for lightlike geodesics presented in Frost \cite{Frost2022,Frost2022b} to unbound timelike geodesics. It was not surprising that for a fixed particle energy the lens maps show the same patterns as the lens maps for light rays. For the spherically symmetric and static Schwarzschild and Reissner-Nordstr\"{o}m spacetimes for fixed particle energies images of different orders from the same quadrant on the two-sphere of particle sources are clearly separated. The critical curves lie on circles marking the boundaries between rings of images of different orders. For the NUT metric and the charged NUT metric the lensing pattern shows a twist. The formerly disconnected areas with images of first and second order of particle sources on the same quadrant of the two-sphere of particle sources connect and we have two sharp boundaries between images of first and second order for timelike geodesics crossing the axes (the same is the case for images of third and fourth order and so on). As for lightlike geodesics \cite{Frost2022,Frost2022b} the critical curves are very likely circles because on one hand the NUT and charged NUT spacetimes have an $SO(3,\mathbb{R})$ symmetry and on the other hand the boundaries between images of first and second, and third and fourth order (and so on) are still circles. Again if we only have particles with a fixed energy or a particle distribution with a very narrow energy peak we have the same problem as for the shadow. We have a degeneracy with respect to the energy $E$, the radius coordinates of the observer and the particle source $r_{O}$ and $r_{P}$, the electric charge $e$, and the gravitomagnetic charge $n$. However, the strength of the twist pattern may allow one to some degree to infer the magnitude of the gravitomagnetic charge $n$. Luckily, an interesting fact came to light when we considered a particle distribution with a broader energy spectrum. For the spherically symmetric and static Schwarzschild and Reissner-Nordstr\"{o}m spacetimes we saw that in this case the images of one specific particle source form a line along a constant longitude. For low-energy particles images of third order can even overlap with images of first order of particles with higher energies from the same particle source. For the NUT metric and the charged NUT metric the twist will very likely lead to the formation of small arclets, whose shape can potentially be used to determine the magnitude of the gravitomagnetic charge. In addition, we observed that the presence of an electric charge leads to an apparent latitudinal broadening of the features in the lens maps when we approach the particle shadow. This effect was, according to our knowledge, first noted by Frost \cite{Frost2022b} in the lens map for light rays. It is also implicitly contained in the results of earlier work (see, e.g., Bozza \cite{Bozza2002}) and seems to be a characteristic feature in the presence of an electric charge and when we combine the image positions for massive particles and light rays it may help to determine the electric charge of a black hole.\\
Finally we discussed two different travel time measures: on one hand the travel time and on the other hand the traveled proper time. The travel time maps generally show the same patterns as for lightlike geodesics. For the spherically symmetric and static Schwarzschild and Reissner-Nordstr\"{o}m spacetimes the travel time maps are rotationally symmetric about the center of the shadow. In the travel time maps for the NUT spacetime and the charged NUT spacetime this symmetry is broken and we observed the same discontinuities for timelike geodesics crossing the Misner string(s) as well as the same spiral-shaped pattern as for lightlike geodesics \cite{Frost2022,Frost2022b}. However, for both types of spacetimes the travel time for massive particles is generally longer than for light rays. The second travel time measure we discussed is the traveled proper time. For all four spacetimes the traveled proper time turned out to be independent of the celestial longitude. It decreased with increasing energy of the particle and increasing electric charge. On the other hand the presence of a gravitomagnetic charge lead to an increase of the traveled proper time.\\
The question in how far the results presented in this paper are of astrophysical relevance is difficult to answer. On one hand the angular resolution of currently existing particle detectors such as Super-Kamiokande \cite{Fukuda2003,Ashie2005} or IceCube \cite{Achterberg2006} is too low to resolve most of the lensing features discussed in this article. Even the resolution of future particle detectors such as Hyper-Kamiokande \cite{TheHKKCollaboration2018,Itow2021} (the angular resolution of Hyper-Kamiokande is expected to be similar to that of Super-Kamiokande \cite{Itow2023}) and PINGU \cite{Aartsen2017} will still be too low to resolve the features we discussed in this article. However, if we assume that we have a particle spectrum with particles with different energies, the detectors may still to some degree be able to indirectly resolve the images of particle sources and their shape via the probability distribution function of the sky localization on the observer's celestial sphere. For the spherically symmetric and static Schwarzschild and Reissner-Nordstr\"{o}m spacetimes in the ideal case the probability distribution function for the sky localization of the spectrum is located on a line along a constant longitude. For the NUT spacetime and the charged NUT spacetime we can expect the probability distribution function of the sky localization to be slightly curved. However, because the electric charge $e$ and the distances of the observer and the particle source to the black hole $r_{O}$ and $r_{P}$ result in a scaling we still have a degeneracy with respect to these parameters. In addition, the discontinuities in the travel time maps also provide a clear indicator for the absence or presence of a gravitomagnetic charge. Furthermore, a thorough discussion of the lens and travel time maps revealed that combining travel time differences between images of particle signals with the same or different energies, between images generated by light rays, and between images generated by particle signals and light rays has the potential to allow the characterization of the spacetime describing an astrophysical black hole and to determine its physical parameters, in the case of the charged NUT metric in particular the determination of the gravitomagnetic charge.\\
Unfortunately, in space, outside the Solar System, most neutrino emissions can either not be directly associated with one particular source or in the case of burst sources with very characteristic particle emission, such as supernova explosions, binary neutron star mergers (for these neutrino emission is expected but has not yet been observed), or tidal disruption events, they tend to be very short-lived. Thus in the near future it will be difficult to measure travel time differences between two neutrino signals emitted by these sources (due to their weak interaction with their environment only neutrinos are suitable candidates). In addition, considering the sensitivity and the angular resolution of current and near future neutrino detectors observing travel time differences between two neutrino signals is still beyond our technological capabilities. On the other hand when we have enough information about the source, e.g., from observations in the full electromagnetic spectrum, and can determine the nature and particle spectrum of the source, decay processes as well as neutrino oscillations along timelike geodesics (for most astrophysical observations most likely only the latter) may allow us to infer the traveled proper time. When the source emits particles which decay along the geodesic along which they travel we can estimate the number of particles emitted by the source and compare it to the number of particles detected by an appropriate particle detector. Assuming that the specific decay laws for these particles are known we can calculate the proper time along the geodesic. A second option for directly observing the traveled proper time is offered by neutrino oscillations. On their path from the source to a detector on Earth neutrinos oscillate between different flavors. When we know the neutrino flavor composition at the source and at the observer we can model the neutrino oscillations along the timelike geodesic and infer the traveled proper time. Note, however, while both approaches in theory sound rather easy because of the complexity of and our limited knowledge about astrophysical particle sources and the complexity of the astrophysical environments along the path of these particles from their source to Earth inferring the traveled proper time from particle decay or neutrino oscillations will be rather complicated.\\
As already mentioned above in nature particle emission by strong burst sources which can be more easily detected and characterized with the currently available detector technology, in particular neutrino sources such as supernova explosions, binary neutron star mergers, or tidal disruption events, tends to be rather rare and short-lived. Therefore, observing all features discussed during our theoretical treatment will be very hard to achieve; however, in the context of a multimessenger approach information gained from the observation of gravitationally lensed particles in combination with information from gravitationally lensed electromagnetic radiation, namely the photon shadow, the positions of multiple images of the source on the observer's celestial sphere, and the travel time differences between these images may help to determine the nature of an astrophysical black hole by constraining the parameters in different black hole models in general relativity. In particular in the presence of circular photon and particle shadows combining these information may help to observe the presence and magnitude of an electric or gravitomagnetic charge. \\
As already pointed out above, treating neutrinos as quantum mechanical particles will very likely allow us to get more information about the background spacetime, in particular when neutrino oscillations are properly considered. First calculations for the Schwarzschild spacetime show already promising results. The first investigations on this topic have already been performed more than 25 years ago. As one of the first Fornengo \emph{et al.} \cite{Fornengo1997} investigated the effect of a curved spacetime on the propagation of neutrinos. In their work they assumed that the neutrinos propagate along the trajectories of light rays in the Schwarzschild spacetime. They found that gravitational effects affect the phase when it is written in terms of the energy and proper distance as measured by a local observer. Their results indicate that gravitational lensing affects the interference pattern resulting from neutrinos traveling along different paths around the black hole and the flavor-changing probability. Cardall and Fuller \cite{Cardall1997} performed a similar analysis and also found that neutrino oscillations are affected in the presence of a Schwarzschild geometry. Since these initial investigations several authors, among them Crocker \emph{et al.} \cite{Crocker2004}, and Alexandre and Clough \cite{Alexandre2018} investigated the effect of gravitational scattering in a Schwarzschild spacetime on neutrino oscillations and found that under the right circumstances when source, lens, and observer are properly aligned an observer on Earth can detect a neutrino pattern which differs from the pattern expected for neutrinos propagating in a flat space vacuum. Dvornikov \cite{Dvornikov2020} investigated the influence of a Schwarzschild geometry on neutrino spin oscillations. He found that close to the critical impact parameter after the scattering initially only left polarized ultrarelativistic neutrinos have a transition probability of about 25\% to right polarized neutrinos. Furthermore, very recent results from Swami \emph{et al.} \cite{Swami2020} indicate that the phase difference of neutrinos traveling along different paths in a Schwarzschild geometry depends on the difference of the squared neutrino masses and also on the neutrino masses themselves. These results let us hope that once more sensitive neutrino detectors are available neutrino lensing will not only serve as supplement but as a fully fledged contributor to multimessenger observations. 

\section*{Acknowledgments}
I thank Volker Perlick, Eva Hackmann, and Oleg Tsupko for their valuable comments and our discussions. In addition, I thank the developers of the programming language Julia and its packages for their continuous effort to provide fast and easily accessible tools for mathematical evaluations. I acknowledge support from the Deutsche Forschungsgemeinschaft (DFG, German Research Foundation) under Germany’s Excellence Strategy EXC-2123 QuantumFrontiers-390837967.

\appendix
\section{ELEMENTARY INTEGRALS}\label{App:EmInt}
While solving the equations of motion for $r$, the time coordinate $t$, and the proper time $\tau$ in Secs.~\ref{Sec:EoMSolr}, \ref{Sec:EoMSolt}, and \ref{Sec:EoMtau} we encountered several elementary integrals. In this appendix we briefly summarize them and their solutions.
\subsection{Radial timelike geodesics}\label{App:EmInt1}
We start with the integrals which occurred for radial timelike geodesics with $K=0$ for the Schwarzschild metric and the Reissner-Nordstr\"{o}m metric. The integrals for the Schwarzschild metric can be rewritten in terms of the following four elementary integrals and their solutions:
\begin{align}
I_{1}=&\int\frac{x\text{d}x}{\sqrt{c_{1}x^2+c_{2}x}}=\frac{\sqrt{(c_{1}x+c_{2})x}}{c_{1}}-\frac{c_{2}}{c_{1}^{\frac{3}{2}}}\ln\left(c_{1}\sqrt{x}+\sqrt{c_{1}(c_{1}x+c_{2})}\right),\label{eq:I1}\\
I_{2}=&\int\frac{\text{d}x}{\sqrt{c_{1}x^2+c_{2}x}}=\frac{2}{\sqrt{c_{1}}}\ln\left(c_{1}\sqrt{x}+\sqrt{c_{1}(c_{1}x+c_{2})}\right),\label{eq:I2}\\
I_{3}=&\int\frac{\text{d}x}{x\sqrt{c_{1}x^2+c_{2}x}}=-\frac{2\sqrt{x(c_{1}x+c_{2})}}{c_{2}x},\label{eq:I3}\\
I_{4}=&\int\frac{\text{d}x}{(x-c_{3})\sqrt{c_{1}x^2+c_{2}x}}=-\frac{2}{\sqrt{(c_{1}c_{3}+c_{2})c_{3}}}\text{arcoth}\left(\sqrt{\frac{(c_{1}c_{3}+c_{2})x}{(c_{1}x+c_{2})c_{3}}}\right),\label{eq:I4}
\end{align}
where in our case we have $c_{1}=E^2-1$ and $c_{2}=c_{3}=2m$. Similarly the integrals for the Reissner-Nordstr\"{o}m metric can be rewritten in terms of the following five elementary integrals and their solutions:
\begin{align}
I_{5}=&\int\frac{x\text{d}x}{\sqrt{c_{1}x^2+c_{2}x+c_{3}}}=\frac{\sqrt{c_{1}x^2+c_{2}x+c_{3}}}{c_{1}}-\frac{c_{2}}{2c_{1}^{\frac{3}{2}}}\ln\left(2\sqrt{c_{1}(c_{1}x^2+c_{2}x+c_{3})}+2c_{1}x+c_{2}\right),\label{eq:I5}\\
I_{6}=&\int\frac{\text{d}x}{\sqrt{c_{1}x^2+c_{2}x+c_{3}}}=\frac{1}{\sqrt{c_{1}}}\ln\left(2\sqrt{c_{1}(c_{1}x^2+c_{2}x+c_{3})}+2c_{1}x+c_{2}\right),\label{eq:I6}\\
I_{7}=&\int\frac{\text{d}x}{x\sqrt{c_{1}x^2+c_{2}x+c_{3}}}=\frac{1}{\sqrt{-c_{3}}}\arcsin\left(\frac{c_{2}x+2c_{3}}{x\sqrt{c_{2}^2-4c_{1}c_{3}}}\right),\label{eq:I7}\\
I_{8}=&\int\frac{\text{d}x}{(x-c_{4})\sqrt{c_{1}x^2+c_{2}x+c_{3}}}=\frac{1}{\sqrt{c_{1}c_{4}^2+c_{2}c_{4}+c_{3}}}\biggl(\ln\left(x-c_{4}\right)\label{eq:I8}\\
&-\ln\left(2(c_{1}c_{4}^2+c_{2}c_{4}+c_{3})+(c_{2}+2c_{1}c_{4})(x-c_{4})+2\sqrt{(c_{1}c_{4}^2+c_{2}c_{4}+c_{3})(c_{1}x^2+c_{2}x+c_{3})}\right)\biggr),\nonumber\\
I_{9}=&\int\frac{\text{d}x}{(x-c_{4})^2\sqrt{c_{1}x^2+c_{2}x+c_{3}}}=-\frac{\sqrt{c_{1}x^2+c_{2}x+c_{3}}}{(c_{1}c_{4}^2+c_{2}c_{4}+c_{3})(x-c_{4})}-\frac{c_{2}+2c_{1}c_{4}}{2(c_{1}c_{4}^2+c_{2}c_{4}+c_{3})^{\frac{3}{2}}}\biggl(\ln\left(x-c_{4}\right)\label{eq:I9}\\
&-\ln\left(2(c_{1}c_{4}^2+c_{2}c_{4}+c_{3})+(c_{2}+2c_{1}c_{4})(x-c_{4})+2\sqrt{(c_{1}c_{4}^2+c_{2}c_{4}+c_{3})(c_{1}x^2+c_{2}x+c_{3})}\right)\biggr),\nonumber
\end{align}
where this time we have $c_{1}=E^2-1$, $c_{2}=2m$, $c_{3}=-e^2$, and $c_{4}$ can be $r_{\text{H}_{\text{o}}}$ [only in (\ref{eq:I8})], $r_{\text{H}_{\text{i}}}$ [only in (\ref{eq:I8})], or $r_{\text{H}}$.

\subsection{Timelike geodesics with $K=V_{E}(r_{\text{pa}_{-}})$}\label{App:EmInt2}
In this subsection we discuss the elementary integrals associated with timelike geodesics with a real double root at $r_{1}=r_{2}=r_{\text{pa}_{-}}$. They do not have turning points in the domain of outer communication and only occur for the NUT metric and the charged NUT metric. In total we have four different elementary integrals. The integrals and their solutions read
\begin{align}
I_{10}=&\int\frac{r\text{d}r}{\sqrt{(R_{3}-r)^2+R_{4}^2}}=\sqrt{(R_{3}-r)^2+R_{4}^2}+R_{3}\text{arsinh}\left(\frac{r-R_{3}}{R_{4}}\right),\label{eq:I10}\\
I_{11}=&\int\frac{\text{d}r}{\sqrt{(R_{3}-r)^2+R_{4}^2}}=\text{arsinh}\left(\frac{r-R_{3}}{R_{4}}\right),\label{eq:I11}\\
I_{12}=&\int\frac{\text{d}r}{(r-r_{a})\sqrt{(R_{3}-r)^2+R_{4}^2}}=-\frac{1}{\sqrt{(R_{3}-r_{a})^2+R_{4}^2}}\text{arsinh}\left(\frac{(r_{a}-R_{3})(r-r_{a})+(R_{3}-r_{a})^2+R_{4}^2}{R_{4}(r-r_{a})}\right),\label{eq:I12}\\
I_{13}=&\int\frac{\text{d}r}{(r-r_{a})^2\sqrt{(R_{3}-r)^2+R_{4}^2}}=\frac{r_{a}-R_{3}}{((R_{3}-r_{a})^2+R_{4}^2)^{\frac{3}{2}}}\text{arsinh}\left(\frac{(r_{a}-R_{3})(r-r_{a})+(R_{3}-r_{a})^2+R_{4}^2}{R_{4}(r-r_{a})}\right)\label{eq:I13}\\
&-\frac{\sqrt{(R_{3}-r)^2+R_{4}^2}}{((R_{3}-r_{a})^2+R_{4}^2)(r-r_{a})},\nonumber
\end{align}
where in (\ref{eq:I12}) the parameter $r_{a}$ can either be $r_{1}=r_{2}=r_{\text{pa}_{-}}$, $r_{\text{H}_{\text{o}}}$, $r_{\text{H}_{\text{i}}}$, or $r_{\text{H}}$, while in (\ref{eq:I13}) $r_{a}$ can only be $r_{\text{H}}$.

\subsection{Timelike geodesics with $K=V_{E}(r_{\text{pa}})$}\label{App:EmInt3}
In this subsection we discuss the elementary integrals associated with timelike geodesics asymptotically coming from or going to an unstable particle sphere at $r_{\text{pa}}$. In total we have two different elementary integrals. After applying the coordinate transformation
(\ref{eq:Case3sub}) they read
\begin{align}
I_{14}=\int\frac{\text{d}y}{(y-y_{a})\sqrt{y-y_{1}}},\label{eq:I14y}\\
I_{15}=\int\frac{\text{d}y}{(y-y_{a})^2\sqrt{y-y_{1}}}.\label{eq:I15y}
\end{align}
Here, in (\ref{eq:I14y}) the parameter $y_{a}$ can take the values $y_{\text{pa}}$, $a_{2}/12$, $y_{\text{H}_{\text{o}}}$, $y_{\text{H}_{\text{i}}}$, or $y_{\text{H}}$ while in (\ref{eq:I15y}) it can only take the values $a_{2}/12$ or $y_{\text{H}}$.
Here, the coordinate transformation (\ref{eq:Case3sub}) relates $y_{1}$, $y_{\text{pa}}$, $y_{\text{H}_{\text{o}}}$, $y_{\text{H}_{\text{i}}}$, and $y_{\text{H}}$ to $r_{4}$, $r_{\text{pa}}$, $r_{\text{H}_{\text{o}}}$, $r_{\text{H}_{\text{i}}}$, and $r_{\text{H}}$, respectively. Now we have to distinguish the cases $y_{a}=a_{2}/12$, $y_{a}=y_{\text{pa}}$ for $r_{\text{H}_{\text{o}}}<r<r_{\text{pa}}$, and $y_{a}=y_{\text{pa}}$ for $r_{\text{pa}}<r$ from all other cases.\\
We start with calculating (\ref{eq:I14y}). We have to distinguish two different cases. When $r_{a}>r$ we have $y>y_{a}$. This also includes the case $y_{a}=a_{2}/12$. In this case we substitute $z=y-y_{a}$. In the second case we have $r>r_{a}$ and thus $y<y_{a}$. In this case we substitute $z=y-y_{1}$. We integrate and rewrite the obtained results in terms of $r$. In total we obtain three structurally different results. They read
\begin{align}
I_{14_{1}}&=-4\sqrt{\frac{r_{3}-r_{4}}{a_{3}}}\text{arcoth}\left(\sqrt{\frac{r-r_{4}}{r-r_{3}}}\right),\label{eq:I141}\\
I_{14_{2}}&=-4\sqrt{\frac{(r_{a}-r_{3})(r_{3}-r_{4})}{a_{3}(r_{a}-r_{4})}}\text{arcoth}\left(\sqrt{\frac{(r-r_{4})(r_{a}-r_{3})}{(r-r_{3})(r_{a}-r_{4})}}\right),\label{eq:I142}\\
I_{14_{3}}&=-4\sqrt{\frac{(r_{a}-r_{3})(r_{3}-r_{4})}{a_{3}(r_{a}-r_{4})}}\text{artanh}\left(\sqrt{\frac{(r-r_{4})(r_{a}-r_{3})}{(r-r_{3})(r_{a}-r_{4})}}\right),\label{eq:I143}
\end{align}
where the first result $I_{14_{1}}$ (\ref{eq:I141}) is only valid for $y_{a}=a_{2}/12$. The second result $I_{14_{2}}$ (\ref{eq:I142}) only occurs for timelike geodesics with $r_{\text{H}_{\text{o}}}<r<r_{\text{pa}}$ and thus the parameter $r_{a}$ only takes the value $r_{\text{pa}}$. In the third result $I_{14_{3}}$ (\ref{eq:I143}) the parameter $r_{a}$ can take four different values. These are $r_{\text{pa}}$, $r_{\text{H}_{\text{o}}}$, $r_{\text{H}_{\text{i}}}$, and $r_{\text{H}}$, however, the integral with $r_{a}=r_{\text{pa}}$ only occurs for timelike geodesics with $r_{\text{pa}}<r$.\\
Analogously we proceed for the second integral (\ref{eq:I15y}). Here, we only have to distinguish two different cases. After the integration and rewriting the obtained results in terms of $r$ we get
\begin{align}
I_{15_{1}}&=\frac{8\sqrt{r_{3}-r_{4}}}{a_{3}^{\frac{3}{2}}}\left((r_{3}-r_{4})\text{arcoth}\left(\sqrt{\frac{r-r_{4}}{r-r_{3}}}\right)-\sqrt{(r-r_{3})(r-r_{4})}\right)
,\label{eq:I151}\\
I_{15_{2}}&=\frac{8}{a_{3}^{\frac{3}{2}}}\left(\frac{(r_{a}-r_{3})^2\sqrt{(r_{3}-r_{4})(r-r_{3})(r-r_{4})}}{(r_{a}-r_{4})(r-r_{a})}+\left(\frac{(r_{3}-r_{4})(r_{a}-r_{3})}{r_{a}-r_{4}}\right)^{\frac{3}{2}}\text{artanh}\left(\sqrt{\frac{(r-r_{4})(r_{a}-r_{3})}{(r-r_{3})(r_{a}-r_{4})}}\right)\right),\label{eq:I152}
\end{align}
where the first result $I_{15_{1}}$ (\ref{eq:I151}) is only valid for $y_{a}=a_{2}/12$ and in the second result $I_{15_{2}}$ (\ref{eq:I152}) the parameter $r_{a}$ can only take the value $r_{\text{H}}$.

\section{ELLIPTIC INTEGRALS}\label{App:EllInt}
In Sec.~\ref{Sec:EoM}, in particular Secs.~\ref{Sec:EoMSolt} and \ref{Sec:EoMtau}, we encountered several general (pseudo)elliptic integrals. All of them can be rewritten in terms of elementary functions and Legendre's elliptic integrals of the first, second, and third kind. Since not all readers might be familiar with Legendre's elliptic integrals in this appendix we will briefly introduce them and then show how we can use them and elementary functions to rewrite nine different nonstandard (pseudo)elliptic integrals in terms of elementary functions and Legendre's elliptic integrals of the first, second, and third kind. \\
Legendre's incomplete elliptic integrals of the first, second, and third kind are defined by
\begin{align}
F_{L}(\chi,k)=&\int_{0}^{\chi}\frac{\text{d}\chi'}{\sqrt{1-k\sin^2\chi'}},\label{eq:LEIF}\\
E_{L}(\chi,k)=&\int_{0}^{\chi}\sqrt{1-k\sin^2\chi'}\text{d}\chi',\label{eq:LEIE}\\
\Pi_{L}(\chi,k,\tilde{n})=&\int_{0}^{\chi}\frac{\text{d}\chi'}{(1-\tilde{n}\sin^2\chi')\sqrt{1-k\sin^2\chi'}},\label{eq:LEIPi}
\end{align}
where $\chi$ is the so-called amplitude, $k$ is the square of the elliptic modulus and $\tilde{n}\in \mathbb{R}$ is an arbitrary real parameter. Note that when we add or subtract an integer multiple $\hat{n}$ of $\pi$ in the limits and transform $\chi \rightarrow \chi\mp\hat{n}\pi$ the integrands remain invariant. When we have $\chi=\pi/2$ the elliptic integrals are commonly referred to as complete elliptic integrals and the amplitude $\chi$ is omitted. In addition, the complete elliptic integral of the first kind is usually written as $K_{L}(k)=F_{L}(\pi/2,k)$. As we can easily read from (\ref{eq:LEIPi}) the denominator of the integrand of the elliptic integral of the third kind leads to a singularity whenever $1\leq \tilde{n}$. In this case we can use (17.7.7) and (17.7.8) in \cite{MilneThomson1972} to rewrite (\ref{eq:LEIPi}) as
\begin{equation}\label{eq:LEIPin}
\Pi_{L}(\chi,k,\tilde{n})=F_{L}(\chi,k)-\Pi_{L}\left(\chi,k,\frac{k}{\tilde{n}}\right)+\frac{1}{2p}\ln\left(\frac{\cos\chi \sqrt{1-k\sin^2\chi}+p\sin\chi}{\left|\cos\chi\sqrt{1-k\sin^2\chi}-p\sin\chi\right|}\right),
\end{equation}
where
\begin{equation}
p=\sqrt{\frac{(\tilde{n}-1)(\tilde{n}-k)}{\tilde{n}}}.
\end{equation}
In addition, in the course of this paper we encountered nine different nonstandard elliptic and pseudoelliptic integrals. They read
\begin{align}
J_{L_{1}}(\chi_{i},\chi,k_{1},\tilde{n})=&\int_{\chi_{i}}^{\chi}\frac{\text{d}\chi'}{(1+\tilde{n}\tan\chi')\sqrt{1-k_{1}\sin^2\chi'}},\label{eq:J1}
\end{align}
\begin{align}
J_{L_{2}}(\chi_{i},\chi,k_{1},\tilde{n})=&\int_{\chi_{i}}^{\chi}\frac{\text{d}\chi'}{(1+\tilde{n}\tan\chi')^2\sqrt{1-k_{1}\sin^2\chi'}},\label{eq:J2}
\end{align}
\begin{align}
J_{L_{3}}(\chi_{i},\chi,k_{1})=&\int_{\chi_{i}}^{\chi}\frac{\tan\chi'\text{d}\chi'}{\sqrt{1-k_{1}\sin^2\chi'}},\label{eq:J3}
\end{align}
\begin{align}
J_{L_{4}}(\chi_{i},\chi,k_{1})=&\int_{\chi_{i}}^{\chi}\frac{\tan^2\chi'\text{d}\chi'}{\sqrt{1-k_{1}\sin^2\chi'}},\label{eq:J4}
\end{align}
\begin{align}
J_{L_{5}}(\chi_{i},\chi,k_{2},\tilde{n})=&\int_{\chi_{i}}^{\chi}\frac{\text{d}\chi'}{(1+\tilde{n}\cos\chi')\sqrt{1-k_{2}\sin^2\chi'}},\label{eq:J5}
\end{align}
\begin{align}
J_{L_{6}}(\chi_{i},\chi,k_{2},\tilde{n})=&\int_{\chi_{i}}^{\chi}\frac{\text{d}\chi'}{(1+\tilde{n}\cos\chi')^2\sqrt{1-k_{2}\sin^2\chi'}},\label{eq:J6}
\end{align}
\begin{align}
J_{L_{7}}(\chi_{i},\chi,k_{2})=&\int_{\chi_{i}}^{\chi}\frac{\text{d}\chi'}{\cos\chi'\sqrt{1-k_{2}\sin^2\chi'}},\label{eq:J7}
\end{align}
\begin{align}
J_{L_{8}}(\chi_{i},\chi,k_{2})=&\int_{\chi_{i}}^{\chi}\frac{\text{d}\chi'}{\cos^2\chi'\sqrt{1-k_{2}\sin^2\chi'}},\label{eq:J8}
\end{align}
\begin{align}
J_{L_{9}}(\chi_{i},\chi,\tilde{k},\tilde{n})=&\int_{\chi_{i}}^{\chi}\frac{\text{d}\chi'}{(1-\tilde{n}\sin^2\chi')^2\sqrt{1-\tilde{k}\sin^2\chi'}}.\label{eq:J9}
\end{align}
In the following we will summarize how we can rewrite them in terms of elementary functions and Legendre's elliptic integrals of the first, second, and third kind. We start with the integrals $J_{L_{1}}(\chi_{i},\chi,k_{1},\tilde{n})$ and $J_{L_{2}}(\chi_{i},\chi,k_{1},\tilde{n})$. In this case $\chi_{i}$ and $\chi$ are related to $r_{i}$ and $r(\lambda)$ (note that here we omit $\lambda$ for $\chi$) by (\ref{eq:Case1NUTachi}) and the square of the elliptic modulus $k_{1}$ is given by (\ref{eq:Case1NUTak}). Note that after the substitution (\ref{eq:Case1NUTasub}) we effectively obtain the forms given by (267.01) and (267.02) in Byrd and Friedman \cite{Byrd1954}. In both integrals we can have $\tilde{n}=g_{0}$ or $\tilde{n}=n_{1}$, where
\begin{equation}
n_{1}=\frac{R_{2}+g_{0}(R_{1}-r_{\tilde{\text{H}}})}{R_{1}-g_{0}R_{2}-r_{\tilde{\text{H}}}},
\end{equation}
where $R$ and $\bar{R}$ are given by (\ref{eq:Case1NUTaCoeff1}) and (\ref{eq:Case1NUTaCoeff2}). $r_{\tilde{\text{H}}}$ can be $r_{\text{H}_{\text{i}}}$, $r_{\text{H}_{\text{o}}}$, or $r_{\text{H}}$. Using elementary functions and Legendre's elliptic integrals of the first, second, and third kind (\ref{eq:J1}) and (\ref{eq:J2}) can now be rewritten as (note that here we omit $\chi_{i}$ in the argument of the functions)
\begin{equation}\label{eq:J1R}
J_{L_{1}}(\chi,k_{1},\tilde{n})=\frac{F_{L}(\chi,k_{1})+\tilde{n}^2\Pi_{L}(\chi,k_{1},1+\tilde{n}^2)}{1+\tilde{n}^2}+\frac{\tilde{n}J_{L}(\chi,k_{1},\tilde{n})}{2\sqrt{(1+\tilde{n}^2)(1-k_{1}+\tilde{n}^2)}},
\end{equation}
\begin{align}\label{eq:J2R}
J_{L_{2}}(\chi,k_{1},\tilde{n})=&\frac{F_{L}(\chi,k_{1})}{(1+\tilde{n}^2)^2}+\frac{\tilde{n}^2}{(1+\tilde{n}^2)(1-k_{1}+\tilde{n}^2)}\left(\tilde{n}+\frac{\sin\chi-\tilde{n}\cos\chi}{\cos\chi+\tilde{n}\sin\chi}\sqrt{1-k_{1}\sin^2\chi}-E_{L}(\chi,k_{1})\right)\\
&+\frac{2(1-k_{1}+\tilde{n}^2)-\tilde{n}^2k_{1}}{(1+\tilde{n}^2)(1-k_{1}+\tilde{n}^2)}\left(\frac{\tilde{n}^2\Pi_{L}(\chi,k_{1},1+\tilde{n}^2)}{1+\tilde{n}^2}+\frac{\tilde{n}J_{L}(\chi,k_{1},\tilde{n})}{2\sqrt{(1+\tilde{n}^2)(1-k_{1}+\tilde{n}^2)}}\right),\nonumber
\end{align}
where in both cases the function $J_{L}(\chi,k_{1},\tilde{n})$ is given by
\begin{equation}\label{eq:JL}
J_{L}(\chi,k_{1},\tilde{n})=\ln\left(\left|\frac{\left(1+\sqrt{\frac{1+\tilde{n}^2}{1-k_{1}+\tilde{n}^2}}\right)\left(1-\sqrt{\frac{1+\tilde{n}^2}{1-k_{1}+\tilde{n}^2}}\sqrt{1-k_{1}\sin^2\chi}\right)}{\left(1-\sqrt{\frac{1+\tilde{n}^2}{1-k_{1}+\tilde{n}^2}}\right)\left(1+\sqrt{\frac{1+\tilde{n}^2}{1-k_{1}+\tilde{n}^2}}\sqrt{1-k_{1}\sin^2\chi}\right)}\right|\right).
\end{equation}
Note that because $1<1+\tilde{n}^2$ we use (\ref{eq:LEIPin}) to avoid the divergence of Legendre's elliptic integral of the third kind. Now we continue with $J_{L_{3}}(\chi_{i},\chi,k_{1})$ and $J_{L_{4}}(\chi_{i},\chi,k_{1})$. We have the same relations for $\chi_{i}$, $\chi$, and $k_{1}$ as for $J_{L_{1}}(\chi_{i},\chi,k_{1},\tilde{n})$ and $J_{L_{2}}(\chi_{i},\chi,k_{1},\tilde{n})$.
Both integrals occur as special cases for $e=\sqrt{m^2+n^2}$, $K=0$, and $\Delta<0$ when we rewrite (\ref{eq:tr2}) in terms of elementary functions and Legendre's elliptic integrals (we can easily show that in this case we have $r_{\text{H}}=m=R_{2}/g_{0}+R_{1}$). Note that in this case $\chi_{\text{H}}=\chi(r=r_{\text{H}})=\pi/2$ and thus we have $\pi/2<\chi_{i},\chi$. In terms of elementary functions and Legendre's elliptic integral of the second kind (\ref{eq:J3}) and (\ref{eq:J4}) now read (we again omit $\chi_{i}$ in the argument and rewrite (\ref{eq:J3}) as antiderivative without integration constant)
\begin{equation}\label{eq:J3R}
J_{L_{3}}(\chi,k_{1})=\frac{1}{\sqrt{1-k_{1}}}\text{arcoth}\left(\sqrt{\frac{1-k_{1}\sin^2\chi}{1-k_{1}}}\right),
\end{equation}
\begin{equation}\label{eq:J4R}
J_{L_{4}}(\chi,k_{1})=\frac{\sqrt{1-k_{1}\sin^2\chi}\tan\chi-E_{L}(\chi,k_{1})}{1-k_{1}}.
\end{equation}
Note that because we can write (\ref{eq:J3R}) in terms of elementary functions (\ref{eq:J3}) is a so-called pseudoelliptic integral. \\ 
Now we turn to $J_{L_{5}}(\chi_{i},\chi,k_{2},\tilde{n})$ and $J_{L_{6}}(\chi_{i},\chi,k_{2},\tilde{n})$. For both integrals $\chi_{i}$ and $\chi$ are related to $r_{i}$ and $r(\lambda)$ (note that we again omit $\lambda$ for $\chi$) by (\ref{eq:Case2chi}) and the square of the elliptic modulus $k_{2}$ is given by (\ref{eq:Case2k}). In this case we have either $\tilde{n}=n_{2}$ or $\tilde{n}=n_{3}$, where $n_{2}$ and $n_{3}$ are given by
\begin{equation}
n_{2}=\frac{\bar{R}+R}{\bar{R}-R},
\end{equation}
\begin{equation}
n_{3}=\frac{(r_{\tilde{\text{H}}}-r_{1})\bar{R}+(r_{\tilde{\text{H}}}-r_{2})R}{(r_{\tilde{\text{H}}}-r_{1})\bar{R}-(r_{\tilde{\text{H}}}-r_{2})R},
\end{equation}
where $R$ and $\bar{R}$ are given by (\ref{eq:Case2Coeff1}) and (\ref{eq:Case2Coeff2}). In $n_{3}$ $r_{\tilde{\text{H}}}$ can be $r_{\text{H}_{\text{i}}}$, $r_{\text{H}_{\text{o}}}$, or $r_{\text{H}}$. For both integrals the integration procedure is straight forward. For (\ref{eq:J5}) we first expand by $1-\tilde{n}\cos\chi'$ and split the integral into two terms. Now it reads (we again drop $\chi_{i}$ in the argument)
\begin{equation}\label{eq:J5I}
J_{L_{5}}(\chi,k_{2},\tilde{n})=\frac{1}{1-\tilde{n}^2}\left(\Pi_{L}\left(\chi,k_{2},\frac{\tilde{n}^2}{\tilde{n}^2-1}\right)-\tilde{n}\int_{0}^{\chi}\frac{\cos\chi'\text{d}\chi'}{\left(1-\frac{\tilde{n}^2}{\tilde{n}^2-1}\sin^2\chi'\right)\sqrt{1-k_{2}\sin^2\chi'}}\right),
\end{equation}
where we already rewrote the first term as Legendre's elliptic integral of the third kind. The second term is an elementary integral. The evaluation of the elementary integral requires several case by case analyses which we will not reproduce here. The final result reads
\begin{equation}\label{eq:J5R}
J_{L_{5}}(\chi,k_{2},\tilde{n})=\frac{\Pi_{L}\left(\chi,k_{2},\frac{\tilde{n}^2}{\tilde{n}^2-1}\right)}{1-\tilde{n}^2}+\frac{\tilde{n}\tilde{J}_{L}(\chi,k_{2},\tilde{n})}{2\sqrt{(\tilde{n}^2-1)(\tilde{n}^2(1-k_{2})+k_{2})}},
\end{equation}
where $\tilde{J}_{L}(\chi,k_{2},\tilde{n})$ is given by (\ref{eq:JTL}) below. For (\ref{eq:J6}) we proceed analogously. First we expand by $(1-\tilde{n}\cos\chi')^2$ and get
\begin{align}\label{eq:J6I}
J_{L_{6}}(\chi,k_{2},\tilde{n})=&\frac{2}{(\tilde{n}^2-1)^2}\left(\int_{0}^{\chi}\frac{\text{d}\chi'}{\left(1-\frac{\tilde{n}^2}{\tilde{n}^2-1}\sin^2\chi'\right)^2\sqrt{1-k_{2}\sin^2\chi'}}\right.\\
&\left.-\tilde{n}\int_{0}^{\chi}\frac{\cos\chi'\text{d}\chi'}{\left(1-\frac{\tilde{n}^2}{\tilde{n}^2-1}\sin^2\chi'\right)^2\sqrt{1-k_{2}\sin^2\chi'}}\right)+\frac{\Pi_{L}\left(\chi,k_{2},\frac{\tilde{n}^2}{\tilde{n}^2-1}\right)}{\tilde{n}^2-1}.\nonumber
\end{align}
This time we got three different terms. The first and the third term are again elliptic integrals. The latter we already rewrote in terms of Legendre's elliptic integral of the third kind. The second term is an elementary integral. Again the evaluation of the elementary integral requires
several case by case analyses, which are too long to be reproduced here. The first term can be rewritten in terms of elementary functions and Legendre's elliptic integrals of the first, second, and third kind using (\ref{eq:J9R}) below. When we evaluate all terms and simplify them the result reads
\begin{align}\label{eq:J6R}
J_{L_{6}}(\chi,k_{2},\tilde{n})&=\frac{\tilde{n}^3\sin\chi\sqrt{1-k_{2}\sin^2\chi}}{(\tilde{n}^2-1)(\tilde{n}^2(1-k_{2})+k_{2})(1+\tilde{n}\cos\chi)}-\frac{\tilde{n}(\tilde{n}^2(1-2k_{2})+2k_{2})\tilde{J}_{L}(\chi,k_{2},\tilde{n})}{2((\tilde{n}^2-1)(\tilde{n}^2(1-k_{2})+k_{2}))^{\frac{3}{2}}}+\frac{F_{L}(\chi,k_{2})}{\tilde{n}^2-1}\\
&-\frac{\tilde{n}^2E_{L}(\chi,k_{2})}{(\tilde{n}^2-1)(\tilde{n}^2(1-k_{2})+k_{2})}+\frac{(\tilde{n}^2(1-2k_{2})+2k_{2})\Pi_{L}\left(\chi,k_{2},\frac{\tilde{n}^2}{\tilde{n}^2-1}\right)}{(\tilde{n}^2-1)^2(\tilde{n}^2(1-k_{2})+k_{2})}.\nonumber
\end{align}
In both, (\ref{eq:J5R}) and (\ref{eq:J6R}) the function $\tilde{J}_{L}(\chi,k_{2},\tilde{n})$ is given by
\begin{equation}\label{eq:JTL}
\tilde{J}_{L}(\chi,k_{2},\tilde{n})=\ln\left(\frac{\sin\chi\sqrt{\frac{\tilde{n}^2(1-k_{2})+k_{2}}{\tilde{n}^2-1}}+\sqrt{1-k_{2}\sin^2\chi}}{\left|\sin\chi\sqrt{\frac{\tilde{n}^2(1-k_{2})+k_{2}}{\tilde{n}^2-1}}-\sqrt{1-k_{2}\sin^2\chi}\right|}\right).
\end{equation}
In both (\ref{eq:J5R}) and (\ref{eq:J6R}), we have $\tilde{n}^2/(\tilde{n}^2-1)>1$ for all $\tilde{n}$ and thus we again use (\ref{eq:LEIPin}) to avoid the divergence of Legendre's elliptic integral of the third kind.\\
Now we proceed to $J_{L_{7}}(\chi_{i},\chi,k_{2})$ and $J_{L_{8}}(\chi_{i},\chi,k_{2})$. The former is a pseudoelliptic integral while the latter is a true elliptic integral. We have the same relations for $\chi_{i}$, $\chi$, and $k_{2}$ as for $J_{L_{5}}(\chi_{i},\chi,k_{2},\tilde{n})$ and $J_{L_{6}}(\chi_{i},\chi,k_{2},\tilde{n})$. Both integrals occur as special cases for $e=\sqrt{m^2+n^2}$, $K=0$, and $\Delta>0$ when we rewrite (\ref{eq:tr2}) in terms of elementary functions and Legendre's elliptic integrals. Note that we can easily show that in this case we have $r_{\text{H}}=m=(r_{1}\bar{R}-r_{2}R)/(\bar{R}-R)$. Note that also in this case we have $\chi_{\text{H}}=\chi(r=r_{\text{H}})=\pi/2$ and thus we have $\pi/2<\chi_{i},\chi$. When we rewrite $J_{L_{7}}(\chi_{i},\chi,k_{2})$ and $J_{L_{8}}(\chi_{i},\chi,k_{2})$ in terms of elementary functions and Legendre's elliptic integrals of the first and second kind they read (again we omit the first argument $\chi_{i}$)
\begin{equation}\label{eq:J7R}
J_{L_{7}}(\chi,k_{2})=\frac{1}{\sqrt{1-k_{2}}}\text{artanh}\left(\frac{\sqrt{1-k_{2}}\sin\chi}{\sqrt{1-k_{2}\sin^2\chi}}\right),
\end{equation}
\begin{equation}\label{eq:J8R}
J_{L_{8}}(\chi,k_{2})=F_{L}(\chi,k_{2})-\frac{E_{L}(\chi,k_{2})}{1-k_{2}}+\frac{\sqrt{1-k_{2}\sin^2\chi}}{1-k_{2}}\tan\chi.
\end{equation}
Note that both integrals diverge when $\chi=\pi/2$.\\
Now the last remaining nonstandard elliptic integral is $J_{L_{9}}(\chi_{i},\chi,\tilde{k},\tilde{n})$. It occurs when we rewrite (\ref{eq:tr2}) in terms of elementary functions and Legendre's elliptic integrals for timelike geodesics which can have turning points at $r_{1}=r_{\text{min}}$ or $r_{2}=r_{\text{max}}$, and when we rewrite (\ref{eq:J6I}) as (\ref{eq:J6R}). Note that for the former $\chi_{i}$ and $\chi$ are related to $r_{i}$ and $r(\lambda)$ (again we omit $\lambda$ for $\chi$) by (\ref{eq:Case4chi1}) or (\ref{eq:Case4chi2}), respectively, and the square of the elliptic modulus $\tilde{k}$ is given by $k_{3}$ (\ref{eq:Case4k}). For the latter $\chi_{i}$ and $\chi$ are related to $r_{i}$ and $r(\lambda)$ by (\ref{eq:Case2chi}) and the square of the elliptic modulus $\tilde{k}$ is given by $k_{2}$ (\ref{eq:Case2k}). After rewriting (\ref{eq:J9}) in terms of elementary functions and Legendre's elliptic integrals of the first, second, and third kind and omitting the first argument $\chi_{i}$ it reads
\begin{align}\label{eq:J9R}
&J_{L_{9}}(\chi,\tilde{k},\tilde{n})=\frac{\tilde{n}^2\sin(2\chi)\sqrt{1-\tilde{k}\sin^2\chi}}{4(\tilde{n}-\tilde{k})(\tilde{n}-1)(1-\tilde{n}\sin^2\chi)}+\frac{F_{L}(\chi,\tilde{k})}{2(\tilde{n}-1)}-\frac{\tilde{n}E_{L}(\chi,\tilde{k})}{2(\tilde{n}-\tilde{k})(\tilde{n}-1)}+\frac{\tilde{n}(\tilde{n}-2)-(2\tilde{n}-3)\tilde{k}}{2(\tilde{n}-\tilde{k})(\tilde{n}-1)}\Pi_{L}(\chi,\tilde{k},\tilde{n}).
\end{align}
For timelike geodesics which can have turning points at $r_{1}=r_{\text{min}}$ or $r_{2}=r_{\text{max}}$ we can use (\ref{eq:J9R}) as is. However, when we use (\ref{eq:J9R}) to rewrite (\ref{eq:J6I}) in terms of elementary functions and Legendre's elliptic integrals of the first, second, and third kind we have to replace $\tilde{n}\rightarrow\tilde{n}^2/(\tilde{n}^2-1)$.

\section{SOLVING DIFFERENTIAL EQUATIONS USING JACOBI'S ELLIPTIC FUNCTIONS}\label{App:SolDiff}
In Sec.~\ref{Sec:EoMSolr} we encountered several differential equations which can only be solved using elliptic functions. For this purpose in general different types of elliptic functions can be used. The most popular are certainly Jacobi's elliptic functions and Weierstra\ss' elliptic $\wp$ function. In this paper we will use Jacobi's elliptic functions. In this appendix we will briefly introduce Jacobi's elliptic functions and show how to use them to solve differential equations of the general type
\begin{equation}\label{eq:BDiff}
\left(\frac{\text{d}z}{\text{d}\lambda}\right)^2=d_{4}z^4+d_{3}z^3+d_{2}z^2+d_{1}z+d_{0},
\end{equation}
where in our case $d_{4}$, $d_{3}$, $d_{2}$, $d_{1}$, and $d_{0}$ are real coefficients and $\lambda$ is an arbitrary variable, which, for now, has no relation to the Mino parameter.\\
Before we proceed to demonstrate how to solve (\ref{eq:BDiff}) using Jacobi's elliptic functions let us briefly introduce them. For this purpose let us start with the differential equation
\begin{equation}\label{eq:BDiffLeg}
\left(\frac{\text{d}\chi}{\text{d}\lambda}\right)^2=1-k\sin^2\chi,
\end{equation}
where $k$ is the square of the elliptic modulus. Now we separate variables and integrate using the initial condition $\chi(\lambda_{i}=0)=\chi_{i}=0$. We get
\begin{equation}\label{eq:LegInt}
\lambda=\int_{0}^{\chi}\frac{\text{d}\chi'}{\sqrt{1-k\sin^2\chi'}},
\end{equation}
where $\chi$ is called the amplitude of $\lambda$ ($\chi=\text{am}\lambda$). We now define the so-called \emph{sinus amplitudinis} and the \emph{cosinus amplitudinis} (also known as Jacobi's elliptic $\text{sn}$ and $\text{cn}$ functions) as
\begin{align}
&\sin\chi=\sin\text{am}\lambda=\text{sn}(\lambda,k),\label{eq:SinAmp}\\
&\cos\chi=\cos\text{am}\lambda=\text{cn}(\lambda,k).\label{eq:CosAmp}
\end{align}
Together with the so-called \emph{delta amplitude}
\begin{equation}
\sqrt{1-k\sin^2\chi}=\sqrt{1-k\text{sn}^2(\lambda,k)}=\text{dn}(\lambda,k)
\end{equation}
they form a set of basic elliptic functions. However, in this paper we will only use Jacobi's elliptic $\text{sn}$ and $\text{cn}$ functions. Using the definitions of the sinus amplitudinis and the cosinus amplitudinis we can also define another elliptic function. It is the elliptic analog of the tangent and reads
\begin{equation}
\tan\chi=\frac{\sin\chi}{\cos\chi}=\frac{\text{sn}(\lambda,k)}{\text{cn}(\lambda,k)}=\text{sc}(\lambda,k).
\end{equation}
It is called Jacobi's elliptic $\text{sc}$ function. However, note that due to its similarity to the tangent in the older literature one can also find the notation $\text{sc}(\lambda,k)=\text{tn}(\lambda,k)$. Jacobi's elliptic $\text{sn}$, $\text{cn}$, and $\text{sc}$ functions have now the unique property to solve (\ref{eq:BDiffLeg}). In the following we will now demonstrate how to use them to solve differential equations of the form (\ref{eq:BDiff}). For this purpose let us first rewrite (\ref{eq:BDiff}) in terms of its roots. We get
\begin{equation}\label{eq:BDiffR}
\left(\frac{\text{d}z}{\text{d}\lambda}\right)^2=d_{4}(z-z_{1})(z-z_{2})(z-z_{3})(z-z_{4}),
\end{equation}
where in our case the four roots $z_{1}$, $z_{2}$, $z_{3}$, and $z_{4}$ can be two pairs of distinct complex conjugate roots, a pair of complex conjugate roots and two distinct real roots, or four distinct real roots. So for applying Jacobi's elliptic functions to solve (\ref{eq:BDiffR}) we require that we do not have real multiple roots [in these cases we can use, e.g., the elementary integrals from Appendix~\ref{App:EmInt} to solve (\ref{eq:BDiff})] or a pair of complex conjugate double roots. In all other cases we now apply coordinate transformations of the form $z=f(\sin\chi)$ (four real roots), $z=f(\cos\chi)$ (two real roots and a pair of complex conjugate roots), or $z=f(\tan\chi)$ (two pairs of complex conjugate roots) to transform (\ref{eq:BDiffR}) into the form
\begin{equation}\label{eq:BDiffEll}
\left(\frac{\text{d}\chi}{\text{d}\lambda}\right)^2=d_{4}C_{L}\left(1-k\sin^2\chi\right),
\end{equation}
where $C_{L}$ is a constant whose exact form depends on the chosen coordinate transformation. We can easily see that the form of (\ref{eq:BDiffEll}) is already very similar to the Legendre form (\ref{eq:BDiffLeg}). Again we separate variables and integrate using the initial conditions $\chi(\lambda_{i})=\chi_{i}$. We get
\begin{equation}\label{eq:LegIntsep}
\lambda-\lambda_{i}=\frac{i_{\chi_{i}}}{\sqrt{d_{4}C_{L}}}\int_{\chi_{i}}^{\chi}\frac{\text{d}\chi'}{\sqrt{1-k\sin^2\chi'}},
\end{equation}
where $i_{\chi_{i}}=\text{sgn}\left(\left.\text{d}\chi/\text{d}\lambda\right|_{\chi=\chi_{i}}\right)$.
We now rewrite the elliptic integral in terms of two incomplete elliptic integrals of the first kind and move all terms containing the initial conditions to the left-hand side. We obtain
\begin{equation}\label{eq:LegIApp}
i_{\chi_{i}}\sqrt{d_{4}C_{L}}\left(\lambda-\lambda_{i}\right)+F_{L}(\chi_{i},k)=\int_{0}^{\chi}\frac{\text{d}\chi'}{\sqrt{1-k\sin^2\chi'}}.
\end{equation}
We now define
\begin{equation}\label{eq:lamtrans}
\tilde{\lambda}(\lambda)=i_{\chi_{i}}\sqrt{d_{4}C_{L}}\left(\lambda-\lambda_{i}\right)+\lambda_{\chi_{i},k},
\end{equation}
where we defined $\lambda_{\chi_{i},k}=F_{L}(\chi_{i},k)$, and get
\begin{equation}\label{eq:LegInttrans}
\tilde{\lambda}(\lambda)=\int_{0}^{\chi}\frac{\text{d}\chi'}{\sqrt{1-k\sin^2\chi'}}.
\end{equation}
Comparing (\ref{eq:LegInttrans}) with (\ref{eq:LegInt}) we now see that in this case we have $\chi=\text{am}\tilde{\lambda}(\lambda)$ and thus the solutions are given by $\text{sn}(\tilde{\lambda}(\lambda),k)$, $\text{cn}(\tilde{\lambda}(\lambda),k)$, and $\text{sc}(\tilde{\lambda}(\lambda),k)$. Using the solutions to (\ref{eq:BDiffEll}) we can now write the solutions to (\ref{eq:BDiff}) as $z(\lambda)=f(\text{sn}(\tilde{\lambda}(\lambda),k))$ (four real roots), $z(\lambda)=f(\text{cn}(\tilde{\lambda}(\lambda),k))$ (two real roots and a pair of complex conjugate roots), or $z(\lambda)=f(\text{sc}(\tilde{\lambda}(\lambda),k))$ (two pairs of complex conjugate roots).

\bibliography{Charged_NUT_Metric_Particle_Lensing.bib}

\begin{thebibliography}{74}%
\makeatletter
\providecommand \@ifxundefined [1]{%
 \@ifx{#1\undefined}
}%
\providecommand \@ifnum [1]{%
 \ifnum #1\expandafter \@firstoftwo
 \else \expandafter \@secondoftwo
 \fi
}%
\providecommand \@ifx [1]{%
 \ifx #1\expandafter \@firstoftwo
 \else \expandafter \@secondoftwo
 \fi
}%
\providecommand \natexlab [1]{#1}%
\providecommand \enquote  [1]{``#1''}%
\providecommand \bibnamefont  [1]{#1}%
\providecommand \bibfnamefont [1]{#1}%
\providecommand \citenamefont [1]{#1}%
\providecommand \href@noop [0]{\@secondoftwo}%
\providecommand \href [0]{\begingroup \@sanitize@url \@href}%
\providecommand \@href[1]{\@@startlink{#1}\@@href}%
\providecommand \@@href[1]{\endgroup#1\@@endlink}%
\providecommand \@sanitize@url [0]{\catcode `\\12\catcode `\$12\catcode
  `\&12\catcode `\#12\catcode `\^12\catcode `\_12\catcode `\%12\relax}%
\providecommand \@@startlink[1]{}%
\providecommand \@@endlink[0]{}%
\providecommand \url  [0]{\begingroup\@sanitize@url \@url }%
\providecommand \@url [1]{\endgroup\@href {#1}{\urlprefix }}%
\providecommand \urlprefix  [0]{URL }%
\providecommand \Eprint [0]{\href }%
\providecommand \doibase [0]{https://doi.org/}%
\providecommand \selectlanguage [0]{\@gobble}%
\providecommand \bibinfo  [0]{\@secondoftwo}%
\providecommand \bibfield  [0]{\@secondoftwo}%
\providecommand \translation [1]{[#1]}%
\providecommand \BibitemOpen [0]{}%
\providecommand \bibitemStop [0]{}%
\providecommand \bibitemNoStop [0]{.\EOS\space}%
\providecommand \EOS [0]{\spacefactor3000\relax}%
\providecommand \BibitemShut  [1]{\csname bibitem#1\endcsname}%
\let\auto@bib@innerbib\@empty
\bibitem [{\citenamefont {Einstein}(1916)}]{Einstein1916}%
  \BibitemOpen
  \bibfield  {author} {\bibinfo {author} {\bibfnamefont {A.}~\bibnamefont
  {Einstein}},\ }\bibfield  {title} {\bibinfo {title} {Die {G}rundlage der
  allgemeinen {R}elativit\"{a}tstheorie},\ }\href
  {https://doi.org/10.1002/andp.19163540702} {\bibfield  {journal} {\bibinfo
  {journal} {Ann. {P}hys. (Berlin)}\ }\textbf {\bibinfo {volume} {354}},\
  \bibinfo {pages} {769} (\bibinfo {year} {1916})}\BibitemShut {NoStop}%
\bibitem [{\citenamefont {Dyson}\ \emph {et~al.}(1920)\citenamefont {Dyson},
  \citenamefont {Eddington},\ and\ \citenamefont {Davidson}}]{Dyson1920}%
  \BibitemOpen
  \bibfield  {author} {\bibinfo {author} {\bibfnamefont {F.~W.}\ \bibnamefont
  {Dyson}}, \bibinfo {author} {\bibfnamefont {A.~S.}\ \bibnamefont
  {Eddington}},\ and\ \bibinfo {author} {\bibfnamefont {C.}~\bibnamefont
  {Davidson}},\ }\bibfield  {title} {\bibinfo {title} {A determination of the
  deflection of light by the {S}un's gravitational field, from observations
  made at the total eclipse of {M}ay 29, 1919},\ }\href
  {http://www.jstor.org/stable/91137} {\bibfield  {journal} {\bibinfo
  {journal} {Phil. {T}rans. {R}. {S}oc. {A}}\ }\textbf {\bibinfo {volume}
  {220}},\ \bibinfo {pages} {291} (\bibinfo {year} {1920})}\BibitemShut
  {NoStop}%
\bibitem [{\citenamefont {{K. Akiyama \emph{et al.} (Event Horizon Telescope
  Collaboration)}}(2019)}]{EHTCollaboration2019a}%
  \BibitemOpen
  \bibfield  {author} {\bibinfo {author} {\bibnamefont {{K. Akiyama \emph{et
  al.} (Event Horizon Telescope Collaboration)}}},\ }\bibfield  {title}
  {\bibinfo {title} {First {M}87 {E}vent {H}orizon {T}elescope results. {I}.
  {T}he shadow of the supermassive black hole},\ }\href
  {https://doi.org/10.3847/2041-8213/ab0ec7} {\bibfield  {journal} {\bibinfo
  {journal} {{A}strophys. {J}. {L}ett.}\ }\textbf {\bibinfo {volume} {875}},\
  \bibinfo {pages} {L1} (\bibinfo {year} {2019})}\BibitemShut {NoStop}%
\bibitem [{\citenamefont {{K. Akiyama \emph{et al.} (Event Horizon Telescope
  Collaboration)}}(2022)}]{EHTCollaboration2022}%
  \BibitemOpen
  \bibfield  {author} {\bibinfo {author} {\bibnamefont {{K. Akiyama \emph{et
  al.} (Event Horizon Telescope Collaboration)}}},\ }\bibfield  {title}
  {\bibinfo {title} {First {S}agittarius $\text{A}^{*}$ {E}vent {H}orizon
  {T}elescope results. {I}. {T}he shadow of the supermassive black hole in the
  center of the {M}ilky {W}ay},\ }\href
  {https://doi.org/10.3847/2041-8213/ac6674} {\bibfield  {journal} {\bibinfo
  {journal} {{A}strophys. {J}. {L}ett.}\ }\textbf {\bibinfo {volume} {930}},\
  \bibinfo {pages} {L12} (\bibinfo {year} {2022})}\BibitemShut {NoStop}%
\bibitem [{\citenamefont {Palanque-Delabrouille}\ \emph
  {et~al.}(2015)\citenamefont {Palanque-Delabrouille}, \citenamefont
  {Y\`{e}che}, \citenamefont {Baur}, \citenamefont {Magneville}, \citenamefont
  {Rossi}, \citenamefont {Lesgourgues}, \citenamefont {Borde}, \citenamefont
  {Burtin}, \citenamefont {LeGoff}, \citenamefont {Rich}, \citenamefont
  {Viel},\ and\ \citenamefont {Weinberg}}]{PalanqueDelabrouille2015}%
  \BibitemOpen
  \bibfield  {author} {\bibinfo {author} {\bibfnamefont {N.}~\bibnamefont
  {Palanque-Delabrouille}}, \bibinfo {author} {\bibfnamefont {C.}~\bibnamefont
  {Y\`{e}che}}, \bibinfo {author} {\bibfnamefont {J.}~\bibnamefont {Baur}},
  \bibinfo {author} {\bibfnamefont {C.}~\bibnamefont {Magneville}}, \bibinfo
  {author} {\bibfnamefont {G.}~\bibnamefont {Rossi}}, \bibinfo {author}
  {\bibfnamefont {J.}~\bibnamefont {Lesgourgues}}, \bibinfo {author}
  {\bibfnamefont {A.}~\bibnamefont {Borde}}, \bibinfo {author} {\bibfnamefont
  {E.}~\bibnamefont {Burtin}}, \bibinfo {author} {\bibfnamefont {J.-M.}\
  \bibnamefont {LeGoff}}, \bibinfo {author} {\bibfnamefont {J.}~\bibnamefont
  {Rich}}, \bibinfo {author} {\bibfnamefont {M.}~\bibnamefont {Viel}},\ and\
  \bibinfo {author} {\bibfnamefont {D.}~\bibnamefont {Weinberg}},\ }\bibfield
  {title} {\bibinfo {title} {Neutrino masses and cosmology with {L}yman-alpha
  forest power spectrum},\ }\href
  {https://doi.org/10.1088/1475-7516/2015/11/011} {\bibfield  {journal}
  {\bibinfo  {journal} {J. Cosmol. Astropart. Phys.}\ }\textbf {\bibinfo
  {volume} {2015}}\bibinfo  {number} { (11)},\ \bibinfo {pages}
  {011}}\BibitemShut {NoStop}%
\bibitem [{\citenamefont {Fukuda}\ \emph {et~al.}(2003)\citenamefont {Fukuda}
  \emph {et~al.}}]{Fukuda2003}%
  \BibitemOpen
\bibfield  {number} {  }\bibfield  {author} {\bibinfo {author} {\bibfnamefont
  {S.}~\bibnamefont {Fukuda}} \emph {et~al.},\ }\bibfield  {title} {\bibinfo
  {title} {The {S}uper-{K}amiokande detector},\ }\href
  {https://doi.org/10.1016/S0168-9002(03)00425-X} {\bibfield  {journal}
  {\bibinfo  {journal} {Nucl. Instrum. Methods Phys. Res., Sect. A}\ }\textbf
  {\bibinfo {volume} {501}},\ \bibinfo {pages} {418} (\bibinfo {year}
  {2003})}\BibitemShut {NoStop}%
\bibitem [{\citenamefont {Ashie}\ \emph {et~al.}(2005)\citenamefont {Ashie}
  \emph {et~al.}}]{Ashie2005}%
  \BibitemOpen
  \bibfield  {author} {\bibinfo {author} {\bibfnamefont {Y.}~\bibnamefont
  {Ashie}} \emph {et~al.},\ }\bibfield  {title} {\bibinfo {title} {Measurement
  of atmospheric neutrino oscillation parameters by {S}uper-{K}amiokande {I}},\
  }\href {https://doi.org/10.1103/PhysRevD.71.112005} {\bibfield  {journal}
  {\bibinfo  {journal} {Phys. {R}ev. {D}}\ }\textbf {\bibinfo {volume} {71}},\
  \bibinfo {pages} {112005} (\bibinfo {year} {2005})}\BibitemShut {NoStop}%
\bibitem [{\citenamefont {Achterberg}\ \emph {et~al.}(2006)\citenamefont
  {Achterberg} \emph {et~al.}}]{Achterberg2006}%
  \BibitemOpen
  \bibfield  {author} {\bibinfo {author} {\bibfnamefont {A.}~\bibnamefont
  {Achterberg}} \emph {et~al.},\ }\bibfield  {title} {\bibinfo {title} {First
  year performance of the {I}ce{C}ube {N}eutrino {T}elescope},\ }\href
  {https://doi.org/10.1016/j.astropartphys.2006.06.007} {\bibfield  {journal}
  {\bibinfo  {journal} {Astropart. {P}hys.}\ }\textbf {\bibinfo {volume}
  {26}},\ \bibinfo {pages} {155} (\bibinfo {year} {2006})}\BibitemShut
  {NoStop}%
\bibitem [{\citenamefont {{IceCube
  Collaboration}}(2013)}]{ICCollaboration2013}%
  \BibitemOpen
  \bibfield  {author} {\bibinfo {author} {\bibnamefont {{IceCube
  Collaboration}}},\ }\bibfield  {title} {\bibinfo {title} {Evidence for
  high-energy extraterrestrial neutrinos at the {I}ce{C}ube detector},\ }\href
  {https://doi.org/10.1126/science.1242856} {\bibfield  {journal} {\bibinfo
  {journal} {Science}\ }\textbf {\bibinfo {volume} {342}},\ \bibinfo {pages}
  {1242856} (\bibinfo {year} {2013})}\BibitemShut {NoStop}%
\bibitem [{\citenamefont {{IceCube
  Collaboration}}(2021)}]{TheICCollaboration2021}%
  \BibitemOpen
  \bibfield  {author} {\bibinfo {author} {\bibnamefont {{IceCube
  Collaboration}}},\ }\bibfield  {title} {\bibinfo {title} {Detection of a
  particle shower at the {G}lashow resonance with {I}ce{C}ube},\ }\href
  {https://doi.org/10.1038/s41586-021-03256-1} {\bibfield  {journal} {\bibinfo
  {journal} {Nature (London)}\ }\textbf {\bibinfo {volume} {591}},\ \bibinfo
  {pages} {220} (\bibinfo {year} {2021})}\BibitemShut {NoStop}%
\bibitem [{\citenamefont {Plebanski}\ and\ \citenamefont
  {Demianski}(1976)}]{Plebanski1976}%
  \BibitemOpen
  \bibfield  {author} {\bibinfo {author} {\bibfnamefont {J.~F.}\ \bibnamefont
  {Plebanski}}\ and\ \bibinfo {author} {\bibfnamefont {M.}~\bibnamefont
  {Demianski}},\ }\bibfield  {title} {\bibinfo {title} {Rotating, charged, and
  uniformly accelerating mass in general relativity},\ }\href
  {https://doi.org/10.1016/0003-4916(76)90240-2} {\bibfield  {journal}
  {\bibinfo  {journal} {Ann. {P}hys. ({N}. {Y}.)}\ }\textbf {\bibinfo {volume}
  {98}},\ \bibinfo {pages} {98} (\bibinfo {year} {1976})}\BibitemShut {NoStop}%
\bibitem [{\citenamefont {Taub}(1951)}]{Taub1951}%
  \BibitemOpen
  \bibfield  {author} {\bibinfo {author} {\bibfnamefont {A.~H.}\ \bibnamefont
  {Taub}},\ }\bibfield  {title} {\bibinfo {title} {Empty space-times admitting
  a three parameter group of motions},\ }\href
  {https://doi.org/10.2307/1969567} {\bibfield  {journal} {\bibinfo  {journal}
  {Ann. {M}ath.}\ }\textbf {\bibinfo {volume} {53}},\ \bibinfo {pages} {472}
  (\bibinfo {year} {1951})}\BibitemShut {NoStop}%
\bibitem [{\citenamefont {Misner}(1963)}]{Misner1963}%
  \BibitemOpen
  \bibfield  {author} {\bibinfo {author} {\bibfnamefont {C.~W.}\ \bibnamefont
  {Misner}},\ }\bibfield  {title} {\bibinfo {title} {The flatter regions of
  {N}ewman, {U}nti, and {T}amburino's generalized {S}chwarzschild space},\
  }\href {https://doi.org/10.1063/1.1704019} {\bibfield  {journal} {\bibinfo
  {journal} {J. {M}ath. {P}hys. ({N}.{Y}.)}\ }\textbf {\bibinfo {volume} {4}},\
  \bibinfo {pages} {924} (\bibinfo {year} {1963})}\BibitemShut {NoStop}%
\bibitem [{\citenamefont {Newman}\ \emph {et~al.}(1963)\citenamefont {Newman},
  \citenamefont {Tamburino},\ and\ \citenamefont {Unti}}]{Newman1963}%
  \BibitemOpen
  \bibfield  {author} {\bibinfo {author} {\bibfnamefont {E.}~\bibnamefont
  {Newman}}, \bibinfo {author} {\bibfnamefont {L.}~\bibnamefont {Tamburino}},\
  and\ \bibinfo {author} {\bibfnamefont {T.}~\bibnamefont {Unti}},\ }\bibfield
  {title} {\bibinfo {title} {Empty-space generalization of the {S}chwarzschild
  metric},\ }\href {https://doi.org/10.1063/1.1704018} {\bibfield  {journal}
  {\bibinfo  {journal} {J. {M}ath. {P}hys. ({N}.{Y}.)}\ }\textbf {\bibinfo
  {volume} {4}},\ \bibinfo {pages} {915} (\bibinfo {year} {1963})}\BibitemShut
  {NoStop}%
\bibitem [{\citenamefont {Brill}(1964)}]{Brill1964}%
  \BibitemOpen
  \bibfield  {author} {\bibinfo {author} {\bibfnamefont {D.~R.}\ \bibnamefont
  {Brill}},\ }\bibfield  {title} {\bibinfo {title} {Electromagnetic fields in a
  homogeneous, nonisotropic universe},\ }\href
  {https://doi.org/10.1103/PhysRev.133.B845} {\bibfield  {journal} {\bibinfo
  {journal} {Phys. {R}ev.}\ }\textbf {\bibinfo {volume} {133}},\ \bibinfo
  {pages} {B845} (\bibinfo {year} {1964})}\BibitemShut {NoStop}%
\bibitem [{\citenamefont {Manko}\ and\ \citenamefont {Ruiz}(2005)}]{Manko2005}%
  \BibitemOpen
  \bibfield  {author} {\bibinfo {author} {\bibfnamefont {V.~S.}\ \bibnamefont
  {Manko}}\ and\ \bibinfo {author} {\bibfnamefont {E.}~\bibnamefont {Ruiz}},\
  }\bibfield  {title} {\bibinfo {title} {Physical interpretation of the {NUT}
  family of solutions},\ }\href {https://doi.org/10.1088/0264-9381/22/17/014}
  {\bibfield  {journal} {\bibinfo  {journal} {Classical {Q}uantum {G}ravity}\
  }\textbf {\bibinfo {volume} {22}},\ \bibinfo {pages} {3555} (\bibinfo {year}
  {2005})}\BibitemShut {NoStop}%
\bibitem [{\citenamefont {Bonnor}(1969)}]{Bonnor1969}%
  \BibitemOpen
  \bibfield  {author} {\bibinfo {author} {\bibfnamefont {W.~B.}\ \bibnamefont
  {Bonnor}},\ }\bibfield  {title} {\bibinfo {title} {A new interpretation of
  the {NUT} metric in general relativity},\ }\href
  {https://doi.org/10.1017/S0305004100044807} {\bibfield  {journal} {\bibinfo
  {journal} {Math. {P}roc. {C}ambridge {P}hilos. {S}oc.}\ }\textbf {\bibinfo
  {volume} {66}},\ \bibinfo {pages} {145} (\bibinfo {year} {1969})}\BibitemShut
  {NoStop}%
\bibitem [{\citenamefont {Sackfield}(1971)}]{Sackfield1971}%
  \BibitemOpen
  \bibfield  {author} {\bibinfo {author} {\bibfnamefont {A.}~\bibnamefont
  {Sackfield}},\ }\bibfield  {title} {\bibinfo {title} {Physical interpretation
  of {N}.{U}.{T}. metric},\ }\href {https://doi.org/10.1017/S0305004100049707}
  {\bibfield  {journal} {\bibinfo  {journal} {Math. {P}roc. {C}ambridge
  {P}hilos. {S}oc.}\ }\textbf {\bibinfo {volume} {70}},\ \bibinfo {pages} {89}
  (\bibinfo {year} {1971})}\BibitemShut {NoStop}%
\bibitem [{\citenamefont {Cl{\' e}ment}\ \emph {et~al.}(2016)\citenamefont
  {Cl{\' e}ment}, \citenamefont {Gal'tsov},\ and\ \citenamefont
  {Guenouche}}]{Clement2016}%
  \BibitemOpen
  \bibfield  {author} {\bibinfo {author} {\bibfnamefont {G.}~\bibnamefont
  {Cl{\' e}ment}}, \bibinfo {author} {\bibfnamefont {D.}~\bibnamefont
  {Gal'tsov}},\ and\ \bibinfo {author} {\bibfnamefont {M.}~\bibnamefont
  {Guenouche}},\ }\bibfield  {title} {\bibinfo {title} {{NUT} wormholes},\
  }\href {https://doi.org/10.1103/PhysRevD.93.024048} {\bibfield  {journal}
  {\bibinfo  {journal} {Phys. {R}ev. {D}}\ }\textbf {\bibinfo {volume} {93}},\
  \bibinfo {pages} {024048} (\bibinfo {year} {2016})}\BibitemShut {NoStop}%
\bibitem [{\citenamefont {Frost}(2022{\natexlab{a}})}]{Frost2022}%
  \BibitemOpen
  \bibfield  {author} {\bibinfo {author} {\bibfnamefont {T.~C.}\ \bibnamefont
  {Frost}},\ }\bibfield  {title} {\bibinfo {title} {Companion paper,
  {G}ravitational lensing in the charged {NUT}-de {S}itter spacetime},\ }\href
  {https://doi.org/10.1103/PhysRevD.105.064064} {\bibfield  {journal} {\bibinfo
   {journal} {Phys. {R}ev. {D}}\ }\textbf {\bibinfo {volume} {105}},\ \bibinfo
  {pages} {064064} (\bibinfo {year} {2022}{\natexlab{a}})}\BibitemShut
  {NoStop}%
\bibitem [{\citenamefont {Castellanos}\ \emph {et~al.}(2018)\citenamefont
  {Castellanos}, \citenamefont {Degollado}, \citenamefont {L\"ammerzahl},
  \citenamefont {Mac\'{i}as},\ and\ \citenamefont {Perlick}}]{Castellanos2018}%
  \BibitemOpen
  \bibfield  {author} {\bibinfo {author} {\bibfnamefont {E.}~\bibnamefont
  {Castellanos}}, \bibinfo {author} {\bibfnamefont {J.~C.}\ \bibnamefont
  {Degollado}}, \bibinfo {author} {\bibfnamefont {C.}~\bibnamefont
  {L\"ammerzahl}}, \bibinfo {author} {\bibfnamefont {A.}~\bibnamefont
  {Mac\'{i}as}},\ and\ \bibinfo {author} {\bibfnamefont {V.}~\bibnamefont
  {Perlick}},\ }\bibfield  {title} {\bibinfo {title} {Bose-{E}instein
  condensates in charged black--hole spacetimes},\ }\href
  {https://doi.org/10.1088/1475-7516/2018/01/043} {\bibfield  {journal}
  {\bibinfo  {journal} {J. Cosmol. Astropart. Phys.}\ }\textbf {\bibinfo
  {volume} {2018}}\bibinfo  {number} { (01)},\ \bibinfo {pages}
  {043}}\BibitemShut {NoStop}%
\bibitem [{\citenamefont {Zimmerman}\ and\ \citenamefont
  {Shahir}(1989)}]{Zimmerman1989}%
  \BibitemOpen
\bibfield  {number} {  }\bibfield  {author} {\bibinfo {author} {\bibfnamefont
  {R.~L.}\ \bibnamefont {Zimmerman}}\ and\ \bibinfo {author} {\bibfnamefont
  {B.~Y.}\ \bibnamefont {Shahir}},\ }\bibfield  {title} {\bibinfo {title}
  {Geodesics for the {NUT} metric and gravitational monopoles},\ }\href
  {https://doi.org/10.1007/BF00758986} {\bibfield  {journal} {\bibinfo
  {journal} {Gen. {R}elativ. {G}ravit.}\ }\textbf {\bibinfo {volume} {21}},\
  \bibinfo {pages} {821} (\bibinfo {year} {1989})}\BibitemShut {NoStop}%
\bibitem [{\citenamefont {Jefremov}\ and\ \citenamefont
  {Perlick}(2016)}]{Jefremov2016}%
  \BibitemOpen
  \bibfield  {author} {\bibinfo {author} {\bibfnamefont {P.~I.}\ \bibnamefont
  {Jefremov}}\ and\ \bibinfo {author} {\bibfnamefont {V.}~\bibnamefont
  {Perlick}},\ }\bibfield  {title} {\bibinfo {title} {Circular motion in {NUT}
  space-time},\ }\href {https://doi.org/10.1088/0264-9381/33/24/245014}
  {\bibfield  {journal} {\bibinfo  {journal} {Classical {Q}uantum {G}ravity}\
  }\textbf {\bibinfo {volume} {33}},\ \bibinfo {pages} {245014} (\bibinfo
  {year} {2016})}\BibitemShut {NoStop}%
\bibitem [{\citenamefont {Kagramanova}\ \emph {et~al.}(2010)\citenamefont
  {Kagramanova}, \citenamefont {Kunz}, \citenamefont {Hackmann},\ and\
  \citenamefont {L{\"a}mmerzahl}}]{Kagramanova2010}%
  \BibitemOpen
  \bibfield  {author} {\bibinfo {author} {\bibfnamefont {V.}~\bibnamefont
  {Kagramanova}}, \bibinfo {author} {\bibfnamefont {J.}~\bibnamefont {Kunz}},
  \bibinfo {author} {\bibfnamefont {E.}~\bibnamefont {Hackmann}},\ and\
  \bibinfo {author} {\bibfnamefont {C.}~\bibnamefont {L{\"a}mmerzahl}},\
  }\bibfield  {title} {\bibinfo {title} {Analytic treatment of complete and
  incomplete geodesics in {T}aub-{NUT} space-times},\ }\href
  {https://doi.org/10.1103/PhysRevD.81.124044} {\bibfield  {journal} {\bibinfo
  {journal} {Phys. {R}ev. {D}}\ }\textbf {\bibinfo {volume} {81}},\ \bibinfo
  {pages} {124044} (\bibinfo {year} {2010})}\BibitemShut {NoStop}%
\bibitem [{\citenamefont {Misner}\ and\ \citenamefont
  {Taub}(1969)}]{Misner1969}%
  \BibitemOpen
  \bibfield  {author} {\bibinfo {author} {\bibfnamefont {C.~W.}\ \bibnamefont
  {Misner}}\ and\ \bibinfo {author} {\bibfnamefont {A.~H.}\ \bibnamefont
  {Taub}},\ }\bibfield  {title} {\bibinfo {title} {A singularity-free empty
  universe},\ }\href {http://jetp.ras.ru/cgi-bin/dn/e_028_01_0122.pdf}
  {\bibfield  {journal} {\bibinfo  {journal} {Sov. {P}hys. {JETP}}\ }\textbf
  {\bibinfo {volume} {28}},\ \bibinfo {pages} {122} (\bibinfo {year}
  {1969})}\BibitemShut {NoStop}%
\bibitem [{\citenamefont {Miller}\ \emph {et~al.}(1971)\citenamefont {Miller},
  \citenamefont {Kruskal},\ and\ \citenamefont {Godfrey}}]{Miller1971}%
  \BibitemOpen
  \bibfield  {author} {\bibinfo {author} {\bibfnamefont {J.~G.}\ \bibnamefont
  {Miller}}, \bibinfo {author} {\bibfnamefont {M.~D.}\ \bibnamefont
  {Kruskal}},\ and\ \bibinfo {author} {\bibfnamefont {B.~B.}\ \bibnamefont
  {Godfrey}},\ }\bibfield  {title} {\bibinfo {title} {Taub-{NUT} ({N}ewman,
  {U}nti, {T}amburino) metric and incompatible extensions},\ }\href
  {https://doi.org/10.1103/PhysRevD.4.2945} {\bibfield  {journal} {\bibinfo
  {journal} {Phys. {R}ev. {D}}\ }\textbf {\bibinfo {volume} {4}},\ \bibinfo
  {pages} {2945} (\bibinfo {year} {1971})}\BibitemShut {NoStop}%
\bibitem [{\citenamefont {Cl{\' e}ment}\ \emph {et~al.}(2015)\citenamefont
  {Cl{\' e}ment}, \citenamefont {Gal'tsov},\ and\ \citenamefont
  {Guenouche}}]{Clement2015}%
  \BibitemOpen
  \bibfield  {author} {\bibinfo {author} {\bibfnamefont {G.}~\bibnamefont
  {Cl{\' e}ment}}, \bibinfo {author} {\bibfnamefont {D.}~\bibnamefont
  {Gal'tsov}},\ and\ \bibinfo {author} {\bibfnamefont {M.}~\bibnamefont
  {Guenouche}},\ }\bibfield  {title} {\bibinfo {title} {Rehabilitating
  space-times with {NUT}s},\ }\href
  {https://doi.org/10.1016/j.physletb.2015.09.074} {\bibfield  {journal}
  {\bibinfo  {journal} {Phys. {L}ett. {B}}\ }\textbf {\bibinfo {volume}
  {750}},\ \bibinfo {pages} {591} (\bibinfo {year} {2015})}\BibitemShut
  {NoStop}%
\bibitem [{\citenamefont {Lynden-Bell}\ and\ \citenamefont
  {Nouri-Zonoz}(1998)}]{LyndenBell1998}%
  \BibitemOpen
  \bibfield  {author} {\bibinfo {author} {\bibfnamefont {D.}~\bibnamefont
  {Lynden-Bell}}\ and\ \bibinfo {author} {\bibfnamefont {M.}~\bibnamefont
  {Nouri-Zonoz}},\ }\bibfield  {title} {\bibinfo {title} {Classical monopoles:
  {N}ewton, {NUT} space, gravomagnetic lensing, and atomic spectra},\ }\href
  {https://doi.org/10.1103/RevModPhys.70.427} {\bibfield  {journal} {\bibinfo
  {journal} {Rev. {M}od. {P}hys.}\ }\textbf {\bibinfo {volume} {70}},\ \bibinfo
  {pages} {427} (\bibinfo {year} {1998})}\BibitemShut {NoStop}%
\bibitem [{\citenamefont {Nouri-Zonoz}\ and\ \citenamefont
  {Lynden-Bell}(1997)}]{NouriZonoz1997}%
  \BibitemOpen
  \bibfield  {author} {\bibinfo {author} {\bibfnamefont {M.}~\bibnamefont
  {Nouri-Zonoz}}\ and\ \bibinfo {author} {\bibfnamefont {D.}~\bibnamefont
  {Lynden-Bell}},\ }\bibfield  {title} {\bibinfo {title} {Gravomagnetic lensing
  by {NUT} space},\ }\href {https://doi.org/10.1093/mnras/292.3.714} {\bibfield
   {journal} {\bibinfo  {journal} {Mon. {N}ot. {R}. {A}stron. {S}oc.}\ }\textbf
  {\bibinfo {volume} {292}},\ \bibinfo {pages} {714} (\bibinfo {year}
  {1997})}\BibitemShut {NoStop}%
\bibitem [{\citenamefont {Halla}\ and\ \citenamefont
  {Perlick}(2020)}]{Halla2020}%
  \BibitemOpen
  \bibfield  {author} {\bibinfo {author} {\bibfnamefont {M.}~\bibnamefont
  {Halla}}\ and\ \bibinfo {author} {\bibfnamefont {V.}~\bibnamefont
  {Perlick}},\ }\bibfield  {title} {\bibinfo {title} {Application of the
  {G}auss-{B}onnet theorem to lensing in the {NUT} metric},\ }\href
  {https://doi.org/10.1007/s10714-020-02766-z} {\bibfield  {journal} {\bibinfo
  {journal} {Gen. {R}elativ. {G}ravit.}\ }\textbf {\bibinfo {volume} {52}},\
  \bibinfo {pages} {112} (\bibinfo {year} {2020})}\BibitemShut {NoStop}%
\bibitem [{\citenamefont {Halla}\ and\ \citenamefont
  {Perlick}(2023)}]{Halla2023}%
  \BibitemOpen
  \bibfield  {author} {\bibinfo {author} {\bibfnamefont {M.}~\bibnamefont
  {Halla}}\ and\ \bibinfo {author} {\bibfnamefont {V.}~\bibnamefont
  {Perlick}},\ }\bibfield  {title} {\bibinfo {title} {Gravitational lensing in
  {B}rill spacetimes},\ }\href {https://doi.org/10.1103/PhysRevD.107.024048}
  {\bibfield  {journal} {\bibinfo  {journal} {Phys. Rev. D}\ }\textbf {\bibinfo
  {volume} {107}},\ \bibinfo {pages} {024048} (\bibinfo {year}
  {2023})}\BibitemShut {NoStop}%
\bibitem [{\citenamefont {Sharif}\ and\ \citenamefont
  {Iftikhar}(2016)}]{Sharif2016}%
  \BibitemOpen
  \bibfield  {author} {\bibinfo {author} {\bibfnamefont {M.}~\bibnamefont
  {Sharif}}\ and\ \bibinfo {author} {\bibfnamefont {S.}~\bibnamefont
  {Iftikhar}},\ }\bibfield  {title} {\bibinfo {title} {Equatorial gravitational
  lensing by accelerating and rotating black hole with {NUT} parameter},\
  }\href {https://doi.org/10.1007/s10509-015-2623-x} {\bibfield  {journal}
  {\bibinfo  {journal} {Astrophys. {S}pace {S}ci.}\ }\textbf {\bibinfo {volume}
  {361}},\ \bibinfo {pages} {36} (\bibinfo {year} {2016})}\BibitemShut
  {NoStop}%
\bibitem [{\citenamefont {Grenzebach}\ \emph {et~al.}(2014)\citenamefont
  {Grenzebach}, \citenamefont {Perlick},\ and\ \citenamefont
  {L\"ammerzahl}}]{Grenzebach2014}%
  \BibitemOpen
  \bibfield  {author} {\bibinfo {author} {\bibfnamefont {A.}~\bibnamefont
  {Grenzebach}}, \bibinfo {author} {\bibfnamefont {V.}~\bibnamefont
  {Perlick}},\ and\ \bibinfo {author} {\bibfnamefont {C.}~\bibnamefont
  {L\"ammerzahl}},\ }\bibfield  {title} {\bibinfo {title} {Photon regions and
  shadows of {K}err-{N}ewman-{NUT} black holes with a cosmological constant},\
  }\href {https://doi.org/10.1103/PhysRevD.89.124004} {\bibfield  {journal}
  {\bibinfo  {journal} {Phys. {R}ev. {D}}\ }\textbf {\bibinfo {volume} {89}},\
  \bibinfo {pages} {124004} (\bibinfo {year} {2014})}\BibitemShut {NoStop}%
\bibitem [{\citenamefont {Grenzebach}(2016)}]{Grenzebach2016}%
  \BibitemOpen
  \bibfield  {author} {\bibinfo {author} {\bibfnamefont {A.}~\bibnamefont
  {Grenzebach}},\ }\href@noop {} {\emph {\bibinfo {title} {The {S}hadow of
  {B}lack {H}oles}}},\ Springer Briefs in Physics\ (\bibinfo  {publisher}
  {Springer},\ \bibinfo {address} {Cham},\ \bibinfo {year} {2016})\BibitemShut
  {NoStop}%
\bibitem [{\citenamefont {Zakharov}(2018)}]{Zakharov2018}%
  \BibitemOpen
  \bibfield  {author} {\bibinfo {author} {\bibfnamefont {A.~F.}\ \bibnamefont
  {Zakharov}},\ }\bibfield  {title} {\bibinfo {title} {Comment on
  '{G}ravitational lensing of massive particles in {S}chwarzschild gravity'},\
  }\href {https://doi.org/10.1088/1361-6382/aa964e} {\bibfield  {journal}
  {\bibinfo  {journal} {Classical {Q}uantum {G}ravity}\ }\textbf {\bibinfo
  {volume} {35}},\ \bibinfo {pages} {028001} (\bibinfo {year}
  {2018})}\BibitemShut {NoStop}%
\bibitem [{\citenamefont {Mielnik}\ and\ \citenamefont
  {Pleba\'{n}ski}(1962)}]{Mielnik1962}%
  \BibitemOpen
  \bibfield  {author} {\bibinfo {author} {\bibfnamefont {B.}~\bibnamefont
  {Mielnik}}\ and\ \bibinfo {author} {\bibfnamefont {J.}~\bibnamefont
  {Pleba\'{n}ski}},\ }\bibfield  {title} {\bibinfo {title} {A study of geodesic
  motion in the field of {S}chwarzschild's solution},\ }\href@noop {}
  {\bibfield  {journal} {\bibinfo  {journal} {Acta Phys. Pol.}\ }\textbf
  {\bibinfo {volume} {21}},\ \bibinfo {pages} {239} (\bibinfo {year}
  {1962})}\BibitemShut {NoStop}%
\bibitem [{\citenamefont {Kobialko}\ \emph {et~al.}(2022)\citenamefont
  {Kobialko}, \citenamefont {Bogush},\ and\ \citenamefont
  {Gal'tsov}}]{Kobialko2022}%
  \BibitemOpen
  \bibfield  {author} {\bibinfo {author} {\bibfnamefont {K.}~\bibnamefont
  {Kobialko}}, \bibinfo {author} {\bibfnamefont {I.}~\bibnamefont {Bogush}},\
  and\ \bibinfo {author} {\bibfnamefont {D.}~\bibnamefont {Gal'tsov}},\
  }\bibfield  {title} {\bibinfo {title} {Geometry of massive particle
  surfaces},\ }\href {https://doi.org/10.1103/PhysRevD.106.084032} {\bibfield
  {journal} {\bibinfo  {journal} {Phys. {R}ev. {D}}\ }\textbf {\bibinfo
  {volume} {106}},\ \bibinfo {pages} {084032} (\bibinfo {year}
  {2022})}\BibitemShut {NoStop}%
\bibitem [{\citenamefont {Claudel}\ \emph {et~al.}(2001)\citenamefont
  {Claudel}, \citenamefont {Virbhadra},\ and\ \citenamefont
  {Ellis}}]{Claudel2001}%
  \BibitemOpen
  \bibfield  {author} {\bibinfo {author} {\bibfnamefont {C.-M.}\ \bibnamefont
  {Claudel}}, \bibinfo {author} {\bibfnamefont {K.~S.}\ \bibnamefont
  {Virbhadra}},\ and\ \bibinfo {author} {\bibfnamefont {G.~F.~R.}\ \bibnamefont
  {Ellis}},\ }\bibfield  {title} {\bibinfo {title} {The geometry of photon
  surfaces},\ }\href {https://doi.org/10.1063/1.1308507} {\bibfield  {journal}
  {\bibinfo  {journal} {J. {M}ath. {P}hys. (N. Y.)}\ }\textbf {\bibinfo
  {volume} {42}},\ \bibinfo {pages} {818} (\bibinfo {year} {2001})}\BibitemShut
  {NoStop}%
\bibitem [{\citenamefont {Zakharov}(1988)}]{Zakharov1988}%
  \BibitemOpen
  \bibfield  {author} {\bibinfo {author} {\bibfnamefont {A.~F.}\ \bibnamefont
  {Zakharov}},\ }\bibfield  {title} {\bibinfo {title} {Effective particle
  capture cross section of a {S}chwarzschild black hole},\ }\href
  {https://articles.adsabs.harvard.edu//full/1988SvA....32..456Z/0000456.000.html}
  {\bibfield  {journal} {\bibinfo  {journal} {Sov. {A}stron.}\ }\textbf
  {\bibinfo {volume} {32}},\ \bibinfo {pages} {456} (\bibinfo {year}
  {1988})}\BibitemShut {NoStop}%
\bibitem [{\citenamefont {Zakharov}(1994)}]{Zakharov1994}%
  \BibitemOpen
  \bibfield  {author} {\bibinfo {author} {\bibfnamefont {A.~F.}\ \bibnamefont
  {Zakharov}},\ }\bibfield  {title} {\bibinfo {title} {Particle capture cross
  sections for a {R}eissner-{N}ordstr\"{o}m black hole},\ }\href
  {https://doi.org/10.1088/0264-9381/11/4/018} {\bibfield  {journal} {\bibinfo
  {journal} {Classical {Q}uantum {G}ravity}\ }\textbf {\bibinfo {volume}
  {11}},\ \bibinfo {pages} {1027} (\bibinfo {year} {1994})}\BibitemShut
  {NoStop}%
\bibitem [{\citenamefont {Accioly}\ and\ \citenamefont
  {Ragusa}(2002)}]{Accioly2002}%
  \BibitemOpen
  \bibfield  {author} {\bibinfo {author} {\bibfnamefont {A.}~\bibnamefont
  {Accioly}}\ and\ \bibinfo {author} {\bibfnamefont {S.}~\bibnamefont
  {Ragusa}},\ }\bibfield  {title} {\bibinfo {title} {Gravitational deflection
  of massive particles in classical and semiclassical general relativity},\
  }\href {https://doi.org/10.1088/0264-9381/19/21/308} {\bibfield  {journal}
  {\bibinfo  {journal} {Classical {Q}uantum {G}ravity}\ }\textbf {\bibinfo
  {volume} {19}},\ \bibinfo {pages} {5429} (\bibinfo {year}
  {2002})}\BibitemShut {NoStop}%
\bibitem [{\citenamefont {Tsupko}(2014)}]{Tsupko2014}%
  \BibitemOpen
  \bibfield  {author} {\bibinfo {author} {\bibfnamefont {O.~Y.}\ \bibnamefont
  {Tsupko}},\ }\bibfield  {title} {\bibinfo {title} {Unbound motion of massive
  particles in the {S}chwarzschild metric: {A}nalytical description in case of
  strong deflection},\ }\href {https://doi.org/10.1103/PhysRevD.89.084075}
  {\bibfield  {journal} {\bibinfo  {journal} {Phys. {R}ev. {D}}\ }\textbf
  {\bibinfo {volume} {89}},\ \bibinfo {pages} {084075} (\bibinfo {year}
  {2014})}\BibitemShut {NoStop}%
\bibitem [{\citenamefont {Liu}\ \emph {et~al.}(2016)\citenamefont {Liu},
  \citenamefont {Yang},\ and\ \citenamefont {Jia}}]{Liu2016}%
  \BibitemOpen
  \bibfield  {author} {\bibinfo {author} {\bibfnamefont {X.}~\bibnamefont
  {Liu}}, \bibinfo {author} {\bibfnamefont {N.}~\bibnamefont {Yang}},\ and\
  \bibinfo {author} {\bibfnamefont {J.}~\bibnamefont {Jia}},\ }\bibfield
  {title} {\bibinfo {title} {Gravitational lensing of massive particles in
  {S}chwarzschild gravity},\ }\href
  {https://doi.org/10.1088/0264-9381/33/17/175014} {\bibfield  {journal}
  {\bibinfo  {journal} {Classical {Q}uantum {G}ravity}\ }\textbf {\bibinfo
  {volume} {33}},\ \bibinfo {pages} {175014} (\bibinfo {year}
  {2016})}\BibitemShut {NoStop}%
\bibitem [{\citenamefont {Crisnejo}\ and\ \citenamefont
  {Gallo}(2018)}]{Crisnejo2018}%
  \BibitemOpen
  \bibfield  {author} {\bibinfo {author} {\bibfnamefont {G.}~\bibnamefont
  {Crisnejo}}\ and\ \bibinfo {author} {\bibfnamefont {E.}~\bibnamefont
  {Gallo}},\ }\bibfield  {title} {\bibinfo {title} {Weak lensing in a plasma
  medium and gravitational deflection of massive particles using the
  {G}auss-{B}onnet theorem. {A} unified treatment},\ }\href
  {https://doi.org/10.1103/PhysRevD.97.124016} {\bibfield  {journal} {\bibinfo
  {journal} {Phys. {R}ev. {D}}\ }\textbf {\bibinfo {volume} {97}},\ \bibinfo
  {pages} {124016} (\bibinfo {year} {2018})}\BibitemShut {NoStop}%
\bibitem [{\citenamefont {Gibbons}\ and\ \citenamefont
  {Werner}(2008)}]{Gibbons2008}%
  \BibitemOpen
  \bibfield  {author} {\bibinfo {author} {\bibfnamefont {G.~W.}\ \bibnamefont
  {Gibbons}}\ and\ \bibinfo {author} {\bibfnamefont {M.~C.}\ \bibnamefont
  {Werner}},\ }\bibfield  {title} {\bibinfo {title} {Applications of the
  {G}auss--{B}onnet theorem to gravitational lensing},\ }\href
  {https://doi.org/10.1088/0264-9381/25/23/235009} {\bibfield  {journal}
  {\bibinfo  {journal} {Classical {Q}uantum {G}ravity}\ }\textbf {\bibinfo
  {volume} {25}},\ \bibinfo {pages} {235009} (\bibinfo {year}
  {2008})}\BibitemShut {NoStop}%
\bibitem [{\citenamefont {Jia}\ and\ \citenamefont {Liu}(2019)}]{Jia2019}%
  \BibitemOpen
  \bibfield  {author} {\bibinfo {author} {\bibfnamefont {J.}~\bibnamefont
  {Jia}}\ and\ \bibinfo {author} {\bibfnamefont {H.}~\bibnamefont {Liu}},\
  }\bibfield  {title} {\bibinfo {title} {Time delay of timelike particles in
  gravitational lensing of the {S}chwarzschild spacetime},\ }\href
  {https://doi.org/10.1103/PhysRevD.100.124050} {\bibfield  {journal} {\bibinfo
   {journal} {Phys. {R}ev. {D}}\ }\textbf {\bibinfo {volume} {100}},\ \bibinfo
  {pages} {124050} (\bibinfo {year} {2019})}\BibitemShut {NoStop}%
\bibitem [{\citenamefont {Pang}\ and\ \citenamefont {Jia}(2019)}]{Pang2019}%
  \BibitemOpen
  \bibfield  {author} {\bibinfo {author} {\bibfnamefont {X.}~\bibnamefont
  {Pang}}\ and\ \bibinfo {author} {\bibfnamefont {J.}~\bibnamefont {Jia}},\
  }\bibfield  {title} {\bibinfo {title} {Gravitational lensing of massive
  particles in {R}eissner--{N}ordstr\"{o}m black hole spacetime},\ }\href
  {https://doi.org/10.1088/1361-6382/ab0512} {\bibfield  {journal} {\bibinfo
  {journal} {Classical {Q}uantum {G}ravity}\ }\textbf {\bibinfo {volume}
  {36}},\ \bibinfo {pages} {065012} (\bibinfo {year} {2019})}\BibitemShut
  {NoStop}%
\bibitem [{\citenamefont {He}\ \emph {et~al.}(2020)\citenamefont {He},
  \citenamefont {Zhou}, \citenamefont {Feng}, \citenamefont {Mu}, \citenamefont
  {Wang}, \citenamefont {Li}, \citenamefont {Pan},\ and\ \citenamefont
  {Lin}}]{He2020}%
  \BibitemOpen
  \bibfield  {author} {\bibinfo {author} {\bibfnamefont {G.}~\bibnamefont
  {He}}, \bibinfo {author} {\bibfnamefont {X.}~\bibnamefont {Zhou}}, \bibinfo
  {author} {\bibfnamefont {Z.}~\bibnamefont {Feng}}, \bibinfo {author}
  {\bibfnamefont {X.}~\bibnamefont {Mu}}, \bibinfo {author} {\bibfnamefont
  {H.}~\bibnamefont {Wang}}, \bibinfo {author} {\bibfnamefont {W.}~\bibnamefont
  {Li}}, \bibinfo {author} {\bibfnamefont {C.}~\bibnamefont {Pan}},\ and\
  \bibinfo {author} {\bibfnamefont {W.}~\bibnamefont {Lin}},\ }\bibfield
  {title} {\bibinfo {title} {Gravitational deflection of massive particles in
  {S}chwarzschild-de {S}itter spacetime},\ }\href
  {https://doi.org/10.1140/epjc/s10052-020-8382-z} {\bibfield  {journal}
  {\bibinfo  {journal} {Eur. {P}hys. {J}. {C}}\ }\textbf {\bibinfo {volume}
  {80}},\ \bibinfo {pages} {835} (\bibinfo {year} {2020})}\BibitemShut
  {NoStop}%
\bibitem [{\citenamefont {Gralla}\ and\ \citenamefont
  {Lupsasca}(2020)}]{Gralla2020}%
  \BibitemOpen
  \bibfield  {author} {\bibinfo {author} {\bibfnamefont {S.~E.}\ \bibnamefont
  {Gralla}}\ and\ \bibinfo {author} {\bibfnamefont {A.}~\bibnamefont
  {Lupsasca}},\ }\bibfield  {title} {\bibinfo {title} {Null geodesics of the
  {K}err exterior},\ }\href {https://doi.org/10.1103/PhysRevD.101.044032}
  {\bibfield  {journal} {\bibinfo  {journal} {Phys. {R}ev. {D}}\ }\textbf
  {\bibinfo {volume} {101}},\ \bibinfo {pages} {044032} (\bibinfo {year}
  {2020})}\BibitemShut {NoStop}%
\bibitem [{\citenamefont {Yang}\ and\ \citenamefont {Wang}(2013)}]{Yang2013}%
  \BibitemOpen
  \bibfield  {author} {\bibinfo {author} {\bibfnamefont {X.}~\bibnamefont
  {Yang}}\ and\ \bibinfo {author} {\bibfnamefont {J.}~\bibnamefont {Wang}},\
  }\bibfield  {title} {\bibinfo {title} {Y{NOGK}: {A} new public code for
  calculating null geodesics in the {K}err spacetime},\ }\href
  {https://doi.org/10.1088/0067-0049/207/1/6} {\bibfield  {journal} {\bibinfo
  {journal} {{A}strophys. {J}. {S}uppl. {S}er.}\ }\textbf {\bibinfo {volume}
  {207}},\ \bibinfo {pages} {6} (\bibinfo {year} {2013})}\BibitemShut {NoStop}%
\bibitem [{\citenamefont {Grenzebach}\ \emph {et~al.}(2015)\citenamefont
  {Grenzebach}, \citenamefont {Perlick},\ and\ \citenamefont
  {L\"ammerzahl}}]{Grenzebach2015}%
  \BibitemOpen
  \bibfield  {author} {\bibinfo {author} {\bibfnamefont {A.}~\bibnamefont
  {Grenzebach}}, \bibinfo {author} {\bibfnamefont {V.}~\bibnamefont
  {Perlick}},\ and\ \bibinfo {author} {\bibfnamefont {C.}~\bibnamefont
  {L\"ammerzahl}},\ }\bibfield  {title} {\bibinfo {title} {Photon regions and
  shadows of accelerated black holes},\ }\href
  {https://doi.org/10.1142/S0218271815420249} {\bibfield  {journal} {\bibinfo
  {journal} {Int. {J}. {M}od. {P}hys. {D}}\ }\textbf {\bibinfo {volume} {24}},\
  \bibinfo {pages} {1542024} (\bibinfo {year} {2015})}\BibitemShut {NoStop}%
\bibitem [{\citenamefont {Perlick}\ and\ \citenamefont
  {Tsupko}(2017)}]{Perlick2017}%
  \BibitemOpen
  \bibfield  {author} {\bibinfo {author} {\bibfnamefont {V.}~\bibnamefont
  {Perlick}}\ and\ \bibinfo {author} {\bibfnamefont {O.~Y.}\ \bibnamefont
  {Tsupko}},\ }\bibfield  {title} {\bibinfo {title} {Light propagation in a
  plasma on {K}err spacetime: {S}eparation of the {H}amilton-{J}acobi equation
  and calculation of the shadow},\ }\href
  {https://doi.org/10.1103/PhysRevD.95.104003} {\bibfield  {journal} {\bibinfo
  {journal} {Phys. {R}ev. {D}}\ }\textbf {\bibinfo {volume} {95}},\ \bibinfo
  {pages} {104003} (\bibinfo {year} {2017})}\BibitemShut {NoStop}%
\bibitem [{\citenamefont {Griffiths}\ and\ \citenamefont {Podolsk{\'
  y}}(2009)}]{Griffiths2009}%
  \BibitemOpen
  \bibfield  {author} {\bibinfo {author} {\bibfnamefont {J.~B.}\ \bibnamefont
  {Griffiths}}\ and\ \bibinfo {author} {\bibfnamefont {J.}~\bibnamefont
  {Podolsk{\' y}}},\ }\href {https://doi.org/10.1017/CBO9780511635397} {\emph
  {\bibinfo {title} {Exact {S}pace-{T}imes in {E}instein's {G}eneral
  {R}elativity}}}\ (\bibinfo  {publisher} {Cambridge {U}niversity {P}ress},\
  \bibinfo {address} {Cambridge, England},\ \bibinfo {year} {2009})\BibitemShut
  {NoStop}%
\bibitem [{\citenamefont {Mino}(2003)}]{Mino2003}%
  \BibitemOpen
  \bibfield  {author} {\bibinfo {author} {\bibfnamefont {Y.}~\bibnamefont
  {Mino}},\ }\bibfield  {title} {\bibinfo {title} {Perturbative approach to an
  orbital evolution around a supermassive black hole},\ }\href
  {https://doi.org/10.1103/PhysRevD.67.084027} {\bibfield  {journal} {\bibinfo
  {journal} {Phys. {R}ev. {D}}\ }\textbf {\bibinfo {volume} {67}},\ \bibinfo
  {pages} {084027} (\bibinfo {year} {2003})}\BibitemShut {NoStop}%
\bibitem [{\citenamefont {Byrd}\ and\ \citenamefont
  {Friedman}(1954)}]{Byrd1954}%
  \BibitemOpen
  \bibfield  {author} {\bibinfo {author} {\bibfnamefont {P.~F.}\ \bibnamefont
  {Byrd}}\ and\ \bibinfo {author} {\bibfnamefont {M.~D.}\ \bibnamefont
  {Friedman}},\ }\href {https://doi.org/10.1007/978-3-642-52803-3} {\emph
  {\bibinfo {title} {Handbook of {E}lliptic {I}ntegrals for {E}ngineers and
  {P}hysicists}}},\ \bibinfo {edition} {1st}\ ed.,\ Die {G}rundlehren der
  {M}athematischen {W}issenschaften\ (\bibinfo  {publisher} {Springer-Verlag,
  Berlin},\ \bibinfo {year} {1954})\BibitemShut {NoStop}%
\bibitem [{\citenamefont {Hancock}(1917)}]{Hancock1917}%
  \BibitemOpen
  \bibfield  {author} {\bibinfo {author} {\bibfnamefont {H.}~\bibnamefont
  {Hancock}},\ }\href@noop {} {\emph {\bibinfo {title} {Elliptic
  {I}ntegrals}}},\ \bibinfo {edition} {1st}\ ed.,\ edited by\ \bibinfo {editor}
  {\bibfnamefont {M.}~\bibnamefont {Merriman}}\ and\ \bibinfo {editor}
  {\bibfnamefont {R.~S.}\ \bibnamefont {Woodward}},\ Mathematical {M}onographs\
  (\bibinfo  {publisher} {John Wiley \& Sons, New York},\ \bibinfo {year}
  {1917})\BibitemShut {NoStop}%
\bibitem [{\citenamefont {Frost}(2022{\natexlab{b}})}]{Frost2022b}%
  \BibitemOpen
  \bibfield  {author} {\bibinfo {author} {\bibfnamefont {T.~C.}\ \bibnamefont
  {Frost}},\ }\emph {\bibinfo {title} {Gravitational {L}ensing in {S}tationary
  and {A}xisymmetric {B}lack {H}ole {S}pacetimes: {A}cceleration and
  {G}ravitomagnetic {C}harge}},\ \href {https://dx.doi.org/10.26092/elib/1673}
  {Ph.D. thesis},\ \bibinfo  {school} {University of Bremen, Faculty 1 Physics
  and Electrical Engineering}, \bibinfo {address} {Bremen, Germany} (\bibinfo
  {year} {2022}{\natexlab{b}})\BibitemShut {NoStop}%
\bibitem [{\citenamefont {Wald}(1984)}]{Wald1984}%
  \BibitemOpen
  \bibfield  {author} {\bibinfo {author} {\bibfnamefont {R.~M.}\ \bibnamefont
  {Wald}},\ }\href@noop {} {\emph {\bibinfo {title} {General {R}elativity}}}\
  (\bibinfo  {publisher} {The {U}niversity of {C}hicago {P}ress, Chicago},\
  \bibinfo {year} {1984})\BibitemShut {NoStop}%
\bibitem [{\citenamefont {Frittelli}\ and\ \citenamefont
  {Newman}(1999)}]{Frittelli1999}%
  \BibitemOpen
  \bibfield  {author} {\bibinfo {author} {\bibfnamefont {S.}~\bibnamefont
  {Frittelli}}\ and\ \bibinfo {author} {\bibfnamefont {E.~T.}\ \bibnamefont
  {Newman}},\ }\bibfield  {title} {\bibinfo {title} {Exact universal
  gravitational lensing equation},\ }\href
  {https://doi.org/10.1103/PhysRevD.59.124001} {\bibfield  {journal} {\bibinfo
  {journal} {Phys. {R}ev. {D}}\ }\textbf {\bibinfo {volume} {59}},\ \bibinfo
  {pages} {124001} (\bibinfo {year} {1999})}\BibitemShut {NoStop}%
\bibitem [{\citenamefont {Perlick}(2004)}]{Perlick2004}%
  \BibitemOpen
  \bibfield  {author} {\bibinfo {author} {\bibfnamefont {V.}~\bibnamefont
  {Perlick}},\ }\bibfield  {title} {\bibinfo {title} {Exact gravitational lens
  equation in spherically symmetric and static spacetimes},\ }\href
  {https://doi.org/10.1103/PhysRevD.69.064017} {\bibfield  {journal} {\bibinfo
  {journal} {Phys. {R}ev. {D}}\ }\textbf {\bibinfo {volume} {69}},\ \bibinfo
  {pages} {064017} (\bibinfo {year} {2004})}\BibitemShut {NoStop}%
\bibitem [{\citenamefont {Bezanson}\ \emph {et~al.}(2017)\citenamefont
  {Bezanson}, \citenamefont {Edelman}, \citenamefont {Karpinski},\ and\
  \citenamefont {Shah}}]{Bezanson2017}%
  \BibitemOpen
  \bibfield  {author} {\bibinfo {author} {\bibfnamefont {J.}~\bibnamefont
  {Bezanson}}, \bibinfo {author} {\bibfnamefont {A.}~\bibnamefont {Edelman}},
  \bibinfo {author} {\bibfnamefont {S.}~\bibnamefont {Karpinski}},\ and\
  \bibinfo {author} {\bibfnamefont {V.~B.}\ \bibnamefont {Shah}},\ }\bibfield
  {title} {\bibinfo {title} {Julia: {A} fresh approach to numerical
  computing},\ }\href {https://doi.org/10.1137/141000671} {\bibfield  {journal}
  {\bibinfo  {journal} {SIAM {R}ev.}\ }\textbf {\bibinfo {volume} {59}},\
  \bibinfo {pages} {65} (\bibinfo {year} {2017})}\BibitemShut {NoStop}%
\bibitem [{\citenamefont {Bohn}\ \emph {et~al.}(2015)\citenamefont {Bohn},
  \citenamefont {Throwe}, \citenamefont {H{\' e}bert}, \citenamefont
  {Henriksson}, \citenamefont {Bunandar}, \citenamefont {Scheel},\ and\
  \citenamefont {Taylor}}]{Bohn2015}%
  \BibitemOpen
  \bibfield  {author} {\bibinfo {author} {\bibfnamefont {A.}~\bibnamefont
  {Bohn}}, \bibinfo {author} {\bibfnamefont {W.}~\bibnamefont {Throwe}},
  \bibinfo {author} {\bibfnamefont {F.}~\bibnamefont {H{\' e}bert}}, \bibinfo
  {author} {\bibfnamefont {K.}~\bibnamefont {Henriksson}}, \bibinfo {author}
  {\bibfnamefont {D.}~\bibnamefont {Bunandar}}, \bibinfo {author}
  {\bibfnamefont {M.~A.}\ \bibnamefont {Scheel}},\ and\ \bibinfo {author}
  {\bibfnamefont {N.~W.}\ \bibnamefont {Taylor}},\ }\bibfield  {title}
  {\bibinfo {title} {What does a binary black hole merger look like?},\ }\href
  {https://doi.org/10.1088/0264-9381/32/6/065002} {\bibfield  {journal}
  {\bibinfo  {journal} {Classical {Q}uantum {G}ravity}\ }\textbf {\bibinfo
  {volume} {32}},\ \bibinfo {pages} {065002} (\bibinfo {year}
  {2015})}\BibitemShut {NoStop}%
\bibitem [{\citenamefont {Bozza}(2002)}]{Bozza2002}%
  \BibitemOpen
  \bibfield  {author} {\bibinfo {author} {\bibfnamefont {V.}~\bibnamefont
  {Bozza}},\ }\bibfield  {title} {\bibinfo {title} {Gravitational lensing in
  the strong field limit},\ }\href {https://doi.org/10.1103/PhysRevD.66.103001}
  {\bibfield  {journal} {\bibinfo  {journal} {Phys. {R}ev. {D}}\ }\textbf
  {\bibinfo {volume} {66}},\ \bibinfo {pages} {103001} (\bibinfo {year}
  {2002})}\BibitemShut {NoStop}%
\bibitem [{\citenamefont {{K. Abe \emph{et al.} (Hyper-Kamiokande
  Proto-Collaboration)}}(2018)}]{TheHKKCollaboration2018}%
  \BibitemOpen
  \bibfield  {author} {\bibinfo {author} {\bibnamefont {{K. Abe \emph{et al.}
  (Hyper-Kamiokande Proto-Collaboration)}}},\ }\bibfield  {title} {\bibinfo
  {title} {Physics potentials with the second {H}yper-{K}amiokande detector in
  {K}orea},\ }\href {https://doi.org/10.1093/ptep/pty044} {\bibfield  {journal}
  {\bibinfo  {journal} {Prog. {T}heor. {E}xp. {P}hys.}\ }\textbf {\bibinfo
  {volume} {2018}},\ \bibinfo {pages} {063C01} (\bibinfo {year}
  {2018})}\BibitemShut {NoStop}%
\bibitem [{\citenamefont {Itow}(2021)}]{Itow2021}%
  \BibitemOpen
  \bibfield  {author} {\bibinfo {author} {\bibfnamefont {Y.}~\bibnamefont
  {Itow}},\ }\bibfield  {title} {\bibinfo {title} {Construction status and
  prospects of the {H}yper-{K}amiokande project},\ }\href
  {https://doi.org/10.22323/1.395.1192} {\bibfield  {journal} {\bibinfo
  {journal} {Proc. Sci. {ICRC}2021}\ }\textbf {\bibinfo {volume} {395}},\
  \bibinfo {pages} {1192} (\bibinfo {year} {2021})}\BibitemShut {NoStop}%
\bibitem [{\citenamefont {Itow}()}]{Itow2023}%
  \BibitemOpen
  \bibfield  {author} {\bibinfo {author} {\bibfnamefont {Y.}~\bibnamefont
  {Itow}},\ }\href@noop {} {\bibinfo {title} {personal
  communication}}\BibitemShut {NoStop}%
\bibitem [{\citenamefont {Aartsen}\ \emph {et~al.}(2017)\citenamefont {Aartsen}
  \emph {et~al.}}]{Aartsen2017}%
  \BibitemOpen
  \bibfield  {author} {\bibinfo {author} {\bibfnamefont {M.~G.}\ \bibnamefont
  {Aartsen}} \emph {et~al.},\ }\bibfield  {title} {\bibinfo {title} {P{INGU}:
  {A} vision for neutrino and particle physics at the {S}outh {P}ole},\ }\href
  {https://doi.org/10.1088/1361-6471/44/5/054006} {\bibfield  {journal}
  {\bibinfo  {journal} {J. {P}hys. {G}}\ }\textbf {\bibinfo {volume} {44}},\
  \bibinfo {pages} {054006} (\bibinfo {year} {2017})}\BibitemShut {NoStop}%
\bibitem [{\citenamefont {Fornengo}\ \emph {et~al.}(1997)\citenamefont
  {Fornengo}, \citenamefont {Giunti}, \citenamefont {Kim},\ and\ \citenamefont
  {Song}}]{Fornengo1997}%
  \BibitemOpen
  \bibfield  {author} {\bibinfo {author} {\bibfnamefont {N.}~\bibnamefont
  {Fornengo}}, \bibinfo {author} {\bibfnamefont {C.}~\bibnamefont {Giunti}},
  \bibinfo {author} {\bibfnamefont {C.~W.}\ \bibnamefont {Kim}},\ and\ \bibinfo
  {author} {\bibfnamefont {J.}~\bibnamefont {Song}},\ }\bibfield  {title}
  {\bibinfo {title} {Gravitational effects on the neutrino oscillation},\
  }\href {https://doi.org/10.1103/PhysRevD.56.1895} {\bibfield  {journal}
  {\bibinfo  {journal} {Phys. {R}ev. {D}}\ }\textbf {\bibinfo {volume} {56}},\
  \bibinfo {pages} {1895} (\bibinfo {year} {1997})}\BibitemShut {NoStop}%
\bibitem [{\citenamefont {Cardall}\ and\ \citenamefont
  {Fuller}(1997)}]{Cardall1997}%
  \BibitemOpen
  \bibfield  {author} {\bibinfo {author} {\bibfnamefont {C.~Y.}\ \bibnamefont
  {Cardall}}\ and\ \bibinfo {author} {\bibfnamefont {G.~M.}\ \bibnamefont
  {Fuller}},\ }\bibfield  {title} {\bibinfo {title} {Neutrino oscillations in
  curved spacetime: {A} heuristic treatment},\ }\href
  {https://doi.org/10.1103/PhysRevD.55.7960} {\bibfield  {journal} {\bibinfo
  {journal} {Phys. {R}ev. {D}}\ }\textbf {\bibinfo {volume} {55}},\ \bibinfo
  {pages} {7960} (\bibinfo {year} {1997})}\BibitemShut {NoStop}%
\bibitem [{\citenamefont {Crocker}\ \emph {et~al.}(2004)\citenamefont
  {Crocker}, \citenamefont {Giunti},\ and\ \citenamefont
  {Mortlock}}]{Crocker2004}%
  \BibitemOpen
  \bibfield  {author} {\bibinfo {author} {\bibfnamefont {R.~M.}\ \bibnamefont
  {Crocker}}, \bibinfo {author} {\bibfnamefont {C.}~\bibnamefont {Giunti}},\
  and\ \bibinfo {author} {\bibfnamefont {D.~J.}\ \bibnamefont {Mortlock}},\
  }\bibfield  {title} {\bibinfo {title} {Neutrino interferometry in curved
  spacetime},\ }\href {https://doi.org/10.1103/PhysRevD.69.063008} {\bibfield
  {journal} {\bibinfo  {journal} {Phys. {R}ev. {D}}\ }\textbf {\bibinfo
  {volume} {69}},\ \bibinfo {pages} {063008} (\bibinfo {year}
  {2004})}\BibitemShut {NoStop}%
\bibitem [{\citenamefont {Alexandre}\ and\ \citenamefont
  {Clough}(2018)}]{Alexandre2018}%
  \BibitemOpen
  \bibfield  {author} {\bibinfo {author} {\bibfnamefont {J.}~\bibnamefont
  {Alexandre}}\ and\ \bibinfo {author} {\bibfnamefont {K.}~\bibnamefont
  {Clough}},\ }\bibfield  {title} {\bibinfo {title} {Black hole interference
  patterns in flavor oscillations},\ }\href
  {https://doi.org/10.1103/PhysRevD.98.043004} {\bibfield  {journal} {\bibinfo
  {journal} {Phys. {R}ev. {D}}\ }\textbf {\bibinfo {volume} {98}},\ \bibinfo
  {pages} {043004} (\bibinfo {year} {2018})}\BibitemShut {NoStop}%
\bibitem [{\citenamefont {Dvornikov}(2020)}]{Dvornikov2020}%
  \BibitemOpen
  \bibfield  {author} {\bibinfo {author} {\bibfnamefont {M.}~\bibnamefont
  {Dvornikov}},\ }\bibfield  {title} {\bibinfo {title} {Spin effects in
  neutrino gravitational scattering},\ }\href
  {https://doi.org/10.1103/PhysRevD.101.056018} {\bibfield  {journal} {\bibinfo
   {journal} {Phys. {R}ev. {D}}\ }\textbf {\bibinfo {volume} {101}},\ \bibinfo
  {pages} {056018} (\bibinfo {year} {2020})}\BibitemShut {NoStop}%
\bibitem [{\citenamefont {Swami}\ \emph {et~al.}(2020)\citenamefont {Swami},
  \citenamefont {Lochan},\ and\ \citenamefont {Patel}}]{Swami2020}%
  \BibitemOpen
  \bibfield  {author} {\bibinfo {author} {\bibfnamefont {H.}~\bibnamefont
  {Swami}}, \bibinfo {author} {\bibfnamefont {K.}~\bibnamefont {Lochan}},\ and\
  \bibinfo {author} {\bibfnamefont {K.~M.}\ \bibnamefont {Patel}},\ }\bibfield
  {title} {\bibinfo {title} {Signature of neutrino mass hierarchy in
  gravitational lensing},\ }\href {https://doi.org/10.1103/PhysRevD.102.024043}
  {\bibfield  {journal} {\bibinfo  {journal} {Phys. {R}ev. {D}}\ }\textbf
  {\bibinfo {volume} {102}},\ \bibinfo {pages} {024043} (\bibinfo {year}
  {2020})}\BibitemShut {NoStop}%
\bibitem [{\citenamefont {Milne-Thomson}(1972)}]{MilneThomson1972}%
  \BibitemOpen
  \bibfield  {author} {\bibinfo {author} {\bibfnamefont {L.~M.}\ \bibnamefont
  {Milne-Thomson}},\ }\bibfield  {title} {\bibinfo {title} {Elliptic
  {I}ntegrals},\ }in\ \href@noop {} {\emph {\bibinfo {booktitle} {Handbook of
  {M}athematical {F}unctions {W}ith {F}ormulas, {G}raphs, and {M}athematical
  {T}ables}}},\ \bibinfo {series and number} {Applied {M}athematics {S}eries},\
  \bibinfo {editor} {edited by\ \bibinfo {editor} {\bibfnamefont
  {M.}~\bibnamefont {Abramowitz}}\ and\ \bibinfo {editor} {\bibfnamefont
  {I.~A.}\ \bibnamefont {Stegun}}}\ (\bibinfo  {publisher} {U. {S}.
  {D}epartment of {C}ommerce, {N}ational {B}ureau of {S}tandards},\ \bibinfo
  {address} {Washington DC},\ \bibinfo {year} {1972})\ \bibinfo {edition}
  {10th}\ ed.,\ pp.\ \bibinfo {pages} {587--607}\BibitemShut {NoStop}%
\end{thebibliography}%

\end{document}